\documentclass[12pt]{report}

\usepackage[T1]{fontenc}
\usepackage[utf8]{inputenc}
\usepackage{pdfpages}

\usepackage{tgtermes}

\usepackage{amsmath,amssymb,amsfonts}
\usepackage{mathtools}
\usepackage{mathrsfs}

\allowdisplaybreaks

\usepackage[
a4paper,
left=30mm,
right=20mm,
top=30mm,
bottom=22mm,
headheight=15pt,
headsep=12mm,
footskip=13mm
]{geometry}

\usepackage{setspace}
\usepackage{array}
\usepackage{longtable}
\usepackage{booktabs}
\usepackage{multirow}
\usepackage{adjustbox}

\newcolumntype{M}[1]{>{\centering\arraybackslash}m{#1}}
\newcolumntype{L}[1]{>{\raggedright\arraybackslash}m{#1}}

\usepackage{graphicx}
\usepackage{caption}
\usepackage{subcaption}
\usepackage{pdflscape}

\usepackage{cite}

\usepackage{textcomp}
\usepackage{newunicodechar}

\newunicodechar{Ω}{$\Omega$}
\newunicodechar{Λ}{$\Lambda$}

\usepackage[dvipsnames]{xcolor}

\usepackage[
colorlinks=true,
linkcolor=blue,
citecolor=magenta,
urlcolor=violet
]{hyperref}

\usepackage{tikz}
\usetikzlibrary{shapes.geometric,arrows,positioning}
\usetikzlibrary{shapes}

\tikzstyle{startstop} =
[
rectangle,
rounded corners,
minimum width=2cm,
minimum height=1cm,
text centered,
text width=4.4cm,
draw=black,
fill=white!30
]

\tikzstyle{io} =
[
trapezium,
trapezium left angle=70,
trapezium right angle=110,
minimum width=2cm,
minimum height=1cm,
text centered,
draw=black
]

\tikzstyle{process} =
[
rectangle,
minimum width=1.5cm,
minimum height=1cm,
text centered,
text width=4cm,
draw=black,
fill=white!30
]

\tikzstyle{decision} =
[
diamond,
minimum width=3cm,
minimum height=1cm,
text centered,
text width=4.2cm,
draw=black,
fill=green!30
]

\tikzstyle{arrow} =
[
thick,
->,
>=stealth
]

\usepackage{placeins}
\usepackage{eso-pic}
\usepackage[framemethod=TikZ]{mdframed}

\usepackage{fancyhdr}

\renewcommand{\chaptermark}[1]
{
\markboth{CHAPTER \thechapter.\ #1}{}
}

\begin{document}

\pagenumbering{roman}
\setcounter{page}{1}

\begin{titlepage}
\centering

{\LARGE\bfseries Realisation of de Sitter space as an excited state in string theory\par}

\vspace{1.5cm}

{\large
A thesis submitted in partial fulfillment of\\
the requirements for the degree of\\[1cm]

\textbf{\Large Doctor of Philosophy}\\[0.5cm]

by\\[0.5cm]

\textbf{\Large Archana Maji}\\[0.0cm]

\textbf{\large (Roll No. 195120042)}\\[0.8cm]

Under the supervision of\\
\textbf{\Large  Prof. P. Ramadevi}\\[1.5cm]
}

\includegraphics[width=2in]{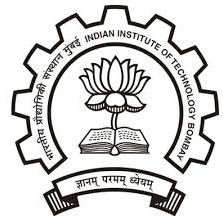}\\[1cm]

{\large
Department of Physics\\
Indian Institute of Technology Bombay\\[0.3cm]

}

\end{titlepage}

\clearpage
\thispagestyle{empty}
\null
\clearpage

\clearpage
\thispagestyle{empty}

\vspace*{\fill}

\begin{center}
{\Large\itshape
Dedicated to my parents
}
\end{center}

\vspace*{\fill}

\clearpage

\clearpage
\thispagestyle{empty}
\null
\clearpage

\thispagestyle{empty}

\includepdf{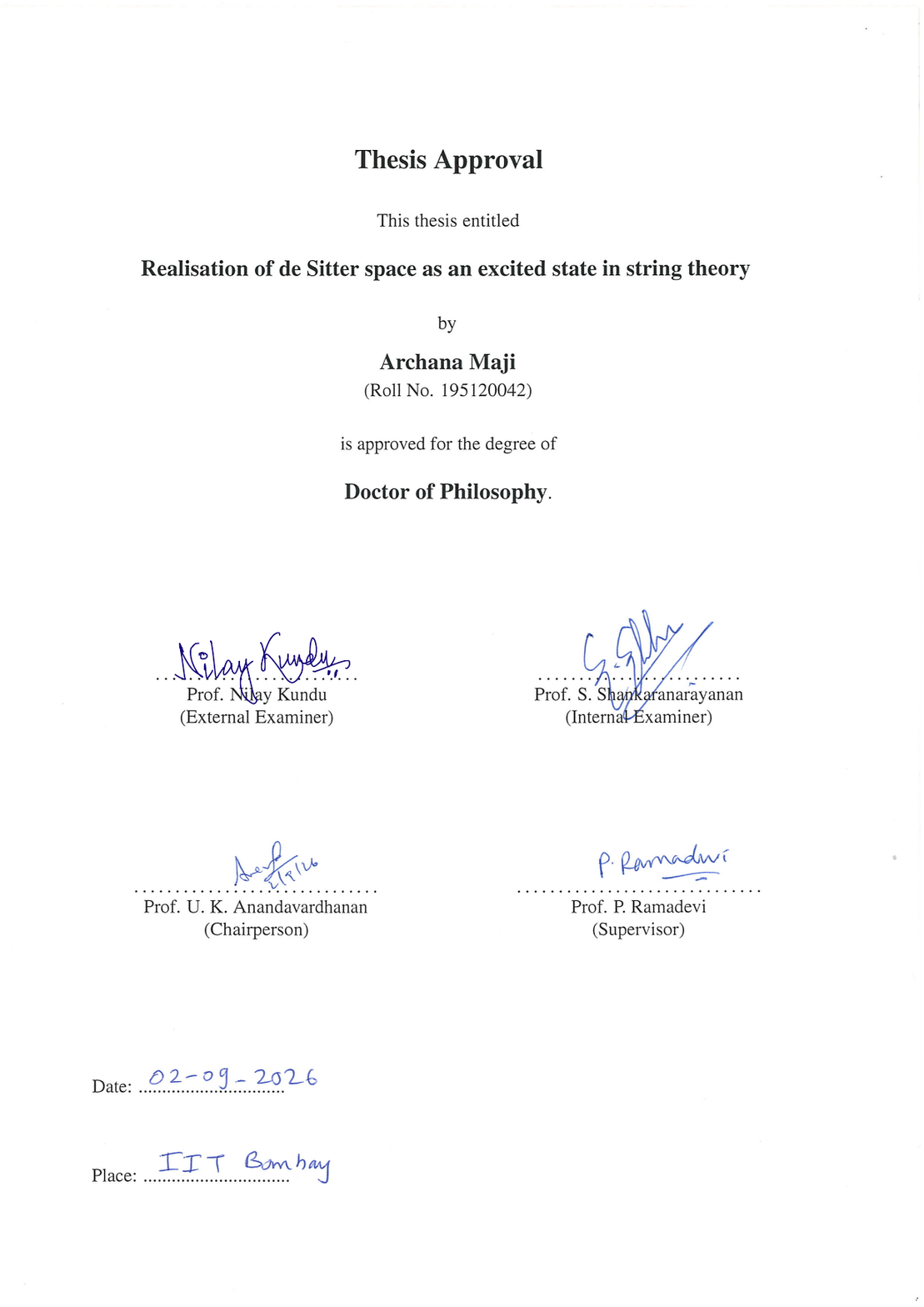}

\includepdf{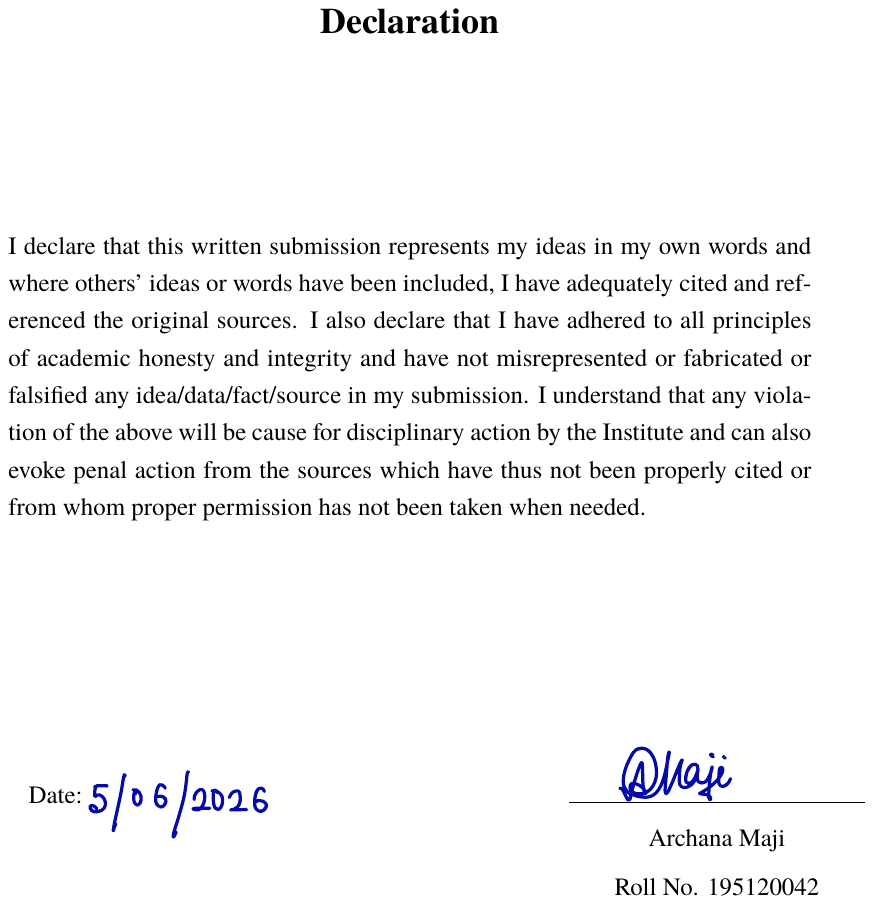}
\thispagestyle{empty}

\clearpage

\chapter*{Acknowledgements}
\addcontentsline{toc}{section}{Acknowledgements}

I would like to begin by thanking my supervisor, P. Ramadevi, for her invaluable guidance, unwavering support, and encouragement throughout my PhD journey. Her thoughtful advice and reassuring presence, both academically and personally, have always been a great source of motivation for me.

I am deeply grateful to Keshav Dasgupta for his generous guidance and support. His immense patience in answering my numerous questions helped me navigate many challenges during the course of our projects. I am especially thankful to him for hosting my academic visit to McGill University---a truly wonderful and enriching experience, during which a significant portion of the work presented in this thesis was completed.

I would also like to express my sincere gratitude to Suvankar Dutta for his encouragement and kind support. I greatly benefited from the academic discussions we had during the early stages of my PhD.

I am deeply thankful to Dileep Jatkar for the insightful academic discussions we have had. His encouragement and support have been truly inspiring.

I would also like to thank my other collaborators---Stephon Alexander, Radu Tatar, Suddhasattwa Brahma, Bohdan Kulinich, Joydeep Chakravarty, Diksha Jain, Kushal Chakraborty, Aditya Dwivedi, and Dmitry Noshchenko---for the engaging discussions, constructive inputs, and kind support.

Many thanks to Joydeep da for his encouraging words throughout this long and challenging journey.

Thanks to my office mates Sachin, Himanshu, Aditya, Lekhika, Zafri, Sadashiv, Ayaz, Ravi, Payel, Amol, Aman, Pravesh, Shamik, and Soumen. Their companionship and support made it much easier to endure many of the challenges.

I am thankful to my hostel mates Arifa, Ishita, Parul, Priyanka, Khushboo, and Dreamsy for all the happiness and memorable moments we shared together.

My thanks also go to all those who have been a part of my Ph.D. journey and contributed, in one way or another, to my academic and personal growth. While not all of them could be acknowledged individually here, I remain deeply grateful for their support, encouragement, and kindness, which have helped me become a stronger and better version of myself.

I would like to sincerely acknowledge the financial support provided by the Prime Minister's Research Fellowship, funded by the Ministry of Education, Government of India. It has been a great privilege to receive this support during the course of my Ph.D. This fellowship afforded me the financial freedom to focus on my research without undue concern and enabled me to travel both internationally and within India for academic activities. These opportunities significantly enriched my research experience and broadened my academic reach.

Finally, I am deeply indebted to my parents and my sister, whose unwavering love, care, and support have sustained me through every challenge along this journey. Their encouragement and belief in me have been a constant source of strength and inspiration.

\newpage

\chapter*{Abstract}
Among currently known frameworks, string theory is widely regarded as the only plausible candidate for a consistent UV-complete theory of quantum gravity. Numerous cosmological observations suggest that our universe is undergoing accelerated expansion, pointing towards the presence of a small positive cosmological constant. Since de Sitter space is the maximally symmetric solution of Einstein's equations with a positive cosmological constant, understanding whether four-dimensional de Sitter space can be realized within string theory has become one of the central questions.

This thesis is devoted to addressing this long-standing question. Following a series of works that realize de Sitter space as an excited state—namely, as a Glauber--Sudarshan (GS) state in type IIB string theory—our goal has been to extend such a construction to heterotic $\mathrm{SO}(32)$ and heterotic $\mathrm{E}_8\times \mathrm{E}_8$ string theories. Realizing de Sitter space as an excited state over a time-independent supersymmetric Minkowski background provides a mechanism for evading the vacuum-based no-go theorems that obstruct de Sitter solutions in string theory. Starting from a generic M-theory configuration, we obtain four-dimensional de Sitter space in the late-time limit of flat-slicing coordinates in both heterotic $\mathrm{SO}(32)$ and heterotic $\mathrm{E}_8 \times \mathrm{E}_8$ theories through appropriate dynamical duality sequences. The corresponding GS state is then constructed by evaluating the expectation value of the metric operator in M-theory using path-integral techniques. We further analyse the conditions required for the existence of a well-defined effective field theory description of the underlying setup and find that these conditions are equivalent to satisfying  the null energy condition in a (3+1)-dimensional Friedmann--Lemaître--Robertson--Walker (FLRW) cosmology. Furthermore, the de Sitter GS states are found to lie well within the temporal bound imposed by the Trans-Planckian Censorship Conjecture (TCC), thereby rendering them as transient states. In the case of heterotic $\mathrm{E}_8\times \mathrm{E}_8$ string theory, we additionally investigate implications arising from axion cosmology. Requiring the time-dependent axion decay constant to obey the astrophysical and cosmological bounds imposes further constraints on the corresponding de Sitter GS state.

\phantomsection
\addcontentsline{toc}{section}{Abstract}
\label{sec:abstract}

 \newpage
\chapter*{List of Publications}
\phantomsection
\addcontentsline{toc}{section}{List of Publications}
\label{sec:publications}
\begin{enumerate}
    \item[$\bullet$] Publications that are included in this thesis
    \begin{itemize}
    \item[(1)] Archana Maji, ``\textbf{A computational companion to transient de sitter and quasi de sitter states in ${\rm SO}(32)$ and
${\rm E}_8 \times {\rm E}_8$ heterotic string theories I: Formalisms}'', \href{https://doi.org/10.1002/prop.70092}{[Fortsch. Phys 74 (2026) 3, e70092]}
        \item[(2)] Suddhasattwa Brahma, Keshav Dasgupta, Bohdan Kulinich, {Archana Maji}, Pichai Ramadevi, Radu Tatar  ``\textbf{de Sitter state in $\mathrm{E}_{8}\times \mathrm{E}_{8}$ string theory}'', 	\href{https://doi.org/10.1002/prop.70047}{[Fortsch. Phys 73 (2025) 12, e70047]}
         \item[(3)]  Stephon Alexander, Keshav Dasgupta, {Archana Maji}, Pichai Ramadevi, Radu Tatar \textbf{``de Sitter State in Heterotic String Theory"}, [\href{https://doi.org/10.1002/prop.202400163}{Fortsch.Phys. 72 (2024) 11, 2400163} ]
    \end{itemize}
    \item[$\bullet$] Other publications and pre-prints (not included in this thesis)
    \begin{enumerate}
     \item[(4)] Keshav Dasgupta, Suddhasattwa Brahma, Bohdan Kulinich, {Archana Maji}, Pichai Ramadevi, Radu Tatar, ``\textbf{Transient de Sitter and Quasi de Sitter States in ${\rm SO}(32)$ and ${\rm E}_8 \times {\rm E}_8$ Heterotic String Theories}'', \href{https://doi.org/10.48550/arXiv.2511.03798}{[arXiv:2511.03798]}
     \item[(5)]  Aditya Dwivedi, Archana Maji, Dmitry Noshchenko, Pichai Ramadevi ``\textbf{Flat connections at infinity on knot surgery manifolds}'', \href{https://arxiv.org/abs/2509.05270}{[arXiv:2509.05270]}     
        \item[(6)] Archana Maji, Kushal Chakraborty, Suvankar Dutta, P. Ramadevi \textbf{{ ``$U(N)$ torus link invariants in the large $N$ limit from the Matrix Model Approach"}},  [\href{https://doi.org/10.1103/PhysRevD.109.065021}{Phys. Rev. D 109 (2024) 6, 065021}]       \item[(7)] Joydeep Chakravarty, Keshav Dasgupta, Dileep Jatkar, Diksha Jain, {Archana Maji}, Radu Tatar \textbf{``{Coherent states in $\mathrm{M}$ theory: A brane scan using the Taub-NUT geometry}"},        [\href{https://journals.aps.org/prd/abstract/10.1103/PhysRevD.108.L081902}{Phys.Rev.D 108 (2023) 8, L081902}]
    \end{enumerate}
\end{enumerate}
\newpage

{
\hypersetup{
  linkcolor=black,
linkbordercolor=red,
 pdfborder={0 0 0.6}
}
\phantomsection
\tableofcontents
}

{
\hypersetup{
  linkcolor=Blue
}
\phantomsection
\listoffigures
\addcontentsline{toc}{section}{List of Figures}
}


\clearpage
\pagenumbering{arabic}
\setcounter{page}{1}
\chapter{Introduction}
\thispagestyle{empty}
\label{Chap:Intro}
 The astronomical observations made towards the end of twentieth century indicate that the cosmological constant of our universe is positive\cite{Riess_1998,Perlmutter_1999}. A spacetime with this feature is termed as a de Sitter space. Motivated by this observational evidence, the search for de Sitter vacua within a consistent quantum theory of gravity -- string theory being a leading candidate -- has attracted considerable interest. However, the construction of de Sitter vacua within string theory has long remained a challenging endeavour.  
 
 For instance, the Maldacena-Nuñez and Gibbons no-go theorems \cite{Maldacena:2000mw,Gibbons_2003} demonstrate the absence of well-defined de Sitter compactifications for a large class of supergravity theories. In particular, a warped Kaluza-Klein compactification of a $\mathrm{D}(>2)$ dimensional supergravity theory to $d(\geq 2)$ dimensions can not yield non-singular Minkowski  or de Sitter solutions under certain assumptions: the $\mathrm{D}$-dimensional supergravity theory contains no higher-curvature corrections, the scalar potential is non-positive, all massless fields have positive kinetic terms, and the $d$ dimensional Newton's constant remains finite\cite{Maldacena:2000mw}. Furthermore, the supergravity theories in $\mathrm{D}=10~$and 11 dimensions obey the strong energy condition \cite{Gibbons_2003} that constrains the energy-momentum tensor $\mathrm{T}_{\mathrm{M}\mathrm{N}}$ as
 \begin{align}
\left(\mathrm{T}_{\mathrm{M} \mathrm{N}}-\frac{1}{\mathrm{D}-2}g_{\mathrm{M} \mathrm{N}}\mathrm{T}^{\mathrm{L}}_{~~\mathrm{L}} \right)\xi^\mathrm{M} \xi^\mathrm{N} \geq 0,
 \end{align}
 where $\xi^\mathrm{M}$ is a timelike vector. This condition continues to hold in any smooth compactification of these theories, whereas an accelerating universe necessarily requires a violation of the strong energy condition. 

Numerous ingenious attempts have been made to circumvent the aforementioned no-go conditions. One of the earliest proposals is based on the flux compactification of a highly warped type IIB string theory, referred to as the KKLT construction\cite{kachru2003sitter}. According to KKLT, metastable de Sitter vacua are obtained by uplifting supersymmetric stabilized Anti-de Sitter (AdS) vacua through the inclusion of anti-D3 branes. Instead of introducing anti-D3 branes there is an alternate mechanism for uplifting the AdS vacuum, as proposed in Ref. \cite{CliffordBurgess_2003}, namely by turning on fluxes of gauge fields on the worldvolume of D7 branes. Following the KKLT model, several works with differing ingredients  appeared soon after \cite{CristinaEscoda_2003,AlexSaltman_2004,FrederikDenef_2004,VijayBalasubramanian_2004}. Another well known framework is the large volume scenario (LVS)\cite{balasubramanian2005systematics,JosephConlon_2005,MicheleCicoli_2008}, studied in the Calabi-Yau compactification of type IIB string theory. In this setup, a non-supersymmetric AdS minimum of the scalar potential is constructed, in which the overall volume of the internal space is stabilized at an exponentially large value (in string units). The resulting AdS vacuum can then be uplifted to a de Sitter vacuum by the inclusion of the anti-D3 branes\cite{kachru2003sitter} or the gauge fluxes on the D7-branes\cite{CliffordBurgess_2003}. Several additional mechanisms for obtaining de Sitter solutions in different corners of string theory/M-theory have also been outlined \cite{BECKER2004223,PhysRevD.70.066008,westphal2007sitter,hertzberg2007inflationary,PhysRevD.77.106006,covi2008sitter,rummel2012sufficient,cicoli2016sitter}. 

Despite these efforts, a parallel set of studies~
\cite{wrase2010classical,shiu2011stability,green2012constraints,Gautason2012-xb, Kutasov:2015eba,Quigley2015-vk,palti2019swampland,heliudsonmirsuddho} have continued to highlight persistent obstructions in constructing de Sitter compactifications within string/M-theory. In string theory, one encounters ingredients such as orientifold planes, which can violate the strong energy condition and allow for higher-derivative interactions. Nevertheless, it has been shown that, in heterotic string theory, the leading strong energy condition violating interactions coming from orientifold planes are insufficient to support an accelerating universe upon compactification~\cite{green2012constraints,Gautason2012-xb}. At the level of classical string theory—incorporating strong curvature effects on the worldsheet beyond the supergravity approximation, as well as non-perturbative effects like world-sheet instantons—macroscopic de Sitter spacetime of dimension four or higher appear to be ruled out, at least in heterotic string theory\cite{Kutasov:2015eba}. There have also been various viewpoints regarding the use of anti-D3 branes in the KKLT setup\cite{PhysRevD.87.106010,Bena2013-el}. In fact, it has been argued that no combination of the localized static sources—including fluxes, Dp-branes, anti Dp-branes, and orientifold planes can lead to de Sitter solution in the 3+1 non-compact directions\cite{evandasgupta2014}. It is therefore fair to say that the existence of a fully controlled classical de Sitter vacuum in these theories remains an open question. 

At this stage, one can ask: is it possible to find de Sitter solutions as excited states which show close resemblance to classical de Sitter vacuum? For instance,
we are familiar with {\it coherent states} $\vert \alpha \rangle$ in one-dimensional quantum mechanical harmonic oscillator obtained by the action of the displacement operator $\hat{\mathrm{D}}(\alpha)$ on a free vacuum $\vert 0\rangle$: $|\alpha\rangle=\hat{\mathrm{D}}(\alpha)\vert 0\rangle$. For these coherent states, the expectation value of position operator match the classical oscillatory solution $x_{cl}(t)$:
\begin{align}
  x_{cl}(t)=  \langle \alpha(t)\vert  \hat x \vert \alpha (t)\rangle= A~\cos (\omega t-\phi),\label{ExpValofxOp}
\end{align}
where $\omega$ is the characteristic frequency of the harmonic oscillator and $\phi$ denotes the phase. 
Such a promising concept has been explored in IIB string theory\cite{DASGUPTA2021115463,Brahma:2020tak,brahma2021four,Bernardo:2021rul}, where the vacuum state is an interacting vacuum $\vert \Omega\rangle$ and the corresponding state obtained by the action of displacement operator $\hat{\mathbb{D}}(\sigma)$ is referred to as {\it Glauber-Sudarshan} (GS) state $|\sigma\rangle = \hat{\mathbb{D}}(\sigma)|\Omega\rangle$. In a similar spirit of Eq. \eqref{ExpValofxOp}, the expectation value of the metric operator computed with respect to $|\sigma\rangle$ is identified with the classical metric solution\cite{Brahma:2020tak,Bernardo:2021rul,brahma2021four}. More details and justifications for the de Sitter solution as GS state in string theory is elaborated in Ref. \cite{Chakravarty_2024}.


In Refs.~\cite{Brahma:2020tak,brahma2021four,Bernardo:2021rul}, it was shown for the first time that a four-dimensional de Sitter space in IIB string theory can be realised as a Glauber–Sudarshan state built over an interacting supersymmetric Minkowski vacuum $\vert \Omega \rangle$ . Related ideas, namely the construction of de Sitter space as a  coherent state in four dimensional quantum gravity had previously appeared in~\cite{Dvali_2017,10.1002/prop.201800094}; however, the key novelty of the works~\cite{Brahma:2020tak,brahma2021four,Bernardo:2021rul} lies in the fact that here the construction is carried out within string theory. In particular, de Sitter isometry in flat slicing is imposed on the four non-compact directions of a type IIB setup. This configuration is then uplifted to M theory via a sequence of dualities involving T duality followed by dimensional uplift. The uplift to M theory simplifies the analysis in several ways; M theory admits a Lagrangian formulation neatly packaging various type IIB fluxes in the 4-form fluxes of M theory, the branes of type IIB get mapped  either to geometry or to M2 and M5 branes. Moreover, the instanton effects and other non-perturbative corrections in type IIB theory become generalised curvature corrections in M theory\cite{Bergshoeff2015-do}. 

In this M theory framework, the expectation value of the metric operator with respect to the GS state is computed by a path integral analysis using the perturbative techniques\cite{Brahma:2022wdl}. This expectation value is then identified with the coefficients of the M-theory line element $ds^2_{11}$:
\begin{align}
    ds^2_{11}&=\langle \hat{\mathbf{g}}_{\mathrm{MN}} \rangle_\sigma~d\mathrm{Y^M}d\mathrm{Y^N},~~~{\rm M,~N}=0,1,...,9,11,\label{MthGSsTaTe}
\end{align}
leading to an emergent de Sitter space in the dual type IIB description:
\begin{align}
    ds^2_{\rm IIB}&=\langle \hat{\bf g}_{\rm AB} \rangle_\rho d{\rm Y}^{\rm A} d{\rm Y}^{\rm B},~~~{\rm A,~B}=0,1,...,9\label{IIBGSstAte}
\end{align}
where $|\rho\rangle$ is the corresponding GS state in type IIB theory.
Note that, the explicit path integral analysis in Ref. \cite{Brahma:2022wdl} has been done in M-theory by considering representative scalars $(\varphi_1,\varphi_2,\varphi_3)$ for the metric $\{\mathbf{g}_{\rm MN}\}$, three-form field $\{\mathbf{C}_{\rm MNP}\}$ and the Rarita-Schwinger fermion $\{\mathbf{\Psi}_{\rm M}\}$ respectively. Since the ultimate goal is to describe a universe undergoing accelerated expansion—closely resembling our own—it becomes necessary to include contributions from all orders in the perturbative expansion.  

It is well known from quantum field theory that, except in rare cases, exact results for observables are difficult to obtain. One generally relies on perturbative analysis, which requires identifying a suitable expansion parameter that can be tuned to be small, allowing the observable of interest to be expressed as a power series in that parameter. Although perturbative methods have had notable success—for instance, in the remarkable precision tests of the anomalous magnetic moment of the electron—they also come with shortcomings. In general, perturbative series are asymptotic with zero radius of convergence. There are two primary sources of such divergences: \textcolor{blue}{(i)} the factorial growth of the number of Feynman diagrams at higher orders in the perturbative expansion due to combinatorics~\cite{Dorigoni:2014hea}, and \textcolor{blue}{(ii)} UV/IR divergences of the so-called renormalon type, arising from the high/low-momentum regions of phase-space integrals~\cite{Beneke_1999}. In the perturbative analysis of the GS states \cite{Brahma:2022wdl}, where the diagrams are not the Feynman diagrams but are referred to as {\it nodal diagrams}, the divergences of the former kind are encountered.

The asymptotic nature of the perturbation series does not represent a fundamental obstacle to computing the observable of interest, but rather signals the necessity of incorporating non-perturbative contributions. These non-perturbative effects arise from non-trivial saddle points of the path integral, known as \textit{instantons}~\cite{Jentschura:2011zza}. The formal series obtained by adding instanton corrections to the original asymptotic expansion is known as a \textit{trans-series}~\cite{Marino:2023epd}. Then one has to employ Borel-\'Ecalle resummation techniques \cite{Dorigoni:2014hea,Beneke_1999,Marino:2023epd,ANICETO20191,serone2024lectures} to resum the trans-series and obtain physically meaningful results for the observables. A detailed resurgence analysis for the de Sitter GS state has been carried out in \cite{Brahma:2022wdl}. 

Realising de Sitter space as a coherent state addresses several issues related to its stability, vacuum energy, and entropy\cite{Brahma:2020tak,brahma2021four}. Most importantly, the underlying spacetime being time-independent Minkowski, the notion of frequency remains well defined, allowing for the implementation of an exact renormalisation group (ERG) procedure. This stands in contrast to a classical de Sitter vacuum configuration, where modes are continuously red-shifted and thereby making the implementation of ERG subtle. Furthermore, since the background Minkowski is supersymmetric in nature, zero-point energies arising due to the bosonic and fermionic degrees of freedom cancel each other. Hence, in this setup the cosmological constant does not get any contribution from the zero-point energy. 

To support the de Sitter excited state configuration, it is essential to incorporate quantum corrections as well as time dependent degrees of freedom on the internal space of the M theory. The importance of such time dependent internal degrees of freedom for the existence of a well defined effective field theory description has been emphasized in Ref. \cite{Dasgupta2019-bp}. In particular, it was shown that the hierarchies associated with the dual type IIA string coupling $g_s$ and the Planck scale $\mathrm{M}_p$ generically break down unless suitable time-dependent degrees of freedom are introduced, for example through time-dependent warp factors on the internal manifold and time-dependent four-form fluxes. These fluxes are further constrained by the flux quantization and anomaly cancellation conditions. 

Interestingly, the existence of a consistent effective field theory (EFT) description of the underlying M-theory background is closely tied to the null energy condition (NEC) for a four-dimensional Friedman-Lema\^{\i}tre-Robertson-Walker (FLRW) cosmology, namely
\begin{align}
\mathrm{T}_{\mu\nu}\zeta^{\mu}\zeta^{\nu}\geq 0,
\label{NECConstraint}
\end{align}
where $\zeta^\mu$ is an arbitrary null vector \cite{Bernardo_2021}. In particular, it was shown in \cite{Bernardo_2021} that requiring the temporal derivative of $g_s$ to scale as a positive power of $g_s$,
\begin{align}
\frac{\partial g_{s}}{\partial t}\propto g_{s}^{+\mathrm{ve}},
\label{EftCriTeriA1}
\end{align}
which is one of the necessary criteria for the validity of the EFT description of the M-theory configuration, is in fact equivalent to imposing the four-dimensional NEC. Another important EFT consistency condition requires that
\begin{align}
g_s<1.\label{EfTCriTeria2}
\end{align}
This, in turn, restricts the conformal time of de Sitter space to a certain interval
within which the de Sitter Glauber-Sudarshan (GS) state description in type IIB remains valid. Remarkably, this time interval coincides with the range implied by the Trans-Planckian Censorship Conjecture (TCC) bound \cite{Bedroya_2020}. The resulting finite domain of validity further implies that the GS states are intrinsically transient in nature. These aspects have been analysed in detail in the context of constructing de Sitter GS states in type IIB string theory \cite{Brahma:2020tak,brahma2021four,Bernardo:2021rul,Chakravarty_2024}. 

The realisation of de Sitter space as a GS state in type IIB string theory naturally raises a deeper question: is this construction specific to type IIB, or does it point to a more general mechanism within string theory? This motivated us to examine, whether de Sitter space, viewed as a GS state, can also be realised in other corners of string theory, rather than being an artefact of a particular setup. Guided by this perspective, we extended the analysis beyond type IIB to heterotic string theory \cite{Alexander:2023qym,10.1002/prop.70047,Dasgupta:2025ypg,Archana_CC2026}, which is particularly compelling due to its phenomenological relevance. This forms the central theme of this thesis, where we detail the emergence of de Sitter space as a GS state in both $\mathrm{SO}(32)$ and $\mathrm{E}_8\times \mathrm{E}_8$ heterotic string theories. 

 In \cite{Alexander:2023qym}, we proposed a novel duality sequence which, starting from a type IIB orientifold, leads to heterotic SO(32) theory, wherein de Sitter isometry is realised in the 3+1-dimensional non-compact directions. We follow the methodology of uplifting the type IIB orientifold configuration to M-theory, as in \cite{Brahma:2020tak,Bernardo:2021rul}, and study the resulting dynamics in M-theory. Interestingly, in both the type IIB and heterotic $\mathrm{SO}(32)$ realisations of de Sitter state, the internal space of M theory is equipped with two time dependent warp factors along with an appropriate manifold structure (with/without orbifolding operation). 
 
 A suitable choice of M theory configuration should, in principle, also allow for the realisation of de Sitter as a GS state in heterotic $\mathrm{E}_8\times \mathrm{E}_8$ string theory. However, it turns out that incorporating only two time dependent warp factors on the internal space of the M theory background is insufficient to accomplish this goal. Instead, one is led to consider a more intricate M-theory orbifold structure that accommodates three time-dependent warp factors on the internal manifold, rendering the analysis significantly more involved.

 In \cite{10.1002/prop.70047}, we developed such a framework to analyse the emergence of a de Sitter GS state in heterotic $\mathrm{E}_8 \times \mathrm{E}_8$ string theory. It is well known that the strong-coupling limit of the ten-dimensional heterotic $\mathrm{E}_8 \times \mathrm{E}_8$ theory is described by eleven-dimensional M-theory compactified on the orbifold $\mathbb{R}^{10} \times \mathbf{S}^1/\mathbb{Z}_2$, as shown by Petr Hořava and Edward Witten \cite{1Horava_1996,2Horava_1996}. Our work \cite{10.1002/prop.70047}, presents a duality sequence that starting from an orbifold in M theory dynamically flows to give heterotic $\mathrm{E}_8\times \mathrm{E}_8$ theory with four dimensional de Sitter isometry in the late time limit. We also show that the four dimensional theory does not violate NEC and  the transient de Sitter phase lies well within the TCC bound. Especially for this $\mathrm{E}_8\times \mathrm{E}_8$ theory we also have to impose constraints coming from axionic cosmology\cite{PeterSvrcek_2006,KeshavDasgupta_2008}. In particular, we obtain a temporally varying axion decay constant  which in turn constrains the late time behavior of the time dependent warp factors\cite{10.1002/prop.70047}.  
 
A detailed analysis of the equations of motion, Bianchi identities, flux quantization conditions, and anomaly cancellation constraints that substantiate the GS state construction in heterotic $\mathrm{SO}(32)$ and $\mathrm{E}_8 \times \mathrm{E}_8$ theories—has been carried out in our recent work \cite{Dasgupta:2025ypg}. The companion paper \cite{Archana_CC2026}, which forms the technical foundation for this thesis, provides all the intermediate steps corresponding to the computations presented in Sections 3 and 4 of Ref. \cite{Dasgupta:2025ypg}. In addition, it also contains a detailed review of the resurgence analysis of the Glauber-Sudarshan state following Ref. \cite{Brahma:2022wdl}.

The structure of the remainder of this thesis is as follows. In the next two sections of this introductory chapter, we review some basic properties of de Sitter space in Poincaré coordinates (also known as flat slicing or planar coordinates) and provide a brief overview of coherent states, with particular emphasis on how Glauber–Sudarshan states differ from the conventional notion of coherent states.

In Chapter \ref{Chapter:2}, we review the realisation of de Sitter space as a GS state in type IIB string theory \cite{Brahma:2020tak,brahma2021four,Bernardo:2021rul,Brahma:2022wdl,Bernardo_2021}. In  \S\ref{typeIIBdSisom}, we begin by presenting a dynamical duality sequence which starts from an M-theory configuration and flows to a type IIB background exhibiting de Sitter isometry in the four non-compact directions in the late-time limit of flat slicing coordinates. In the subsequent \S\ref{sec:PIinMth}, we discuss in detail the computation of the expectation value of the metric operator with respect to the GS state in M-theory using perturbative techniques. This computation leads to a perturbative series which, being asymptotic in nature, requires Borel resummation, as discussed in \S\ref{BorelREsum}. We then analyse, in \S\ref{sec:NECIIB}, the null energy condition, which is necessary for the existence of a well-defined effective field theory description of the excited state configuration.

Chapter \ref{Chap:3} is based on the works \cite{Alexander:2023qym,Archana_CC2026} and focuses on the construction of de Sitter space as a Glauber–Sudarshan state in heterotic $\mathrm{SO}(32)$ string theory. In  \S\ref{dSinHetSO(32)}, we present a duality sequence in which an M-theory orbifold configuration dynamically leads, in the late-time limit, to a four-dimensional de Sitter background in heterotic $\mathrm{SO}(32)$ string theory upon compactification.  \S\ref{sec:NEC} is devoted to the analysis of the null energy condition in this setup. In  \S\ref{sec:TCCconsTraints}, we derive the Trans-Planckian Censorship Conjecture (TCC) bound, which constrains the temporal regime over which the effective field theory description remains valid.

Chapter \ref{Chap:4} builds on the works \cite{10.1002/prop.70047,Archana_CC2026}. In  \S\ref{E8XE8ThrdistWarpf}, we present a novel duality sequence starting from an M-theory orbifold with three time-dependent warp factors on the internal manifold, which dynamically evolves to a four-dimensional de Sitter background in heterotic $\mathrm{E}_8 \times \mathrm{E}_8$ string theory in the late-time limit of flat-slicing coordinates. In \S\ref{sec:NECE8XE8}, we examine the null energy condition in this framework.  \S\ref{sec:AxionsinE8Thoery} is devoted to the study of constraints arising from axion cosmology in heterotic $\mathrm{E}_8 \times \mathrm{E}_8$ string theory, with particular emphasis on how observational bounds on the axion decay constant restrict the allowed time-dependent warp factors. 

Finally, in Chapter \ref{sec:Summary&Con}, we summarise the main results and discuss their implications. We also highlight some open problems and outline possible directions for future research.

The appendices—\ref{DualitYRulES}, \ref{FIntApproachPhi4}, and \ref{sec:AppendixD}—contain technical material that complements the main text. In Appendix \ref{DualitYRulES}, we review various duality transformations in string theory, including T-duality, S-duality, dimensional uplift, and dimensional reduction. Appendix \ref{FIntApproachPhi4} presents details of the functional integral approach to computing correlation functions in quantum field theory. In Appendix \ref{sec:AppendixD}, we compute the volume of an internal manifold , demonstrating that it is independent of the Kaluza--Klein gauge fields that describe the twisting of a two-dimensional fiber over the four-dimensional base manifold.  

To set the notation for the rest of the thesis, in the following section we discuss some basic features of de Sitter space in Poincar\'e coordinates. 
\section{de Sitter space in {Poincar\'e} coordinates}
\label{dSspacePoincareCo}
 The Einstein's equation of motion in the presence of energy momentum tensor ($\mathrm{T}_{\mu\nu}$) and cosmological constant ($\Lambda$) is given as 
 \begin{align}
      \mathrm{R}_{\mu\nu}-\frac{1}{2}g_{\mu\nu}\mathrm{R}+\Lambda g_{\mu\nu}=8\pi \mathrm{G}_{\mathrm{N}}\mathrm{T}_{\mu\nu}.\label{EEofmoTion}
 \end{align}
 de Sitter space is a solution to the above equation of motion for vanishing $\mathrm{T}_{\mu\nu}$ and $\Lambda>0$. An alternate way of realising de Sitter space is from the embedding space point of view: a $d$ dimensional de Sitter space $d\mathrm{S}^d$ of length scale $l_d$ is a set of points $(\mathrm{X}^0,\mathrm{X}^1,\cdots,\mathrm{X}^d)$ in a flat $d+1$ dimensional Minkowski spacetime $\mathbb{R}^{d,1}$ satisfying 
\begin{align}
        -(\mathrm{X}^0)^2+(\mathrm{X}^1)^2+\cdots+(\mathrm{X}^d)^2&=l_{d}^2.\label{ConstdeSitter}
\end{align}
 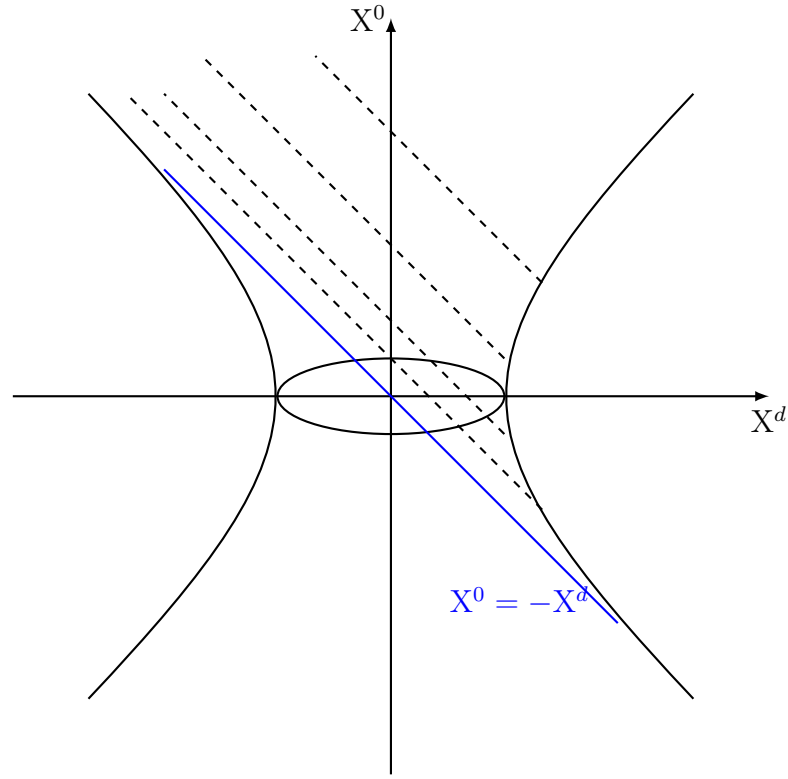
\begin{figure}[!ht]
    \centering
 \begin{tikzpicture}
 \draw[-latex,thick] (-5,0) --(5,0);
 \draw[-latex,thick] (0,-5) --(0,5);
 \draw[thick] (0,0) ellipse (1.5 and 0.5);
 \draw[thick] (-4,4) .. controls (-0.7,0.5) and (-0.7,-0.5)  .. (-4,-4);
 \draw[thick] (4,4) .. controls (0.7,0.5) and (0.7,-0.5)  .. (4,-4);
 \draw[blue, thick] (3,-3) --(-3,3);
 \node at (5,-0.3) {$\mathrm{X}^d$};
 \node at (-0.3,5) {$\mathrm{X}^0$};
 \node at (1.7,-2.7) {$\textcolor{blue}{\mathrm{X}^0=-\mathrm{X}^d}$};
 \draw[dashed, thick] (2,-1.5) --(-3.5,4);
 \draw[dashed, thick] (1.5,-0.5) -- (-3,4);
 \draw[dashed, thick] (1.5,0.5) -- (-2.5,4.5);
 \draw[dashed, thick] (2,1.5) -- (-1,4.5);
 \end{tikzpicture}
    \caption{The region of embedding space $d\mathrm{S}^d$ covered by the Poincar\'e coordinates $\mathrm{X}^0\geq -\mathrm{X}^d$ (shown using dashed lines). The constant $\mathrm{X}^0$ slices are $\mathrm{S}^{d-1}$.}
    \label{FlatSlicing}
\end{figure}
At a given value of $\mathrm{X}^0$, the $d$ coordinates $(\mathrm{X}^1,\cdots, \mathrm{X}^d)$ form a sphere $\mathrm{S}^{d-1}$ 
\begin{align}
    (\mathrm{X}^1)^2+(\mathrm{X}^2)^2+\cdots+(\mathrm{X}^d)^2&=\left(\sqrt{l_{d}^2+(\mathrm{X}^0)^2}\right)^2,
\end{align}
of radius $\sqrt{l_{d}^2+(\mathrm{X}^0)^2}$. As the coordinate $\mathrm{X}^0$ goes from $-\infty$ to $\infty$, the radius of the sphere starts out infinitely large, shrinks to a minimum value $l_d$ at $\mathrm{X}^0=0$ and then expands again to infinity.  Thus, de Sitter space has the topology of $\mathbb{R}\times \mathrm{S}^{d-1}.$   It can be parameterised using different coordinate systems: static coordinates, global coordinates, Poincar\'e coordinates, Kruskal  coordinates etc. We will be working with the Poincar\'e coordinates (also known as  flat slicing coordinates). So, in the following we elaborate on this. A $d$ dimensional de Sitter space can be parameterised with Poincar\'e coordinates $(\tau,x^1,...,x^{d-1})$ as
\begin{equation}
\begin{aligned}
\mathrm{X}^0&=l_d\sinh\left( \frac{\tau}{l_d}\right)+\frac{(x^i)^2}{2l_d}e^{\tau/l_d},\\
    \mathrm{X}^i&=x^i e^{\tau/l_d},~~~~~(i=1,...,d-1)\\
    \mathrm{X}^d&=l_d\cosh\left( \frac{\tau}{l_d}\right)-\frac{(x^i)^2}{2l_d}e^{\tau/l_d},\label{FSlicingPar}
\end{aligned}
\end{equation}
where $\tau\in (-\infty,\infty)$ and $x^i\in (-\infty,\infty)$. 
The metric of de Sitter space $d\mathrm{S}^d$, obtained upon substituting (\ref{FSlicingPar}) in the metric of $d+1$ dimensional embedding Minkowski spacetime 
\begin{align}
    ds^2_{\mathbb{R}^{d,1}}&=-(d\mathrm{X}^0)^2+(d\mathrm{X}^1)^2+\cdots+(d\mathrm{X}^d)^2,
\end{align}
is given by
\begin{align}
  ds^2_{d\mathrm{S}^d}    &=-d\tau^2+e^{2\tau/l_d}\left[(dx^1)^2+\cdots+(dx^{d-1})^2\right],\label{PlanarMet}  
\end{align}
where the spatial slices $\sum_{i=1}^{d-1}(dx^i)^2$ are flat Euclidean spaces $\mathbb{E}^{d-1}$.  Therefore, for a four dimensional de Sitter space the Poincar\'e metric becomes 
\begin{align}
ds^2_{d\mathrm{S}^4}&=-d\tau^2+e^{2\tau/l_4}\left[(dx^1)^2+(dx^2)^2+(dx^3)^2\right].\label{4DPoinCareMetric}
\end{align}
The above metric along with the Einstein's equation of motion (\ref{EEofmoTion}) (with $\mathrm{T}_{\mu\nu}$ set equal to zero) allows us to fix the de Sitter length scale $l_4$ in terms of the cosmological constant: 
\begin{align}
\mathrm{R}_{11}-\frac{g_{11}}{2}\mathrm{R}+\Lambda g_{11}&=0\nonumber\\
\implies    \frac{3}{l_4^2}e^{2\tau/l_4}-\frac{e^{2\tau/l_4}}{2}\left(\frac{12}{l_4^2}\right)+\Lambda (e^{2\tau/l_4})&=0\nonumber\\
     \implies l_4=\sqrt{\frac{3}{\Lambda}}&.
    \end{align}
With the conformal time $t$, defined as 
\begin{align}
t= -l_4~e^{-\tau/l_4} ,\quad\quad -\infty<t < 0,
\end{align}
 the metric \eqref{4DPoinCareMetric} becomes conformally flat
\begin{align}
ds^{2}_{d\mathrm{S}^4}&=\frac{1}{l_4^{-2} t^2}(-dt^2+(dx^1)^2+(dx^2)^2+(dx^3)^2)\nonumber\\
&=\frac{3}{\Lambda t^2}(-dt^2+(dx^1)^2+(dx^2)^2+(dx^3)^2).\label{dSFlAtSlic}
\end{align}
Note that, the Poincar\'e coordinates (\ref{FSlicingPar}) cover only half of the global de Sitter space (see Fig. \ref{FlatSlicing})
\begin{align}
\mathrm{X}^0+\mathrm{X}^d&=l_de^{\tau/l_d}\geq 0 \implies \mathrm{X}^0 \ge -\mathrm{X}^d.
\end{align} 
{In a similar manner, the other half of the de Sitter space can be spanned by an appropriate modification of the coordinates (\ref{FSlicingPar}).} 

As mentioned earlier, our goal is to realise de Sitter space as an excited state, more specifically as a Glauber–Sudarshan state in string theory. We now illustrate the key differences between the Glauber–Sudarshan states that we aim to construct and the well-known coherent states.

\section{Coherent states vs. Glauber--Sudarshan states}
\label{GS state}

The importance of coherent states for the study of photon statistics of radiation fields is widely known \cite{PhysRev.131.2766}. The property that makes these quantum states of considerable interest is their well-defined classical limit. In fact, they are typically known as the \textit{minimum uncertainty states} --- the expectation value of any observable computed with respect to such states follows the classical trajectory with minimum uncertainty. Even though the coherent states are not orthogonal to each other, they form a complete set, or more precisely, an overcomplete set. These states are characterised as the eigenstates $|z\rangle$ of the annihilation operator $\hat{a}$ in the context of a quantum harmonic oscillator:
\begin{align}
    \hat{a}|z\rangle = z |z\rangle,
\end{align}
where $z$ is the corresponding eigenvalue. Since $\hat{a}$ is not a Hermitian operator, $z = |z| e^{i\varphi} \in \mathbb{C}$ is generically a complex number. Let us express the state $|z\rangle$ as a linear combination of the basis states $\{|n\rangle,~n=0,1,\dots\}$, namely the eigenstates of the number operator of the harmonic oscillator:
\begin{align}
    |z\rangle
    &= \sum_{n=0}^{\infty} c_n |n\rangle = c_0|0\rangle + c_1|1\rangle + c_2|2\rangle + c_3|3\rangle + \cdots .
\end{align}
Then the action of $\hat{a}$ on the above expression gives 
\begin{align}
    \hat{a}|z\rangle
    &= c_1|0\rangle + \sqrt{2}c_2|1\rangle + \sqrt{3}c_3|2\rangle + \cdots \nonumber\\
    &= z(c_0|0\rangle + c_1|1\rangle + c_2|2\rangle + \cdots),
\end{align}
where we have used $\hat{a}|n\rangle = \sqrt{n}|n-1\rangle$. Comparing both sides, we see that the coefficients satisfy the recurrence relation
\begin{align}
    c_{n+1} = \frac{z}{\sqrt{n+1}} c_n.
\end{align}
Hence, the coherent state is given by
\begin{align}
    |z\rangle = c_0 \sum_{n=0}^{\infty} \frac{z^n}{\sqrt{n!}} |n\rangle .
\end{align}
We can fix the overall constant $c_0$ by demanding $|z\rangle$ to be normalised:
\begin{align}
    \langle z|z\rangle
    &= |c_0|^2 \sum_{m,n} \frac{(z^\star)^m z^n}{\sqrt{m!n!}} \langle m|n\rangle \implies |c_0|
    = \exp\left(-\frac{|z|^2}{2}\right).
\end{align}
Upon expressing the $n$-th excited state in terms of the vacuum state, the coherent state
\begin{align}
    |z\rangle
    &= \exp\left(-\frac{|z|^2}{2}\right)
    \sum_{n=0}^{\infty}
    \frac{z^n}{\sqrt{n!}}
    \frac{(\hat{a}^\dagger)^n}{\sqrt{n!}}
    |0\rangle \equiv \hat{\mathrm{D}}(z)|0\rangle ,
    \label{CoHeREntStatE}
\end{align}
can be interpreted as a state obtained by displacing the vacuum configuration $|0\rangle$, where the displacement operator
\begin{align}
    \hat{\mathrm{D}}(z)
    &= \exp\left(-\frac{|z|^2}{2}\right)
    \exp\left(z\hat{a}^\dagger\right)
    \exp\left(-z^\star \hat{a}\right) \nonumber\\
    &= \exp\left(z\hat{a}^\dagger - z^\star \hat{a}\right),
    \label{DispOp}
\end{align}
is a unitary operator satisfying
\begin{align}
    \hat{\mathrm{D}}(z)\hat{\mathrm{D}}^\dagger(z)
    =
    \hat{\mathrm{D}}^\dagger(z)\hat{\mathrm{D}}(z)
    = 1.
\end{align}
In $3+1$ dimensional quantum field theory we have one harmonic oscillator for each momentum mode $k_n$, in other words we have an infinite set of annihilation and creation operators $\hat a_{k_n},~\hat{a}^\dagger_{k_n}$, satisfying the canonical commutation relations
\begin{align}
    [\hat a_{k_n},\hat{a}^\dagger_{k^\prime_n}]&=\delta_{k_{n},k^\prime_n}.\label{CanonicalCommFTheory}
\end{align}
Thus, we can  express the coherent state (\ref{CoHeREntStatE}) in field theory as \cite{bookKlauder,zhang1999coherent} 
\begin{align}
    |\{z_{k}\}\rangle &=\exp\left(\frac{1}{\mathrm{V}}\sum_{n} \left(z_{k_n} \hat{a}^{\dagger}_{k_n}-z^{\star}_{k_n}\hat{a}_{k_n}\right)\right)|0\rangle= \hat{\rm D}(\{z_k\})\vert 0\rangle,\label{DispInFTheory}
\end{align}
where we have considered the four dimensional space-time to be a square lattice of volume $\mathrm{V}=L^{4}$ such that $k_{n}^{\mu}=\frac{2\pi n^{\mu}}{L}$, $n^\mu$ are integers.    
Now, the state that we construct differs from the usual coherent states. We spell out the differences in the following. 
\begin{itemize}
    \item[(a)]  Rather than a free vacuum $|0\rangle$, the state of our consideration is obtained by displacing an \textit{interacting vacuum} $|\Omega\rangle$. Suppose we consider a double well potential of the form $V(x)=(x^2-a^2)^2$. 
\begin{figure}[!ht]
    \centering
\includegraphics[width=0.80\textwidth]{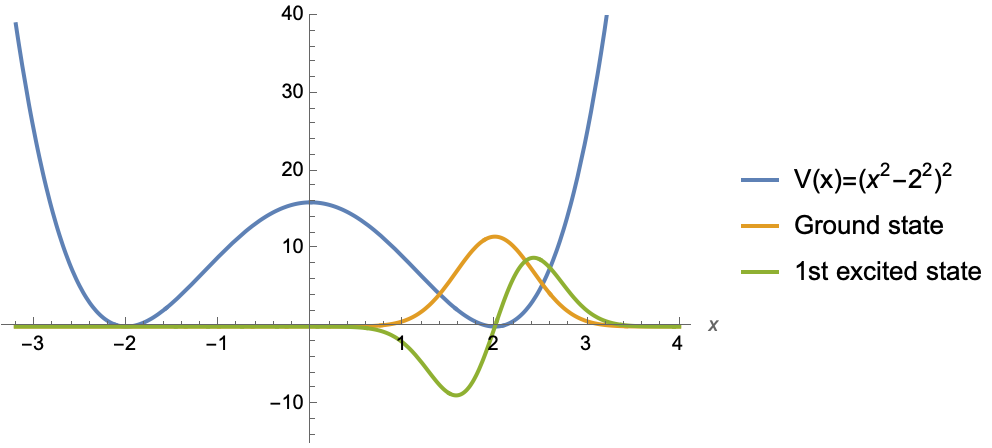}
    \caption{Double well potential $V(x)=(x^2-2^2)^2$ with ground state and 1st excited state around the minima at $x=+2$.}
    \label{DWellPot}
\end{figure}
It contains two minima at $x=\pm a$. The ground state corresponding to any of these two minima we refer to as the free vacuum. In Fig. \ref{DWellPot}, the free vacuum corresponding to the minima at $x=+a$ is shown by the orange curve. 
\begin{figure}[!ht]
    \centering
\includegraphics[width=0.80\textwidth]{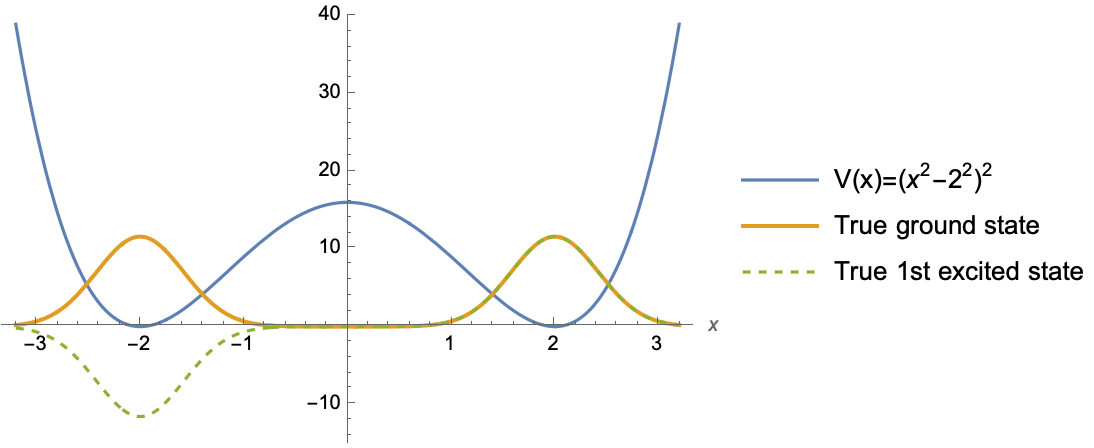}
    \caption{Double well potential $V(x)=(x^2-2^2)^2$ with true ground state and 1st excited state}
    \label{DWPotTGS}
\end{figure}
The true ground state of the double well potential is depicted by the orange curve in Fig. \ref{DWPotTGS}. We refer to this particular state as the interacting vacuum.  The interacting vacuum can be expressed using the total Hamiltonian $H=H_{0}+H_{\mathrm{int}}$ of a system as \cite{Peskin:1995ev} 
\begin{align}
    |\Omega \rangle &=\lim_{{T}\rightarrow \infty (1-i\epsilon)}\left(e^{-iE_{0}T}\langle \Omega |0 \rangle \right)^{-1} e^{-iHT}|0\rangle ,
\end{align}
where $E_0= \langle \Omega | H | \Omega \rangle$ is the true ground state energy.

\item[(b)] For such an interacting vacuum configuration, the form of the corresponding displacement operator need not take the same form as (\ref{DispOp}). To quantify this situation, we introduce effective creation and annihilation operators $\hat{a}^\dagger_{\mathrm{eff},k}$ and $\hat{a}_{\mathrm{eff},k}$ respectively. The interacting vacuum $|\Omega\rangle$ can be defined by the condition that it is annihilated by $\hat{a}_{\mathrm{eff},k}$ \cite{Brahma:2020tak} 
\begin{align}
\hat{a}_{\mathrm{eff},k}|\Omega\rangle&=0.
\end{align} 
Note that, the effective operators $\hat{a}_{\mathrm{eff},k}$ and $\hat{a}^\dagger_{\mathrm{eff},k}$ do not satisfy the conventional canonical commutation relation (\ref{CanonicalCommFTheory}). Hence it in not that straightforward to write the displacement operator $\hat{\mathbb{D}}(\sigma)$  analogous to Eq. (\ref{DispInFTheory}).
In fact a possible form for such an operator has been given in Ref. \cite{Brahma:2022wdl} for an interacting scalar theory: 
\begin{align}
   \hat{\mathbb{D}}(\sigma)&=\exp\left(\frac{1}{\mathrm{V}}\sum_{i} \left(\sigma(k_{i}) \tilde{\varphi}^\star (k_i)+\sigma^{\star}(k_{i})\tilde{\varphi}(k_{i})  \right) \right)\nonumber\\
    &=\exp\left(\frac{2}{\mathrm{V}}\sum_{i}\left(\mathbf{Re}~\sigma(k_i)~\mathbf{Re}~\tilde{\varphi}(k_i)+\mathbf{Im}~\sigma(k_i)~\mathbf{Im}~\tilde{\varphi}(k_i)~\right) \right),\label{DOpIntVAccUm}
\end{align}
where $\tilde{\varphi}(k_i)$ are the Fourier modes of the scalar field $\varphi(x)$. 
\end{itemize}

To make distinction from coherent states (\ref{DispInFTheory}), the states obtained by acting $\hat{\mathbb{D}}(\sigma)$ on an interacting vacuum $|\Omega\rangle:$
\begin{align} |\sigma\rangle=\hat{\mathbb{D}}(\sigma)|\Omega\rangle,
\end{align}
are referred to as \textit{Glauber-Sudarshan} (GS) states.

In this introductory chapter, we have provided a brief overview of de Sitter space in Poincaré coordinates and introduced the formalism of the Glauber--Sudarshan state. In the next chapter, we review the realization of de Sitter space as a Glauber--Sudarshan state in type IIB string theory \cite{Brahma:2020tak,Bernardo:2021rul,brahma2021four}, which lays the groundwork for the heterotic constructions developed in this thesis.

\chapter{de Sitter as an excited state in type IIB string theory}
\label{Chapter:2}
A first step towards realising de Sitter space as an excited state in string theory was taken in the context of type IIB string theory\cite{Brahma:2020tak,brahma2021four,Bernardo:2021rul,Brahma:2022wdl}. In this chapter, we elaborate on a duality sequence from a generic M theory metric configuration that leads to de Sitter isometry $d\mathrm{S}_4$ within type IIB string theory. In fact, as we will see the duality sequence proceeds dynamically to give rise to de Sitter as the endpoint of that sequence in the late time limit i.e. in the $t\rightarrow 0^-$ limit of the Poincar\'e coordinates. We emphasize that the metric configuration at each stage of the sequence is to be understood as obtained by taking an expectation value of the metric operator with respect to the corresponding Glauber-Sudarshan state (\ref{MthGSsTaTe},  \ref{IIBGSstAte}).   
\section{de Sitter isometry in Type \texorpdfstring{$\mathrm{IIB}$}{IIB}  string theory} 
\label{typeIIBdSisom}
The goal of \cite{Brahma:2020tak,Bernardo:2021rul, brahma2021four} was to see whether the following time dependent background in type $\mathrm{IIB}$ string theory  
\begin{align}
ds^2_{\mathrm{IIB}} &=\langle  \hat{\mathbf{g}}_{\rm AB}\rangle_\rho d\mathrm{Y}^{\rm A}d\mathrm{Y}^{\rm B}\nonumber\\
    &=\frac{ {a}^{2}(t)}{ {\rm H}^2(y)}\left[-dt^2 +  \left(dx^1\right)^2 +\left(dx^2\right)^2 + \left(dx^3\right)^2\right] + {\rm H}^2(y)\big[\mathrm{F}_{1}(t)g_{\alpha\beta}dy^{\alpha}dy^{\beta}\nonumber\\
    &+\mathrm{F}_{2}(t)g_{mn} dy^m dy^n \big],\label{tyPeIIBMetrIc}
\end{align}
with the coupling constant 
\begin{align}
    g_{b}=1,\label{tyPeIIbCoupLConstaNT}
\end{align}
can be realised as a Glauber-Sudarshan state \eqref{IIBGSstAte} with  four dimensional de Sitter isometry in flat-slicing provided the scale factor $a(t)$ takes the form
\begin{align}
  {a^{2}(t)}=\frac{1}{\Lambda t^2} ~~~\text{where}~~ -\infty \le t<0 .\label{TheScaLeFactOr}  
\end{align} 
The type IIB coupling $g_b=1$ means axio-dilaton, $\tau:=C_{0}+i e^{-\varphi}$ which is a combination of the {R-R} field $C_{0}$ (viz. axion) and the dilaton field $\varphi$, vanishes.
The type $\mathrm{IIB}$ coupling constant is expressed in terms of the axio-dilaton field as $g_{b}=1/\mathrm{Im}(\tau)$. By vanishing axio-dilaton it means that $C_{0}=\varphi=0$. 
 Such a configuration is supported at a constant coupling point in $\mathrm{F}$-theory\cite{Sen_1996,DASGUPTA1996125}.

In Eq. (\ref{tyPeIIBMetrIc}),  the compact six dimensional internal space is taken to be the product of a 2-manifold, $\mathcal{M}_{2}$ and a 4-manifold, $\mathcal{M}_{4}$. Both of these manifolds are non-k$\ddot{\text{a}}$hler and non-complex in nature. The corresponding coordinates are denoted as $y^{\alpha}\in \mathcal{M}_{2}$, $y^{m}\in \mathcal{M}_{4}$ and $y=(y^\alpha,y^m)$. Note that, $\mathrm{H}(y),~\mathrm{F}_1(t)$ and $\mathrm{F}_2(t)$ are the warp factors, where the latter two are depicting time dependence of the internal manifolds. The unwarped metric components for the manifolds $\mathcal{M}_2$ and $\mathcal{M}_4$ are denoted as $g_{\alpha \beta}(y)$ and $g_{mn}(y)$ respectively. 
One has to ensure that the time dependent warp factors satisfy the following constraint 
\begin{align}
    \mathrm{F}_{1}(t)\mathrm{F}^{2}_{2}(t)=1.\label{eq:IIBconst}
\end{align}
 The above constraint arises by demanding time independence of the volume of the internal space which is a necessary condition for the time independence of the four-dimensional Newton's constant. {Note that, the four dimensional Newton's constant $\mathbb{G}^{(4)}$ is given in terms of the ten dimensional Newton's constant $\mathbb{G}^{(10)}$ and the volume $\mathcal{V}_{6}$ of the six dimensional internal space as $\mathbb{G}^{(4)}=\mathbb{G}^{(10)}/\mathcal{V}_{6}$. Since $\mathbb{G}^{(10)}$ is time independent to begin with, time independence of $\mathbb{G}^{(4)}$ necessitates $\mathcal{V}_{6}$ to be time independent}. 

The construction of a Glauber-Sudarshan state  is technically challenging from this type $\mathrm{IIB}$ string theory with $g_{b}=1$ \cite{Brahma:2020tak,brahma2021four,Bernardo:2021rul}. The way out is to dualize to $\mathrm{M}$-theory and then construct the state therein. There are many advantages with the M theory uplift. As is evident from the duality rules elaborated in Appendix \ref{DualitYRulES}, the various flux degrees of freedom of type $\mathrm{IIB}$ theory get converted to four form fluxes in M theory. Furthermore, the branes of type IIB get mapped to geometric spaces or to M2 and M5-branes.  

We now obtain the corresponding metric configuration of $\rm{M}$ theory. In order to do so we first perform a $\mathrm{T}$ duality (see Appendix \ref{DualitYRulES} for the duality rules) along $x_{3}$ direction of type $\mathrm{IIB}$ theory (\ref{tyPeIIBMetrIc}) which gives type $\mathrm{IIA}$ theory with the following metric configuration
\begin{align}
ds^{2}_{\mathrm{IIA}}&=\frac{a^{2}(t)}{{\rm H}^2(y)}\left[-dt^2 + \sum_{i = 1}^2 \left(dx^i\right)^2 \right] +\frac{\mathrm{H}^{2}(y)}{a^{2}(t)}(dx^{3})^2 + {\rm H}^2(y)\big[\mathrm{F}_{1}(t)g_{\alpha\beta}dy^{\alpha}dy^{\beta}\nonumber\\
&~~~+\mathrm{F}_{2}(t)g_{mn} dy^m dy^n \big]. \label{Type IIA MetRIc}
\end{align}
 The type IIA coupling constant equals
\begin{align}
    {g}_{s}=\frac{\mathrm{H}(y)}{a(t)}.\label{TyPeIIA coUplInG}
\end{align}
Demanding de Sitter isometry (\ref{TheScaLeFactOr}) in type $\mathrm{IIB}$ theory (\ref{tyPeIIBMetrIc}) implies that the above coupling constant becomes
\begin{align}
    g_{s}&=\mathrm{H}(y)|t|\sqrt{\Lambda}.\label{IIAcouPlingdeSitterIIB}
\end{align}
\noindent
 Upon dimensional uplift (as elaborated in Appendix \ref{DualitYRulES}), the type IIA theory (\ref{Type IIA MetRIc}) becomes the eleven-dimensional M-theory that has the following metric structure
\begin{align}
&ds^{2}_{11}\equiv \langle  \hat{\mathbf{g}}_{\mathrm{MN}}\rangle_\sigma ~d\mathrm{Y^M}d\mathrm{Y^N}\nonumber\\
&=g_{s}^{-\frac 8 3}\left[-dt^2 + \sum_{i = 1}^2 \left({dx^i}\right)^2 \right] +g_{s}^{-\frac 2 3} {\rm H}^2(y)\left[\mathrm{F}_{1}(t)g_{\alpha\beta}dy^{\alpha}dy^{\beta}+\mathrm{F}_{2}(t)g_{mn} dy^m dy^n \right]+g_{s}^{\frac 4 3}\delta_{ab}d\omega^{a}d\omega^{b}
\label{MthEorYmeTric}
\end{align}
where $\omega^{a}\equiv (x^3,x^{11})$ are the coordinates of $\mathrm{M}$-theory {torus} ${\mathbb{T}_{3,11}^{2}}/{\mathcal{G}}$, $\mathcal{G}$ is a group action without any fixed points. An important thing to note here is that the $\mathrm{M}$ theory torus is of finite size. This is precisely because a well defined effective field theory description of the underlying $\mathrm{M}$ theory action requires the type $\mathrm{IIA}$ coupling constant (\ref{IIAcouPlingdeSitterIIB}) to satisfy an upper bound $\frac {g_{s}}{\mathrm{H}(y)}<1$. This in turn implies that the de Sitter isometry in type $\mathrm{IIB}$ theory exists only for a finite time domain 
\begin{align}
    -\frac{1}{\sqrt{\Lambda}}<t<0.\label{TCCinTypeIIBStrINgTh}
\end{align}
This temporal domain is surprisingly identical to that of the Trans-Planckian Cosmic Censorship (TCC) bound\cite{Bedroya_2020,PhysRevD.101.103502,Bernardo_2021}. In the next chapter, we will discuss more on this.
\begin{figure}[!ht]
\centering
\begin{tikzpicture}[font=\small,thick]
\node[draw, rounded corners,fill=teal!10,
    minimum width=1.5cm,
    minimum height=1cm](block1) { \textcolor{violet}{
$\substack{ \textstyle \mathrm{M}~\textbf{theory} \\ \\\textstyle \textbf{on}\\ \textstyle\textbf{$\mathbb{R}^{2,1}\times\mathcal{ M}_{2}\times \mathcal{M}_{4}\times \frac{\mathbb{T}^{2}_{3,11}}{\mathcal{G}} $}}$
}};
\node[draw, rounded rectangle, fill=teal!10,
    right=of block1,
    minimum width=1.5cm,
    minimum height=1cm] (block2) { \textcolor{violet}{ 
$\substack{\textstyle\textbf{Type}~\mathrm{IIA}~\textbf{theory}\\ \\ \textstyle\textbf{on} \\\textstyle \mathbb{R}^{2,1}\times  \mathcal{M}_{2}\times \mathcal{M}_{4}\times {\mathbf{S}^{1}_{3}}}$
}};
\node[draw, rounded corners, fill=teal!10,
    right =of block2,
    minimum width=1.5cm,
    minimum height=1.5cm] (block3) { \textcolor{violet}{
$\substack{\textstyle\textbf{Type}~\mathrm{IIB}~\textbf{theory}\\ \\ \textstyle \textbf{on} \\\textstyle \mathbb{R}^{3,1}\times \mathcal{M}_{2}\times \mathcal{M}_{4}}$
     }};
\draw[-latex] (block1) edge (block2)
    (block2) edge (block3);
\node at (2.9,.8) {${\substack{\textstyle\text{Dimensional}\\ \textstyle \text{reduction} \\ \textstyle \text{along}\\\textstyle x^{11} }}$} ; 
\node at (8.4,0.7) { $\substack{\textstyle\mathrm{T}~\text{duality}\\ \textstyle \text{along}\\\textstyle~x^3}$};
\end{tikzpicture}
    \caption{Duality sequence from the $\mathrm{M}$ theory (\ref{MthEorYmeTric}) to $\mathrm{IIB}$ (\ref{tyPeIIBMetrIc}) having de Sitter isometry. } 
    \label{Type IIB tree diagram}
\end{figure}
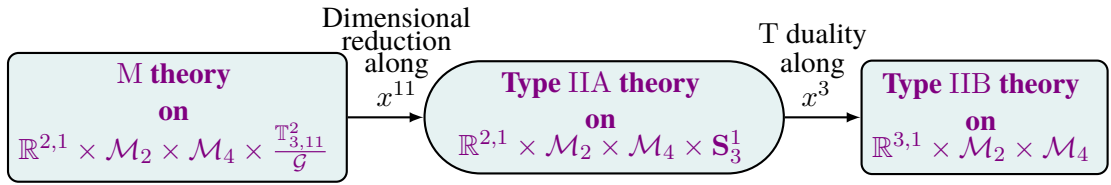

\subsubsection{Dynamical Duality Sequence:}
Starting with the $\mathrm{M}$ theory set-up (\ref{MthEorYmeTric}) we can reach type $\mathrm{IIB}$ theory (\ref{typeIIBdSisom}) \textit{dynamically}. To implement this, one has to specify generic scalings of the time dependent warp factors as
\begin{equation}
\begin{aligned}
    &\mathrm{F}_{1}(t)\equiv\mathrm{F}_{1}\left(\frac{g_{s}}{\mathrm{H}(y)}\right)=\sum_{k=0}^{\infty}{\rm A}_{k}\left(\frac{g_{s}}{\mathrm{H}(y) }\right)^{\beta_{o}+\frac{2k}{3}},\\
    &\mathrm{F}_{2}(t)\equiv \mathrm{F}_{2}\left(\frac{g_{s}}{\mathrm{H}(y)}\right)=\sum_{k=0}^{\infty}{\rm B}_{k}\left(\frac{g_{s}}{\mathrm{H}(y) }\right)^{\alpha_{o}+\frac{2k}{3}} \label{ScalingOfFi's}
\end{aligned}
\end{equation}
where $\mathrm{A}_k,~\mathrm{B}_k$ are non-zero coefficients that get constrained by the condition \eqref{eq:IIBconst} and the dominant scalings $\alpha_{o}~,~\beta_{o}$ have to satisfy the upper bound
\begin{align}
    \alpha_{o}<+\frac{2}{3},~~~\beta_{o}<+\frac{2}{3}.\label{UbdofdomFactor}
\end{align}
This bound is required so that only the $\mathrm{M}$ theory torus along 3,11 in Eq. (\ref{MthEorYmeTric})  shrinks to zero size 
taking us dynamically to the dual type $\mathrm{IIB}$ configuration (\ref{tyPeIIBMetrIc}). In fact, for the case at consideration we want the $\mathrm{IIB}$ configuration to be the final stage of our duality sequence having de Sitter isometry.  For this to happen it is necessary to set 
\begin{align}
\alpha_o=\beta_o=0.\label{TheValofalphAoBetAo}
\end{align}
The dynamical duality sequence is pictorially depicted in Fig. \ref{Type IIB tree diagram}.

So far we have illustrated the duality sequence in order to realise de Sitter space as a GS state in type IIB theory (\ref{tyPeIIBMetrIc}). Such a realisation requires one to compute the expectation value of the metric operator with respect to the corresponding GS state in the M theory background (\ref{MthEorYmeTric}). 
In the next section, we will review the path integral analysis to perform this computation following Ref. \cite{Brahma:2022wdl}. This analysis has also been detailed to some extent in our more recent paper \cite{Archana_CC2026}.
\section{Path integral analysis }
\label{sec:PIinMth}
In the low energy limit, the effective action of M-theory becomes identical to that of the 11 dimensional supergravity theory. The field content of this theory consists of the graviton $ \{\mathrm{g}_{\mathrm{M}\mathrm{N}} \}$, which transforms as a symmetric rank two tensor of the little group $SO(9)$ and possesses 44 physical degrees of freedom; the three form field $ \{\mathrm{C}_{\mathrm{MNP}} \}$ with 84 physical degrees of freedom; and the spin-$\frac{3}{2}$ Rarita-Schwinger field, $\{\Psi^{\mathrm{M}}_{a~} \}$, with 128 physical degrees of freedom. Here $a$ denotes the spinor index, while $\mathrm{M,~N,~P}=0,...,11$ are eleven-dimensional spacetime indices associated with the M-theory background $\mathbb{R}^{2,1}\times \mathcal{M}_{2}\times \mathcal{M}_{4}\times  \frac{\mathbb{T}_{3,11}^2}{\mathcal{G}}$ appearing in Eq. (\ref{MthEorYmeTric}). Let the GS state be denoted as $|\sigma\rangle=|{\alpha},{\beta},{\gamma}\rangle$, where $|{\alpha}\rangle \equiv |{\alpha}_{\mathrm{MN}}\rangle$, $|{\beta} \rangle\equiv |{\beta}_{\mathrm{MNP}}\rangle$ and $|{\gamma}\rangle\equiv |{\gamma}_{\mathrm{M}}\rangle$ are the GS states associated to the graviton, three-form field and the Rarita-Schwinger fermion degrees of freedom respectively. We express the expectation value of the metric operator in M-theory with respect to the aforementioned state as
\begin{align}
\langle \hat{\mathbf{g}}_{\mathrm{MN}}\rangle_{\sigma}
&=
\frac{
\langle \sigma |\hat{\mathbf{g}}_{\mathrm{MN}}|\sigma \rangle
}{
\langle \sigma|\sigma\rangle
}
=
\frac{
\langle \Omega |
\hat{\mathbb{D}}^{\dagger}(\sigma)\,
\hat{\mathbf{g}}_{\mathrm{MN}}\,
\hat{\mathbb{D}}(\sigma)
|\Omega\rangle
}{
\langle \Omega |
\hat{\mathbb{D}}^{\dagger}(\sigma)
\hat{\mathbb{D}}(\sigma)
|\Omega \rangle
}
\nonumber\\[4pt]
&=
\frac{
\int [{\cal D} {\rm g}_{\mathrm{MN}}]
[{\cal D} {\rm C}_{\mathrm{MNP}}]
[{\cal D}\Psi_{\mathrm{N}}]\,
\exp\!\left(-\mathbf{S}_{\mathrm{tot}}\right)\,
\mathbb{D}^{\star}(\alpha,\beta,\gamma)\,
{\rm g}_{\mathrm{MN}}\,
\mathbb{D}(\alpha,\beta,\gamma)
}{
\int [{\cal D} {\rm g}_{\mathrm{MN}}]
[{\cal D} {\rm C}_{\mathrm{MNP}}]
[{\cal D}\Psi_{\mathrm{N}}]\,
\exp\!\left(-\mathbf{S}_{\mathrm{tot}}\right)\,
\mathbb{D}^{\star}(\alpha,\beta,\gamma)\,
\mathbb{D}(\alpha,\beta,\gamma)
}.
\label{ExpofMetOp}
\end{align}
where the total action $\mathbf{S}_{\rm tot}$, involving the kinetic ${\bf S}_{\rm kin}$, interaction ${\bf S}_{\rm int}$, ghost ${\bf S}_{\rm gh}$ and gauge-fixing ${\bf S}_{\rm g-f}$ terms can be schematically written as 
\begin{align}
    {\bf S}_{\rm tot} \equiv {\bf S}_{\rm kin} + {\bf S}_{\rm int} + 
{\bf S}_{\rm gh} + {\bf S}_{\rm g-f}.\label{TotaLActIonSchematic}
\end{align}
Since the exact evaluation of the above path integral by keeping the tensorial structure of $\{\mathrm{g}_{\rm MN}\}$, $\{\mathrm{C}_{\rm MNP}\}$ and the spinorial structure of $\{\mathrm{\Psi}^{\rm M}_a \}$ is, in general, difficult to perform, a toy model is at consideration without any loss of generality\cite{Brahma:2022wdl}. In particular,  three scalar fields are taken into account $(\varphi_{1},\varphi_{2},\varphi_{3})$ each representing one of the components of graviton, three-form field and the Rarita-Schwinger fermion respectively. Correspondingly, the path integral (\ref{ExpofMetOp}) becomes:
\begin{align}
    \langle \hat \varphi_1\rangle_{\sigma}&=\frac{ \int [\mathcal{D}\varphi_1][\mathcal{D}\varphi_2][\mathcal{D}\varphi_3] e^{-\mathbf{S}_{\rm tot}} \mathbb{D}^\star(\alpha,\beta,\gamma) \varphi_1 \mathbb{D}(\alpha,\beta,\gamma) }{\int [\mathcal{D}\varphi_1][\mathcal{D}\varphi_2][\mathcal{D}\varphi_3] e^{-\mathbf{S}_{\rm tot}} \mathbb{D}^\star(\alpha,\beta,\gamma)  \mathbb{D}(\alpha,\beta,\gamma)}.\label{ScalarPIntAnA}
\end{align}
Note that, for the scalar analysis, the ghost and gauge-fixing parts of the total action (\ref{TotaLActIonSchematic}) can be ignored. Further, a generic  interaction term can have the following form 
\begin{align}
\mathbf{S}_{\text{int}}&\equiv \int d^{11}\mathrm{Y} \sum_{n,...,s}c_{nmpqrs}\partial^{n} \varphi^{q}_{1}(\mathrm{Y})\partial^{m}\varphi_{2}^{r}(\mathrm{Y})\partial^{p}\varphi_{3}^{s}(\mathrm{Y}),\label{SinTeraCtiOn}
\end{align}
where $n+m+p\in 2 \mathbb{Z}_{+}$ 
and $(n,m,p,q,r,s)\in \mathbb{Z}_{+}$. 
Replacing the fields $\varphi_{i}(\mathrm{Y})$ in terms of their Fourier transform:
\begin{align}
\varphi_i(\mathrm{Y})&=\int_{-\infty}^{\infty} d^{11}k ~~e^{ik\mathrm{Y} }\tilde{\varphi}_i(k),
\end{align}
$\mathbf{S}_{\text{int}}$ takes the following form
{
\begin{align}
    \mathbf{S}_{\text{int}}&\equiv \int d^{11}\mathrm{Y} \sum_{n,...,s}c_{nmpqrs}~\partial^{n}\left(\int_{-\infty}^{\infty} d^{11}k~e^{ik\mathrm{Y} }~\widetilde{\varphi}_{1}(k)\right)^{q}\partial^{m}\left(\int_{-\infty}^{\infty} d^{11}l~e^{il\mathrm{Y} }~\widetilde{\varphi}_{2}(l)\right)^{r}\nonumber\\
    &~~~~~~~~~~~
   \times \partial^{p}\left(\int_{-\infty}^{\infty} d^{11}f~e^{if\mathrm{Y} }~\widetilde{\varphi}_{3}(f)\right)^{s}\nonumber\\
   &=\int d^{11}\mathrm{Y} \sum_{n,...,s}c_{nmpqrs} (i)^{n+m+p}\int \prod_{i=1}^{q}d^{11}k_{i} \prod_{j=1}^{r}d^{11}l_{j}\prod_{t=1}^{s}d^{11}f_{t}~\left(\sum_{i=1}^{q}k_{i} \right)^n \left(\sum_{j=1}^{r}l_{j} \right)^{m}\nonumber\\
   &\times \left(\sum_{t=1}^{s}f_{t} \right)^{p} \prod_{i=1}^{q}\widetilde{\varphi}_{1}(k_{i})\prod_{j=1}^{r}\widetilde{\varphi}_{2}(l_{j})\prod_{t=1}^{s}\widetilde{\varphi}_{3}(f_{t})\exp\left[i\left(\sum_{i=1}^{q}k_{i}+\sum_{j=1}^{r}l_{j}+\sum_{t=1}^{s}f_{t}\right)\mathrm{Y} \right]\nonumber\\
   &=\sum_{n,...,s}c_{nmpqrs} (-1)^{(n+m+p)/2}\int \prod_{i=1}^{q}d^{11}k_{i} \prod_{j=1}^{r}d^{11}l_{j}\prod_{t=1}^{s}d^{11}f_{t}~\left(\sum_{i=1}^{q}k_{i} \right)^n \left(\sum_{j=1}^{r}l_{j} \right)^{m}\nonumber\\
   &\times \left(\sum_{t=1}^{s}f_{t} \right)^{p} \prod_{i=1}^{q}\widetilde{\varphi}_{1}(k_{i})\prod_{j=1}^{r}\widetilde{\varphi}_{2}(l_{j})\prod_{t=1}^{s}\widetilde{\varphi}_{3}(f_{t})~\delta^{(11)}\left(\sum_{i=1}^{q}k_{i}+\sum_{j=1}^{r}l_{j}+\sum_{t=1}^{s}f_{t}\right)\label{DelTaFunCTionn}\\
   &= \sum_{n,...,s}c_{nmpqrs} (-1)^{(n+m+p)/2}\int \prod_{i=1}^{q}d^{11}k_{i} \prod_{j=1}^{r}d^{11}l_{j}\prod_{t=1}^{s-1}d^{11}f_{t}~\left(\sum_{i=1}^{q}k_{i} \right)^n \left(\sum_{j=1}^{r}l_{j} \right)^{m}\nonumber\\
   \times \bigg(&-\sum_{i=1}^{q}k_{i}-\sum_{j=1}^{r}l_{j} \bigg)^{p} \prod_{i=1}^{q}\widetilde{\varphi}_{1}(k_{i})\prod_{j=1}^{r}\widetilde{\varphi}_{2}(l_{j})\prod_{t=1}^{s-1}\widetilde{\varphi}_{3}(f_{t})\widetilde{\varphi}_{3}\left(-\sum_{i=1}^{q}k_{i}-\sum_{j=1}^{r}l_{j}-\sum_{t=1}^{s-1}f_{t}\right).\label{SinTerActiONinTegrAlREp}
\end{align}
}
Since the fields $\varphi_i(\mathrm{Y})$ are \textit{real} valued:
\begin{align}
    \varphi_{i}^{\star}(\mathrm{Y} )=\int_{-\infty}^{\infty}d^{11}k ~\widetilde{\varphi}_{i}^{\star}(k)&e^{-ik\mathrm{Y} }=\int_{-\infty}^{\infty}d^{11}k ~\widetilde{\varphi}_{i}^{\star}(-k)e^{ik\mathrm{Y} }=\varphi_{i}(\mathrm{Y}),
\end{align}
the corresponding Fourier modes satisfy $\widetilde{\varphi}_{i}^{\star}(k)=\widetilde{\varphi}_{i}(-k)$. Therefore, $\mathbf{S}_{\rm{int}}$ can be expressed as 
\begin{align}
\mathbf{S}_{\text{int}}&= \sum_{n, ..., s} c_{nmpqrs} 
(-1)^{(n + m + 3p)/2} \int \prod_{i = 1}^q d^{11} {k}_i \prod_{j = 1}^r d^{11} {l}_j \prod_{t = 1}^{s - 1} d^{11} {f}_t~ \left(\sum_{i = 1}^q {k}_i\right)^n \left(\sum_{j = 1}^r {l}_j\right)^m\nonumber\\
\times  &\left(\sum_{i = 1}^q {k}_i + \sum_{j = 1}^ r {l}_j\right)^p
 \prod_{i = 1}^q \widetilde{\varphi}_1({k}_i) \prod_{j = 1}^r \widetilde{\varphi}_2({l}_j) \prod_{t = 1}^{s - 1} \widetilde{\varphi}_3({f}_t)
~\widetilde\varphi^\ast_3\left(\sum_{i = 1}^q {k}_i + \sum_{j = 1}^ r {l}_j + \sum_{t = 1}^{s-1} {f}_t\right)\nonumber\\
\equiv & \sum_{n, ..., s} c_{nmpqrs} 
(-1)^{(n + m + 3p)/2} \frac{1}{\mathrm{V}^{q+r+s-1}}\sum_{\left\{{k_{u_i}}\right\},\left\{{l_{v_j}}\right\},\left\{{f_{w_t}}\right\}}~ \left(\sum_{i = 1}^q {k}_{u_i}\right)^n \left(\sum_{j = 1}^r {l}_{v_j}\right)^m\nonumber\\
\times  \bigg(\sum_{i = 1}^q &{k}_{u_i} + \sum_{j = 1}^ r {l}_{v_j}\bigg)^p
 \prod_{i = 1}^q \widetilde{\varphi}_1({k}_{u_i}) \prod_{j = 1}^r \widetilde{\varphi}_2({l}_{v_j}) \prod_{t = 1}^{s - 1} \widetilde{\varphi}_3({f}_{w_t})
~\widetilde\varphi^\ast_3\left(\sum_{i = 1}^q {k}_{u_i} + \sum_{j = 1}^ r {l}_{v_j} + \sum_{t = 1}^{s-1} {f}_{w_t}\right).\label{FormOfSint}
\end{align}
To arrive at the final step we have replaced
\begin{align}
    \int d^{11}k\rightarrow \frac{1}{\rm V} \sum_{k},\label{DisCreteLiMiT}
\end{align}
where $\mathrm{V}$ denotes volume of the 11-dimensional discretised spacetime. Likewise, the kinetic part of the action expressed in terms of the fourier modes of the fields is as follows
\begin{align}
\mathbf{S}_{\mathrm{kin}}&=\int_{-\infty}^{\infty}d^{11}\mathrm{Y}~\big( \partial_{\rm M}\varphi_{1}\partial^{\rm M}\varphi_{1}+\partial_{\rm M}\varphi_{2}\partial^{\rm M}\varphi_{2}+\partial_{\rm M}\varphi_{3}\partial^{\rm M}\varphi_{3}\big)\nonumber\\
&=\int_{-\infty}^{\infty}d^{11}k~ k^{2}~|\widetilde{\varphi}_{1}(k)|^2+\int_{-\infty}^{\infty}d^{11}l~ l^{2}~|\widetilde{\varphi}_{2}(l)|^2+\int_{-\infty}^{\infty}d^{11}f~ f^{2}~|\widetilde{\varphi}_{3}(f)|^2.
\end{align}
There is another ingredient that we require in order to evaluate the path integral (\ref{ScalarPIntAnA}), viz. the displacement operator corresponding to the GS state. The form of the displacement operator  $\hat{\mathbb{D}}(\sigma)$ (\ref{DOpIntVAccUm}) will now take the following form 
\begin{align}
\mathbb{D}(\sigma\equiv\{\alpha,\beta,\gamma\})&=\exp\bigg(\frac{2}{\mathrm{V}}\bigg(\sum_{i}\mathbf{Re}~{\alpha}(k_{i}) ~\mathbf{Re}~\widetilde{\varphi}_{1}(k_{i})+\mathbf{Im}~{\alpha}(k_{i})~\mathbf{Im}~\widetilde{\varphi}_{1}(k_{i})\nonumber\\
&~~~~~~~~~~~~~~+\sum_{j}\mathbf{Re}~{\beta}(l_{j}) ~\mathbf{Re}~\widetilde{\varphi}_{2}(l_{j})+\mathbf{Im}~{\beta}(l_{j})~\mathbf{Im}~\widetilde{\varphi}_{2}(l_{j})\nonumber\\
&~~~~~~~~~~~~~~+\sum_{t}\mathbf{Re}~{\gamma}(f_{t}) ~\mathbf{Re}~\widetilde{\varphi}_{3}(f_{t})+\mathbf{Im}~{\gamma}(f_{t})~\mathbf{Im}~\widetilde{\varphi}_{3}(f_{t})\bigg)\bigg).   
\end{align}
Now, we can write down the numerator of the path integral (\ref{ScalarPIntAnA}) as   
\begin{align}\label{NumPInT}
 &{\rm Num}\left[\langle  \hat{\varphi}_1 \rangle_{{\sigma}}\right]\equiv 
\prod_{i, j, t}\int d\left({\bf Re}~\widetilde{\varphi}_1(k_i)\right) d\left({\bf Im}~\widetilde{\varphi}_1(k_i)\right) d\left({\bf Re}~\widetilde{\varphi}_2(l_j)\right) d\left({\bf Im}~\widetilde{\varphi}_2(l_j)\right)
\nonumber\\
&\times  d\left({\bf Re}~\widetilde{\varphi}_3(f_t)\right) d\left({\bf Im}~\widetilde{\varphi}_3(f_t)\right)
\cdot \frac{1}{ {\rm V}} \sum_{k_{i}^{''}} \psi_{{\bf k}''_i}({\bf x}, y, z) e^{-ik''_{i,0} t} 
\left({\bf Re}~\widetilde{\varphi}_1(k''_i) + i {\bf Im}~\widetilde{\varphi}_1(k''_i)\right) \nonumber\\
&\times  {\rm exp}\Big[\frac{4}{ {\rm V}}\sum_{i, j, t}\Big({\bf Re}~{\alpha}(k'_i)
~{\bf Re}~\widetilde{\varphi}_1(k'_i) + {\bf Im}~{\alpha}(k'_i)
~{\bf Im}~\widetilde{\varphi}_1(k'_i) + {\bf Re}~{\beta}(l'_j)
~{\bf Re}~\widetilde{\varphi}_2(l'_j) \nonumber\\
& ~~~~~+  {\bf Im}~{\beta}(l'_j)
~{\bf Im}~\widetilde{\varphi}_2(l'_j) + {\bf Re}~{\gamma}(f'_t)
~{\bf Re}~\widetilde{\varphi}_3(f'_t) + {\bf Im}~{\gamma}(f'_t)
~{\bf Im}~\widetilde{\varphi}_3(f'_t) \Big)\Big] \nonumber\\ 
& \times  {\rm exp}\Big[-\frac{1}{ {\rm V}} \Big(\sum_{i} k_i^2 \vert \widetilde\varphi_1(k_i)\vert^2 + 
\sum_{j} l_j^2 \vert \widetilde\varphi_2(l_j)\vert^2 + 
\sum_{t} f_t^2 \vert \widetilde\varphi_3(f_t)\vert^2\Big)\Big] \nonumber\\
&\times  {\rm exp}\Big[ -\sum_{\mathcal{S}^\prime}\frac{1}{ {\rm V}^u} c_{nmpqrs} (-1)^{(n + m + 3p)/2}
\left(\sum_{i = 1}^q {k}_{u_i}\right)^n \left(\sum_{j = 1}^r {l}_{v_j}\right)^m
\left(\sum_{i = 1}^q {k}_{u_i} + \sum_{j = 1}^ r {l}_{v_j}\right)^p\nonumber\\
& \times 
\prod_{i, j, t = 1}^{q, r, s-1}\Big({\bf Re}~\widetilde{\varphi}_1(k_{u_i}) + i {\bf Im}~\widetilde{\varphi}_1(k_{u_i})\Big) 
\Big({\bf Re}~\widetilde{\varphi}_2(l_{v_j}) + i {\bf Im}~\widetilde{\varphi}_2(l_{v_j})\Big) 
\Big({\bf Re}~\widetilde{\varphi}_3(f_{w_t}) + i {\bf Im}~\widetilde{\varphi}_3(f_{w_t})\Big) 
\nonumber\\
&\times 
\Big({\bf Re}~\widetilde{\varphi}_3\Big(\sum_{i = 1}^q {k}_{u_i} + \sum_{j = 1}^ r {l}_{v_j} + \sum_{t = 1}^{s-1}{f}_{w_t}\Big) - i {\bf Im}~\widetilde{\varphi}_3\Big(\sum_{i = 1}^q {k}_{u_i} + \sum_{j = 1}^ r {l}_{v_j} + \sum_{t = 1}^{s-1}{f}_{w_t}\Big)\Big)\Big], 
\end{align}
where $u=q+r+s-1$ and ${\cal S}^\prime= (\left\{{k_{u_i}}\right\},\left\{{l_{v_j}}\right\},\left\{{f_{w_t}}\right\},n,m,p,q,r,s)$. The insertion of $\varphi_1$ field in the path integral is expressed as
\begin{align}
    \varphi_1(\mathrm{Y})=\frac{1}{\rm V}\sum_{k_i}\psi_{\mathbf{k}_i}(\mathrm{X})e^{-ik_{i,0}t}\widetilde{\varphi}_{1}(k_i)=\frac{1}{\rm V}\sum_{k_i}\psi_{\mathbf{k}_i}(\mathrm{X})e^{-ik_{i,0}t}\left(\mathbf{Re}\widetilde{\varphi}_{1}(k_i)+i\mathbf{Im}\widetilde{\varphi}_{1}(k_i)\right),\label{TheInSertIonField}
\end{align}
where the coordinate $\mathrm{X}\in\mathbb{R}^{2}\times  \mathcal{M}_2\times\mathcal{M}_4\times \frac{\mathbb{T}^2_{3,11}}{\mathcal{G}}$.
For further simplification, we set the imaginary parts of all the Fourier modes in Eq. (\ref{NumPInT}) to zero. Introducing the variables
\begin{equation}
\begin{aligned} (a_{j},b_{j},c_{j})\equiv & (a(k_j),b(l_j),c(f_j))=\frac{1}{\mathrm{V}}(k_{j}^{2},l_{j}^{2},f_{j}^{2}),\\ (\overline{\alpha}_{i},\overline{\beta}_{i},\overline{\gamma}_{i})\equiv &(\overline{\alpha}(k_{i}),\overline{\beta}(l_{i}),\overline{\gamma}(f_{i}))=\frac{2}{\mathrm{V}}(\alpha(k_i),\beta(l_i),\gamma(f_i)), \label{abcDef}
\end{aligned}    
\end{equation}
we can rewrite (\ref{NumPInT}) in the following way
\begin{align}
     &{\rm Num}[\langle \hat{\varphi}_1\rangle_{\overline\sigma}]  =  \mathbb{C}(\overline\alpha_i, a_j, {\rm V}, ..) 
\int d\widetilde\varphi_1(k_1) 
~{\rm exp}\Big[-a_1\Big(\widetilde\varphi_1(k_1) - \frac{\overline{\alpha}_1}{a_1}  \Big)^2\Big]
~\cdots\nonumber\\ 
& ~~~~ \int d\widetilde\varphi_2(l_1) 
~{\rm exp}\Big[-b_1\Big(\widetilde\varphi_2(l_1) - \frac{\overline{\beta}_1}{b_1} \Big)^2\Big] \;\int d\widetilde\varphi_2(l_2) 
~{\rm exp}\Big[-b_2\Big(\widetilde\varphi_2(l_2) - \frac{\overline{\beta}_2}{b_2} \Big)^2\Big]\cdots\nonumber\\
& ~~~~\int d\widetilde\varphi_3(f_1) 
~{\rm exp}\Big[-c_1\Big(\widetilde\varphi_3(f_1) - \frac{\overline{\gamma}_1}{c_1} \Big)^2\Big]\int d\widetilde\varphi_3(f_2) 
~{\rm exp}\Big[-c_2\Big(\widetilde\varphi_3(f_2) - \frac{\overline{\gamma}_2}{c_2} \Big)^2\Big]\cdots \nonumber \\
& \times   \frac{1}{ {\rm V}}\Big(\widetilde\varphi_1(k_1)\psi_{{\bf k}_1}({\rm X})e^{-ik_{0, 1}t} + 
\widetilde\varphi_1(k_2)\psi_{{\bf k}_2}({\rm X})e^{-ik_{0, 2}t} + 
\widetilde\varphi_1(k_3)\psi_{{\bf k}_3}({\rm X})e^{-ik_{0, 3}t} +\cdots \Big)\nonumber \\
&\times \Big(1 - \sum_{{\cal S}}{\rm V}^{-u} c_{nmpqrs} (-1)^{v}
\begin{pmatrix} p \\ e_o \end{pmatrix}(k_{u_1} + k_{u_2} +\cdots + k_{u_q})^{n + e_o}(l_{v_1} + l_{v_2} +\cdots + l_{v_r})^{m + p - e_o}\nonumber\\
 &\times \widetilde\varphi_1(k_{u_1})\widetilde\varphi_1(k_{u_2})... \widetilde\varphi_1(k_{u_q}) \widetilde\varphi_2(l_{v_1})
\widetilde\varphi_2(l_{v_2})... \widetilde\varphi_2(l_{v_r}) \widetilde\varphi_3(f_{w_1}) \widetilde\varphi_3(f_{w_2}) ... \widetilde\varphi_3(f_{w_{s-1}})
\widetilde\varphi_3(f_{w_s})\nonumber\\
& ~~ + {\cal O}(c^2_{nmpqrs})\Big), 
\label{NumErAtorOfExpValuE}
\end{align}
where the variable $v=\frac{1}{2}(n+m+3p)$, the quantity $f_{w_s}$ is defined to be

\begin{align}
f_{w_{s}}=\sum_{i=1}^{q}k_{u_{i}}+\sum_{j=1}^{r}l_{v_{j}}+\sum_{t=1}^{s-1}f_{w_{t}},\label{fwsRelation}\end{align}
  the summation over $\mathcal{S}$ denotes
\begin{align}
{\cal S}&= \left(\left\{{k_{u_i}}\right\},\left\{{l_{v_j}}\right\},\left\{{f_{w_t}}\right\},n,m,p,q,r,s,e_o\right),
\end{align}
and $\mathbb{C}(\overline{\alpha}_{i},a_{i},\mathrm{V},...)$ is an overall factor that gets canceled out with a similar factor from the denominator of the path integral. 
Note that, in (\ref{NumErAtorOfExpValuE}) we have called $$\mathbf{Re}~{\xi}:=\xi~~ \text{for}~~ \xi=(\widetilde{\varphi}_{1},\widetilde{\varphi}_{2},\widetilde{\varphi}_{3},\overline{\alpha}_{i},\overline{\beta}_{i},\overline{\gamma}_{i}).$$ In the following subsections we will analyse in detail the path integral \eqref{NumErAtorOfExpValuE} order by order in the coupling $c_{nmpqrs}$. We follow the functional integral approach for computing such path integrals as elaborated in Appendix \ref{FIntApproachPhi4} for the scalar $\varphi^4$ theory in 3+1 dimensional quantum field theory. 
\subsection{Contribution to the path integral at the tree level}
\label{subsec:TreeLevel}
Taking $c_{nmpqrs}$ to be zero we obtain  the tree level contribution to the path integral (\ref{NumErAtorOfExpValuE}) as
\begin{align}
{\rm Num}\left[\langle  \hat{\varphi}_1 \rangle_{{\sigma}}\right]\bigg|_{\mathrm{Tree~level}}&=  \frac{1}{\mathrm{V}}\sum_{k_i}\int d\widetilde\varphi_1(k_i) 
~{\rm exp}\Big[-a_i\Big(\widetilde\varphi_1(k_i) - \frac{\overline\alpha_i}{ a_i}\Big)^2\Big]
\widetilde\varphi_1(k_i)\psi_{{\bf k}_i}({\rm X})e^{-ik_{0, i}t}\nonumber\\
\prod_{j \ne i} 
\int d\widetilde\varphi_1(k_j) 
~&{\rm exp}\Big[-a_j\Big(\widetilde\varphi_1(k_j) - \frac{\overline\alpha_j}{ a_j}\Big)^2\Big]\times \prod_{u} 
\int d\widetilde\varphi_2(l_u) 
~{\rm exp}\Big[-b_u\Big(\widetilde\varphi_2(l_u) - \frac{\overline\beta_u}{ b_u}\Big)^2\Big] \nonumber\\
&~~\times \prod_{v} 
\int d\widetilde\varphi_3(f_v) 
~{\rm exp}\Big[-c_v\Big(\widetilde\varphi_3(f_v) - \frac{\overline\gamma_v}{ c_v}\Big)^2\Big]  .
\label{TrEEleVeldiAgNum}
\end{align}
Using the standard Gaussian integration results:
\begin{equation}
\begin{aligned}
    \int d\widetilde{\varphi}_{1}(k_{j})\exp\Big[-a_j\Big(\widetilde\varphi_1(k_j) - \frac{\overline\alpha_j}{ a_j}\Big)^2\Big]&= \left(\frac{\pi}{a_{j}}\right)^{1/2},\\
\int d\widetilde\varphi_1(k_i) 
~{\rm exp}\Big[-a_i\Big(\widetilde\varphi_1(k_i) - \frac{\overline\alpha_i}{ a_i}\Big)^2\Big]\widetilde{\varphi}_{1}(k_{i})&=\frac{\overline{\alpha}_{i}}{a_{i}}\left(\frac{\pi}{a_{i}}\right)^{1/2},\label{GInteGrals1}
\end{aligned}
\end{equation}
and taking the continuum limit (i.e. the reverse limit of (\ref{DisCreteLiMiT})) we arrive at 
\begin{align}
{\rm Num}\left[\langle  \hat{\varphi}_1 \rangle_{{\sigma}}\right]\bigg|_{\mathrm{Tree~level}}&=\left(\prod_{j} \Big(\frac{\pi_j}{ a_j}\Big)^{1/2} 
\prod_{u} \Big(\frac{\pi_u}{ b_u}\Big)^{1/2} \prod_{v} \Big(\frac{\pi_v}{ c_v}\Big)^{1/2} \right)\int d^{11}k ~\frac{\overline\alpha(k)}{ a(k)}~\psi_{{\bf k}}({\rm X})e^{-ik_{0}t}.\label{TrEeNumPInt}
\end{align}
where $\pi_{j}=\pi_{u}=\pi_{v}=\pi$. Note that, a similar tree level contribution for the denominator of the path integral (\ref{ScalarPIntAnA}) is given by
\begin{align}
   {\rm Den}\left[\langle  \hat{\varphi}_1 \rangle_{{\sigma}}\right]\bigg|_{\mathrm{Tree~level}}&= \prod_{j} \Big(\frac{\pi_j}{ a_j}\Big)^{1/2} 
\prod_{u} \Big(\frac{\pi_u}{ b_u}\Big)^{1/2} \prod_{v} \Big(\frac{\pi_v}{ c_v}\Big)^{1/2},\label{TrEeDenPathInt}
\end{align}
which implies the overall factor gets cancelled out when we take into account the tree level result of the denominator as well. That is, at the tree level we get:
\begin{align}
\langle  \hat{\varphi}_1 \rangle_{{\sigma}}\bigg\vert_{\rm Tree~level }&=\int d^{11}k ~\frac{\overline\alpha(k)}{ a(k)}~\psi_{{\bf k}}({\rm X})e^{-ik_{0}t}\equiv \mathbb{T}.\label{TrEELeVelAnsWerr}
\end{align}
Next, we evaluate the quantum corrections to this tree level result.
\subsection{First order quantum correction}
We now analyze the contribution to the path integral (\ref{NumErAtorOfExpValuE}) at first order in $c_{nmpqrs}$ due to the three field sectors ${\varphi}_{1},~{\varphi}_{2}$ and $~{\varphi}_{3}$ individually. Let us begin with the $\varphi_{2}$ sector.
\subsection*{First order quantum correction due to \texorpdfstring{$\varphi_{2}$}{varphi2} field sector} The contribution to (\ref{NumErAtorOfExpValuE}) at $\mathcal{O}(c_{nmpqrs})$ due to $\varphi_{2}$ field is given by 
\begin{align}
{\rm Num}\left[\langle  \hat{\varphi}_1 \rangle_{{\sigma}}\right]\bigg|_{\mathcal{O}(c_{nmpqrs}),\varphi_2}&=\int d\widetilde{\varphi}_{2}(l_1){\rm exp}\Big[-b_1\Big(\widetilde\varphi_2(l_1) - \frac{\overline{\beta}_1}{b_1} \Big)^2\Big] \nonumber\\
    \times\int d\widetilde{\varphi}_{2}(l_2){\rm exp}\Big[-b_2&\Big(\widetilde\varphi_2(l_2) - \frac{\overline{\beta}_2}{b_2} \Big)^2\Big]\int d\widetilde{\varphi}_{2}(l_3){\rm exp}\Big[-b_3\Big(\widetilde\varphi_2(l_3) - \frac{\overline{\beta}_3}{b_3} \Big)^2\Big] \cdots \nonumber\\
\times\frac{1}{\mathrm{V}^{r}}\sum_{\left\{{l_{v_i}}\right\}}(l_{v_1}+l_{v_2}+&l_{v_3}+\cdots+l_{v_r})^{m+p-e_o}\widetilde\varphi_2(l_{v_1})
\widetilde\varphi_2(l_{v_2})\widetilde\varphi_2(l_{v_3})\cdots \widetilde\varphi_2(l_{v_r}).\label{O(c)phi2Case1}
\end{align}
To proceed further we consider various cases for the momenta $l_{v_i}$ of the $r$ copies of the field $\widetilde{\varphi}_{2}(l_{v_i})$ appearing in $\mathbf{S}_{\mathrm{int}}$ (\ref{FormOfSint}). 
\vskip.2in
\noindent{\textbf{Case 1:} {$\boldsymbol{l_{v_1} = l_{v_2} = l_{v_3} = \cdots = l_{v_r} \equiv l_i}$}}
\vskip.1in
This is the case where all the $r$ momentum values are considered to be identical. Thus, from Eq. (\ref{O(c)phi2Case1}) we get 
\begin{align}
{\rm Num}\left[\langle  \hat{\varphi}_1 \rangle_{{\sigma}}\right]\bigg|^{\mathrm{Case~1}}_{\mathcal{O}(c_{nmpqrs}),\varphi_2}&=\frac{1}{\mathrm{V}^r}\sum_{l_i}(rl_{i})^{m+p-e_o}\int d\widetilde{\varphi}_{2}(l_{i})\exp\left[-b_i\Big(\widetilde\varphi_2(l_i) - \frac{\overline{\beta}_i}{b_i} \Big)^2\right]\widetilde{\varphi}_{2}^{r}(l_i)\nonumber\\
   & ~~~~\prod_{j\neq i}\int d\widetilde{\varphi}_{2}(l_j){\rm exp}\Big[-b_j\Big(\widetilde\varphi_2(l_j) - \frac{\overline{\beta}_j}{b_j} \Big)^2\Big].
\end{align}
To evaluate the first Gaussian integration we make a change of variable $W=\widetilde\varphi_2(l_i) - \frac{\overline{\beta}_i}{b_i}$ and obtain the following 
\begin{align}
&\int_{-\infty}^{\infty} d\widetilde{\varphi}_{2}(l_{i})\exp\left[-b_i\Big(\widetilde\varphi_2(l_i) - \frac{\overline{\beta}_i}{b_i} \Big)^2\right](rl_{i})^{m+p-e_o}\widetilde{\varphi}_{2}^{r}(l_i)\nonumber\\
    &=\int_{-\infty}^{\infty} d W e^{-b_i W^2} (rl_i)^{m + p - e_o} \; \left(W + \frac{\overline{\beta}_{i}}{b_i} \right)^r \nonumber\\
    &= \int_{-\infty}^{\infty} dW e^{-b_{i}W^2} \; (rl_{i})^{m+p-e_o} \sum_{e_{p}\in 2\mathbb{Z}_{+}}^{r}\begin{pmatrix}
    r\\
    e_{p}\end{pmatrix} \bigg(\frac{\overline{\beta}_{i}}{b_{i}}\bigg)^{r-e_{p}} W^{e_{p}}\nonumber\\
    &= \sqrt{\frac{\pi}{b_{i}}}   \sum_{e_p \in 2\mathbb{Z}_+}^r \left(\begin{matrix} r \\ e_p \\ \end{matrix}\right) 
\left(\frac{\overline\beta_i}{ b_i}\right)^{r - e_p} \frac{(rl_i)^{m+p-e_o}(e_p - 1)!!}{ (2b_i)^{e_p/2}}.\label{GInteGrals2}
\end{align}
Therefore, in the continuum limit this becomes 
\begin{align}
{\rm Num}&\left[\langle  \hat{\varphi}_1 \rangle_{{\sigma}}\right]\bigg|^{\mathrm{Case~1}}_{\mathcal{O}(c_{nmpqrs}),\varphi_2}\nonumber\\
        &=\prod_{j} \Big(\frac{\pi_j}{ b_j}\Big)^{1/2} \frac{1}{\mathrm{V}^{r-1}}\sum_{e_p \in 2\mathbb{Z}_+}^r \left(\begin{matrix} r \\ e_p \\ \end{matrix}\right)
\int d^{11} l ~(rl)^{m+p-e_o}  
\left(\frac{\overline\beta(l)}{ b(l)}\right)^{r - e_p} \frac{(e_p - 1)!!}{ \left(2b(l)\right)^{e_p/2}},
\end{align}
where $\pi_{j}=\pi,\forall ~j$\footnote{This notation should be understood from now on.}. The reason for not taking the continuum limit of the overall factor $ \prod_{j}\left(\frac{\pi_{j}}{b_{j}}\right)^{1/2}$ is that they cancel out when we take into account the denominator of the path integral. Note that, this case has led to a volume suppression of the form $\frac{1}{\mathrm{V}^{r-1}}$. 

Let us now elaborate the computation when one of the $r$ momentum values differ from the rest. 
\vskip.2in
\noindent {\textbf{Case 2}: $\boldsymbol{l_{v_1} = l_{v_2} = l_{v_3} = ..... = l_{v_{r-1}} = l_i, l_{v_r} = l_j,~\text{with}~ l_i \neq l_j }$}
\vskip.1in
For this case (\ref{O(c)phi2Case1}) evaluates to 
\begin{align}
{\rm Num}\left[\langle  \hat{\varphi}_1 \rangle_{{\sigma}}\right]\bigg|^{\mathrm{Case~2}}_{\mathcal{O}(c_{nmpqrs}),\varphi_2}&=\frac{1}{\mathrm{V}^{r}}\sum_{{l_i,l_j}}\int d \widetilde{\varphi}_{2}(l_{i})\exp\bigg[-b_i \left( \widetilde{\varphi}_{2}(l_i)-\frac{\overline{\beta}_{i}}{b_{i}}\right)^2\bigg]\widetilde{\varphi}_{2}^{r-1}(l_{i})\nonumber\\
    \times \int d &\widetilde{\varphi}_{2}(l_{j})\exp\bigg[-b_j \left( \widetilde{\varphi}_{2}(l_j)-\frac{\overline{\beta}_{j}}{b_{j}}\right)^2\bigg]\widetilde{\varphi}_{2}(l_{j})\left((r-1)l_{i}+l_{j} \right)^{m+p-e_o}\nonumber\\
     &~~\times\prod_{k\neq i,j}\int d\widetilde{\varphi}_{2}(l_k){\rm exp}\Big[-b_k\Big(\widetilde\varphi_2(l_k) - \frac{\overline{\beta}_k}{b_k} \Big)^2\Big]\nonumber\\
     =\frac{1}{\mathrm{V}^r}\sum_{e_1=0}^{m + p - e_o} \left(\begin{matrix} m + p - e_o \\ e_1 \\ \end{matrix}\right) & (r - 1)^{e_1} \sum_{l_i}l^{e_1}_i\int d\widetilde{\varphi}_{2}(l_{i}) \exp\bigg[-b_i \left( \widetilde{\varphi}_{2}(l_i)-\frac{\overline{\beta}_{i}}{b_{i}}\right)^2\bigg]\widetilde{\varphi}_{2}^{r-1}(l_{i}) \nonumber\\
     \times\sum_{l_j}l^{m+p-e_o-e_1}_j&\int d\widetilde{\varphi}_{2}(l_{j}) \exp\bigg[-b_j \left( \widetilde{\varphi}_{2}(l_j)-\frac{\overline{\beta}_{j}}{b_{j}}\right)^2\bigg]\widetilde{\varphi}_{2}(l_{j})\nonumber\\
    \times\prod_{k\neq i,j}&\int d\widetilde{\varphi}_{2}(l_k){\rm exp}\Big[-b_k\Big(\widetilde\varphi_2(l_k) - \frac{\overline{\beta}_k}{b_k} \Big)^2\Big]\nonumber\\
    =\frac{1}{\mathrm{V}^r}\sum_{e_1=0}^{m + p - e_o} \left(\begin{matrix} m + p - e_o \\ e_1 \\ \end{matrix}\right)&  (r - 1)^{e_1} \sum_{l_i}l^{e_1}_i\left( \frac{\pi_{i}}{b_{i}}\right)^{1/2}\sum_{e_{p}\in 2\mathbb{Z}_{+}}^{r-1} \left(\begin{matrix} r-1 \\ e_p \\ \end{matrix}\right) 
\left(\frac{\overline\beta_i}{ b_i}\right)^{r-1 - e_p} \frac{(e_p - 1)!!}{ (2b_i)^{e_p/2}} \nonumber\\
     \times\sum_{l_j}l^{m+p-e_o-e_1}_j&\left(\frac{\pi_{j}}{b_{j}}\right)^{1/2}\frac{\overline{\beta}_{j}}{b_{j}}\prod_{k\neq i,j}\left(\frac{\pi_{k}}{b_{k}}\right)^{1/2}.
\end{align}
To arrive at the last step we have used the Gaussian integration results (\ref{GInteGrals1}) and (\ref{GInteGrals2}). Now, taking the continuum limit we get 
\begin{align}
    {\rm Num}&\left[\langle  \hat{\varphi}_1 \rangle_{{\sigma}}\right]\bigg|^{\mathrm{Case~2}}_{\mathcal{O}(c_{nmpqrs}),\varphi_2}\nonumber\\
    &=\prod_{j}\left( \frac{\pi_{j}}{b_{j}}\right)^{1/2}\frac{1}{\mathrm{V}^{r-2}}\sum_{e_1=0}^{m + p - e_o} \left(\begin{matrix} m + p - e_o \\ e_1 \\ \end{matrix}\right)  (r - 1)^{e_1}\sum_{e_{p}\in 2\mathbb{Z}_{+}}^{r-1} \left(\begin{matrix} r-1 \\ e_p \\ \end{matrix}\right)\frac{(e_{p}-1)!!}{2^{e_{p}/2}}\nonumber\\
&~~~~\times \int d^{11} l \frac{l^{e_1}}{b^{e_p/2}(l)} \left( \frac{\overline{\beta}(l)}{b(l)} \right)^{r-1-e_{p}} \; \int d^{11}  l^{\prime} \;{l^{\prime }}^{m+p-e_o-e_1}   \frac{\overline{\beta}(l^{\prime} )}{b(l^{\prime} )} 
\end{align}
This result is again volume suppressed due to the presence of the factor $\frac{1}{\mathrm{V}^{r-2}}$. In fact, such a volume suppression will always be there unless all the $r$ momentum values $l_{v_{i}}$ differ from each other. Let us now analyse such a case.
\vskip.2in
\noindent {\textbf{Case 3}: $\boldsymbol{l_{v_1} \neq l_{v_2} \neq l_{v_3} \neq ..... \neq l_{v_{r-1}}\neq l_{v_r} }$}
\vskip.1in
When all the $r$ momentum values of the fields $\widetilde{\varphi}_2(l_{v_i})$ differ from each other, Eq. (\ref{O(c)phi2Case1}) evaluates to
\begin{align}
    & {\rm Num}\left[\langle  \hat{\varphi}_1 \rangle_{{\sigma}}\right]\bigg|^{\mathrm{Case~3}}_{\mathcal{O}(c_{nmpqrs}),\varphi_2}\nonumber\\
    &=\frac{1}{\mathrm{V}^r} \sum_{e_1,...,e_{r-1}}\begin{pmatrix}
    m+p-e_o\\
    e_{1}\end{pmatrix}\begin{pmatrix}
    m+p-e_o-e_{1}\\
e_{2}\end{pmatrix}\cdots\begin{pmatrix}
    m+p-e_o-\sum_{i=1}^{r-2}e_{i}\\
    e_{r-1}\end{pmatrix} \nonumber\\
    &\times\sum_{l_1,...,l_r}l_{1}^{e_{1}}l_{2}^{e_2}\cdots l_{r}^{e_r}\int d\widetilde{\varphi}_{2}(l_{1})\exp\bigg[-b_{1}\left( \widetilde{\varphi}_{2}(l_1)-\frac{\overline{\beta}_{1}}{b_{1}}\right)^2\bigg]\widetilde{\varphi}_{2}(l_1)\nonumber\\
    &\times\int d\widetilde{\varphi}_{2}(l_{2})\exp\bigg[-b_{2}\left( \widetilde{\varphi}_{2}(l_2)-\frac{\overline{\beta}_{2}}{b_{2}}\right)^2\bigg]\widetilde{\varphi}_{2}(l_2)\cdots \nonumber\\
&\times\int d\widetilde{\varphi}_{2}(l_{r})\exp\bigg[-b_{r}\left( \widetilde{\varphi}_{2}(l_r)-\frac{\overline{\beta}_{r}}{b_{r}}\right)^2\bigg]\widetilde{\varphi}_{2}(l_r)\nonumber\\
&\times\prod_{j\neq 1,...,r}\int d\widetilde{\varphi}_{2}(l_{j})\exp\bigg[-b_{j}\left( \widetilde{\varphi}_{2}(l_j)-\frac{\overline{\beta}_{j}}{b_{j}}\right)^2\bigg]\nonumber\\
    &=\frac{1}{\mathrm{V}^r} \sum_{e_1,...,e_{r-1}}\begin{pmatrix}
    m+p-e_o\\
    e_{1}\end{pmatrix}\begin{pmatrix}
    m+p-e_o-e_{1}\\
e_{2}\end{pmatrix}\cdots\begin{pmatrix}
    m+p-e_o-\sum_{i=1}^{r-2}e_{i}\\
    e_{r-1}\end{pmatrix} \nonumber\\
    &\times\sum_{l_{1},...,l_r}l_{1}^{e_{1}}l_{2}^{e_2}\cdots l_{r}^{e_r}\frac{\overline{\beta}_{1}}{b_{1}}\left( \frac{\pi}{b_1}\right)^{1/2}\frac{\overline{\beta}_{2}}{b_{2}}\left( \frac{\pi}{b_2}\right)^{1/2}\cdots \frac{\overline{\beta}_{r}}{b_{r}} \left( \frac{\pi}{b_r}\right)^{1/2}\prod_{j\neq 1,...,r}\left(\frac{\pi_{j}}{b_{j}} \right)^{1/2},
\end{align}
where $e_{r}=m+p-e_o-\sum_{i=1}^{r-1}e_{i}$. Therefore, if we take the continuum limit we arrive at the following result
\begin{align}
     & {\rm Num}\left[\langle  \hat{\varphi}_1 \rangle_{{\sigma}}\right]\bigg|^{\mathrm{Case~3}}_{\mathcal{O}(c_{nmpqrs}),\varphi_2}\nonumber\\
     &=\prod_{i}\left( \frac{\pi_{i}}{b_{i}}\right)^{1/2} \sum_{e_1,...,e_{r-1}}\begin{pmatrix}
    m+p-e_o\\
    e_{1}\end{pmatrix}\begin{pmatrix}
    m+p-e_o-e_{1}\\
e_{2}\end{pmatrix}\cdots\begin{pmatrix}
    m+p-e_o-\sum_{i=1}^{r-2}e_{i}\\
    e_{r-1}\end{pmatrix} \nonumber\\
    &~~~~~~~~~~~~~~~~~~~~\times\int d^{11}l_1 \frac{\overline{\beta}_{1}}{b_{1}}l_{1}^{e_{1}} \int d^{11}l_2 \frac{\overline{\beta}_{2}}{b_{2}}l_{2}^{e_{2}}\cdots\int d^{11}l_r \frac{\overline{\beta}_{r}}{b_{r}}l_{r}^{e_{r}} . \label{DomVarphi2}
\end{align}
So this does not have any volume suppression as per our expectation. Only such dominant contributions are taken into account in order to evaluate the expectation value of the metric operator with respect to the GS state\cite{Brahma:2022wdl,Archana_CC2026}. In Table \ref{tab:ThreeCasesPhi2}, we summarise the above studied three cases:
\begin{table}[htbp]\centering\begin{adjustbox}{width=\textwidth}\begin{tabular}{||>{\centering\arraybackslash}m{0.16\textwidth}|>{\centering\arraybackslash}m{0.22\textwidth}|>{\centering\arraybackslash}m{0.26\textwidth}|>{\centering\arraybackslash}m{0.30\textwidth}||}\hline
 &\textbf{Case 1}& \textbf{Case 2}& \textbf{Case 3}
 \\\hline\hline
Momentum values of the $r\geq 4$ fields $\{\widetilde{\varphi}_{2}(l_{v_j})\}$ & $l_{v_1}=l_{v_2}=\cdots=l_{v_r}$& $l_{v_1}=l_{v_2}=\cdots=l_{v_{r-1} }=l_{i},~l_{v_r}=l_j$ with $l_{i}\neq l_j$& $l_{v_1}\neq l_{v_2}\neq \cdots \neq l_{v_r}$\\
\hline
Volume suppression&   $\frac{1}{\mathrm{V}^{r-1}}$  & $\frac{1}{\mathrm{V}^{r-2}}$ & $1$
\\
\hline
{Relevance to the path integral}& Will be ignored & Will be ignored & Dominant contribution
\\
\hline
\end{tabular}
\end{adjustbox}
\caption{Numerator of the path integral (\ref{NumErAtorOfExpValuE}) at $\mathcal{O}(c_{nmpqrs})$ due to the $\varphi_2$ field sector}
\label{tab:ThreeCasesPhi2}
\end{table}

\noindent
Let us now analyse $\varphi_{3}$ sector before getting onto the source sector $\varphi_{1}$.
\subsection*{First order quantum correction due to \texorpdfstring{$\varphi_{3}$}{varphi3} field sector}
The contribution to the path integral (\ref{NumErAtorOfExpValuE}) at $\mathcal{O}(c_{nmpqrs})$ due to $\varphi_{3}$ field sector is given by 
\begin{align}
       {\rm Num}\left[\langle  \hat{\varphi}_1 \rangle_{{\sigma}}\right]\bigg|_{\mathcal{O}(c_{nmpqrs}),\varphi_3}&=\int d\widetilde{\varphi}_{3}(f_1){\rm exp}\Big[-c_1\Big(\widetilde\varphi_3(f_1) - \frac{\overline{\gamma}_1}{c_1} \Big)^2\Big] \nonumber\\
    \times\int d\widetilde{\varphi}_{3}(f_2){\rm exp}\Big[-c_2&\Big(\widetilde\varphi_3(f_2) - \frac{\overline{\gamma}_2}{c_2} \Big)^2\Big]\int d\widetilde{\varphi}_{3}(f_3){\rm exp}\Big[-c_3\Big(\widetilde\varphi_2(f_3) - \frac{\overline{\gamma}_3}{c_3} \Big)^2\Big] \cdots \nonumber\\
\times\frac{1}{\mathrm{V}^{s-1}}\sum_{\left\{{f_{w_t}}\right\}}&\widetilde\varphi_3(f_{w_1})
\widetilde\varphi_3(f_{w_2})\widetilde\varphi_3(f_{w_3})\cdots  \widetilde\varphi_3(f_{w_{s-1}}) \widetilde\varphi_3(f_{w_s}).\label{O(c)phi3Case1}
\end{align}
Recall that, the momentum $f_{w_s}$ depends on the momenta values of the other two field sectors via the relation (\ref{fwsRelation}). This will make the following analysis a bit different from that of the $\varphi_{2}$ sector analysed previously. For the purpose of illustration, we first study the case in which all the $s-1$ momentum values $\{f_{w_t};~t=1,...,s-1 \}$ of the Fourier modes $\widetilde{\varphi}_3(f_{w_{t}})$ are identical. 
\vskip.2in
\noindent{\textbf{Case 1:} {$\boldsymbol{f_{w_1} = f_{w_2} = f_{w_3} = \cdots = f_{w_{s-1}} \equiv f_k}$}}
\vskip.1in
This case implies:
\begin{align}   f_{w_s}&=\sum_{i=1}^{q}k_{u_i}+\sum_{j=1}^{r}l_{v_j}+(s-1)f_{k}.
\end{align}
As an example we consider $k_{u_1}=k_{u_{2}}=\cdots=k_{u_{q}}= k_{i}$ and $l_{v_1}=l_{v_2}=\cdots=l_{v_{r}}= l_{j}$ which leads to $f_{w_s}=qk_{i}+rl_{j}+(s-1)f_{k}$. Thus from (\ref{O(c)phi3Case1}) we obtain 
\begin{align}
 {\rm Num}&\left[\langle  \hat{\varphi}_1 \rangle_{{\sigma}}\right]\bigg|^{\mathrm{Case~1}}_{\mathcal{O}(c_{nmpqrs}),\varphi_3}=\frac{1}{\mathrm{V}^{s-1}}\sum_{f_k}\int d\widetilde{\varphi}_{3}(f_{k})\exp\bigg[-c_k\left( \widetilde{\varphi}_{3}(f_k)-\frac{\overline{\gamma}_{k}}{c_{k}}\right)^2\bigg]\widetilde{\varphi}^{s-1}_{3}(f_{k})\nonumber\\
    &\times\int d\widetilde{\varphi}_{3}(f_{w_s})\exp\bigg[-c_{w_s}\left( \widetilde{\varphi}_{3}(f_{w_s})-\frac{\overline{\gamma}_{w_s}}{c_{w_s}}\right)^2\bigg]\widetilde{\varphi}_{3}(f_{w_s})\nonumber\\
    &~~~~\times\prod_{t\neq k,w_{s}}\int d\widetilde{\varphi}_{3}(f_{t})\exp\bigg[-c_t\left( \widetilde{\varphi}_{3}(f_t)-\frac{\overline{\gamma}_{t}}{c_{t}}\right)^2\bigg]\nonumber\\
    &=\frac{1}{\mathrm{V}^{s-1}}\sum_{f_k}\left( \frac{\pi}{c_{k}}\right)^{1/2}\sum_{e_{p}\in 2\mathbb{Z}_{+}}\begin{pmatrix}
        s-1\\
        e_{p}
    \end{pmatrix} \left(\frac{\overline{\gamma}_{k}}{c_{k}} \right)^{s-1-e_{p}}\frac{(e_{p}-1)!!}{(2c_{k})^{e_p/2}}\frac{\overline{\gamma}_{w_{s}}}{c_{w_s}} \left( \frac{\pi}{c_{w_s}}\right)^{1/2}\prod_{t\neq k,w_s}\left( \frac{\pi_{t}}{c_{{t}}}\right)^{1/2}\label{Case1phi3Sec}
\end{align}
The factor $\frac{\overline{\gamma}_{w_s}}{c_{w_{s}}}$ appearing in the above expression can be written more explicitly by using the condition (\ref{fwsRelation}). From (\ref{abcDef}) we know that $c_{w_{s}}=\frac{1}{\rm V} f_{w_s}^2$. Thus we get
\begin{align}
    \frac{1}{c_{w_s}}= \frac{\rm V}{f_{w_s}^2}&=\frac{\rm V}{\left(\sum_{i=1}^{q}k_{u_i}+\sum_{j=1}^r l_{v_j}+\sum_{t=1}^{s-1}f_{w_t} \right)^2}\nonumber\\
    &=\frac{\rm V}{\left(\sum_{t=1}^{s-1}f_{w_t}\right)^2} \left(1+\frac{\sum_{i=1}^{q}k_{u_i}+\sum_{j=1}^{r}l_{v_{j}}}{\sum_{t=1}^{s-1}f_{w_t}}\right)^{-2}.\label{Inverseofcws}
\end{align}
If we can impose the following condition for the values of the momentum 
\begin{align}
    \zeta\equiv \sum_{t=1}^{s-1}f_{w_{t}}>> \sum_{i=1}^{q}k_{u_{i}}+\sum_{j=1}^{r}l_{v_j},\label{MomentumConst}
\end{align}
then (\ref{Inverseofcws}) can be simplified further as
\begin{align}
    \frac{1}{c_{w_s}}&\sim\frac{\rm V}{\left(\sum_{t=1}^{s-1}f_{w_t}\right)^2} \left(1-2\frac{\sum_{i=1}^{q}k_{u_i}+\sum_{j=1}^{r}l_{v_{j}}}{\sum_{t=1}^{s-1}f_{w_t}}\right)\nonumber\\
    &=\frac{\rm V}{\left(\sum_{t=1}^{s-1}f_{w_t}\right)^2} \sum_{n_1=0}^{1}\begin{pmatrix}
        2\\
        n_1
    \end{pmatrix}(-1)^{n_1}\left(\frac{\sum_{i=1}^{q}k_{u_i}+\sum_{j=1}^{r}l_{v_{j}}}{\sum_{t=1}^{s-1}f_{w_t}}\right)^{n_1}\nonumber\\
    &={\rm V}\sum_{n_1=0}^{1}\begin{pmatrix}
        2\\
        n_1
    \end{pmatrix}(-1)^{n_1}\sum_{e_{n_1}=0}^{n_1}\begin{pmatrix}
        n_1\\
        e_{n_1}
    \end{pmatrix}\left( \sum_{i=1}^{q}k_{u_i}\right)^{e_{n_1}}\left(\sum_{j=1}^{r}l_{v_j} \right)^{n_{1}-e_{n_1}}\frac{1}{ \zeta^{2+n_1}}.\label{1/cws}
\end{align}
 Taylor expansion of $\overline{\gamma}(f_{w_s})$ around $\zeta$ gives
\begin{align}
 &\overline{\gamma} ( f_{\omega_s}) =\overline{\gamma}(\zeta)+(f_{w_s}-\zeta)\; \overline{\gamma}^{'}(\zeta)+\frac{\left(f_{w_{s}}-\zeta\right)^2}{2!}\overline{\gamma}^{''}(\zeta)+\cdots\nonumber\\
&=\overline{\gamma}(\zeta) + \left(\sum_{i = 1}^q k_{u_i} + \sum_{j = 1}^r l_{v_j} \right) \frac{\partial \overline{\gamma}}{\partial \zeta} (\zeta) + \frac{1}{2!} \left(\sum_{i = 1}^q k_{u_i} + \sum_{j = 1}^r l_{v_j} \right)^2 \frac{\partial^2 \overline{\gamma}}{\partial \zeta^2} ( \zeta) + \cdots\nonumber
\\
& = \sum_{n_2=0}^{\infty} \frac{1}{n_2 !} \left(\sum_{i = 1}^q k_{u_i} + \sum_{j = 1}^r l_{v_j} \right)^{n_2} \frac{\partial^{n_2} \overline{\gamma}}{\partial \zeta^{n_2}} (\zeta) \nonumber
\\ 
& = \sum_{n_2=0}^{\infty} \frac{1}{n_2 !} \sum_{e_{n_2}=0}^{n_2}  \left(\begin{matrix} n_2 \\ e_{n_2} \\ \end{matrix}\right) \left(\sum_{i = 1}^q k_{u_i} \right)^{e_{n_2}} \left ( \sum_{j = 1}^r l_{v_j} \right)^{n_2 - e_{n_2}} \frac{\partial^{n_2} \overline{\gamma}}{\partial \zeta^{n_2}} (\zeta).\label{gammafws}
\end{align}
From (\ref{1/cws}) and (\ref{gammafws}) we note that $\frac{\overline{\gamma}_{w_s}}{c_{w_s}}$ involves factors of the form 
\begin{align}
\sum_{n_1=0}^{1}\sum_{n_2=0}^{\infty} \sum_{e_{n_1}=0}^{n_1}\sum_{e_{n_2}=0}^{n_2} \bigg(\cdots\bigg) \left(\sum_{i = 1}^q k_{u_i} \right)^{e_{n_1}+e_{n_2}}
 \left ( \sum_{j = 1}^r l_{v_j} \right)^{n_1+n_2  -e_{n_1}- e_{n_2}} \frac{\rm V}{\left(\zeta\right)^{2+n_1}} \frac{\partial^{n_2} \overline{\gamma}}{\partial \zeta^{n_2}} (\zeta).
 \end{align}
This indicates that if the assumption (\ref{MomentumConst}) holds, then the factors $(\sum_{i=1}^{q}k_{u_{i}})^{n+e_{o}}$ and 

\noindent
$(\sum_{j=1}^{r}l_{v_j})^{m+p-e_o}$ in Eq. (\ref{NumErAtorOfExpValuE}), associated to $\varphi_1$ and $\varphi_2$ sectors respectively  get modified as:
\begin{align}
\left(\sum_{i=1}^{q}k_{u_{i}}\right)^{n+e_{o}}&\longrightarrow \left(\sum_{i=1}^{q}k_{u_{i}}\right)^{n+e_{o}+e_{n_1}+e_{n_2}}=\left(\sum_{i=1}^{q}k_{u_{i}}\right)^{\widetilde n}\label{widEtilDen}\\
\text{and},~~~ \left(\sum_{j=1}^{r}l_{v_j}\right)^{m+p-e_o}&\longrightarrow \left(\sum_{j=1}^{r}l_{v_j}\right)^{m+p+n_1+n_2-e_o-e_{n_1}-e_{n_2}}=\left(\sum_{j=1}^{r}l_{v_j}\right)^{\widetilde m}.\label{WidEtiLDem}
\end{align}
  After incorporating (\ref{1/cws}) and (\ref{gammafws}) we write down (\ref{Case1phi3Sec}) in the following way
\begin{align}
   & {\rm Num}\left[\langle  \hat{\varphi}_1 \rangle_{{\sigma}}\right]\bigg|^{\mathrm{Case~1}}_{\mathcal{O}(c_{nmpqrs}),\varphi_3}=\prod_{t}\left( \frac{\pi_{t}}{c_{t}}\right)^{1/2}\frac{1}{\mathrm{V}^{s-1}}\sum_{e_p\in 2\mathbb{Z}_{+}}\begin{pmatrix}
        s-1\\
        e_p
\end{pmatrix}\frac{(e_p-1)!!}{2^{e_p/2}}\sum_{f_{k}}\frac{1}{c_k^{e_p/2}}\left(\frac{\overline{\gamma}_{k}}{c_{k}}\right)^{s-1-e_{p}}\nonumber\\
&~~~~~~~~~~~~~~~~~~~~~~~~~~~~~~~~~~~~\times\sum_{n_1=0}^{1}(-1)^{n_1}\begin{pmatrix}
    2\\
    n_1
\end{pmatrix}\sum_{n_2=0}^{\infty}\frac{1}{n_2!}\frac{\rm V}{\left((s-1)f_{k} \right)^{2+n_1}}\frac{\partial^{n_2}\overline{\gamma}((s-1)f_{k})}{\partial \zeta^{n_2}},\nonumber
\end{align}
where we have ignored the combinatorial factors involving $e_{n_1},~\text{and}~e_{n_2}$ as appearing in (\ref{1/cws}), and (\ref{gammafws}). Now,
substituting $\zeta=(s-1)f_{k}$ and then taking the continuum limit  we arrive at 
\begin{align}
    {\rm Num}\left[\langle  \hat{\varphi}_1 \rangle_{{\sigma}}\right]\bigg|^{\mathrm{Case~1}}_{\mathcal{O}(c_{nmpqrs}),\varphi_3} =&\prod_{t}\left( \frac{\pi_{t}}{c_{t}}\right)^{1/2}\frac{1}{\mathrm{V}^{s-2}}\sum_{e_p\in 2\mathbb{Z}_{+}}\begin{pmatrix}
        s-1\\
        e_p   \end{pmatrix}\frac{(e_p-1)!!}{2^{e_p/2}(s-1)^{2+n_1+n_2}}\nonumber\\
        \times\sum_{n_1=0}^{1}(-1)^{n_1}\begin{pmatrix}
            2\\
           n_{1}
           \end{pmatrix}&\sum_{n_2=0}^{\infty}\frac{1}{n_2!}\int d^{11}f\frac{1}{(c(f))^{e_p/2}}\left(\frac{\overline{\gamma}(f)}{c(f)}\right)^{s-1-e_{p}}\frac{\rm V}{f^{2+n_1}}\frac{\partial^{n_2}\overline{\gamma}((s-1)f)}{\partial f^{n_2}}.
\end{align}
As one could infer, the above result is volume suppressed due to the presence of the factor $\frac{1}{\mathrm{V}^{s-2}}$. Next one may analyse ``\textbf{Case 2}'', in which any one of the $s-1$ momentum values $\{f_{w_t}\}$ differs from the rest. However, that would again lead to volume suppression. We therefore proceed directly to ``\textbf{Case 3}'', where all the $s-1$ momentum values are distinct. 
\vskip.2in
\noindent{\textbf{Case 3:} {$\boldsymbol{f_{w_1} \neq f_{w_2} \neq f_{w_3} \neq \cdots \neq f_{w_{s-1}} }$}}
\vskip.1in
For this case (\ref{O(c)phi3Case1}) evaluates to
\begin{align}
& {\rm Num}\left[\langle  \hat{\varphi}_1 \rangle_{{\sigma}}\right]\bigg|^{\mathrm{Case~3}}_{\mathcal{O}(c_{nmpqrs}),\varphi_3}\nonumber\\
&=\frac{1}{\mathrm{V}^{s-1}}\sum_{f_{1},...,f_{s-1}}\int d\widetilde{\varphi}_{3}(f_{1})\exp\bigg[-c_{1}\left( \widetilde{\varphi}_{3}(f_1)-\frac{\overline{\gamma}_{1}}{c_{1}}\right)^2\bigg]\widetilde{\varphi}_{3}(f_{1})\nonumber\\
&\times\int d\widetilde{\varphi}_{3}(f_{2})\exp\bigg[-c_{2}\left( \widetilde{\varphi}_{3}(f_2)-\frac{\overline{\gamma}_{2}}{c_{2}}\right)^2\bigg]\widetilde{\varphi}_{3}(f_{2})\cdots \nonumber\\
&\times\int d\widetilde{\varphi}_{3}(f_{w_s})\exp\bigg[-c_{w_s}\left( \widetilde{\varphi}_{3}(f_{w_s})-\frac{\overline{\gamma}_{w_s}}{c_{w_s}}\right)^2\bigg]\nonumber\\
&\times\prod_{t\neq {1},...,{s-1}}\int d\widetilde{\varphi}_{3}(f_{t})\exp\bigg[-c_{t}\left( \widetilde{\varphi}_{3}(f_{t})-\frac{\overline{\gamma}_{t}}{c_{t}}\right)^2\bigg]\widetilde{\varphi}_{3}(f_{t})\nonumber\\
&=\prod_{i}\left( \frac{\pi_{i}}{c_{i}}\right)^{1/2}\frac{1}{\mathrm{V}^{s-1}}\sum_{f_1,...,f_{s-1}}\frac{\overline{\gamma}_{1}}{c_1}\cdots \frac{\overline{\gamma}_{s-1}}{c_{s-1}}\frac{\overline{\gamma}_{w_{s}}}{c_{w_s}}.
\end{align}
Incorporating the relevant factors from (\ref{1/cws}) and (\ref{gammafws}) we can express the above expression as
\begin{align}
   {\rm Num}&\left[\langle  \hat{\varphi}_1 \rangle_{{\sigma}}\right]\bigg|^{\mathrm{Case~3}}_{\mathcal{O}(c_{nmpqrs}),\varphi_3}  =\prod_{i}\left( \frac{\pi_{i}}{c_{i}}\right)^{1/2}\frac{1}{\mathrm{V}^{s-1}}\sum_{f_1,...,f_{s-1}}\frac{\overline{\gamma}_{1}}{c_1}\cdots \frac{\overline{\gamma}_{s-1}}{c_{s-1}}\sum_{n_1=0}^{1}(-1)^{n_1}\begin{pmatrix}
       2\\
       n_{1}   \end{pmatrix}\nonumber\\
       &~~~~~~~~~~~~~~~~~~~~~~~~~~~~~~~\times\sum_{n_2=0}^{\infty}\frac{1}{n_2!}\frac{\rm V}{\zeta^{2+n_1}}\frac{\partial^{n_{2}}\overline{\gamma}}{\partial \zeta^{n_2}}(\zeta)\nonumber\\
  &=\prod_{i}\left( \frac{\pi_{i}}{c_{i}}\right)^{1/2} \frac{1}{\mathrm{V}^{s-1}}\sum_{f_{1},...,f_{s-1}}\frac{\overline{\gamma}_{1}}{c_1}\cdots \frac{\overline{\gamma}_{s-1}}{c_{s-1}}\sum_{n_1=0}^{1}(-1)^{n_1}\begin{pmatrix}
      2\\
      n_1
  \end{pmatrix}\frac{1}{(f_{1}+\cdots+f_{s-1})^{2+n_1}}\nonumber\\
&~~~~~~~~~~~~~~~~~\times\sum_{n_2=0}^{\infty}\frac{1}{n_2!}\left(\frac{\partial}{\partial f_{1}}+\cdots+\frac{\partial}{\partial f_{s-1}}\right)^{n_{2}}\mathrm{V}\overline{\gamma}(f_{1}+\cdots+f_{s-1})\nonumber\\
&=\prod_{i}\left( \frac{\pi_{i}}{c_{i}}\right)^{1/2} \int d^{11}f_1\frac{\overline{\gamma}_{1}}{c_1}\cdots\int d^{11}f_{s-1} \frac{\overline{\gamma}_{s-1}}{c_{s-1}}\sum_{n_1=0}^{1}(-1)^{n_1}\begin{pmatrix}
    2\\
    n_1
\end{pmatrix}\frac{1}{(f_{1}+\cdots+f_{s-1})^{2+n_1}}\nonumber\\
&~~~~~~~~~~~~~~~~~\times\sum_{n_2=0}^{\infty}\frac{1}{n_2!}\left(\frac{\partial}{\partial f_{1}}+\cdots+\frac{\partial}{\partial f_{s-1}}\right)^{n_{2}}{\rm V}\overline{\gamma}(f_{1}+\cdots+f_{s-1}),\label{CAsE3VarPhi3ord1}
\end{align}
where to arrive at the last line we have taken the continuum limit. Note that, the denominator of the path integral will give identical result  for the $\widetilde{\varphi}_2$ and $\widetilde{\varphi}_3$ sectors.

\vskip0.1in
\noindent
Now we turn to the analysis of the $\varphi_1$ sector.
\subsection*{First order quantum correction due to \texorpdfstring{$\varphi_{1}$}{varphi1} field sector}
In the path integral (\ref{NumErAtorOfExpValuE}), the contribution due to $\varphi_{1}$ sector  at $\mathcal{O}(c_{nmpqrs})$ is as follows
\begin{align}
&     {\rm Num}\left[\langle  \hat{\varphi}_1 \rangle_{{\sigma}}\right]\bigg|_{\mathcal{O}(c_{nmpqrs}),\varphi_1}=\int d\widetilde{\varphi}_{1}(k_1){\rm exp}\Big[-a_1\Big(\widetilde\varphi_1(k_1) - \frac{\overline{\alpha}_1}{a_1} \Big)^2\Big] \nonumber\\
   & \times\int d\widetilde{\varphi}_{1}(k_2){\rm exp}\Big[-a_2\Big(\widetilde\varphi_1(k_2) - \frac{\overline{\alpha}_2}{a_2} \Big)^2\Big]\int d\widetilde{\varphi}_{1}(k_3){\rm exp}\Big[-a_3\Big(\widetilde\varphi_1(k_3) - \frac{\overline{\alpha}_3}{a_3} \Big)^2\Big] \cdots \nonumber\\
    &\times   \frac{1}{ {\rm V}}\Big(\widetilde\varphi_1(k_1)\psi_{{\bf k}_1}({\rm X})e^{-ik_{0, 1}t} + 
\widetilde\varphi_1(k_2)\psi_{{\bf k}_2}({\rm X})e^{-ik_{0, 2}t} + 
\widetilde\varphi_1(k_3)\psi_{{\bf k}_3}({\rm X})e^{-ik_{0, 3}t} +\cdots \Big)\nonumber\\
&
\times\frac{1}{\mathrm{V}^{q}}\sum_{\left\{{k_{u_i}}\right\}}(k_{u_1}+k_{u_2}+k_{u_3}+\cdots+k_{u_q})^{n+e_o}\widetilde\varphi_1(k_{u_1})
\widetilde\varphi_1(k_{u_2})\widetilde\varphi_1(k_{u_3})\cdots \widetilde\varphi_1(k_{u_q}).\label{O(c)phi1Case1}
\end{align}
We have seen for the $\varphi_2$ and $\varphi_3$ analysis, the dominant contribution arises only when all the momentum values of the internal fields differ from each other. However, for the present case we also have an external field insertion. Let us first consider the case where the momentum of the insertion field also differs from all the distinct $q$ momentum values $\{k_{u_i}\}\equiv (k_{1},k_2,...,k_q)$ of the internal fields $\{\widetilde{\varphi}_1(k_{u_i})\}$.

\vskip.2in
\noindent{\textbf{Case 3:} {$\boldsymbol{k_{u_1} \neq k_{u_2} \neq k_{u_3} \neq \cdots \neq k_{u_{q-1}}\neq k_{u_q}\neq k_i~}$} where $k_i$ denotes the momentum of the external field insertion.}

For this case (\ref{O(c)phi1Case1}) evaluates to
\begin{align}
     {\rm Num}&\left[\langle  \hat{\varphi}_1 \rangle_{{\sigma}}\right]\bigg|^{\mathrm{Case}~3}_{\mathcal{O}(c_{nmpqrs}),\varphi_1}\nonumber\\
    &=\frac{1}{\mathrm{V}^q}\sum_{e_{1},...,e_{q-1}}\begin{pmatrix}
        n+e_o\\
        e_1
    \end{pmatrix}\begin{pmatrix}
        n+e_o-e_1\\
        e_2
    \end{pmatrix}\cdots\begin{pmatrix}
        n+e_o-\sum_{i=1}^{q-1}e_i\\
        e_{q-1}
    \end{pmatrix}\nonumber\\
&~~~~~~\times\sum_{k_1,...,k_q}k_{1}^{e_1}k_{2}^{e_2}\cdots k_{q}^{e_q}\int d\widetilde{\varphi}_{1}(k_1){\rm exp}\Big[-a_1\Big(\widetilde\varphi_1(k_1) - \frac{\overline{\alpha}_1}{a_1} \Big)^2\Big] \widetilde{\varphi}_{1}(k_1)\nonumber\\
    \times\int &d\widetilde{\varphi}_{1}(k_2){\rm exp}\Big[-a_2\Big(\widetilde\varphi_1(k_2) - \frac{\overline{\alpha}_2}{a_2} \Big)^2\Big]\widetilde{\varphi}_{1}(k_2)\cdots\int d\widetilde{\varphi}_{1}(k_q){\rm exp}\Big[-a_q\Big(\widetilde\varphi_1(k_q) - \frac{\overline{\alpha}_q}{a_q} \Big)^2\Big] \widetilde{\varphi}_{1}(k_{q}) \nonumber\\
    &\times   \frac{1}{ {\rm V}}\sum_{k_i\neq k_1,...,k_q}\int d\widetilde{\varphi}_{1}(k_{i})\exp\bigg[-a_{i}\left( \widetilde{\varphi}_{1}(k_{i})-\frac{\overline{\alpha}_{i}}{a_{i}}\right)^2\bigg]\Big(\widetilde\varphi_1(k_i)\psi_{{\bf k}_i}({\bf x}, y, z)e^{-ik_{0, i}t}\Big)\nonumber\\
    &~~~~~~~~~~\times\prod_{j\neq i,1,...,q}\int d\widetilde{\varphi}_{1}(k_{j})\exp\bigg[-a_{j}\left( \widetilde{\varphi}_{1}(k_{j})-\frac{\overline{\alpha}_{j}}{a_{j}}\right)^2\bigg]\nonumber\\
    &=\frac{1}{\mathrm{V}^{q}}\sum_{e_{1},...,e_{q-1}}\begin{pmatrix}
        n+e_o\\
        e_1
    \end{pmatrix}\begin{pmatrix}
        n+e_o-e_1\\
        e_2
    \end{pmatrix}\cdots\begin{pmatrix}
        n+e_o-\sum_{i=1}^{q-1}e_i\\
        e_{q-1}
    \end{pmatrix}\nonumber\\
&~~~~~~\times\sum_{k_1,...,k_q}k_{1}^{e_1}k_{2}^{e_2}\cdots k_{q}^{e_q} \frac{\overline{\alpha}_{1}}{a_{1}}\left(\frac{\pi}{a_{1}} \right)^{1/2}\frac{\overline{\alpha}_{2}}{a_{2}}\left(\frac{\pi}{a_{2}} \right)^{1/2}\cdots \frac{\overline{\alpha}_{q}}{a_{q}}\left(\frac{\pi}{a_{q}} \right)^{1/2}\nonumber\\
    &~~~~~~~~~\times   \frac{1}{ {\rm V}}\sum_{k_i\neq k_1,...,k_q}\frac{\overline{\alpha}_{i}}{a_{i}}\left( \frac{\pi}{a_{i}}\right)^{1/2}\psi_{{\bf k}_i}({\bf x}, y, z)e^{-ik_{0, i}t}~~\times\prod_{j\neq i,1,...,q}\left( \frac{\pi_{j}}{a_{j}}\right)^{1/2},
\end{align}
where $e_{q}=n+e_{o}-\sum_{i=1}^{q-1}e_{i}$. Taking the continuum limit we get 
\begin{align}
     {\rm Num}&\left[\langle  \hat{\varphi}_1 \rangle_{{\sigma}}\right]\bigg|^{\mathrm{Case}~3}_{\mathcal{O}(c_{nmpqrs}),\varphi_1}\nonumber\\
    &=\prod_{i}\left(\frac{\pi_{i}}{a_{i}} \right)^{1/2}\sum_{e_{1},...,e_{q-1}}\begin{pmatrix}
        n+e_o\\
        e_1
    \end{pmatrix}\begin{pmatrix}
        n+e_o-e_1\\
        e_2
    \end{pmatrix}\cdots\begin{pmatrix}
        n+e_o-\sum_{i=1}^{q-1}e_i\\
        e_{q-1}
\end{pmatrix}\nonumber\\
&~~\times\int d^{11}k_{1} \frac{\overline{\alpha}_{1}}{a_{1}} k_{1}^{e_1}\int d^{11}k_{2}\frac{\overline{\alpha}_{2}}{a_{2}}k_{2}^{e_2}\cdots\int d^{11}k_{q}\frac{\overline{\alpha}_{q}}{a_{q}}k_{q}^{e_q}\int d^{11}k \frac{\overline{\alpha}(k)}{a(k)}\psi_{\mathbf{k}}(\mathrm{X})e^{-ik_{0}t}\nonumber\\
&\equiv [{\rm dom}]_{\varphi_1}\cup {\mathbb{T}}.\label{DomVaRphi1}
\end{align}

As elaborated in \cite{Brahma:2022wdl}, for the $\varphi_1$ sector, in addition to the above studied most dominant contribution,  it is important to consider the case where the external momentum  match with at least one of the internal momentum values. Otherwise, the end result obtained by incorporating the denominator of the path integral becomes identical to that of the tree level answer as we will see. So, in the following we are going to present such a case which we dub as ``\textbf{Case 4}''.

\vskip0.2in
\noindent{\textbf{Case 4:} {$\boldsymbol{k_{u_1} \neq k_{u_2} \neq k_{u_3} \neq \cdots \neq k_{u_{q-1}}\neq k_{u_q}~\text{and}~k_{u_q}=k_{i}}$} where $k_i$ denotes the momentum of the external field insertion}
\vskip.1in
For this case (\ref{O(c)phi1Case1}) evaluates to
\begin{align}
     {\rm Num}&\left[\langle  \hat{\varphi}_1 \rangle_{{\sigma}}\right]\bigg|^{\mathrm{Case}~4}_{\mathcal{O}(c_{nmpqrs}),\varphi_1}\nonumber\\
    &=\frac{1}{\mathrm{V}^q}\sum_{e_{1},...,e_{q-1}}\begin{pmatrix}
        n+e_o\\
        e_1
    \end{pmatrix}\begin{pmatrix}
        n+e_o-e_1\\
        e_2
    \end{pmatrix}\cdots\begin{pmatrix}
        n+e_o-\sum_{i=1}^{q-1}e_i\\
        e_{q-1}
 \end{pmatrix}\nonumber\\
&\times\sum_{k_1,...,k_{q-1} }k_{1}^{e_1}k_{2}^{e_2}\cdots k_{q-1}^{e_{q-1} } \int d\widetilde{\varphi}_{1}(k_1){\rm exp}\Big[-a_1\Big(\widetilde\varphi_1(k_1) - \frac{\overline{\alpha}_1}{a_1} \Big)^2\Big] \widetilde{\varphi}_{1}(k_1)\nonumber\\
 &   \times\int d\widetilde{\varphi}_{1}(k_2){\rm exp}\Big[-a_2\Big(\widetilde\varphi_1(k_2) - \frac{\overline{\alpha}_2}{a_2} \Big)^2\Big]\widetilde{\varphi}_{1}(k_2)\cdots\nonumber\\
    &\times \int d\widetilde{\varphi}_{1}(k_{q-1}){\rm exp}\Big[-a_{q-1}\Big(\widetilde\varphi_1(k_{q-1}) - \frac{\overline{\alpha}_{q-1} }{a_{q-1} } \Big)^2\Big] \widetilde{\varphi}_{1}(k_{q-1}) \nonumber\\
    &\times   \frac{1}{ {\rm V}}\sum_{k_i\neq k_1,...,k_{q-1} }k_{i}^{e_q}\int d\widetilde{\varphi}_{1}(k_{i})\exp\bigg[-a_{i}\left( \widetilde{\varphi}_{1}(k_{i})-\frac{\overline{\alpha}_{i}}{a_{i}}\right)^2\bigg]\Big(\widetilde\varphi_1^2(k_i)\psi_{{\bf k}_i}({\bf x}, y, z)e^{-ik_{0, i}t}\Big)\nonumber\\
    &\times\prod_{k_j\neq k_i,k_1,...,k_{q-1}}\int d\widetilde{\varphi}_{1}(k_{j})\exp\bigg[-a_{j}\left( \widetilde{\varphi}_{1}(k_{j})-\frac{\overline{\alpha}_{j}}{a_{j}}\right)^2\bigg]\nonumber\\
    &=\frac{1}{\mathrm{V}^{q}}\sum_{e_{1},...,e_{q-1}}\begin{pmatrix}
        n+e_o\\
        e_1
    \end{pmatrix}\begin{pmatrix}
        n+e_o-e_1\\
        e_2
    \end{pmatrix}\cdots\begin{pmatrix}
        n+e_o-\sum_{i=1}^{q-1}e_i\\
        e_{q-1}
    \end{pmatrix}\nonumber\\
&~~~~~~\times\sum_{k_1,...,k_{q-1} }k_{1}^{e_1}k_{2}^{e_2}\cdots k_{q-1}^{e_{q-1} } \frac{\overline{\alpha}_{1}}{a_{1}}\left(\frac{\pi}{a_{1}} \right)^{1/2}\frac{\overline{\alpha}_{2}}{a_{2}}\left(\frac{\pi}{a_{2}} \right)^{1/2}\cdots \frac{\overline{\alpha}_{q-1}}{a_{q-1}}\left(\frac{\pi}{a_{q-1}} \right)^{1/2}\nonumber\\
    &\times   \frac{1}{ {\rm V}}\sum_{k_i\neq k_1,...,k_{q-1} }\left( \frac{\pi}{a_{i}}\right)^{1/2}k_{i}^{e_q}\left[\left(\frac{\overline{\alpha}_{i}}{a_{i}}\right)^2+\frac{1}{2a_i}\right]\psi_{{\bf k}_i}({\bf x}, y, z)e^{-ik_{0, i}t}\times\prod_{j\neq i,1,...,q-1}\left( \frac{\pi_{j}}{a_{j}}\right)^{1/2},
\end{align}
where $e_{q}=n+e_{o}-\sum_{i=1}^{q-1}e_{i}$. Taking the continuum limit we get 
\begin{align}
     {\rm Num}\left[\langle  \hat{\varphi}_1 \rangle_{{\sigma}}\right]\bigg|^{\mathrm{Case}~4}_{\mathcal{O}(c_{nmpqrs}),\varphi_1}&=\prod_{i}\left(\frac{\pi_{i}}{a_{i}} \right)^{1/2}\frac{1}{\rm V}\sum_{e_{1},...,e_{q-1}}\begin{pmatrix}
        n+e_o\\
        e_1
\end{pmatrix}\begin{pmatrix}
        n+e_o-e_1\\
        e_2
\end{pmatrix}\cdots\nonumber\\
\times\begin{pmatrix}
        n+e_o-\sum_{i=1}^{q-1}e_i\\
        e_{q-1}
\end{pmatrix}&~\int d^{11}k_{1} \frac{\overline{\alpha}_{1}}{a_{1}} k_{1}^{e_1}\int d^{11}k_{2}\frac{\overline{\alpha}_{2}}{a_{2}}k_{2}^{e_2}\cdots\int d^{11}k_{q-1}\frac{\overline{\alpha}_{q-1}}{a_{q-1}}k_{q-1}^{e_{q-1}}\nonumber\\
&~\times\int d^{11}k \left[ \left(\frac{\overline{\alpha}(k)}{a(k)}\right)^2+\frac{1}{2a(k)}\right]k^{e_q}\psi_{\mathbf{k}}(\mathrm{X})e^{-ik_{0}t}\nonumber\\
&\equiv [{\rm dom}]_{\varphi_1}\cup \overset{\curvearrowleft}{\mathbb{T}}.\label{DomVaRphi1Case4}
\end{align}
Now, the numerator of $\langle \hat{\varphi}_1\rangle_{\sigma}$ (\ref{NumErAtorOfExpValuE}) at $\mathcal{O}(c_{nmpqrs})$ can be written using ``\textbf{Case 3}'' of the $\varphi_2$ (\ref{DomVarphi2}) and $\varphi_3$ (\ref{CAsE3VarPhi3ord1}) sectors, and ``\textbf{Case 3}'' (\ref{DomVaRphi1}) as well as the ``\textbf{Case 4}'' (\ref{DomVaRphi1Case4}) of $\varphi_1$ sector  as
\begin{align}
  &~~~~  \mathrm{Num}[\langle \hat{\varphi}_1\rangle_{\sigma}]\bigg\vert_{\mathcal{O}(c_{nmpqrs})}\nonumber\\
  &=-(-1)^v \left(\prod_{i}\left( \frac{\pi_{i}}{a_{i}}\cdot\frac{\pi_i}{b_i}\cdot\frac{\pi_i}{c_i}\right)^{1/2}\right)\sum_{\substack{e_o}}\sum_{n_1=0}^{1}\begin{pmatrix}
      2\\
      n_1
  \end{pmatrix} (-1)^{n_1}\sum_{n_2=0}^{\infty}\frac{1}{n_2!} \sum_{e_{n_1}=0}^{n_1}\begin{pmatrix}
      n_1\\
      e_{n_1}
  \end{pmatrix} \sum_{e_{n_2}=0}^{n_2}\begin{pmatrix}
      n_2\\
      e_{n_2}
  \end{pmatrix}\nonumber\\
  &~\times\sum_{e_1,...,e_{q-1}}\mathbb{C}^{\varphi_1}(\widetilde{n},e_1,...,e_{q-1})\left(\prod_{j=1}^{q-1} \int d^{11}k_{j}\frac{\overline{\alpha}(k_j)}{a(k_j)}k_{j}^{e_j}\right) \sum_{e_1^\prime,...,e_{r-1}^\prime}\mathbb{C}^{\varphi_2}(\widetilde{m},e^\prime_1,...,e^\prime_{r-1})\nonumber\\
  &\times \left(\prod_{i=1}^{r}\int d^{11}l_{i}\frac{\overline{\beta}(l_i)}{b(l_i)}l_{i}^{e^\prime_i} \right)\left( \prod_{t=1}^{s-1}\int d^{11}f_{t}\frac{\overline{\gamma}(f_t)}{c(f_{t})}\right)\nonumber\\
  &\times\frac{1}{(f_1+\cdots+f_{s-1})^{2+n_1}}\left( \frac{\partial}{\partial f_1}+\cdots+\frac{\partial}{\partial f_{s-1}} \right)^{n_2}\mathrm{V}\overline{\gamma}(f_1+\cdots+f_{s-1})\nonumber\\
  &\times\left(\int d^{11}k_{q}\frac{\overline{\alpha}(k_q)}{a(k_q)}k_{q}^{e_q}\int d^{11}k \frac{\overline{\alpha}(k)}{a(k)}+\int d^{11}k \frac{1}{\rm V}\left[\left(\frac{\overline{\alpha}(k)}{a(k)}\right)^2+\frac{1}{2a(k)} \right]k^{e_q}\right)\psi_{\bf k}({\rm X})e^{-ik_0t}\nonumber\\
  &\equiv \left([\mathrm{dom}]_{\varphi_1}\cup \mathbb{T} \right)\cup [\mathrm{dom}]_{\varphi_2}\cup [\mathrm{dom}]_{\varphi_3}+\left([\mathrm{dom}]_{\varphi_1}\cup \overset{\curvearrowleft}{\mathbb{T}} \right)\cup [\mathrm{dom}]_{\varphi_2}\cup [\mathrm{dom}]_{\varphi_3} \label{NumaTord1PathInt}
\end{align}
where $v=\frac{1}{2}(n+m+3p)$, $\widetilde{n}$ and $\widetilde{m}$ are as defined in (\ref{widEtilDen}) and (\ref{WidEtiLDem}). The parameters $e_{q}=\widetilde{n}-\sum_{i=1}^{q-1}e_{i},~\text{and}~e^\prime_{r}=\widetilde{m}-\sum_{i=1}^{r-1}e^\prime_{i}$.  

Note that, the above procedure can be similarly followed to obtain the denominator of the path integral (\ref{ScalarPIntAnA}) at $\mathcal{O}(c_{nmpqrs})$. The result can be quoted as 
\begin{align}
  &\mathrm{Den}[\langle \hat{\varphi}_1\rangle_{\sigma}]\bigg\vert_{\mathcal{O}(c_{nmpqrs})}\nonumber\\
  &=-(-1)^v \left(\prod_{i}\left( \frac{\pi_{i}}{a_{i}}\cdot\frac{\pi_i}{b_i}\cdot\frac{\pi_i}{c_i}\right)^{1/2}\right)\sum_{\substack{e_o}}\sum_{n_1=0}^{1}\begin{pmatrix}
      2\\
      n_1
  \end{pmatrix} (-1)^{n_1}\sum_{n_2=0}^{\infty}\frac{1}{n_2!} \sum_{e_{n_1}=0}^{n_1}\begin{pmatrix}
      n_1\\
      e_{n_1}
  \end{pmatrix} \sum_{e_{n_2}=0}^{n_2}\begin{pmatrix}
      n_2\\
      e_{n_2}
  \end{pmatrix}\nonumber\\
  &\times\sum_{e_1,...,e_{q-1}}\mathbb{C}^{\varphi_1}(\widetilde{n},e_1,...,e_{q-1})\left(\prod_{j=1}^{q} \int d^{11}k_{j}\frac{\overline{\alpha}(k_j)}{a(k_j)}k_{j}^{e_j}\right) \sum_{e_1^\prime,...,e_{r-1}^\prime}\mathbb{C}^{\varphi_2}(\widetilde{m},e^\prime_1,...,e^\prime_{r-1})\nonumber\\
  &\times\left(\prod_{i=1}^{r}\int d^{11}l_{i}\frac{\overline{\beta}(l_i)}{b(l_i)}l_{i}^{e^\prime_i} \right) \left( \prod_{t=1}^{s-1}\int d^{11}f_{t}\frac{\overline{\gamma}(f_t)}{c(f_{t})}\right)\nonumber\\
  &\times\frac{1}{(f_1+\cdots+f_{s-1})^{2+n_1}}\left( \frac{\partial}{\partial f_1}+\cdots+\frac{\partial}{\partial f_{s-1}} \right)^{n_2}\mathrm{V}\overline{\gamma}(f_1+\cdots+f_{s-1})\nonumber\\
  &\equiv [\mathrm{dom}]_{\varphi_1}\cup [\mathrm{dom}]_{\varphi_2}\cup [\mathrm{dom}]_{\varphi_3}.
\label{DenOrd1PathInTeg}
\end{align}
Since ``\textbf{Case 4}'' for the $\varphi_1$ sector (\ref{DomVaRphi1Case4}) is $\frac{1}{\mathrm{V}}$ suppressed, one might wonder about the analogous $\frac{1}{\mathrm{V}}$-suppressed contributions arising from the $\varphi_2$ and $\varphi_3$ sectors, which we denote as $[\mathrm{sdom}]_{\varphi_2}$ and $[\mathrm{sdom}]_{\varphi_3}$ respectively, as well as from the configurations in the $\varphi_1$ sector where any two of the internal field momenta are identical, denoted as $[\mathrm{sdom}]_{\varphi_1}\cup{\mathbb{T}}$. It turns out that even after incorporating all such contributions, the final result remains unchanged. Suppose we express $\langle \hat{\varphi}_1\rangle_{\sigma}$ up to $\mathcal{O}(c_{nmpqrs})$ while including all the aforementioned $\frac{1}{\mathrm{V}}$-suppressed corrections, then we obtain
\begin{align}
\langle \hat{\varphi}_1\rangle_{\sigma}&=\frac{\substack{\mathbb{T}+ \sum_{n,...,s}c_{nmpqrs}\bigg[ \left([{\rm dom}]_{\varphi_1}\cup \mathbb{T}\right)  \cup[{\rm dom}]_{\varphi_2}\cup[{\rm dom}]_{\varphi_3}+\left([{\rm sdom}]_{\varphi_1} \cup \mathbb{T}\right)\cup[{\rm dom}]_{\varphi_2}\cup[{\rm dom}]_{\varphi_3}\bigg.\\ \bigg.+([{\rm dom}]_{\varphi_1}\cup\mathbb{T})\cup[{\rm sdom}]_{\varphi_2}\cup[{\rm dom}]_{\varphi_3}+\left([{\rm dom}]_{\varphi_1}\cup\mathbb{T}\right)\cup[{\rm dom}]_{\varphi_2}\cup[{\rm sdom}]_{\varphi_3} +\left([\mathrm{dom}]_{\varphi_1}\cup \overset{\curvearrowleft}{\mathbb{T}} \right)\cup [\mathrm{dom}]_{\varphi_2}\cup [\mathrm{dom}]_{\varphi_3} \bigg]} }{  \substack{ \mathbb{I}+ \sum_{n,...,s}c_{nmpqrs}\bigg[ [{\rm dom}]_{\varphi_1}  \cup[{\rm dom}]_{\varphi_2}\cup[{\rm dom}]_{\varphi_3}+[{\rm sdom}]_{\varphi_1} \cup[{\rm dom}]_{\varphi_2}\cup[{\rm dom}]_{\varphi_3}+[{\rm dom}]_{\varphi_1} \cup[{\rm sdom}]_{\varphi_2}\cup[{\rm dom}]_{\varphi_3} \bigg. \\ \bigg. +[{\rm dom}]_{\varphi_1}\cup[{\rm dom}]_{\varphi_2}\cup[{\rm sdom}]_{\varphi_3} \bigg] }  }\nonumber\\
&=\mathbb{T}+\frac{\sum_{n,...,s}c_{nmpqrs} \left([\mathrm{dom}]_{\varphi_1}\cup \overset{\curvearrowleft}{\mathbb{T}} \right)\cup [\mathrm{dom}]_{\varphi_2}\cup [\mathrm{dom}]_{\varphi_3} }{\substack{\textstyle \mathbb{I}+ \sum_{n,...,s}c_{nmpqrs}\bigg[ [{\rm dom}]_{\varphi_1}  \cup[{\rm dom}]_{\varphi_2}\cup[{\rm dom}]_{\varphi_3}+[{\rm sdom}]_{\varphi_1} \cup[{\rm dom}]_{\varphi_2}\cup[{\rm dom}]_{\varphi_3} \bigg. \\ \bigg. \textstyle +[{\rm dom}]_{\varphi_1} \cup[{\rm sdom}]_{\varphi_2}\cup[{\rm dom}]_{\varphi_3}+[{\rm dom}]_{\varphi_1}\cup[{\rm dom}]_{\varphi_2}\cup[{\rm sdom}]_{\varphi_3} \bigg] }}\label{UptoOrdercExpValUe}
\end{align}
Note that the numerator of the second term in \eqref{UptoOrdercExpValUe} is $\frac{1}{\rm V}$ suppressed. As a result, when the denominator is expanded binomially the sub dominant terms in it give rise to $\mathcal{O}\left(\frac{1}{\rm V^2}\right)$ volume suppression in the numerator. Thus, apart from the ``\textbf{Case 4}'' of $\varphi_1$ sector (\ref{DomVaRphi1Case4}), we can safely ignore all other $\frac{1}{\rm V}$ corrections. Finally, from (\ref{UptoOrdercExpValUe}) we get
\begin{align}
 \langle \hat{\varphi}_1\rangle_{\sigma}   &=\mathbb{T}+\sum_{n,...,s}
c_{nmpqrs} \left([\mathrm{dom}]_{\varphi_1}\cup \overset{\curvearrowleft}{\mathbb{T}} \right)\cup [\mathrm{dom}]_{\varphi_2}\cup [\mathrm{dom}]_{\varphi_3}+\mathcal{O}(c_{nmpqrs}^2).
\end{align}
\subsection{The perturbative series}
The full path integral (\ref{ScalarPIntAnA}) for a fixed choice of $c_{nmpqrs}$, call $g$ can be schematically expressed as 
\begin{align}
\langle\hat{\varphi}_1\rangle_{\sigma}&=\mathbb{T}+\sum\limits_{\rm N=1}^{\infty}\frac{g^{\rm N}}{\rm N!} \left(([{\rm dom}]_{\varphi_1})^{\rm N}\right)\cup ([\mathrm{dom}]_{\varphi_2})^{\rm N}\cup ([\mathrm{dom}]_{\varphi_3})^{\rm N} \cup \overset{\curvearrowleft}{\mathbb{T}} , \label{TheFPinTeGraL}
\end{align}
where the ${\rm N}$th term is given as 
\begin{align}
 & \frac{1}{\rm V}  \frac{g^{\rm N}}{\rm N!} \prod_{i}\left( \frac{\pi_{i}}{a_{i}}\right)^{1/2}\sum_{\{e_i\}}\mathbb{C}^{\varphi_1}\left(e_1,...e_{q-1},e_{q+1},..., e_{{\rm N}q-1}\right) \mathbb{D}_{\varphi_2}({\rm N}r)\mathbb{D}_{\varphi_3}({\rm N}(s-1))\nonumber\\
 &\times \int d^{11}k_1~\frac{\overline\alpha(k_1) k_1^{e_1}}{ a(k_1)} \cdots \int d^{11}k_q~\frac{\overline\alpha(k_q) k_q^{e_q}}{ a(k_q)}\int d^{11}k_{q+1}~\frac{\overline\alpha(k_{q+1}) k_{q+1}^{e_{q+1}}}{ a(k_{q+1})}\cdots\int d^{11}k_{2q}~\frac{\overline\alpha(k_{2q}) k_{2q}^{e_{2q}}}{ a(k_{2q})}\nonumber\\
&\times  \int d^{11}k_{{\rm 2}q+1}~\frac{\overline\alpha(k_{{\rm 2}q+1}) k_{{\rm 2}q+1}^{e_{{\rm 2}q+1}}}{ a(k_{{\rm 2}q+1})}\cdots \int d^{11}k_{{\rm N}q-1}~\frac{\overline\alpha(k_{{\rm N}q-1}) k_{{\rm N}q-1}^{e_{{\rm N}q-1}}}{ a(k_{{\rm N}q-1})}\nonumber\\
&\times\int d^{11}k \left[\frac{\overline{\alpha}^2(k)}{a^2(k)}+\frac{1}{2a(k)}\right]k^{e_{{\rm N}q}}\psi_{\mathbf{k}}(\mathrm{X})e^{-ik_{0}t}.\label{NthTerMinThePerExpAns}
\end{align}
Here, the coefficient $\mathbb{C}^{\varphi_{1}}\left(e_1,...e_{q-1},e_{q+1},..., e_{{\rm N}q-1}\right)$ encodes the combinatorial factors involved due to the ${\varphi}_1$ sector and $e_{\mathrm{N}q}=\widetilde{n}-\sum^{\mathrm{N}q-1}_{i=(\mathrm{N}-1)q+1}e_{i}$. The factors $\mathbb{D}_{\varphi_2}({\rm N}r)$ and $\mathbb{D}_{\varphi_3}({\rm N}(s-1))$ contain terms arising due to the $\varphi_2$ and $\varphi_3$ sectors. 

With the assumption that the combinatorial factors associated with each of the three field sectors grow as ${\rm N}!$ in the large-${\rm N}$ limit \cite{Brahma:2022wdl}, the perturbative series (\ref{TheFPinTeGraL}) can be expressed as
\begin{align}
    \langle \hat{\varphi}_1\rangle_{\sigma}&=\mathbb{T}+\frac{1}{\rm V}\sum_{{\rm N}=1}^{\infty}g^{\rm N}({\rm N}!)^2\sum_{ \{e_i\} }\mathcal{A}^{\rm N}(\{e_i\})\int d^{11}k \left[\frac{ \overline{\alpha}^2(k) }{a^2(k)}+\frac{1}{2a(k)}\right]k^{e_{{\rm N} q}} \psi_{\bf k}({\rm X})e^{-ik_0 t},\label{PerTSeriESAsymp}
\end{align}
where $\mathcal{A}^{\rm N}(\{e_i\})$ contains the rest of the integrals in Eq. (\ref{NthTerMinThePerExpAns}).  For some $e^{\star}$, if the following bound holds true \cite{Brahma:2022wdl}
\begin{align}
    \sum_{\{e_i\}}\mathcal{A}^{\rm N}(\{e_i\})k^{e_{{\rm N}q}}\leq \mathcal{A}_{\rm eff}^{\rm N}k^{e^\star},\quad\quad \forall {\rm N}\in \mathbb{Z}_{+}
\end{align}
then we can rewrite (\ref{PerTSeriESAsymp}) as
\begin{align}
    \langle \hat{\varphi}_1\rangle_{\sigma}&=\mathbb{T}+\frac{1}{\rm V}\sum_{{\rm N}=1}^{\infty}g^{\rm N}({\rm N}!)^2\mathcal{A}^{\rm N}_{\rm eff}\int d^{11}k \left[\frac{ \overline{\alpha}^2(k) }{a^2(k)}+\frac{1}{2a(k)}\right]k^{e^{\star}} \psi_{\bf k}({\rm X})e^{-ik_0 t}.
\end{align}
Adding and subtracting the ${\rm N}=0$ term, and substituting $\mathbb{T}$ from Eq. \eqref{TrEELeVelAnsWerr}  we get 
\begin{align}
    \langle \hat{\varphi}_1\rangle_{\sigma}&=\int d^{11}k \left[\frac{\overline{\alpha}(k)}{a(k)}-\frac{k^{e^\star}}{\rm V}\left(\frac{\overline{\alpha}^2(k)}{a^2(k)} +\frac{1}{2a(k)}\right)\right]\psi_{\bf k}({\rm X})e^{-ik_0 t}\nonumber\\
    &+\frac{1}{\rm V}\sum_{{\rm N}=0}^{\infty} {g^{\rm N }} ({\rm N }!)^2  \mathcal{A}_{\rm eff}^{\rm N}\int d^{11}k \left[\frac{ \overline{\alpha}^2(k) }{a^2(k)}+\frac{1}{2a(k)}\right]k^{e^{\star}} \psi_{\bf k}({\rm X})e^{-ik_0 t}.\label{TheeXpValUeofVarPhi1}
\end{align}
It was shown in \cite{Brahma:2022wdl}, that the first integral in \eqref{TheeXpValUeofVarPhi1} can be set to zero:
\begin{align}
    \int d^{11}k \left[\frac{\overline{\alpha}(k)}{a(k)} -\frac{k^{e^\star}}{\rm V}\left(\frac{\overline{\alpha}^2(k) }{a^2(k)}+\frac{1}{2a(k)} \right) \right]\psi_{\bf k}({\rm X})e^{-ik_0 t}=0,\label{SettingDelTto0}
\end{align}
which in turn allows us to express the expectation value $\langle \hat{\varphi}_1\rangle_{\sigma}$ in the form of the following series:  
\begin{align}
\langle\hat{\varphi}_1\rangle_{\sigma}
&=\sum_{\rm N=0}^{\infty}g^{\rm N}({\rm N}!)^2 \mathcal{A}^{\rm N}\int d^{11}k \frac{\overline{\alpha}(k)}{a(k)} \mathbf{Re}\left(\psi_{\bf k}({\rm X})e^{-ik_0t}\right).\label{TheExpValofVaRPhi1}
\end{align}
Note that, we have replaced $\psi_{\mathbf{k}}({\rm X})e^{-ik_0 t}$ by its real part, i.e.\ $\mathbf{Re}\!\left(\psi_{\mathbf{k}}({\rm X})e^{-ik_0 t}\right)$. This is because we have set the imaginary parts of the Fourier modes to zero in the above analysis starting from Eq. (\ref{NumErAtorOfExpValuE}). Hence, the replacement $\psi_{\mathbf{k}}({\rm X})e^{-ik_0 t}\rightarrow \mathbf{Re}\!\left(\psi_{\mathbf{k}}({\rm X})e^{-ik_0 t}\right)$ ensures that the insertion field $\varphi_{1}(\mathrm{Y})$  \eqref{TheInSertIonField} remains real.

As the resulting series (\ref{TheExpValofVaRPhi1}) is asymptotic with zero radius of convergence, the next step is to employ resummation techniques to get a sensible answer. This will be the focus of the following section. 
\section{Resummation of the asymptotic series}
\label{BorelREsum}
Generically, for each component of the metric operator one can write Eq. (\ref{TheExpValofVaRPhi1}) as 
\begin{align}
\langle \hat{\mathbf{g}}_{\rm MN}\rangle_{\overline{\sigma}}&=\sum_{\mathrm{N}=0}^{\infty}g^{\mathrm{N}} \mathcal{A}^{\mathrm{N}}(\mathrm{N}!)^{l}\int  d^{11}k \frac{\overline{\alpha}_{\rm MN}(k)}{a(k)}\mathbf{Re}(\psi_{\mathbf{k}}(\mathrm{X})e^{-ik_0 t}),\label{ExpValMOp}
\end{align}
where $l$ denotes one less than the total number of field sectors. 
As the $\mathrm{N}$th coefficient of the series grows as $(\mathrm{N}!)^{l}$, this is  termed as a Gevrey-$l$ series. It is important to note that the above series with $(\mathrm{N}!)^{l}$ growth can equivalently be expressed as a series with $(l \mathrm{N})!$ growth. This can be understood in the  following manner: using Stirlings approximation in the limit $\mathrm{N}\rightarrow \infty$ we get
\begin{align}
    \log (\mathrm{N}!)^{l}&\approx l (\mathrm{N}~\log \mathrm{N}-\mathrm{N})=l \mathrm{N}\log(l \mathrm{N})-l \mathrm{N}-l \mathrm{N}\log l=\log \left(\frac{(l \mathrm{N})!}{l^{l \mathrm{N}}} \right)\nonumber\\
    \therefore~&~~~(\mathrm{N}!)^{l}\sim \frac{(l \mathrm{N})!}{l^{l \mathrm{N}}}.
\end{align}
Hence we have a series
\begin{align}
\mathscr{F}(g)&= \sum_{\mathrm{N}=0}^{\infty}g^{\mathrm{N}}a_{\mathrm{N}},~~~~\mathrm{where}~~a_{\mathrm{N}}\sim \mathcal{A}^{\mathrm{N}}(l \mathrm{N})! .
\label{SeRiesphI(g)}
\end{align}
Here we have absorbed the constant $l^l$ into the definition of the parameter $g$. To resum this series  we introduce a factor of $(l\mathrm{N})!$ in its numerator and denominator as:
 \begin{align}
   \mathscr{F}(g)&= \sum_{\mathrm{N}=0}^{\infty}g^{\mathrm{N}}a_{\mathrm{N}}\frac{\int_{0}^{\infty} d\mathrm{S}\exp\left(-\mathrm{S}\right)\mathrm{S}^{l \mathrm{N}}}{(l \mathrm{N})!}~.
 \end{align}
 Interchanging the order of summation and integration we get 
 \begin{align}
     \mathscr{F}(g)&\rightarrow \int_{0}^{\infty}d\mathrm{S}\exp(-\mathrm{S})\left[\sum_{\mathrm{N}=0}^{\infty}g^{\mathrm{N}}a_{\mathrm{N}}\frac{\mathrm{S}^{l\mathrm{N}}}{(l \mathrm{N})!}\right].
 \end{align}
The quantity inside the square bracket is termed as \textit{Borel transform} of  $\mathscr{F}(g)$ 
 \begin{align}
\mathcal{B}[\mathscr{F}](g\mathrm{S})&=\sum_{\mathrm{N}=0}^{\infty}g^{\mathrm{N}}a_{\mathrm{N}}\frac{\mathrm{S}^{l\mathrm{N}}}{(l \mathrm{N})!},
 \end{align}
which in fact has a finite radius of convergence. Thus, the \textit{Borel resummation} of the formal power series $\mathscr{F}(g)$ is basically the Laplace transform of the Borel transform 
 \begin{align}
\mathfrak{s}(\mathscr{F})(g)&= \int_{0}^{\infty}d\mathrm{S}\exp \left(-\mathrm{S} \right) \mathcal{B}[\mathscr{F}] (g\mathrm{S}).
 \end{align}
 Further simplification of this result yields
 \begin{align}
     \mathfrak{s}(\mathscr{F})(g)&= \int_{0}^{\infty}d\mathrm{S}~\exp\left(-\mathrm{S}\right) \sum_{\mathrm{N}=0}^{\infty}g^{\mathrm{N}}\mathcal{A}^{\mathrm{N}}\mathrm{S}^{l \mathrm{N}}\nonumber\\
     &= \left(\frac{1}{g^{1/l}}\right)\int_{0}^{\infty}d\mathrm{S}~\exp\left(-\frac{\mathrm{S}}{g^{1/l}} \right)\sum_{\mathrm{N}=0}^{\infty}\mathcal{A}^{\mathrm{N}}\mathrm{S}^{l \mathrm{N}}\nonumber\\
     &= \left(\frac{1}{g^{1/l}} \right)\int_{0}^{\infty} d\mathrm{S}\exp\left(-\frac{\mathrm{S}}{g^{1/l}} \right) \frac{1}{1-\mathcal{A}\mathrm{S}^{l}},\label{BoReLreSumMedAnsWer}
 \end{align}
 where to go from 1st line to the 2nd line we have made a change of integration variable $\mathrm{S}\rightarrow \frac{\mathrm{S}}{g^{1/l}}$ and to arrive at the last line we have performed the sum over $\mathrm{N}$. We note that the Borel resummation (\ref{BoReLreSumMedAnsWer}) has led to non-perturbative terms of the form $\exp \left(-\frac{\mathrm{S}}{g^{1/l}}\right)$. The zeros of the following polynomial  
\begin{align}
    \mathrm{S}^{l}-\frac{1}{\mathcal{A}}&\equiv \prod_{i=1}^{l} \left(\mathrm{S}-a_{i}\right)
\end{align}
are nothing but the poles in the Borel plane. 
Substituting the Borel resummed series (\ref{BoReLreSumMedAnsWer}) in (\ref{ExpValMOp}) we arrive at 
\begin{align}
\langle \hat{\mathbf{g}}_{\rm MN}\rangle_{\overline{\sigma}}&=\left[\frac{1}{g^{1/l}} \int_{0}^{\infty} d\mathrm{S}\exp\left(-\frac{\mathrm{S}}{g^{1/l}} \right) \frac{1}{1-\mathcal{A}\mathrm{S}^{l}}\right]_{\mathrm{P.V.}}\int  d^{11}k \frac{\overline{\alpha}_{\rm MN}(k)}{a(k)}\mathbf{Re}(\psi_{\mathbf{k}}(\mathrm{X})e^{-ik_0 t}).\label{BORELRSummed}
\end{align}
When ${\rm M,~N}\equiv\{\mu,~\nu\}\in \mathbb{R}^{2,1}$, then the aforementioned result has to be equated to the coefficient of $\mathbb{R}^{2,1}$ in the M-theory metric (\ref{MthEorYmeTric}) along with the value of $g_s$ as obtained in (\ref{IIAcouPlingdeSitterIIB}). That is,
\begin{align}
   \left[\frac{1}{g^{1/l}} \int_{0}^{\infty} d\mathrm{S}\exp\left(-\frac{\mathrm{S}}{g^{1/l}} \right) \frac{1}{1-\mathcal{A}\mathrm{S}^{l}}\right]_{\mathrm{P.V.}}\int  d^{11}k \frac{\overline{\alpha}_{\rm \mu\nu}(k)}{a(k)}\mathbf{Re}(\psi_{\mathbf{k}}(\mathrm{X})e^{-ik_0 t})&=(\sqrt{\Lambda}\mathrm{H}(y)|t|)^{-\frac 8 3}\eta_{\mu\nu}.\label{R21AnsweRforGSstate}
\end{align}
This allows to write the following non-perturbative definition of the cosmological constant 
\begin{align}
    \frac{1}{\Lambda^{4/3}}&=\left[\frac{1}{g^{1/l}} \int_{0}^{\infty} d\mathrm{S}\exp\left(-\frac{\mathrm{S}}{g^{1/l}} \right) \frac{1}{1-\mathcal{A}\mathrm{S}^{l}}\right]_{\mathrm{P.V.}} .\label{NpeRDefCosMCons}
\end{align}
Equating the remaining terms of both sides of (\ref{R21AnsweRforGSstate}) we get
\begin{align}
   \int  d^{11}k \frac{\overline{\alpha}_{\rm \mu\nu}(k)}{a(k)}\mathbf{Re}(\psi_{\mathbf{k}}(\mathrm{X})e^{-ik_0 t})&= (\mathrm{H}(y)|t|)^{-\frac 8 3}\eta_{\mu\nu}.
\end{align}
Ideally, one should perform an inverse Fourier transform of this equation to extract $\overline{\alpha}_{\mu\nu}(k)$. Nevertheless, in \cite{Brahma:2022wdl}, authors have taken a specific Hermite polynomial ansatz for $\overline{\alpha}(k)$ to validate the aforementioned equation.

Thus, we have demonstrated that the metric solution in M-theory (\ref{MthEorYmeTric}) may be realized as the expectation value of the metric operator evaluated with respect to a Glauber--Sudarshan (GS) state. This analysis further enabled us to express the cosmological constant in a compact form as given in Eq. (\ref{NpeRDefCosMCons}).  

A natural question that arises at this stage is whether the above construction admits a well-defined effective field theory description. Addressing this issue turns out to be equivalent to a careful analysis of the null energy condition (NEC) within the context of a \(3+1\)-dimensional FLRW cosmology, as discussed in \cite{Bernardo_2021}. In the following section, we shall briefly review this aspect. 
\section{NEC in type \texorpdfstring{$\mathrm{IIB}$}{IIB} string theory}
\label{sec:NECIIB}
There are various energy conditions that put restrictions on the stress energy tensor. For instance, according to the null energy condition, the stress energy tensor has to satisfy the constraint (\ref{NECConstraint}). For a perfect fluid, the stress energy tensor is of the form 
\begin{align}
\mathrm{T}_{\mu\nu}&=(\rho +p)u_{\mu}u_{\nu}+pg_{\mu \nu}\label{StressEnergyTenPF}
\end{align}
where $\rho$ denotes the energy density, $p$ denotes the momentum density and $u_{\mu}$ is the velocity 4-vector. Substituting (\ref{StressEnergyTenPF}) in (\ref{NECConstraint}) we obtain 
\begin{align}
   \mathrm{T}_{\mu \nu}\zeta^{\mu }\zeta^{\nu}&=( \rho +p)u_{\mu }\zeta^\mu u_{\nu}\zeta^\nu +pg_{\mu\nu}\zeta^\mu \zeta^\nu \geq 0\nonumber\\
   & \implies (\rho+ p)(u_{\mu}\zeta^\mu)^2\geq 0.
\end{align}
Since the factor $(u_{\mu}\zeta^{\mu})^2$ is non-negative, the NEC for a perfect fluid boils down to 
\begin{align}
    (\rho+p)\geq 0.\label{NECforPF} 
\end{align} 
For a flat 3+1 dimensional FLRW spacetime with metric configuration
\begin{align}
    ds^2=-d\tau^2+ a^2(\tau) ((dx^1)^2+(dx^2)^2+(dx^3)^2),
\end{align}
the Friedman equations take the form
\begin{align}
 (\mathrm{H}(\tau))^2&=\frac{8 \pi G_N}{3}\rho,~~~~\dot{\mathrm{H}}(\tau)=-4\pi G_{N}(\rho + p),
\end{align}
where $\mathrm{H}(\tau)=\dot{a}(\tau)/a(\tau)$ is the Hubble parameter and $\tau\in (-\infty,\infty)$ is the time coordinate. Thus, the NEC in Eq. (\ref{NECforPF}) implies $\dot{\rm H}(\tau)\leq 0$. If we assume $a(\tau)\propto \tau^\delta$ then the null energy condition becomes $\delta \geq 0$. In terms of the conformal time coordinate $t$, the FLRW metric takes the form 
\begin{align}
    ds^2&=a^2(t)\left(-dt^2+ (dx^1)^2+(dx^2)^2+(dx^3)^2\right),
\end{align}
where the coordinates $\tau$ and $t$ are related as 
\begin{align}
    \int\tau^{-\delta} d\tau &= \int dt \nonumber\\
    \implies \tau =( (&1-\delta)t)^{{1}/{(1-\delta)}}.
\end{align}
Therefore, the scale factor in conformal time is given as 
\begin{align}
    a(t)\propto t^{\delta/(1-\delta)}:=t^n.\label{ScaleFacFLRW}
\end{align}
Consequently, the null energy condition $\delta \geq 0$ becomes 
\begin{align}
  \frac{1}{n}= \frac{1}{\delta}-1\geq -1.\label{NECforFLRWcos}
\end{align}
Now, a well defined effective field theory description of the excited state configuration requires us to put the constraint (\ref{EftCriTeriA1}) on the type IIA coupling constant.
Let us analyze this for the case of realising de Sitter isometry in type IIB theory.
The time dependent part of the 3+1 dimensional non-compact spacetime (\ref{tyPeIIBMetrIc}) can be equated to 
\begin{align}
 a^2(t)&=(\Lambda t^2)^n,\label{TypeIIBNEcConDitIOn}
\end{align}
where the parameter $n$ is defined analogously to that appearing in Eq.~(\ref{ScaleFacFLRW}). The dual type IIA coupling constant (\ref{TyPeIIA coUplInG}) becomes 
\begin{align}
\overline{g}_{s}\equiv\frac{g_s}{\mathrm{H}(y)}&=\frac{1}{a(t)}=\frac{1}{(\Lambda t^2)^{n/2}}.\label{ValUeofgsForNECinTyPeIIB}
\end{align}
Correspondingly, the EFT criterion (\ref{EftCriTeriA1}) implies 
\begin{align}
    \frac{\partial \overline{g}_s}{\partial t }&=-\frac{n}{t}\overline{g}_{s}=-n{\Lambda^{1/2}\overline{g}_{s}^{1/n} }\overline{g}_{s}\propto \overline{g}_s^{(1+1/n)\geq 0}\nonumber\\
   &~~~~~~~~~~~~ \implies \frac{1}{n}\geq -1.\label{NECinTypeIIBThry}
\end{align}
This is precisely, the null energy condition for a flat 3+1 dimensional FLRW cosmology as appearing in Eq. (\ref{NECforFLRWcos}). Thus, for the de Sitter configuration in type IIB theory (\ref{tyPeIIBMetrIc}) \textit{the existence of well defined effective field theory description boils down to satisfying the null energy condition}.

Having reviewed the realization of de Sitter space as a Glauber--Sudarshan state in type IIB string theory, we now turn to the central theme of this thesis: the realization of such time-dependent backgrounds in phenomenologically viable sectors of string theory.  We shall focus in the following chapter on the heterotic $\mathrm{SO}(32)$ theory and explore how an analogous construction may be achieved in this framework.
\chapter{de Sitter as an excited state in heterotic \texorpdfstring{$\mathrm{SO}(32)$}{SO(32)}
string theory}
\label{Chap:3}
The possible realisation of de Sitter isometry in type IIB string theory naturally prompts the question of whether such a construction can be achieved in other string theories. In particular, one may ask whether a suitable duality sequence can be devised to realise de Sitter space as an excited state in heterotic string theory. This chapter is devoted to addressing this problem specifically in the context of heterotic $\mathrm{SO}(32)$ string theory\cite{Alexander:2023qym,Archana_CC2026}.      
\section{de Sitter isometry in heterotic \texorpdfstring{$\mathrm{SO}(32)$}{SO(32)} string theory}\label{dSinHetSO(32)}

Just as in the realization of de Sitter space in type IIB theory discussed in \S\ref{typeIIBdSisom}, our aim here is to present a duality sequence which, starting from an $\mathrm{M}$-theory background, leads to de Sitter isometry in heterotic $\mathrm{SO}(32)$ theory. The very first question that one has to address is the choice of compactification in $\mathrm{M}$ theory that allows the duality chain to proceed dynamically and eventually realize four dimensional de Sitter space in the late-time limit of the Poincar\'e patch. 

Exploring various possibilities, we found a particular compactification in which the $\mathrm{M}$-theory background is an orbifold of the following form
\begin{align}
\mathbb{R}^{2,1}\times \frac{\mathbb{T}_{4,5}^{2}}{\mathcal{I}_{4}\mathcal{I}_{5}}\times \mathcal{M}_{4}\times \frac{\mathbb{T}^{2}_{3,11}}{\mathcal{I}_{3}\mathcal{I}_{11}}.\label{MThCompcTifedForSO(32)}
\end{align}
Note that, the two-manifold $\mathcal{M}_2$, which was left arbitrary in the construction of the de Sitter GS state in type IIB theory (see Fig.~\ref{Type IIB tree diagram}), is now taken to be the toroidal orbifold
$\mathcal{M}_2\equiv\frac{\mathbb{T}^2_{4,5}}{\mathcal{I}_4\mathcal{I}_5}$.
In addition, unlike in the type IIB case, the M-theory torus along 3 and 11 is now orbifolded.
 The $\mathbb{Z}_{2}$ action $\mathcal{I}_{a}$ on the coordinate $x^{a}$ of a circle $\mathrm{S}^{1}_{a}$ identifies $x^{a}\sim -x^{a}$, thereby converting the circle into an interval. Consequently, the orbifold $\mathrm{S}^{1}_{a}/\mathcal{I}_{a}$ possesses two fixed points corresponding to the endpoints of the interval. Such an orbifold structure is essential for reaching the theory of interest---namely, the heterotic $\mathrm{SO}(32)$ theory---through the duality symmetries of string theory. 

In the following, we describe the complete duality sequence and present the metric configuration, fluxes, and coupling constants associated with each intermediate theory for clarity.

    [$\bullet$] The orbifold action not being reflected at the level of the metric, the $\mathrm{M}$ theory metric remains same as in Eq. (\ref{MthEorYmeTric})
    \begin{align}
ds^{2}_{11}&=\langle \hat{\mathbf{g}}_{\rm MN}\rangle_\sigma d\mathrm{Y}^{\rm M}d\mathrm{Y}^{\rm N}\nonumber\\
&=g_{s}^{-8/3}\left(-dt^2 + \sum_{i = 1}^2 \left({dx^i}\right)^2 \right) +g_{s}^{-2/3} {\rm H}^2(y)\left[\mathrm{F}_{1}(t)\delta_{\alpha\beta}dy^{\alpha}dy^{\beta}+\mathrm{F}_{2}(t)g_{mn} dy^m dy^n \right]\nonumber\\
&~~~~~~~~+g_{s}^{4/3}\delta_{ab}d\omega^{a}d\omega^{b},\label{MthHetSO32}
\end{align}
where $y^{\alpha}\in \frac{\mathbb T^2_{4,5}}{\mathcal{I}_4\mathcal{I}_5},~y^{m}\in \mathcal{M}_{4}$ and $\omega^a\equiv (x^3,x^{11})\in \frac{\mathbb{T}^2_{3,11}}{\mathcal{I}_3\mathcal{I}_{11}}$.
   However, there is one modification that we have incorporated viz. considered the unwarpped metric of the toroidal orbifold along $4,5$ to be $\delta_{\alpha\beta}$ as opposed to an arbitrary $g_{\alpha\beta}$, reason being that the $\mathcal{M}_2$ is now a toroidal orbifold $ \frac{\mathbb{T}^2_{4,5}}{\mathcal{I}_4\mathcal{I}_5}$. Furthermore, the warp factor $\mathrm{H}(y)$ depends only on the coordinates $y\equiv y^m$ of $\mathcal{M}_{4}$.

   [$\bullet$]  Upon dimensional reduction of 11th direction we get the type $\mathrm{IIA}$ theory with metric  
   \begin{equation}
\begin{aligned}
ds^{2}_{\mathrm{IIA}}&=\frac{a^{2}(t)}{{\rm H}^2(y)}\left(-dt^2 + \sum_{i = 1}^2 \left(dx^i\right)^2 \right) +\frac{\mathrm{H}^{2}(y)}{a^{2}(t)}(dx^{3})^2 + {\rm H}^2(y)\mathrm{F}_{1}(t)\delta_{\alpha\beta}dy^{\alpha}dy^{\beta}\\
    &~~~~~~~~~~~~+{\rm H}^2(y)\mathrm{F}_{2}(t)g_{mn} dy^m dy^n .
\end{aligned}
\end{equation}
 A $\mathrm{T}$ duality along $x^{3}$ of {{type} $\mathrm{IIA}$ theory} leads to type $\mathrm{IIB}$ theory having the metric configuration as in Eq. (\ref{tyPeIIBMetrIc}) (with $g_{\alpha\beta}$ replaced by $\delta_{\alpha\beta}$).  Note that, the solitonic structures of both type $\mathrm{IIA}$ and $\mathrm{IIB}$ backgrounds  as shown in Fig. \ref{Het SO(32) tree diagram} differ from what we had in the previous case of realising de Sitter isometry in $\mathrm{IIB}$ theory (see Fig. \ref{Type IIB tree diagram}). In particular, shrinking the toroidal orbifold along $3,11$ of the $\mathrm{M}$ theory configuration to zero size has led to a type $\mathrm{IIB}$ orientifold $\mathbb{R}^{3,1}\times \frac{\mathbb{T}^2_{4,5}}{\Omega (-1)^{\mathrm{F}_{\mathrm{L}}}\mathcal{I}_{4}\mathcal{I}_{5}}\times \mathcal{M}_{4}$. Here, $\Omega$ is the world-sheet parity transformation operator which reverses the orientation of the world sheet coordinate $$\Omega: \sigma \rightarrow -\sigma.$$ It exchanges the left and right moving modes of the worldsheet fields. Since the left moving and right moving modes of the world sheet fields of type IIA theory have opposite space time chirality, the $\mathrm{IIA}$ theory is chiral on world sheet. In other words, $\Omega$ is not a symmetry of type $\mathrm{IIA}$ theory. However, the simultaneous identification of the space time coordinate $x \sim -x$ via the $\mathbb{Z}_2$ operation restores the symmetry of the theory. For type $\mathrm{IIB}$ theory both the left moving and right moving modes transform under the same space time chirality and hence $\Omega$ is a world sheet symmetry of type $\mathrm{IIB}$ theory. Further, the operator $(-1)^{\mathrm{F_{L}}}$ changes the sign of all the left moving Ramond sector states. The type $\mathrm{IIB}$ theory on $\mathbb{T}^{2}_{4,5}$ modded out by the $\mathbb{Z}_{2}$ operation $\Omega (-1)^{\mathrm{F_{L}}} \mathcal{I}_{4}\mathcal{I}_{5} $ is an orientifold and this is related to type $\mathrm{I}$ theory via $\mathrm{T}$ dualities \cite{Sen_1996}. 
 
 Note that, the type $\mathrm{IIA}$ coupling constant 
\begin{align}
    g_{s}&=\frac{\mathrm{H}(y)}{a(t)}\label{TypeIIAcoUpliNg}
\end{align}
can no longer be equated to (\ref{IIAcouPlingdeSitterIIB}). Instead it will be determined by demanding de Sitter isometry at the final stage of our duality sequence i.e. heterotic $\mathrm{SO(32)}$ theory for the case at consideration.

Because of the orientifolding operation of the form  $\mathcal{I}_{4}\mathcal{I}_{5}$ on the type $\mathrm{IIB}$ solitonic configuration we can only switch on \textbf{NS-NS} and \textbf{R-R} fields with one leg along the toroidal directions $\mathbb{T}^{2}_{4,5}$ and another along $\mathcal{M}_{4}$. We denote these fields as $\mathcal{B}^{(1)}_{\alpha m}$ and ${\mathcal{B}}^{(2)}_{\alpha m}$ respectively where $y^{\alpha}\in \frac{\textstyle \mathbb{T}^{2}_{4,5}}{\textstyle \mathcal{I}_{4}\mathcal{I}_{5}}$ and $y^{m}\in \mathcal{M}_{4}$. There are $\mathrm{O}7$ planes and $\mathrm{D}7$ branes in $\mathrm{IIB}$ theory located at the four fixed points of the toroidal orbifold and extended along $\mathbb{R}^{3,1}\times \mathcal{M}_{4}$ directions. 

[$\bullet$]  A $\mathrm{T}$ duality along $y^{4}$ of $\mathrm{IIB}$ metric (\ref{tyPeIIBMetrIc}) leads to type $\mathrm{IIA}$ theory on  $\mathbb{R}^{3,1}\times \frac{\textstyle \mathbf{S}^{1}_{5}}{\textstyle \Omega \mathcal{I}_{5}}\times \left(\mathcal{M}_{4}\rtimes \mathbf{S}^{1}_{4}\right)$ $\left(\text{or, in other words type}~\mathrm{IA}~\text{theory on}~\mathbb{R}^{3,1}\times \frac{\textstyle \mathbf{S}^{1}_{5}}{ \textstyle\mathcal{I}_{5}}\times \left(\mathcal{M}_{4}\rtimes \mathbf{S}^{1}_{4}\right) \right)$ with the following metric configuration
\begin{align}
    d{s}^{2}_{\mathrm{IIA}}&= g_{s}^{-2}\left( -dt^{2}+\sum_{i=1}^{2}(dx^{i})^2 +(dx^{3})^{2}\right)+\frac{1}{\mathrm{H}^{2}(y)\mathrm{F}_{1}(t)}\left( dy^4\right)^2+\mathrm{H}^{2}(y)\mathrm{F}_{1}(t)\left( dy^{5}\right)^2\nonumber\\
    &+\left[\mathrm{H}^{2}(y)\mathrm{F}_{2}(t)g_{mn}+\frac{\mathcal{B}^{(1)}_{4m}\mathcal{B}^{(1)}_{4n}}{\mathrm{H}^{2}(y)\mathrm{F}_{1}(t)}\right]dy^{m}dy^{n}+\frac{2{\mathcal{B}}^{(1)}_{4m}}{\mathrm{H}^{2}(y)\mathrm{F}_{1}(t)}dy^{4}dy^{m}\nonumber\\
    &=g_{s}^{-2}\left(-dt^2+\sum_{i=1}^{2}(dx^{i})^2+(dx^{3})^{2} \right)+\frac{1}{\mathrm{H}^{2}(y)\mathrm{F}_{1}(t)}\left( dy^{4}+\mathcal{B}^{(1)4}_{~~~m}dy^{m}\right)^2\nonumber\\
    &~~~~~~~~~~+\mathrm{H}^{2}(y)\mathrm{F}_{1}(t)(dy^{5})^{2}+\mathrm{H}^{2}(y)\mathrm{F}_{2}(t)g_{mn}dy^{m}dy^{n},\label{SecIIAcoNfiG}
\end{align}
and the corresponding coupling constant is given as
\begin{align}
    \tilde{g}_{s}&=\frac{1}{\mathrm{H}(y)\sqrt{\mathrm{F}_{1}(t)}}.\label{TypEIIAcc}
\end{align}
The \textbf{NS-NS} fluxes on this theory are $\mathrm{B}^{}_{5m}=\mathcal{B}^{(1)}_{5m}$ and the \textbf{R-R} fluxes are ${A}_{m}, ~{C}_{45m}$. There are $\mathrm{O}8$ planes and $\mathrm{D}8$ branes located at the two fixed points of the orbifold $\frac{\mathbf{S}^1_{5}}{\Omega \mathcal{I}_{5}}$ and these are extended  along $\mathbb{R}^{3,1}\times \mathcal{M}_{4}\times \tilde{\mathbf{S}}^{1}_{4}$.

[$\bullet$] 
A further $\mathrm{T}$ duality along $y^{5}$ leads to type $\mathrm{I}$ theory on $\mathbb{R}^{3,1}\times   \mathcal{M}_{4}\rtimes \left( \widetilde{\mathbf{S}}_{4}^{1}\times \widetilde{\mathbf{S}}_{5}^{1}\right)$, metric configuration of the same being 
\begin{align}
&ds^{2}_{\mathrm{I}}=g_{s}^{-2}\left(-dt^2+\sum_{i=1}^{2}\left( dx^{i}\right)^2+\left( dx^3\right)^2 \right)+\frac{1}{\mathrm{H}^2(y)\mathrm{F}_{1}(t)}\left( dy^4\right)^2\nonumber\\
&~+\frac{1}{\mathrm{H}^{2}(y)\mathrm{F}_{1}(t)}\left(dy^5 \right)^2+\left[\mathrm{H}^2(y)\mathrm{F}_{2}(t)g_{mn}+\frac{\mathcal{B}^{(1)}_{4m}\mathcal{B}^{(1)}_{4n}}{\mathrm{H}^2(y)\mathrm{F}_{1}(t)}+\frac{\mathcal{B}^{(1)}_{5m}\mathcal{B}^{(1)}_{5n}}{\mathrm{H}^2(y)\mathrm{F}_{1}(t)}\right]dy^m dy^n\nonumber\\
    &~~~~~~~~~~+\frac{2\mathcal{B}^{(1)}_{4m}}{\mathrm{H}^2(y)\mathrm{F}_{1}(t)}dy^4dy^m+\frac{2\mathcal{B}^{(1)}_{5m}}{\mathrm{H}^2(y)\mathrm{F}_{1}(t)}dy^5dy^m\nonumber\\
    &=g_{s}^{-2}\left(-dt^2+\sum_{i=1}^{2}(dx^{i})^2+(dx^{3})^{2} \right)+\frac{1}{\mathrm{H}^{2}(y)\mathrm{F}_{1}(t)}\left( dy^{4}+\mathcal{B}^{(1)4}_{~~~m}dy^{m}\right)^2\nonumber\\
    & ~~~~~~+\frac{1}{\mathrm{H}^2(y)\mathrm{F}_{1}(t)}\left( dy^5+\mathcal{B}^{(1)5}_{~~~m}dy^m\right)^2+\mathrm{H}^{2}(y)\mathrm{F}_{2}(t)g_{mn}dy^{m}dy^{n},\label{typeImetriC}
\end{align}
and the type $\mathrm{I}$ coupling constant is given by 
\begin{align}
g^{\mathrm{I}}&=\frac{1}{\mathrm{H}^{2}(y)\mathrm{F}_{1}(t)}.\label{TypeICouplinG}
\end{align}

{This theory has $\tilde{\mathcal{B}}^{(2)}_{\alpha m}$ \textbf{R-R} fluxes  as well as space-time filling orientifold plane and D-brane.}

[$\bullet$] The $\mathrm{S}$ dual of the type $\mathrm{I}$ theory is our designated heterotic $\mathrm{SO(32)}$ theory 
\begin{align}
ds^{2}_{\mathrm{Het~SO}(32)}&=g_{s}^{-2}\mathrm{H}^2(y)\mathrm{F}_{1}(t)\left(-dt^2+\sum_{i}\left(dx^{i}\right)^2+\left( dx^3\right)^2 \right)+\mathrm{H}^{4}(y)\mathrm{F}_{1}(t)\mathrm{F}_{2}(t)g_{mn}dy^{m}dy^{n} \nonumber\\
    &\quad\quad+\delta_{\alpha\beta}\left(dy^{\alpha}+\mathcal{B}^{(1)\alpha}_{~~~m} dy^{m}\right)\left(dy^{\beta}+\mathcal{B}^{(1)\beta}_{~~~n} dy^{n}\right)\nonumber\\
    &=\mathrm{F}_{1}(t)a^2(t)\left(-dt^2+\sum_{i=1}^3\left(dx^{i}\right)^2 \right) +\mathrm{H}^{4}(y)\mathrm{F}_{1}(t)\mathrm{F}_{2}(t)g_{mn}dy^{m}dy^{n}\nonumber\\
&\quad\quad+\delta_{\alpha\beta}\left(dy^{\alpha}+\mathcal{B}^{(1)\alpha}_{~~~m} dy^{m}\right)\left(dy^{\beta}+\mathcal{B}^{(1)\beta}_{~~~n} dy^{n}\right),\label{HeTerotiCSO(32)MetrIc}
\end{align}
with the heterotic coupling 
\begin{align}
g_{\text{Het}}&=\frac{\textstyle 1}{\textstyle g_{\mathrm{I}}}=\mathrm{H}^{2}(y)\mathrm{F}_{1}(t).\label{HeterCoupling}
\end{align}
The fluxes in this theory are $B_{\alpha m}$. Because of the presence of Wilson lines the gauge group of this theory is not $\mathrm{SO}(32)$ but is broken down to four copies of $\mathrm{SO}(8)$ group \cite{Sen_1996}.

Clearly, if we want to realise de Sitter isometry in \textit{flat slicing} on the heterotic theory, we have to demand 
\begin{align}
    \mathrm{F}_{1}(t)a^{2}(t)=\mathrm{F}_{1}(t)g
_{s}^{-2} \mathrm{H}^2(y)&=\frac{1}{\Lambda t^{2}}~~~\text{with}~~-\infty \le t <0 ,\label{dSIsometryHetSO}
\end{align}
along with 
\begin{align}
    \mathrm{F}_{1}(t)\mathrm{F}_{2}(t)=1, \label{ConstrF1F2}
\end{align}
to ensure time independence of the six dimensional internal space. Note that the internal manifold in \eqref{HeTerotiCSO(32)MetrIc} also contains the time dependent two forms $\mathcal{B}^{(1)}_{\alpha m}$ but they do not contribute to its volume for reasons as elaborated in Appendix \ref{sec:AppendixD}. Correspondingly, the type $\mathrm{IIA}$ coupling constant (\ref{TypeIIAcoUpliNg}) becomes equal to 
\begin{align}
    g_{s}&=\mathrm{H}(y)|t|\sqrt{\Lambda\mathrm{F}_{1}(t)},\label{TYpe IIA HetSO(32)}
\end{align}
which also depends on the time dependent warp factor $\mathrm{F}_{1}(t)$ unlike the case of realising de Sitter in type IIB theory (\ref{IIAcouPlingdeSitterIIB}). 
\begin{figure}[ht!]
\centering
\begin{tikzpicture}[scale=.85,font=\small,thick]
\node at (-5.1,0) [draw, rounded corners,fill=teal!10,
    minimum width=4.5cm,
    minimum height=1cm](block1) { \textcolor{violet}{
$\substack{ \textstyle \mathrm{M}~\textbf{theory} \\ \\\textstyle \textbf{on}\\ \textstyle\textbf{$\mathbb{R}^{2,1}\times \frac{\mathbb{T}_{4,5}^{2}}{\mathcal{I}_{4}\mathcal{I}_{5}}\times \mathcal{M}_{4}\times \frac{\mathbb{T}^{2}_{3,11}}{\mathcal{I}_{3}\mathcal{I}_{11}}$}}$
}};
\node at (5.1,0) [draw, rounded rectangle, fill=teal!10,
    minimum width=2.5cm,
    minimum height=1cm] (block2) { \textcolor{violet}{ 
$\substack{\textstyle\textbf{Type}~\mathrm{IIA}~\textbf{theory}\\ \\ \textstyle\textbf{on} \\\textstyle \mathbb{R}^{2,1}\times  \frac{\mathbb{T}_{4,5}^{2}}{\mathcal{I}_{4}\mathcal{I}_{5}}\times \mathcal{M}_{4}\times \frac{\mathbf{S}^{1}_{3}}{\Omega\mathcal{I}_{3}}}$
}};
\node at (0,-3.3) [draw, rounded corners, fill=teal!10,
    minimum width=2.5cm,
    minimum height=1.5cm] (block3) { \textcolor{violet}{
$\substack{\textstyle\textbf{Type}~\mathrm{IIB}~\textbf{theory}\\ \\ \textstyle \textbf{on} \\\textstyle \mathbb{R}^{3,1}\times  \frac{\mathbb{T}_{4,5}^{2}}{\Omega(-1)^{\mathrm{F_{L}}}\mathcal{I}_{4}\mathcal{I}_{5}}\times \mathcal{M}_{4}
}$}};
\node at (0,-6.6) [draw, rounded corners, fill=teal!10,
    minimum width=2.5cm,
    minimum height=1.5cm] (block4) { \textcolor{violet}{
$\substack{\textstyle\textbf{Type}~\mathrm{IIA}~\textbf{theory}\\ \\ \textstyle \textbf{on} \\\textstyle \mathbb{R}^{3,1}\times  \frac{\mathbf{S}^{1}_{5}}{\Omega\mathcal{I}_{5}}\times \left(\mathcal{M}_{4}\rtimes \widetilde{\mathbf{S}}_{4}^{1} \right)
}$}};
\node at (5.1, -9.9) [draw, rounded rectangle, fill=teal!10,
    minimum width=2.5cm,
    minimum height=1.5cm] (block5) { \textcolor{violet}{
$\substack{\textstyle\textbf{Type}~\mathrm{I}~\textbf{theory}\\ \\ \textstyle \textbf{on} \\\textstyle \mathbb{R}^{3,1}\times   \mathcal{M}_{4}\rtimes \left( \widetilde{\mathbf{S}}_{4}^{1}\times \widetilde{\mathbf{S}}_{5}^{1} \right)
}$}};
\node at (-5.1, -9.9) [draw, rounded corners, fill=teal!10,
    minimum width=2.5cm,
    minimum height=1.5cm] (block6) { \textcolor{violet}{
$\substack{\textstyle\textbf{Heterotic}~\mathrm{SO(32)}~\textbf{theory}\\ \\ \textstyle \textbf{on} \\ \textstyle \mathbb{R}^{3,1}\times   \mathcal{M}_{4}\rtimes \left( \widetilde{\mathbf{S}}_{4}^{1}\times \widetilde{\mathbf{S}}_{5}^{1} \right)
}$}};
\draw[-latex] 
(block1) -- (block2)
(block2) edge (block3)
    (block3) edge (block4)
    (block4) edge (block5)
    (block5) edge (block6);
\node at (0,.5) {${\substack{\textstyle\text{Dimensional reduction } \\ \textstyle \text{along $x^{11}$} }}$} ; 
\node at (3.7,-2.0) { $\substack{\textstyle\mathrm{T}~\text{duality}\\ \textstyle \text{along}~x^3}$};
\node at (1.5,-4.9) { $\substack{\textstyle\mathrm{T}~\text{duality}\\ \textstyle \text{along}~y^4}$};
\node at (3.9,-8.1) { $\substack{\textstyle\mathrm{T}~\text{duality}\\ \textstyle \text{along}~y^5}$};
\node at (0,-9.4) {$\mathrm{S}$ duality};
\end{tikzpicture}
    \caption{Duality sequence from $\mathrm{M}$ theory (\ref{MthEorYmeTric}) to heterotic $\mathrm{SO(32)}$ theory (\ref{HeTerotiCSO(32)MetrIc})}
    \label{Het SO(32) tree diagram}
\end{figure}

\vspace{5pt}
\subsubsection{Dynamical duality sequence:}

The above duality sequence leading to heterotic $\mathrm{SO(32)}$ theory can in fact proceed dynamically starting with the $\mathrm{M}$ theory set up. To elaborate this statement we consider the scalings of the temporal warp factors as defined in Eq. (\ref{ScalingOfFi's})  with the dominant scalings $\alpha_{o}$ and $\beta_{o}$ satisfying the similar upper bound as in (\ref{UbdofdomFactor})
\begin{align}
    \alpha_{o}<+\frac{2}{3},~~~\beta_{o}<+\frac{2}{3}.
\end{align}
Just to reiterate, the reason for such upper bound is that we require only the $\mathrm{M}$ theory torus along 3,11  to shrink to zero size taking us to the dual type $\mathrm{IIB}$ configuration.  However, we no longer want the type $\mathrm{IIB}$ theory to be the final stage of our duality sequence, instead two further $\mathrm{T}$ dualities should take place along both the cycles of the toroidal orbifold $\frac{\mathrm{T}^2_{4,5}}{\mathcal{I}_4\mathcal{I}_{5}}$ of type $\mathrm{IIB}$ configuration. Hence $\beta_{o}$ must be greater than zero, implying it lies in the range 
\begin{align}
    0<\beta_{o}<+\frac{2}{3},\label{beta0Range}
\end{align}
and the condition in (\ref{ConstrF1F2}) implies $
    \alpha_{o}=-\beta_{o}$. With the above choice of $\beta_{o}$ (\ref{beta0Range}), the temporal warp factor $\mathrm{F}_{1}(t)$ appearing along the toroidal orbifold $\frac{\textstyle \mathbb{T}^{2}_{4,5}}{\textstyle \mathcal{I}_{4}\mathcal{I}_{5}}$ of type $\mathrm{IIB}$ metric (\ref{tyPeIIBMetrIc}) tends to zero in the late time limit as $t\rightarrow 0^-$. Consequently, two $\mathrm{T}$ dualities along these two cycles take place dynamically and we obtain type $\mathrm{I}$ theory with coupling (\ref{TypeICouplinG}). The strongly coupled type $\mathrm{I}$ theory goes to the $\mathrm{S}$ dual picture and gives weakly coupled (\ref{HeterCoupling}) heterotic $\mathrm{SO(32)}$ theory with de Sitter isometry (\ref{dSIsometryHetSO}) and time independent internal space (\ref{ConstrF1F2}) in the late time limit. 

Now, from Eq. (\ref{ScalingOfFi's}) the generic scaling of the warp factor $\mathrm{F}_{1}(t)$ upon substituting (\ref{TYpe IIA HetSO(32)}) takes the form 
\begin{align}
    \mathrm{F}_1(t)&=(t\sqrt{\Lambda\mathrm{F}_1(t)})^{\beta_o}\sum_{k=0}^{\infty} \mathrm{A}_{k}(t\sqrt{\Lambda \mathrm{F}_1(t)})^{2k/3}.
\end{align}
Since the coefficients $\mathrm{A}_k$ can only be determined by solving the equations of motion, which is a highly non-trivial exercise beyond the scope of this thesis, we shall restrict ourselves to a simple case. We take $\mathrm{A}_0>>\mathrm{A}_k,~\forall~k$ and set $\mathrm{A}_0=1$ to obtain:
\begin{align}
    \mathrm{F}_1(t)&=(\Lambda t^2)^{\frac{\beta_o}{2-\beta_o}}.\label{VAlF1inHetSO32Thry}
\end{align}
Now, identifying the result obtained for the expectation value of the metric operator in Eq.~\eqref{BORELRSummed} with the M-theory metric coefficient in Eq.~\eqref{MthHetSO32}, together with the appropriate value of the string coupling $g_s$ \eqref{TYpe IIA HetSO(32)}, we obtain
\begin{align}
    \left[\frac{1}{g^{1/l}} \int_{0}^{\infty} d\mathrm{S}\exp\left(-\frac{\mathrm{S}}{g^{1/l}} \right) \frac{1}{1-\mathcal{A}\mathrm{S}^{l}}\right]_{\mathrm{P.V.}}\int  &d^{11}k \frac{\overline{\alpha}_{\rm \mu\nu}(k)}{a(k)}\mathbf{Re}\left(\psi_{\mathbf{k}}(\mathrm{X})e^{-ik_0 t}\right)\nonumber\\
      &=(\sqrt{\Lambda \mathrm{F}_1(t)}\mathrm{H}(y)|t|)^{-\frac 8 3}\eta_{\mu\nu}\nonumber\\ 
      &=\Lambda^{-\frac{8}{3(2-\beta_o)}}~\mathrm{H}^{-\frac{8}{3}}(y)~|t|^{-\frac{16}{3(2-\beta_o)}}~\eta_{\mu\nu},
\end{align}
where $\eta_{\mu\nu}=\mathrm{diag}(-1,1,1).$ To arrive at the last step we have substituted the value of $\mathrm{F}_1(t)$ from Eq. \eqref{VAlF1inHetSO32Thry}. 
Therefore, the form of the cosmological constant is given by 
\begin{align}
    \frac{1}{\Lambda^{\kappa}}&=      \left[\frac{1}{g^{1/l}} \int_{0}^{\infty} d\mathrm{S}\exp\left(-\frac{\mathrm{S}}{g^{1/l}} \right) \frac{1}{1-\mathcal{A}\mathrm{S}^{l}}\right]_{\mathrm{P.V.}}\quad\quad \text{where},\quad \kappa=\frac{8}{3(2-\beta_o)},
\end{align}
whereas, the temporal and $y\in \mathcal{M}_4$ dependent pieces are obtained as   
\begin{align}
    \int d^{11}k\frac{\overline{\alpha}_{\mu\nu}(k)}{a(k)}\mathbf{Re}\left(\psi_{\mathbf{k}}(\mathrm{X})e^{-ik_0t}\right)&=\mathrm{H}^{-\frac{8}{3}}(y)~|t|^{-\frac{16}{3(2-\beta_o)}}\eta_{\mu\nu},\label{TheFModeYandt}
\end{align}
where $\mathrm{X}\in \mathbb{R}^{2}\times \frac{\mathbb{T}^2_{4,5}}{\mathcal{I}_4\mathcal{I}_5}\times \mathcal{M}_4\times \frac{\mathbb{T}^2_{3,11}}{\mathcal{I}_3\mathcal{I}_{11}}$ denotes the spatial and the internal space coordinates of the M-theory background (\ref{MthHetSO32}).
An inverse Fourier transformation of this equation can be performed to determined $\overline{\alpha}_{\mu\nu}(k)$\cite{Alexander:2023qym}.

Now, we will elaborate on the construction of de Sitter GS state in the Einstein frame, and compare with the string frame results discussed so far. Note that, in the case of realizing de Sitter space as a GS state in type IIB theory (\ref{tyPeIIBMetrIc}), the Einstein frame metric coincides with the string frame metric. This is because the coupling constant of the IIB theory is $g_b=1$.
\subsection{The string frame versus the Einstein frame}
\label{StrFraMEinstFramSO32}
In ten dimensions, the Einstein frame metric is related to the string frame metric as follows
\begin{align}
\mathrm{g}_{\mu\nu}^{(\mathrm{E})}&=e^{-\frac{\phi}{2}}\mathrm{g}_{\mu\nu}^{(S)},\label{EinsStriFRAme}
\end{align}
where $e^{\phi}$ is the coupling of the theory at consideration. Therefore, in Einstein frame, the metric configuration of heterotic $\mathrm{SO}(32)$ theory (\ref{HeTerotiCSO(32)MetrIc}) becomes\cite{Archana_CC2026}
\begin{align}
ds^2_{\mathrm{Het~SO(32),~E}}&=(\mathrm{H}^2(y)\mathrm{F}_{1}(t))^{-1/2}\bigg(\mathrm{F}_{1}(t)a^2(t)\big(-dt^2+\sum_{i=1}^2\left(dx^{i}\right)^2+\left( dx^3\right)^2 \big) \nonumber\\
    ~+\mathrm{H}^{4}&(y)\mathrm{F}_{1}(t)\mathrm{F}_{2}(t)g_{mn}dy^{m}dy^{n}+\delta_{\alpha\beta}\left(dy^{\alpha}+\mathcal{B}^{(1)\alpha}_{~~~m} dy^{m}\right)\left(dy^{\beta}+\mathcal{B}^{(1)\beta}_{~~~n} dy^{n}\right)\bigg)\nonumber\\
    &=\frac{\sqrt{\mathrm{F}_{1}(t)}a^2(t)}{\mathrm{H}(y)}\left(-dt^2+\sum_{i=1}^3\left(dx^{i}\right)^2 \right)+\mathrm{H}^{3}(y)\sqrt{\mathrm{F}_{1}(t)}\mathrm{F}_{2}(t)g_{mn}dy^{m}dy^{n} \nonumber\\
    ~&~~~~~~~+\frac{1}{\mathrm{H}(y)\sqrt{\mathrm{F}_{1}(t)}}\delta_{\alpha\beta}\left(dy^{\alpha}+\mathcal{B}^{(1)\alpha}_{~~~m} dy^{m}\right)\left(dy^{\beta}+\mathcal{B}^{(1)\beta}_{~~~n} dy^{n}\right),\label{EinsFrameSO32}
\end{align}
where we have substituted $e^{\phi}=g_{\rm Het}$ (\ref{HeterCoupling}). 
Thus, if we demand de Sitter isometry in Einstein frame then the following condition must hold
\begin{align}
\sqrt{\mathrm{F}_{1}(t)}a^2(t)\equiv g_{s}^{-2}\mathrm{H}^2(y)\sqrt{\mathrm{F}_{1}(t)}&=\frac{1}{\Lambda t^2}.\label{CCinEinsFrameHetSO32thr}
\end{align}
Moreover, the time independence of the volume of the six dimensional internal space $\sqrt{\mathrm{F}_{1}(t)\mathrm{F}_{2}^4(t)}$, implies the dominant scalings of $\mathrm{F}_{1}$ and $\mathrm{F}_{2}$ (\ref{ScalingOfFi's}) must satisfy $\alpha_{o}=-\frac{\beta_o}{4}$ while maintaining $0<\beta_o<\frac{2}{3}$ for the duality sequence to proceed appropriately. However, the volume of the internal four manifold $\sqrt{\mathrm{F}_1^2(t)\mathrm{F}_2^4(t)}=\left(\frac{g_{s}}{\mathrm{H}(y)}\right)^{\beta_o+2\alpha_o}=\left(\frac{g_{s}}{\mathrm{H}(y)}\right)^{\beta_o/2}$ implies that $\mathcal{M}_{4}$ shrinks to zero size as $t\rightarrow 0^-$ showing late time singularity whereas the volume of the toroidal manifold being $\frac{1}{\sqrt{\mathrm{F}_{1}(t)}}=\left(\frac{g_{s}}{\mathrm{H}(y)}\right)^{-\beta_o/2}$, it grows in the late time limit. Thus we can not get a  four dimensional de Sitter configuration in both the string frame and Einstein frame simultaneously.  To summarize, there are two possibilities: 
\begin{itemize}
    \item[$\bullet$] We can either have de Sitter isometry in string frame (\ref{HeTerotiCSO(32)MetrIc}) 
    \begin{align}
&g_{s}^{-2}\mathrm{H}^2(y)\mathrm{F}_{1}(t)=\frac{1}{\Lambda t^2}\implies \left(\frac{g_s}{\mathrm{H}(y)}\right)=(\Lambda t^2)^{\frac{1}{2-\beta_o}},\label{StringCCInSF}  
\end{align}
    with time-independent six dimensional internal space and correspondingly time independent four dimensional Newton's constant 
\begin{align}
    \mathrm{F}_{1}(t)\mathrm{F}_{2}(t)&=1~~~~\implies ~~\alpha_{o}=-\beta_{o},~~\mathrm{with}~~0<\beta_o<\frac{2}{3}.\label{TimeIndpIntSp}
\end{align}
    
However, it is dual to a non de Sitter configuration in Einstein frame (\ref{EinsFrameSO32}) since 
    \begin{align}
g_{s}^{-2}\mathrm{H}^2(y)\sqrt{\mathrm{F}_{1}(t)} \neq \frac{1}{\Lambda t^2}.
    \end{align}
    Moreover, in Einstein frame the six dimensional internal space and hence the four dimensional Newton’s constant become time-dependent. In fact, the internal six dimensional manifold shows late time decompactification as
    \begin{align}
        \sqrt{\mathrm{F}_{1}(t)\mathrm{F}_{2}^4(t)}=\mathrm{F}_{2}^{{3}/ {2}}(t)=& \left(\frac{g_{s}}{\mathrm{H}(y)}\right)^{-\frac{3}{2}\beta_o}\quad\quad (\mathrm{Using~Eq.}~(\ref{TimeIndpIntSp}))\nonumber\\
        &\rightarrow\infty ,\quad \text{as}\quad t\rightarrow 0.
    \end{align}
    \item Or, we can have de Sitter isometry in Einstein frame i.e.
    \begin{align}
g_{s}^{-2}\mathrm{H}^2(y)\sqrt{\mathrm{F}_{1}(t)}=\frac{1}{\Lambda t^2}\implies \left(\frac{g_s}{\mathrm{H}(y)}\right)=(\Lambda t^2)^{\frac{2}{4-\beta_o}},\label{EFStringC}
    \end{align}
    with time-independent six dimensional internal space and correspondingly time independent four dimensional Newton’s constant 
    \begin{align}
        {\mathrm{F}_{1}(t)\mathrm{F}^4_{2}(t)}=1\implies \alpha_o=-\frac{\beta_o}{4},~~\mathrm{with}~~0<\beta_o<\frac{2}{3}\label{TimeIndpEinsFrm}
    \end{align}
    along with the internal four manifold $\mathcal{M}_{4}$ showing late time singularity
    \begin{align}
\sqrt{\mathrm{F}_{1}^2(t)\mathrm{F}_{2}^4(t)}=\left(\frac{g_{s}}{\mathrm{H}(y)}\right)^{\beta_o/2}\rightarrow 0 \label{VolOfM4Mani}
    \end{align}
    and the internal two manifold $\mathcal{M}_{2}$ showing late time decompactification 
\begin{align}
\frac{1}{\sqrt{\mathrm{F}_{1}(t)}}=\left(\frac{g_{s}}{\mathrm{H}(y)}\right)^{-\beta_o/2}\rightarrow \infty.
\end{align}
This is dual to a non-de Sitter configuration in string frame since 
\begin{align}
g_{s}^{-2}\mathrm{H}^2(y)\mathrm{F}_1(t)\neq \frac{1}{\Lambda t^2}.
\end{align}
Moreover, in string frame the four dimensional internal manifold $\mathcal{M}_{4}$ shows late time singularity
    \begin{align}
        &\sqrt{\mathrm{F}_{1}(t)\mathrm{F}_{2}(t)}\nonumber\\
        &=\mathrm{F}_{2}^{-3/2}(t)=\left(\frac{g_{s}}{\mathrm{H}(y)}\right)^{\frac{3}{8}\beta_o}\rightarrow 0,~~~(\mathrm{Using ~Eq.}~(\ref{TimeIndpEinsFrm})). 
    \end{align}
\end{itemize}
From the above discussion it seems that string frame is preferred (as it has no pathologies) over the Einstein frame when we want to realise de Sitter isomtery in the heterotic SO(32) string theory. On the other hand, suppose we demand
\begin{align}
{\rm F}_1(t) = \begin{cases} \left(\Lambda t^2\right)^{ \frac{v\beta_o} { 2v -\beta_o}},~~~~~~ -{ \frac{1}{ \sqrt{\Lambda}}} < t \leq -\epsilon \\
~~~ \\
~~ 1~~,~~~~~~~~~~~~~~~ -\epsilon <  t < 0
\end{cases}\label{F1inHetSo32Th}
\end{align}
where $v=1,2$ for the string and the Einstein frame respectively, $\epsilon$ is a temporal point. Then in the regime $-\epsilon <t<0$, $\mathrm{F}_{1}(t)$ and hence $\mathrm{F}_{2}(t)$ become a constant, thus we get a consistent de Sitter isometry in either frame. To implement Eq. (\ref{F1inHetSo32Th}) we define 
\begin{align}
&\mathrm{F}_1(t)\equiv(\Lambda t^2)^{\frac{v\beta(t)}{2v-\beta(t)}},~~~
\mathrm{where}~~\beta(t)=\begin{cases}
       {\beta_o},~~~~~~ -{ \frac{1}{ \sqrt{\Lambda}}} < t \leq -\epsilon \\
 0,~~~~~~~~~~ -\epsilon < t < 0
    \end{cases},\label{F1(t)withbeta(t)}
\end{align}
with $0<\beta_o<\frac{2}{3}$. Correspondingly, the IIA coupling constant in string frame (\ref{StringCCInSF}) and Einstein frame (\ref{EFStringC}) can be expressed as 
\begin{align}
    \frac{g_{s}}{\mathrm{H}(y)}&=\left(\sqrt{\Lambda} |t|\right)^{\frac{2v}{2v-\beta(t)}}\label{StringCCSFEF}
\end{align}
where $v=1,2$ for the string frame, the Einstein frame as defined earlier.  The value of $\epsilon$ can be estimated by demanding that at $|t|=\epsilon$, the Einstein frame volume of the four manifold $\mathcal{M}_{4}$ (\ref{VolOfM4Mani}) attains size $\mathrm{V}_{1}>>\alpha^{\prime 2}$. If  $\mathrm{V}_o(-1/\sqrt{\Lambda})$ is its stabilized volume at $|t|= 1/\sqrt{\Lambda}$ then we obtain
\begin{align}
\frac{\left(\frac{g_s}{{\rm H}(y)}\right)^{\beta_o/2}\bigg|_{|t|=\epsilon}}{\left(\frac{g_s}{{\rm H}(y)}\right)^{\beta_o/2}\bigg|_{|t|= 1/\sqrt{\Lambda}}}
    &=\frac{\mathrm{V}_1}{\mathrm{V}_o(-1/\sqrt{\Lambda})}\nonumber\\
    \implies \left(\sqrt{\Lambda} \epsilon\right)^{\frac{2\beta_o}{4-\beta_o}}&=\frac{\rm V_1}{{\mathrm{V}}_o(-1/\sqrt{\Lambda})}~~~~~~~(\mathrm{Using~Eq.}~ (\ref{StringCCSFEF})~\mathrm{with}~v=2)\nonumber\\
    \implies \epsilon=\frac{1}{\sqrt{\Lambda}}&\left[\frac{\mathrm{V}_1}{\mathrm{V}_o(-1/\sqrt{\Lambda})}\right]^{\frac{4-\beta_o}{2\beta_o}}.\label{ValueofEpsilonHetSO}
\end{align}
The other time dependent warp factor is given by 
\begin{align}
    \mathrm{F}_{2}(t)&=\left( \frac{g_{s}}{\mathrm{H}(y)}\right)^{\alpha(t)}=\left(\Lambda t^2\right)^{\frac{v\alpha(t)}{2v-\beta(t)}}\label{F2tinEinstFrSO(32)}
\end{align}
where 
\begin{align}
    \alpha(t)&=-\frac{\beta(t)}{v^2}.
\end{align}
Our next goal is to investigate whether the GS state in heterotic ${\rm SO}(32)$  theory (\ref{HeTerotiCSO(32)MetrIc}),(\ref{EinsFrameSO32}) satisfy the effective field theory constraint. This is equivalent to satisfying the null energy condition as elaborated on \S\ref{sec:NECIIB}. 
\section{NEC in heterotic \texorpdfstring{$\mathrm{SO}(32)$}{SO(32)} string theory}
\label{sec:NEC} 
We now examine the null energy condition in heterotic $\mathrm{SO}(32)$ string theory, following the approach developed in \S\ref{sec:NECIIB}. From the string frame metric (\ref{HeTerotiCSO(32)MetrIc}), the non-compact $3+1$ dimensional spacetime metric can be read off as
\begin{align}
\label{SO32HetMetrICStrinG}
g_{\mu\nu}^{\mathrm{Het~SO}(32)}&=\mathrm{\mathrm{F}_{1}(t)}a^2(t)\eta_{\mu\nu},
\end{align}
where $\mu,\nu=0,1,2,3$ and $\eta_{\mu\nu}=\mathrm{diag}(-1,1,1,1).$ Similar to the  Eq. (\ref{TypeIIBNEcConDitIOn}), we now demand the above metric coefficient to be
\begin{align}
\mathrm{F}_1(t)a^2(t)&=(\Lambda t^2)^n.
\end{align}
where $n$ is analogous to that appearing in (\ref{ScaleFacFLRW}). Substituting the dominant contribution of $\mathrm{F}_1(t)$ from Eq. (\ref{ScalingOfFi's}) and the scale factor $a(t)$ from Eq. (\ref{TypeIIAcoUpliNg}) we get 
\begin{align}
    \left(\frac{g_s}{\mathrm{H}(y)}\right)^{\beta_{o}-2}\equiv   \left({\overline{g}_s}\right)^{\beta_{o}-2}&=(\Lambda t^2)^n
    \implies {\overline{g}_s}=\left(\Lambda t^2\right)^{\frac{n}{\beta_o -2}}.\label{SFrameCoup}
\end{align}
Recall in Einstein frame the metric along the non-compact $3+1$ dimensional spacetime  (\ref{EinsFrameSO32})  is given as  
\begin{align}
g_{\mu\nu}^{\mathrm{Het~SO}(32),~\mathrm{E}}&=\frac{\sqrt{\mathrm{F}_{1}(t)}a^2(t)}{\mathrm{H}(y)}\eta_{\mu\nu}.
\end{align}
As a result, we now demand the time-dependent part to be equated to
\begin{align}  \sqrt{\mathrm{F}_1(t) }a^2(t)&=(\Lambda t^2)^n\implies \left(\overline{g}_s\right)^{\frac{\beta_o }{2}-2}  =(\Lambda t^2)^n\implies {\overline{g}_s}=(\Lambda t^2)^{\frac{2n}{\beta_o -4}}.\label{EinFrameCoup}
\end{align}
Combining Eqs. (\ref{SFrameCoup}) and (\ref{EinFrameCoup}) we can write
\begin{align}
    \overline{g}_s&=(\sqrt{\Lambda} |t|)^{\frac{2 nv}{\beta_o -2v}},\label{typeIIANECHetSO32}
\end{align}
where $v=1,2$ for the string frame, the Einstein frame respectively. Computing its derivative with respect to $t$ we get 
\begin{align}
  \frac{\partial \overline{g}_s}{\partial t}&= \left( \frac{2 n v}{\beta_o -2v}\right) \frac{\overline{g}_s}{t}\nonumber\\
  &=-\left( \frac{2 n v}{\beta_o -2v}\right)\Lambda^{1/2}\overline{g}_s^{\frac{2v-\beta_o}{2nv}} {\overline{g}_s}\propto \overline{g}_s^{\left(1+ \frac{2v-\beta_o}{2nv}\right)\geq 0}\nonumber\\
  \implies ~~& 1+\frac{2v-\beta_o}{2nv}\geq 0 \implies \frac{1}{n}\geq -\frac{1}{1-\frac{ \beta_o }{2v}},\label{NECinHetSO32Thry}
\end{align}
with $0< \beta_o < \frac{2}{3}$. Since, the bound should be independent of the frame at consideration as well as the  parameter $\beta_o$ of the theory, one has to take the intersection of all the allowed ranges of $\frac{1}{n}$ for $v=1,~2$, and for all values of $\beta_o$ lying within its range of validity. This gives $\frac{1}{n}\geq -1,$ which is identical to the NEC (\ref{NECforFLRWcos}).

If we further consider the temporal variation of the parameter, i.e. $\beta(t)$ as in Eq. (\ref{F1(t)withbeta(t)}) then the type IIA coupling (\ref{typeIIANECHetSO32}) gets modified as 
\begin{align}
    \overline{g}_s&=(\Lambda t^2)^{\frac{n v}{\beta(t)-2v}}. \label{gsbarToObtainNEC}
\end{align}
Taking its derivative with respect to $t$ gives 
\begin{align}
\log \overline{g}_s&=\left(\frac{2nv}{\beta(t)-2v}\right) \log (\sqrt{\Lambda} |t|)\nonumber\\
    \implies \frac{1}{\overline{g}_s}\frac{\partial \overline{g}_s}{\partial t }&= -\frac{2nv}{(\beta(t)-2v)^2}\dot{\beta}(t)\log( \sqrt{\Lambda} |t|)-\left(\frac{2nv}{\beta(t)-2v}\right) \frac{1 }{ |t|}\nonumber\\   &=\left(\frac{\dot{\beta}(t) }{2v-\beta(t)}\right) \log \overline{g}_s + \frac{2nv\sqrt{\Lambda}}{(2v-\beta(t))}\overline{g}_s^{\frac{2v-\beta(t)}{2nv}}\nonumber\\
 \implies   \frac{\partial \overline{g}_s }{\partial t}&=\overline{g}_s^{1+\mathrm{B}(t)}+2nv \sqrt{\Lambda}~ \overline{g}_s^{1+\frac{1}{n}-\frac{\beta(t)}{2nv}+\mathrm{C}(t)}\label{tderofgsbar}
\end{align}
where 
\begin{align}
    \overline{g}_s^{\mathrm{B}(t)}&=\left(\frac{\dot{\beta}(t) }{2v-\beta(t)}\right) \log \overline{g}_s \implies \mathrm{B}(t)=\frac{\log\left[\left(\frac{\dot{\beta}(t) }{2v-\beta(t)}\right) \log \overline{g}_s\right]}{\log \overline{g}_s},\label{ExpofB(t)}\\ \mathrm{and}~~~\overline{g}_s^{\mathrm{C}(t)}&=\frac{1}{(2v-\beta(t))}~~\implies C(t)=-\frac{\log(2v-\beta(t))}{\log\overline{g}_s}.\label{ExpofC(t)}
\end{align}
Since $\overline{g}_s<1$ \eqref{EfTCriTeria2}, and $2v-\beta(t)>1$ for $v=(1,2)$ and  $0\leq \beta(t)<\frac{2}{3}$ (\ref{F1(t)withbeta(t)}) we find that
\begin{align}
\text{sign}~\mathrm{C}(t)=\text{sign}\left(\frac{\log(2v-\beta(t))}{|\log \overline{g}_s|} \right)>0.
\end{align}
Now, requiring the exponent of the second term on the right-hand side of Eq.~(\ref{tderofgsbar}) to be positive semi-definite, we obtain
\begin{align}
  1+\frac{1}{n}-\frac{\beta(t)}{2nv}+\mathrm{C}(t) \geq 0\implies  \frac{1}{n}\geq - \frac{1+C(t)}{1-\frac{\beta(t)}{2v}}.\label{PosI2ndTermNEC}
\end{align}
Again, taking the intersection of the allowed regions for $\frac{1}{n}$ corresponding to all admissible values of the time-dependent parameters of the theory yields the null energy condition \eqref{NECforFLRWcos},
\begin{align}
    \frac{1}{n}\geq -1.\label{NuEnConDiSO32TimDep}
\end{align} 
Note that, for $\mathrm{B}(t)$ in Eq. (\ref{ExpofB(t)}) to be a well-defined function, $\beta(t)$ must be a decreasing function satisfying $\dot{\beta}(t)<0$, which in fact is consistent with the form of $\beta(t)$ given in Eq. (\ref{F1(t)withbeta(t)}). With this information, we can write $\mathrm{B}(t)$ as 
\begin{align}
    \mathrm{B}(t)&=-\frac{\log\left[\frac{|\dot{\beta}(t)|}{(2v-\beta(t))}|\log\overline{g}_s| \right] }{|\log\overline{g}_s|}.
\end{align}
However, to determine the sign of $\mathrm{B}(t)$ we need to know the precise value of $|\dot{\beta}(t)|.$ 
\begin{enumerate}
    \item Suppose we have
\begin{align}
\frac{|\dot{\beta}(t)|}{(2v-\beta(t))}<\frac{1}{|\log \overline{g}_s|}.\label{EFTCondItionForSO32Het}
\end{align}
It is evident that for this case $\mathbf{sgn}~\mathrm{B}(t)>0.$ As a result, the first term on the right hand side of Eq. (\ref{tderofgsbar}) satisfies the effective field theory criterion (\ref{EftCriTeriA1}).
\item If instead we have $|\dot{\beta}(t)|>>1,$ then 
\begin{align}
\text{sign}~\mathrm{B}(t)=\text{sign}\left(\frac{\log(2v-\beta(t))-\log|\log \overline{g}_s|-\log|\dot{\beta}(t)|}{|\log\overline{g}_s |} \right)<0.
\end{align}
This follows from the fact that $1 < 2v - \beta(t) \leq 4$, and therefore the sum
\begin{equation}
\log\lvert \log \overline{g}_s \rvert + \log \lvert \dot{\beta}(t) \rvert
\end{equation}
dominates over $\log\big(2v - \beta(t)\big)$. In this case, we cannot make a definitive statement about the validity of the effective field theory, since this would require detailed knowledge of the magnitude  $\lvert \mathrm{B}(t) \rvert$. Note, however, that in the late-time limit as $\overline{g}_s \to 0$, it is reasonable to expect that
$
\lvert \log \overline{g}_s \rvert > \log\!\big(2v - \beta(t)\big), 
\lvert \log \overline{g}_s \rvert > \log \lvert \log \overline{g}_s \rvert ~
\text{along~with}~\log \lvert \log \overline{g}_s \rvert > \log\!\big(2v - \beta(t)\big).
$ This implies  $|\mathrm{B}(t)|$ is less than 1 provided  
\begin{align}
    |\log \overline{g}_s|&>\log|\dot{\beta}(t)|\nonumber\\
    \implies -\log \overline{g}_s&>\log|\dot{\beta}(t)|\nonumber\\
    \implies \log ~\overline{g}_s|\dot{\beta}&(t)|<0\nonumber\\
    \implies \overline{g}_s|\dot{\beta}(t)|&<1.\label{ConstEFTinHetSo32}
\end{align} 
As a result $1+\mathrm{B}(t)>0$ holds true, satisfying the effective field theory constraint (\ref{EftCriTeriA1}) for the case of heterotic $\mathrm{SO(32)}$ theory with time dependent parameter $\beta(t).$  
\end{enumerate}

\noindent

In the next section we will discuss the other effective field theory criteria (\ref{EfTCriTeria2}) that constrains the temporal regime of the GS state. Further we also investigate how the time dependence of $\beta(t)$ and the constraint (\ref{ConstEFTinHetSo32}) affects this temporal regime. 
\section{Trans-Planckian Censorship Conjecture bound}
\label{sec:TCCconsTraints}
\begin{figure}[!ht]
    \centering
\includegraphics[width=0.65\linewidth]{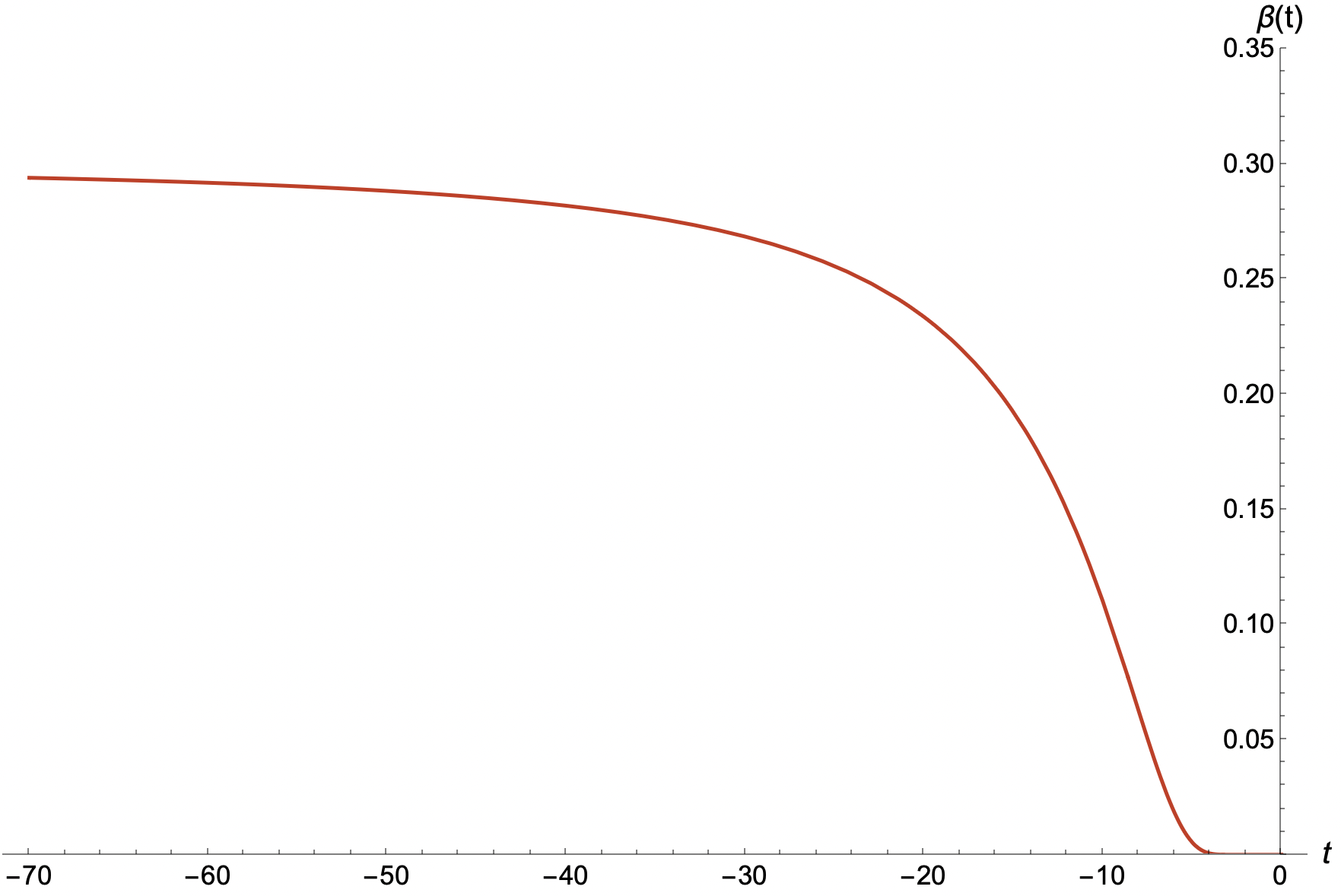}
    \caption{Plot of $\beta(t)$ as in Eq. \eqref{2ndbetatAnsatZ} with $\sigma=60,$ $\beta_o=0.3,~\mathrm{and}~\epsilon=\frac{1}{6}$. }
\label{2ndBetatPlot}
\end{figure}

According to the Trans-Planckian Censorship Conjecture, in any consistent quantum theory of gravity a sub-Planckian mode can never exit the Hubble horizon and freeze \cite{Bedroya_2020}. This gives rise to a bound on the duration of the accelerating phase. Surprisingly, the constraint $\overline{g}_s<1$, which is necessary for a well defined effective field theory description\cite{Brahma:2020tak,brahma2021four,Bernardo:2021rul,Chakravarty_2024}, gives us a temporal domain which is identical to the Trans-Planckian Censorship Conjecture (TCC) bound. Recall, that the above constraint for $\overline{g}_s$ (\ref{IIAcouPlingdeSitterIIB}) in the case of de Sitter realisation in type IIB led to (\ref{TCCinTypeIIBStrINgTh})
\begin{align}
\overline{g}_s\equiv \frac{g_s}{\mathrm{H}(y)}=\sqrt{\Lambda }|t|<1\implies -\frac{1}{\sqrt{\Lambda}}<t<0.\label{TCCBounD}
\end{align}
Likewise, the expression of $\overline{g}_s$ (\ref{StringCCSFEF}) in the case of de Sitter realisation in heterotic $\mathrm{SO}(32)$ theory give us the same temporal domain 
\begin{align}
    \overline{g}_s&=(\sqrt{\Lambda}|t|)^{\frac{ 2v }{2v-\beta(t)}}<1\implies -\frac{1}{\sqrt{\Lambda}}<t<0.\label{TCCBdinSO(32)HeTThy}
\end{align}
However, the additional EFT constraint $\overline{g}_s|\dot{\beta}(t)|<1$ \eqref{ConstEFTinHetSo32} can effect this temporal domain. To investigate this let us consider the following functional form of $\beta(t)$ 
\begin{align}
\beta(t)&=\frac{\beta_o\exp \left[\frac{1}{\sqrt{\Lambda }}\left(\frac{1}{|\epsilon|}-\frac{1}{|t|} \right) \right] }{1+\exp \left[\frac{1}{\sqrt{\Lambda }}\left(\frac{1}{|\epsilon|}-\frac{1}{|t|} \right) \right]},\label{AnsatzofBetat1}
\end{align}
where $0<\beta_o< \frac{2}{3}$ and $\epsilon$ is as given in Eq. (\ref{ValueofEpsilonHetSO}). 
\begin{figure}
    \centering
\includegraphics[width=0.69\linewidth]{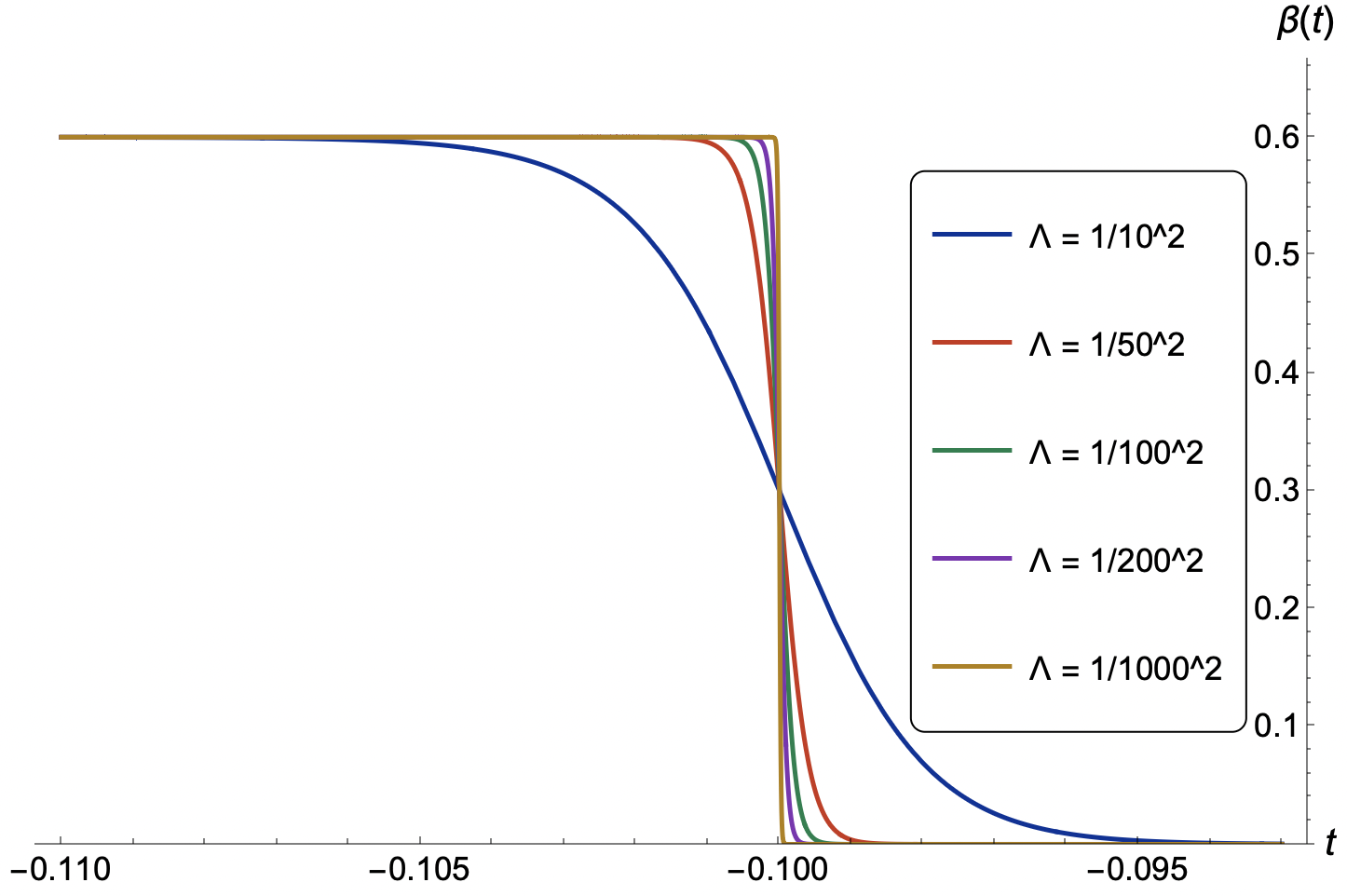}
    \caption{ $\beta(t)$ (\ref{AnsatzofBetat1}) vs. $t$ for different $\Lambda$ values  with $\beta_o=0.6,~\epsilon =0.1$}
\label{betat1}
\end{figure}
Plots of this function for different values of $\Lambda$ appear in Fig. \ref{betat1}. Computing its derivative with respect to $t$ we get 
\begin{align}
    \dot{\beta}(t)&=-\frac{\beta_o}{\sqrt{\Lambda}|t|^2}\frac{\exp \left[\frac{1}{\sqrt{\Lambda }}\left(\frac{1}{|\epsilon|}-\frac{1}{|t|} \right) \right] }{\left[1+\exp \left[\frac{1}{\sqrt{\Lambda }}\left(\frac{1}{|\epsilon|}-\frac{1}{|t|} \right) \right]\right]^2}.\label{Ansatzdotbetat1}
\end{align}
\begin{figure}[ht!]
    \centering
\includegraphics[width=0.74\linewidth]{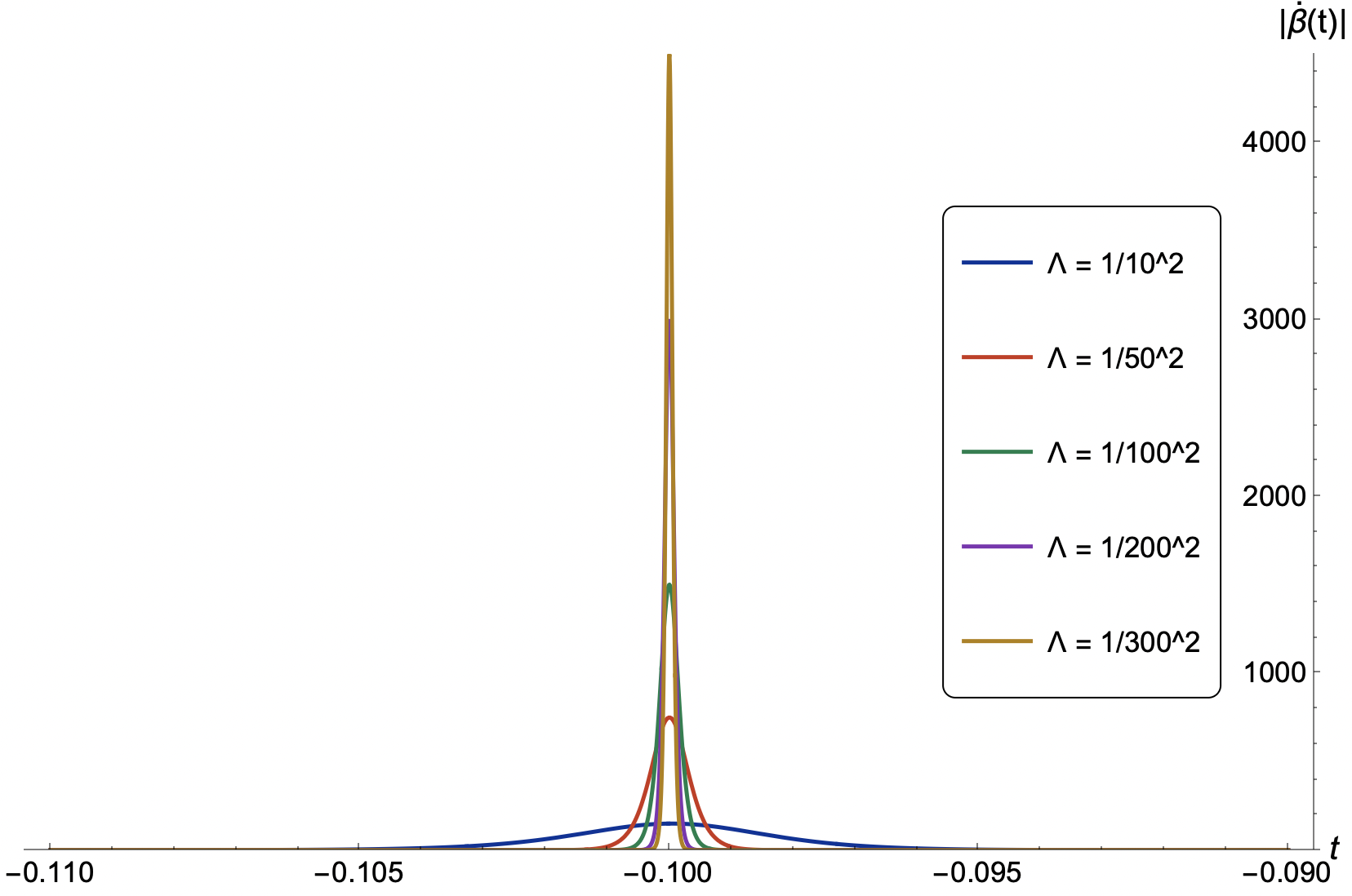}
    \caption{$|\dot{\beta}(t)|$ (\ref{Ansatzdotbetat1}) vs. $t$ for different $\Lambda$ values with $\beta_o=0.6,~|\epsilon|=0.1$}
\label{dotbetat1}
\end{figure}
The absolute values of $\dot{\beta}(t)$, for different values of $\Lambda$, have been plotted in Fig. \ref{dotbetat1}. We notice that these are sharply peaked at $|t|=|\epsilon|$, with $\dot{\beta}(t)\bigg\vert_{t=-\epsilon}=-\frac{\beta_o}{4|\epsilon|^2\sqrt{\Lambda}}.$ 
\begin{figure}[!ht]
    \centering
\includegraphics[width=0.74\linewidth]{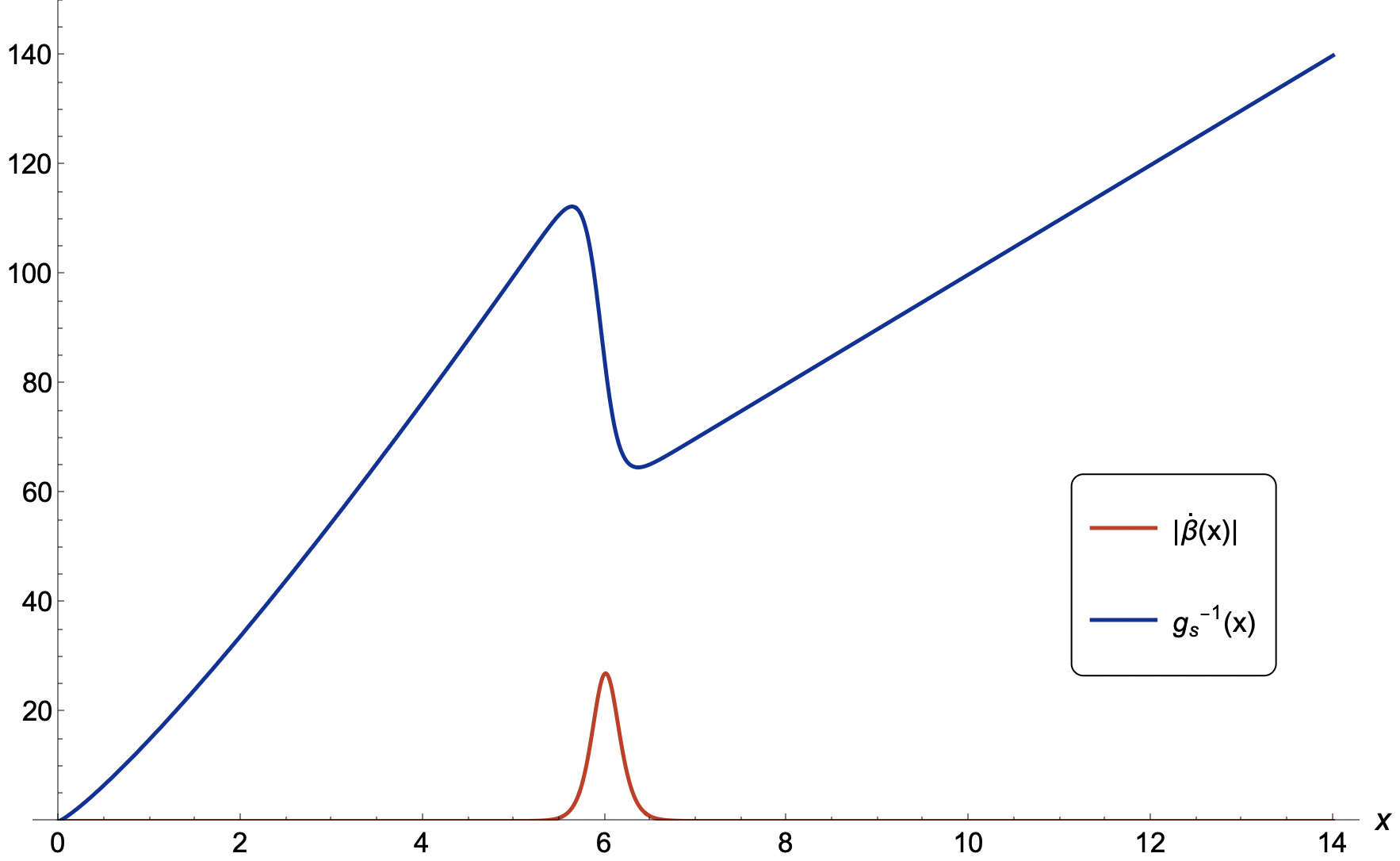}
    \caption{ $|\dot{\beta}(x)|$ and $\overline{g}_{s}^{-1}(x)=(\sqrt{\Lambda} x^{-1})^{-\frac{2v}{(2v-\beta(x))}}$ with $\beta(x)$ as in Eq. (\ref{bEtaInVarx}) and $\Lambda=0.01,$ $\beta_o=0.3,~v=1,~|\epsilon|=\frac{1}{6}$. Here $x:=\frac{1}{|t|}\in (0,\infty)$. }
\label{betadOt&gsinV}
\end{figure}
\begin{figure}[!ht]
    \centering
\includegraphics[width=0.65\linewidth]{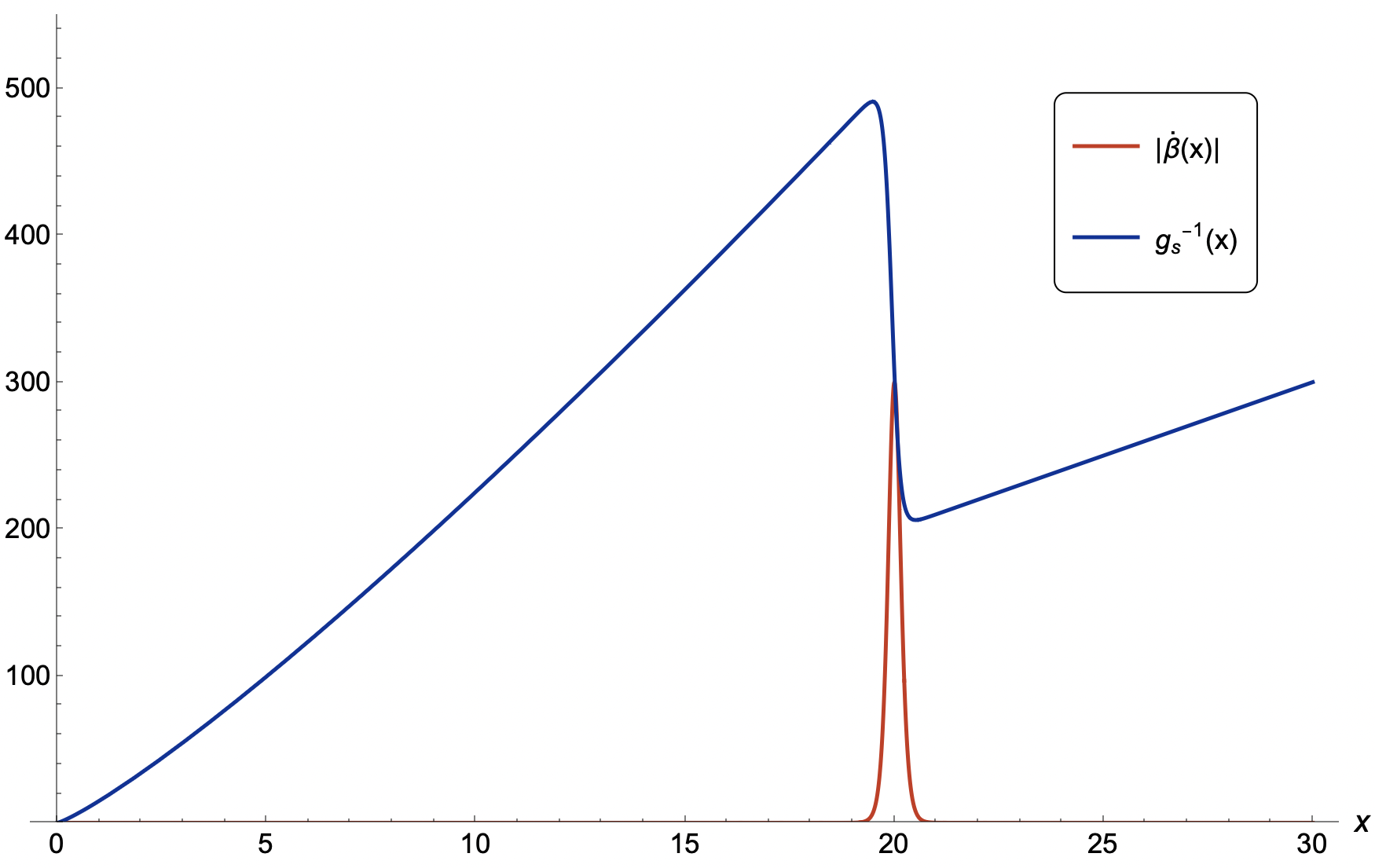}
    \caption{ $|\dot{\beta}(x)|$ and $\overline{g}_{s}^{-1}(x)=(\sqrt{\Lambda} x^{-1})^{-\frac{2v}{(2v-\beta(x))}}$ with $\beta(x)$ as in Eq. (\ref{bEtaInVarx}) and $\Lambda=0.01,$ $\beta_o=0.3,~v=1,~|\epsilon|=\frac{1}{20}$. Here $x:=\frac{1}{|t|}\in (0,\infty)$. The two curves intersect at $x=20.01$ and $x=20.08$.}
\label{betadOt&gsinVmin20}
\end{figure}
 Defining $\frac{1}{|t|}\equiv x\in(0,\infty)$ we can rewrite (\ref{AnsatzofBetat1}) as 
\begin{align}
    \beta(x)&=\frac{\beta_o \exp\left[ \frac{1}{\sqrt{\Lambda}}\left(\frac{1}{|\epsilon|}-x \right) \right] }{1+\exp\left[ \frac{1}{\sqrt{\Lambda}}\left(\frac{1}{|\epsilon|}-x\right) \right]}.\label{bEtaInVarx}
\end{align}
In Fig. \ref{betadOt&gsinV} we have plotted $|\dot{\beta}(x)|$ and $\overline{g}_{s}^{-1}(x)$ in the string frame with $\Lambda=0.01,~\beta_o=0.3$ and $|\epsilon|=\frac{1}{6}$. We see that the two curves do not intersect at any value of $x$. For this case, it is easy to check numerically that $\overline{g}_s(x)|\dot{\beta}(x)|<1$ is satisfied for all $x$ lying in the range $\sqrt{\Lambda}<x<\infty$. If instead we choose the value of $|\epsilon|$ to be $\frac{1}{20}$, then the two curves $|\dot{\beta}(x)|$ and $\overline{g}_s^{-1}(x)$ intersect close to $x=\frac{1}{|\epsilon|}=20$ (see Fig. \ref{betadOt&gsinVmin20}). Numerically we find that close to the intersection points the constraint $\overline{g}_s(x)|\dot{\beta}(x)|<1$ ceases to hold. In other words, we no longer have a well-defined effective field theory description in the entire domain $\sqrt{\Lambda }<x<\infty$. Instead, we can break this domain into two parts $\sqrt{ \Lambda}<x<\frac{1}{|\epsilon|}$ and $\frac{1}{|\epsilon|}<x<\infty$. Since the latter corresponds to the late-time regime and is essential for realizing de Sitter isometry, it is reasonable to say that the effective field theory is valid only in $-\epsilon<t<0$.  One might think that decreasing the value of $\Lambda$ can circumvent this issue. But we have checked that we can always find a suitable value of $\epsilon$ for which there is a breakdown of the effective field theory constraint (\ref{ConstEFTinHetSo32}) near $x=\frac{1}{|\epsilon|}$. This implies that the functional ansatz of $\beta(t)$ (\ref{Ansatzdotbetat1}) is not suitable and hence we look for a smoother function having the following form
\begin{align}
    \beta(t)&=\beta_o \exp\left( -\frac{\sigma^2 |\epsilon|^2}{|t|^2}\right),\label{2ndbetatAnsatZ}
\end{align}
where $\sigma>>1$ is a dimensionless constant. The plot of this function with $\sigma=60$ and $|\epsilon|=\frac{1}{6}$ appears in Fig. \ref{2ndBetatPlot}. As $t\rightarrow 0^{-},$ $\beta(t)\rightarrow \beta_{o}\exp(-\infty)\approx 0$. At $|t|=t_{\star}=\sigma |\epsilon|,$ $\beta(t_{\star})=0.37\beta_o$. Thus, in the temporal domain $-\epsilon < t<0$, $\beta(t)= 0$ satisfying our requirement as in Eq. \eqref{F1(t)withbeta(t)}. Further, as $t\rightarrow -\frac{1}{\sqrt{\Lambda}},$ $\beta (t)\rightarrow \beta_o \exp\left(-\sigma^2|\epsilon|^2 \Lambda \right)\approx \beta_o$ provided $\sigma |\epsilon|\sqrt{\Lambda}<<1$.
The derivative of $\beta(t)$ \eqref{2ndbetatAnsatZ} with respect to $t$ gives 
\begin{align}
    \dot{\beta}(t)= -\frac{ 2 \beta_o \sigma^2 |\epsilon|^2}{|t|^3}\exp\left( -\frac{\sigma^2 |\epsilon|^2}{|t|^2}\right),\label{DotBeTAt2nd}
\end{align}
\begin{figure}[!ht]
    \centering
\includegraphics[width=0.65\linewidth]{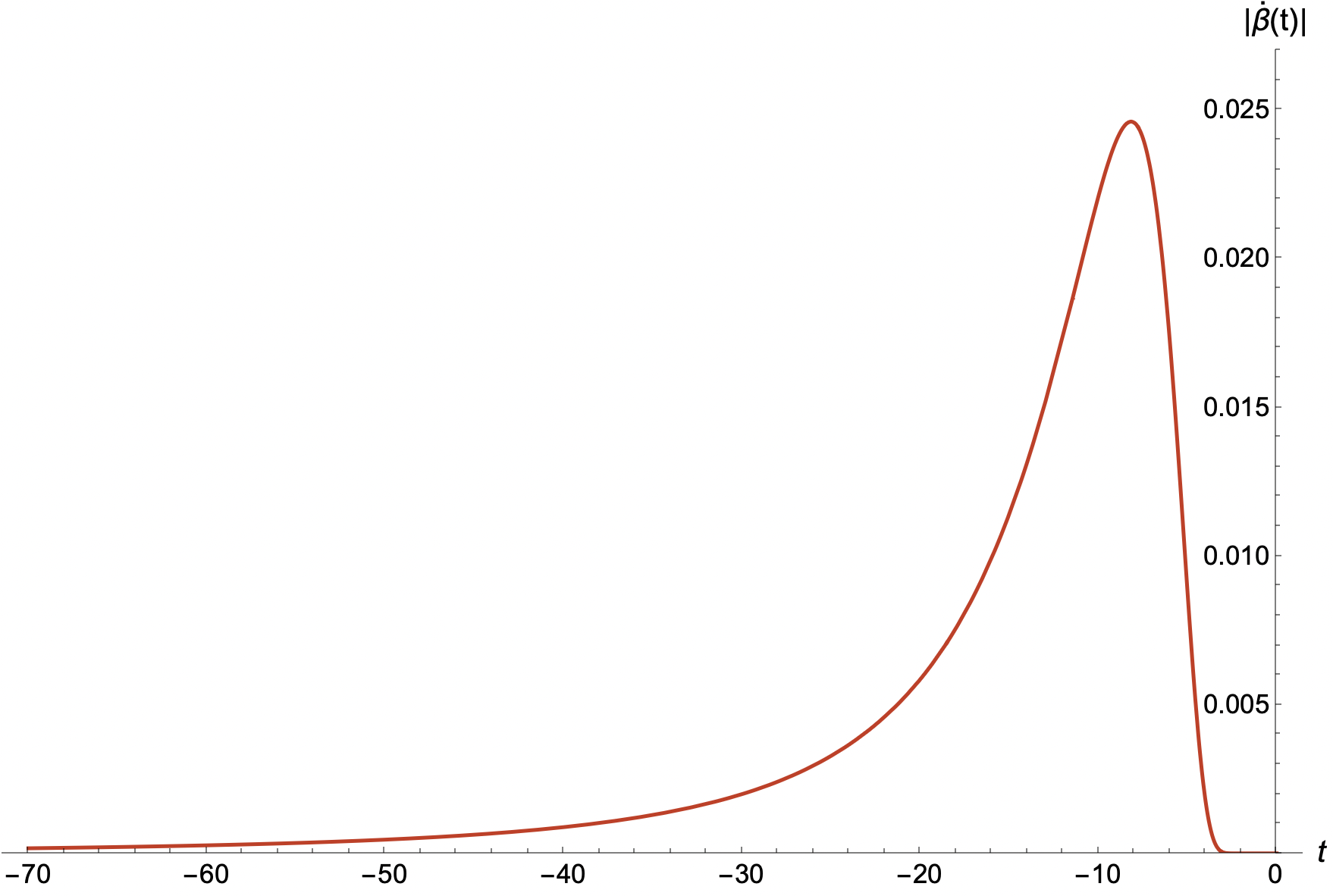}
    \caption{Plot of $|\dot\beta(t)|$ as in Eq. \eqref{DotBeTAt2nd} with $\sigma=60,$ $\beta_o=0.3,~\mathrm{and}~\epsilon=\frac{1}{6}$. }
\label{2ndBetadotPlot}
\end{figure}
\noindent
which peaks at 
\begin{align}
    \ddot{\beta}(t)\big|_{|t|}&=\left(-\frac{ 6 \beta_o \sigma^2|\epsilon|^2 }{|t|^4}+ \frac{ 4\beta_o \sigma^4|\epsilon|^4 }{|t|^6} \right)\exp\left(- \frac{\sigma^2 |\epsilon|^2 }{|t|^2}\right)=0\nonumber\\
    \implies &|t|=\sqrt{ \frac{2\sigma^2|\epsilon|^2 }{3}}<<\sqrt{ \frac{2}{3 \Lambda } }.
\end{align}
The plot of $|\dot{\beta}(t)|$ as in Fig. \ref{2ndBetadotPlot} shows that this is not as sharply peaked as the ones in Fig. \ref{dotbetat1}.  
\begin{figure}[!ht]
    \centering
\includegraphics[width=0.75\linewidth]{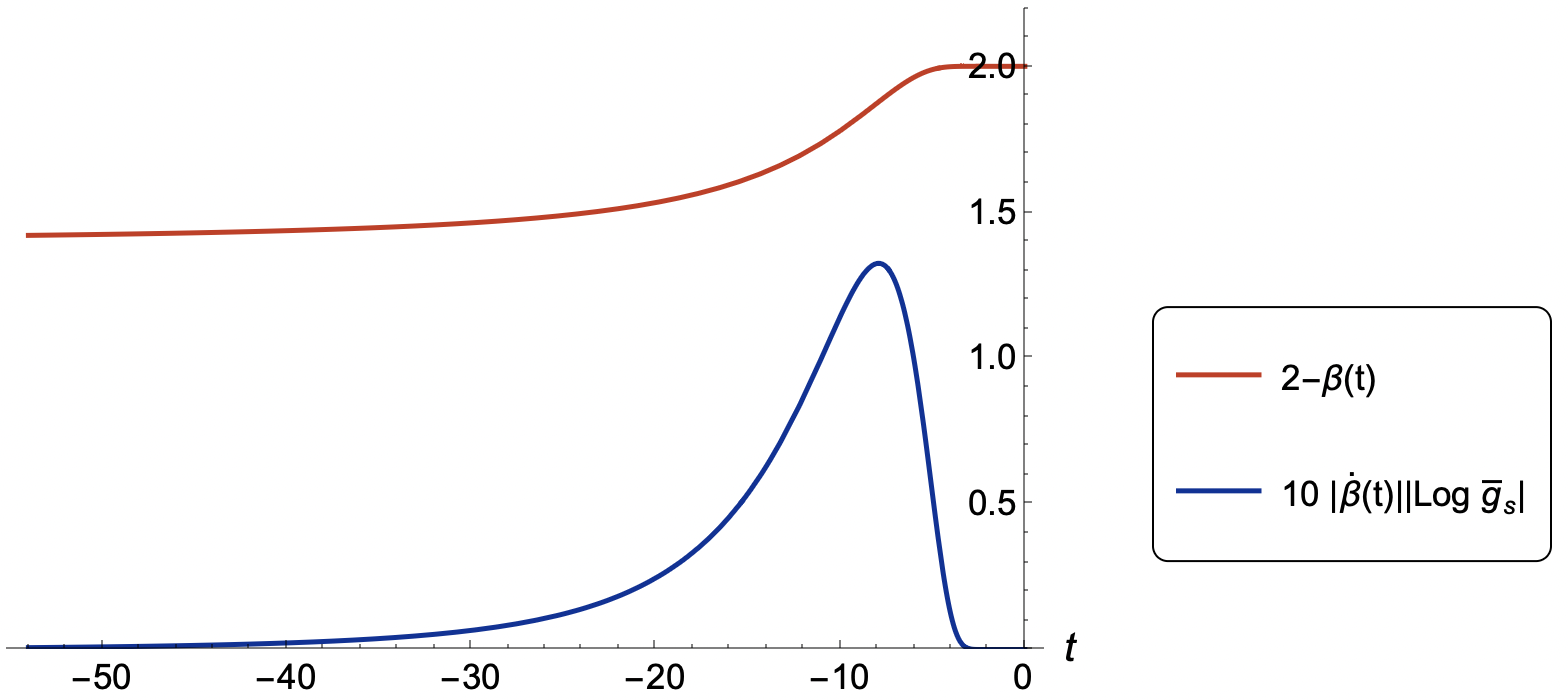}
\caption{Plots of $2-\beta(t)$ and $10 |\dot{\beta}(t)||\log\overline{g}_s|$ with $\beta(t),\dot{\beta}(t)~\mathrm{and}~\overline{g}_s$ as in Eqs. \eqref{2ndbetatAnsatZ}, \eqref{DotBeTAt2nd} and \eqref{StringCCSFEF}. Various parameters have been taken to be  $\sigma=60,$ $\beta_o=0.6,~|\epsilon|=\frac{1}{6},~\Lambda=10^{-4}$. }
\label{Plotsof2minbetat&DotBetaLoggs}
\end{figure}
In Fig. \ref{Plotsof2minbetat&DotBetaLoggs} we have plotted $2-\beta(t)$ and $10 |\dot{\beta}(t)||\log \overline{g}_s|$. It shows that for all values of $t$ the condition
\begin{align}
    2-\beta(t)> |\dot{\beta}(t)||\log \overline{g}_s|
\end{align}
holds true. From Eq. (\ref{EFTCondItionForSO32Het}), it implies that there is a well defined effective field theory description in the entire temporal domain $-\frac{1}{\sqrt{\Lambda}}<t<0$.  

\noindent
\begin{table}[htbp]
\centering
\small
\begin{adjustbox}{width=\textwidth}
\begin{tabular}{||>{\centering\arraybackslash}m{0.10\textwidth}|
>{\centering\arraybackslash}m{0.38\textwidth}|
>{\centering\arraybackslash}m{0.38\textwidth}||}
\hline
\vspace{5pt}
$\beta(t)$ &
$\frac{\beta_o\exp \left[\frac{1}{\sqrt{\Lambda }}\left(\frac{1}{|\epsilon|}-\frac{1}{|t|} \right) \right] }{1+\exp \left[\frac{1}{\sqrt{\Lambda }}\left(\frac{1}{|\epsilon|}-\frac{1}{|t|} \right) \right]}$ 
\vspace{1mm}
&
$\beta_o \exp\left( -\frac{\sigma^2 |\epsilon|^2}{|t|^2}\right)$
\\
\hline\hline

$|\dot{\beta}(t)|$
\vspace{1mm}
&
 Sharply peaked ($|\dot{\beta}(\epsilon)|>>1$)
\vspace{1mm}
&
Not sharply peaked
\vspace{1mm}
\\
\hline

EFT constraint 
&
$\overline{g}_s|\dot{\beta}(t)|<1$ fails to hold $\forall t\in \left(-\frac{1}{\sqrt{\Lambda}},0\right)$
&
$\frac{|\dot{\beta}(t)|}{(2v-\beta(t))}<\frac{1}{|\log \overline{g}_s|}$ holds true $\forall t\in \left(-\frac{1}{\sqrt{\Lambda}},0\right)$
\\
\hline
Temporal domain of validity
&
$-\epsilon<t<0$
&
$-\frac{1}{\sqrt{\Lambda}}<t<0$

\\
\hline

\end{tabular}
\end{adjustbox}

\caption{Comparison between the two functional ansatze for $\beta(t)$ (\ref{AnsatzofBetat1}) and \eqref{2ndbetatAnsatZ}}
\label{tab:ComPbeTEnTwoFuctAnsAtz}
\end{table}

Table  \ref{tab:ComPbeTEnTwoFuctAnsAtz} presents a comparison between the two functional ansätze for $\beta(t)$, given in (\ref{AnsatzofBetat1}) and \eqref{2ndbetatAnsatZ}. While the former fails to provide a well-defined effective field theory description throughout the interval $-\frac{1}{\sqrt{\Lambda}}<t<0$, the latter successfully admits such a description. This suggests that, provided $\beta(t)$ is a smooth function whose derivative does not exhibit sharp peaks, the TCC bound remains unchanged.  

\vspace{5pt}
\vspace{5pt}

In this chapter we have presented a suitable compactification of M-theory and a dynamical duality sequence leading to de Sitter space as an excited state in heterotic $\mathrm{SO}(32)$ string theory. This analysis naturally leads to the next question: can it be further extended to the case where the gauge group of the heterotic theory is $\mathrm{E}_8\times \mathrm{E}_8$?  The following chapter is devoted to addressing realisation of de Sitter space in heterotic $\mathrm{E}_8\times \mathrm{E}_8$ string theory.
\chapter{de Sitter as an excited state in heterotic \texorpdfstring{$\mathrm{E}_8\times \mathrm{E}_8$}{E8} string theory}
\label{Chap:4}
An interesting M-theoretic interpretation of the heterotic $\mathrm{E}_8\times \mathrm{E}_8$ theory was developed by Horava and Witten \cite{1Horava_1996,2Horava_1996}. In particular, they showed that M-theory compactified on the orbifold $\mathbb{R}^{10}\times \mathbf{S}^1/\mathbb{Z}_2$ reduces, in the low energy limit, to the heterotic $\mathrm{E}_8\times \mathrm{E}_8$ theory.  

In this chapter we present a novel dynamical duality sequence to realise four dimensional de Sitter spacetime within the heterotic $\mathrm{E}_8\times \mathrm{E}_8$ framework. By analysing various compactifications of the M theory maifold, we found that obtaining $\mathrm{E}_8 \times \mathrm{E}_8$ gauge group at the final stage of the duality chain requires introduction of three time-dependent warp factors in the six-dimensional internal space\cite{Alexander:2023qym,Archana_CC2026}. This feature stands in contrast with the constructions used for realising de Sitter space in type IIB \S\ref{typeIIBdSisom} and heterotic $\mathrm{SO}(32)$ \S\ref{dSinHetSO(32)} string theories, where the corresponding six-dimensional internal space of the parent M-theory configuration (\ref{MthEorYmeTric}), (\ref{MthHetSO32}) involves only two time-dependent warp factors, namely $\mathrm{F}_1(t)$ and $\mathrm{F}_2(t)$.

\section{de Sitter isometry in heterotic \texorpdfstring{$\mathrm{E}_{8}\times \mathrm{E}_{8}$}{E8XE8} string theory}
\label{E8XE8ThrdistWarpf}
The two-manifold $\mathcal{M}_2$, which was taken to be a toroidal orbifold $
\frac{\mathbb{T}^2_{4,5}}{\mathcal{I}_4\mathcal{I}_5}$ at the starting point of the duality sequence leading to the heterotic $\mathrm{SO}(32)$ realization of de Sitter space (see~\eqref{MThCompcTifedForSO(32)}), will now be replaced by an orbifold of the form
\begin{equation}
\mathcal{M}_2=\mathbf{S}^1_{\theta_1}\times \frac{\mathbf{S}^1_{\theta_2}}{\mathcal{I}_{\theta_2}}.
\end{equation}
The two directions $\theta_1$ and $\theta_2$ are endowed with distinct time-dependent warp factors, $\mathrm{F}_1(t)$ and $\mathrm{F}_3(t)$, respectively. The time-dependent warp factor associated with $\mathcal{M}_4$ continues to be denoted by $\mathrm{F}_2(t)$, as in the previous constructions. 

[$\bullet$] Therefore, we begin with the following metric configuration in $\mathrm{M}$ theory 
\begin{align}
ds^{2}_{11}&=g_{s}^{-8/3} \left(-dt^2+ \sum_{i=1}^{2}\left(dx^{i}\right)^{2}\right)+g_{s}^{-2/3}\mathrm{H}^{2}(y)\bigg[\mathrm{F}_{1}(t)(d\theta_{1})^2+\mathrm{F}_{3}(t)(d\theta_{2})^2 \nonumber\\
&~~~~~+\mathrm{F}_{2}(t)~g_{mn}dy^m dy^n\bigg]+g_{s}^{4/3}\delta_{ab}d\omega^a d\omega^b\label{MthThreewarpFac}
\end{align} 
with the compactification being of the form 
\begin{align}
\mathbb{R}^{2,1}\times \mathbf{S}^{1}_{\theta_{1}}\times \frac{\textstyle \mathbf{S}^{1}_{\theta_{2}}}{\textstyle \mathcal{I}_{\theta_{2}}}\times\mathcal{M}_{4}\times \frac{\textstyle \mathbb{T}^{2}_{3,11}}{\textstyle\mathcal{G}},\label{CompInMTheoRY}    
\end{align}
 here $\mathcal{M}_{4}$ is spanned by the coordinates $(y^6,y^7,y^8,y^9)$ and $\omega^{a}\equiv (x^3,x^{11})$ parameterises the $\mathrm{M}$ theory torus $\frac{\mathbb{T}^2_{3,11}}{\mathcal{G}}$. $\mathcal{G}$ is a group action without any fixed points. There exists two Horava-Witten walls at the boundaries of the orbifold $\mathbf{S}^{1}_{\theta_2}/\mathcal{I}_{\theta_2}$ \cite{1Horava_1996, 2Horava_1996}.

From the solitonic structure of the above $\mathrm{M}$ theory configuration one might think that the dimensional reduction along $\theta_{2}$ immediately gives rise to heterotic $\mathrm{E}_{8}\times \mathrm{E}_{8}$ theory with  four dimensional de Sitter isometry. However, we encounter several subtleties  when we try to perform such a naive dimensional reduction \cite{Dasgupta:2025ypg,Archana_CC2026}.

  Below we elaborate on a duality sequence (depicted on Fig. \ref{Het E8XE8 tree diagram}) that is devoid of any such shortcomings. 

[$\bullet$] 
Dimensional reduction of the $x^{11}$ direction of $\mathrm{M}$ theory (\ref{MthThreewarpFac}) leads to 
type IIA theory on $\mathbb{R}^{2,1}\times \mathbf{S}^{1}_{\theta_{1}}\times \frac{\textstyle \mathbf{S}_{\theta_{2}}^{1}}{\textstyle \Omega \mathcal{I}_{\theta_{2}}}\times \mathcal{M}_{4}\times \mathbf{S}^{1}_{3}$ with metric: 
\begin{align}
ds^{2}_{\mathrm{IIA}}&=g_{s}^{-2} \left[-dt^2+ \sum_{i=1}^{2}\left(dx^{i}\right)^2\right]+\mathrm{H}^{2}(y)\bigg[\mathrm{F}_{1}(t)~(d\theta_{1})^2+\mathrm{F}_{3}(t)~(d\theta_{2})^2+\mathrm{F}_{2}(t)g_{mn}dy^m dy^n\bigg]\nonumber\\
&~~~+{g_{s}^{2}}~(dx^3)^2
\end{align}
and  $g_{s}=\frac{\mathrm{H}(y)}{a(t)}$ is the corresponding type $\mathrm{IIA}$ coupling constant. 
This theory has $\mathrm{O}8$ planes and $\mathrm{D}8$ branes located at the fixed points of the orientifold $\mathbf{S}^{1}_{\theta_2}/\Omega \mathcal{I}_{\theta_2}.$

[$\bullet$] $\mathrm{T}$ duality along $x^{3}$ direction gives rise to type $\mathrm{IIB}$ theory on $\mathbb{R}^{2,1}\times  \frac{\textstyle \mathbf{R}_{3}\times\mathbf{S}^{1}_{\theta_1}\times \mathbf{S}^{1}_{\theta_2}}{\textstyle\Omega(-1)^{\mathrm{F_{L}}}\mathcal{I}_{3}\mathcal{I}_{\theta_2}}\times \mathcal{M}_{4}
 $ with the following metric configuration
\begin{align}
ds^{2}_{\mathrm{IIB}}&=\frac{a^2(t)}{{\rm H}^2(y)} \left(-dt^2+ \sum_{i=1}^{3}\left(dx^{i}\right)^2\right)+\mathrm{H}^{2}(y)\bigg[\mathrm{F}_{1}(t)~(d\theta_{1})^2+\mathrm{F}_{3}(t)~(d\theta_{2})^2+\mathrm{F}_{2}(t)~g_{mn}dy^m dy^n\bigg].\label{DualIIBE8XE8}
\end{align}
The type $\rm{IIB}$ coupling $g_{b}=1$.
Clearly, the vector bundles of this theory are $\mathrm{O}7$ planes and $\mathrm{D}7$ branes along $\mathbb{R}^{2,1}\times \mathbf{S}^{1}_{\theta_1}\times\mathcal{M}_{4}$. 

[$\bullet$] A further $\mathrm{T}$ duality along $\theta_{1}$ leads to type $\mathrm{IIA}$ theory on $\mathbb{R}^{2,1}\times  \frac{\textstyle \mathbf{R}_{3}\times\tilde{\mathbf{S}}^{1}_{\theta_{1}}\times\mathbf{S}^{1}_{\theta_{2}}}{\textstyle \Omega\mathcal{I}_{3}\mathcal{I}_{\theta_{1}}\mathcal{I}_{\theta_{2}}}\times \mathcal{M}_{4}$ whose metric is as follows
\begin{align}
d\tilde{s}^{2}_{\mathrm{IIA}}&=g_{s}^{-2} \left(-dt^2+ \sum_{i=1}^{3}\left(dx^{i}\right)^2\right)+\bigg[\frac{1}{\mathrm{H}^2(y)\mathrm{F}_{1}(t)}(d\theta_{1})^2+\mathrm{H}^{2}(y)\mathrm{F}_{3}(t)(d\theta_{2})^2\nonumber\\
&+\mathrm{H}^{2}(y)\mathrm{F}_{2}(t)g_{mn}dy^m dy^n\bigg].\label{TypeIIA2MetRic} 
\end{align}
The coupling constant of this theory is given by  
\begin{align}
    \tilde{g}_{s}=\frac{1}{{\rm H}(y)\sqrt{{\rm F}_{1}(t)~}}.\label{IIA2CouplinG}
\end{align}
There are $\mathrm{O}6$ planes and $\mathrm{D}6$ branes along $\mathbb{R}^{2,1}\times \mathcal{M}_{4}.$

[$\bullet$] Now, we perform a dimensional uplift to obtain
$\mathrm{M}$ theory on $\mathbb{R}^{2,1}\times  \frac{\textstyle\mathbf{R}_{3}\times\tilde{\mathbf{S}}^{1}_{\theta_{1}}\times\mathbf{S}^{1}_{\theta_{2}}}{\textstyle \mathcal{I}_{3}\mathcal{I}_{\theta_{1}}\mathcal{I}_{\theta_{2}}}\times \mathcal{M}_{4}\times \tilde{\mathbf{S}}^{1}_{11}$. This is nothing but the Taub-NUT geometry where the Taub-NUT base is the orbifold $\frac{\mathbf{R}_{3}\times\tilde{\mathbf{S}}^{1}_{\theta_1}\times\mathbf{S}^1_{\theta_2}}{\mathcal{I}_{3}\mathcal{I}_{\theta_1}\mathcal{I}_{\theta_2}}$ and $\tilde{\mathbf{S}}^1_{11}$ is the circle fibration. 
{
\begin{align}
ds^{2}_{11}&=(\tilde{g}_{s})^{-2/3}g_{s}^{-2} \left(-dt^2+ \sum_{i=1}^{3}\left(dx^{i}\right)^2\right)+(\tilde{g}_{s})^{-2/3}\bigg[\frac{1}{\mathrm{H}^2~\mathrm{F}_{1}(t)~}(d\theta_{1})^2+\mathrm{H}^{2}(y)~\mathrm{F}_{3}(t)~ (d\theta_{2})^2\nonumber\\
&~~+\mathrm{H}^{2}(y)\mathrm{F}_{2}(t)~g_{mn}dy^m dy^n\bigg] +\tilde{g}_{s}^{4/3}(d\tilde{x}_{11})^2\nonumber\\
&=\frac{{\rm H}^{2/3}(y)~{\rm F}_{1}^{1/3}(t)}{g_{s}^{2}}\left(-dt^2+ \sum_{i=1}^3\left(dx^{i}\right)^2\right)+\bigg[\frac{{\rm F}_{1}^{-2/3}(t)}{{\rm H}^{4/3}(y)} (d\theta_1)^2+{\rm H}^{8/3}(y){\rm F}_{1}^{1/3}(t){\rm F}_{3}(t)(d\theta_2)^2\nonumber\\
&+{\rm H}^{8/3}(y) {\rm F}_{1}^{1/3}(t){\rm F}_{2}(t)g_{mn}dy^{m}dy^{n}\bigg]+{\rm H}^{-4/3}(y)  {\rm F}_{1}^{-2/3}(t) (d\tilde{x}_{11})^2.\label{Mth2ndE8E8dS}
\end{align}}

[$\bullet$] Next step is to perform two blow-ups on the  solitonic structure of the above $\mathrm{M}$ theory configuration (\ref{Mth2ndE8E8dS})
\begin{align}
    \frac{\textstyle \mathbf{R}_{3}}{\textstyle \mathcal{I}_{3}}\rightarrow \hat{\mathbf{R}}_{3},~~~\frac{\textstyle \tilde{\mathbf{S}}^{1}_{\theta_{1}}}{\textstyle \mathcal{I}_{\theta_{1}}}\rightarrow \hat{\mathbf{S}}^1_{\theta_{1}}.
\end{align}
This give rise to an $\mathrm{M}$ theory  on $\mathbb{R}^{3,1}\times\hat{\mathbf{S}}^1_{\theta_{1}}\times\frac{\mathbf{S}_{\theta_2}^1}{\mathcal{I}_{\theta_2}}\times \mathcal{M}_{4}\times\tilde{\mathbf{S}}^{1}_{11}$. Consequently, there are two Horava-Witten walls at the two fixed points of $\frac{\textstyle \mathbf{S}^{1}_{\theta_{2}}}{\textstyle \mathcal{I}_{\theta_{2}}}$\cite{1Horava_1996,2Horava_1996}. This theory admits fluxes of the form $\mathbf{C}_{\theta_2 \mathrm{PQ}}$, where $\mathrm{(P,Q)}\in\mathbb{R}^{3,1}$ or $\mathrm{(P,Q)}\in \mathcal{M}_4\times \left(\hat{\mathbf{S}}^1_{\theta_1}\times\widetilde{\mathbf{S}}^1_{11}\right) $. These are the only flux components that do not get projected out by the orbifold action along $\theta_2$. 

\begin{figure}[!ht]
\centering
\begin{tikzpicture}[scale=0.94,font=\small,thick]
\node at (-5.1,0) [draw, rounded rectangle,fill=teal!10,
    minimum width=4.5cm,
    minimum height=1cm](block1) { \textcolor{violet}{
$\substack{ \textstyle \mathrm{M}~\textbf{theory} \\ \\\textstyle \textbf{on}\\ \textstyle\textbf{$\mathbb{R}^{2,1}\times \mathbf{S}^{1}_{\theta_1}\times\frac{\mathbf{S}^{1}_{\theta_2}}{\mathcal{I}_{\theta_{2}}}\times \mathcal{M}_{4}\times \frac{\mathbb{T}^{2}_{3,11}}{\mathcal{G}}$}}$
}};
\node at (5.1,0) [draw, rounded corners, fill=teal!10,
    minimum width=2.5cm,
    minimum height=1cm] (block2) { \textcolor{violet}{ 
$\substack{\textstyle\textbf{Type}~\mathrm{IIA}~\textbf{theory}\\ \\ \textstyle\textbf{on} \\\textstyle \mathbb{R}^{2,1}\times  \mathbf{S}^{1}_{\theta_{1}}\times \frac{\mathbf{S}^{1}_{\theta_2}}{\Omega \mathcal{I}_{\theta_2}}\times \mathcal{M}_{4}\times {\mathbf{S}^{1}_{3}}}$
}};
\node at (2.0,-3.3) [draw, rounded corners, fill=teal!10,
    minimum width=2.5cm,
    minimum height=1.5cm] (block3) { \textcolor{violet}{
$\substack{\textstyle\textbf{Type}~\mathrm{IIB}~\textbf{theory}\\ \\ \textstyle \textbf{on} \\\textstyle \mathbb{R}^{2,1}\times  \frac{\mathbf{R}_{3}\times\mathbf{S}^{1}_{\theta_1}\times \mathbf{S}^{1}_{\theta_2}}{\Omega(-1)^{\mathrm{F_{L}}}\mathcal{I}_{3}\mathcal{I}_{\theta_2}}\times \mathcal{M}_{4}
}$}};

\node at (-0.7,-6.6) [draw, rounded corners, fill=teal!10,
    minimum width=2.5cm,
    minimum height=1.5cm] (block4) { \textcolor{violet}{
$\substack{\textstyle\textbf{Type}~\mathrm{IIA}~\textbf{theory}\\ \\ \textstyle \textbf{on} \\\textstyle \mathbb{R}^{2,1}\times  \frac{\textstyle\mathbf{R}_{3}\times\tilde{\mathbf{S}}^{1}_{\theta_{1}}\times\mathbf{S}^{1}_{\theta_{2}}}{\textstyle\Omega\mathcal{I}_{3}\mathcal{I}_{\theta_{1}}\mathcal{I}_{\theta_{2}}}\times \mathcal{M}_{4} 
}$}};

\node at (1.8, -9.9) [draw, rounded corners, fill=teal!10,
    minimum width=2.5cm,
    minimum height=1.5cm] (block5) { \textcolor{violet}{
$\substack{\textstyle \mathrm{M}~\textbf{theory}\\ \\ \textstyle \textbf{on} \\\textstyle \mathbb{R}^{2,1}\times  \frac{\mathbf{R}_{3}\times\tilde{\mathbf{S}}^{1}_{\theta_{1}}\times\mathbf{S}^{1}_{\theta_{2}}}{\mathcal{I}_{3}\mathcal{I}_{\theta_{1}}\mathcal{I}_{\theta_{2}}}\times \mathcal{M}_{4}\times \tilde{\mathbf{S}}^{1}_{11}}$}};

\node at (4.9, -13.2) [draw, rounded corners, fill=teal!10,
    minimum width=2.5cm,
    minimum height=1.5cm] (block6) { \textcolor{violet}{
$\substack{\textstyle \mathrm{M}~\textbf{theory}\\ \\ \textstyle \textbf{on} \\\textstyle \mathbb{R}^{2,1}\times  \hat{\mathbf{R}}_{3}\times\hat{\mathbf{S}}^{1}_{\theta_{1}}\times\frac{\mathbf{S}^{1}_{\theta_{2}}}{\mathcal{I}_{\theta_{2}}}\times \mathcal{M}_{4}\times \tilde{\mathbf{S}}^{1}_{11}}$}};

\node at (-5.1, -13.2) [draw, rounded rectangle, fill=teal!10,
    minimum width=2.5cm,
    minimum height=1.5cm] (block7) { \textcolor{violet}{
$\substack{\textstyle\textbf{Heterotic}~\mathrm{E}_{8}\times \mathrm{E}_{8}~\textbf{theory}\\ \\ \textstyle \textbf{on} \\ \textstyle \mathbb{R}^{3,1}\times   \hat{\mathbf{S}}^{1}_{\theta_1}\times\mathcal{M}_{4}\times \tilde{\mathbf{S}}_{11}^{1}
}$}};
\draw[-latex] 
(block1) -- (block2)
(block2) edge (block3)
    (block3) edge (block4)
    (block4) edge (block5)
    (block5) edge (block6)
    (block6) edge (block7);
\node at (0.2,0.5) {${\substack{\textstyle\text{Dimensional reduction } \\ \textstyle \text{along $x^{11}$} }}$} ; 
\node at (2.3,-1.7) { $\substack{\textstyle\mathrm{T}~\text{duality}\\ \textstyle \text{along}~x^3}$};
\node at (-0.4,-4.9) { $\substack{\textstyle\mathrm{T}~\text{duality}\\ \textstyle \text{along}~\theta_{1}}$};
\node at (-.8,-8.3) { $\substack{\textstyle\text{Dimensional}\\ \textstyle \text{uplift}}$};
\node at (2.3,-11.5) {Blow-ups};
\node at (5.6,-11.5) {$\left(\frac{\mathbf{R}_{3}}{\mathcal{I}_{3}}\rightarrow \hat{\mathbf{R}}_{3},~\frac{\widetilde{\mathbf{S}}^{1}_{\theta_1}}{\mathcal{I}_{\theta_1}}\rightarrow \hat{\mathbf{S}}^{1}_{\theta_{1}} \right)$};
\node at (-0.3,-12.7) {$\substack{\textstyle\text{Dimensional reduction}\\ \textstyle \text{along}~\theta_{2}} $ };
\end{tikzpicture}
    \caption{Duality sequence from $\mathrm{M}$ theory (\ref{MthThreewarpFac}) to heterotic $\mathrm{E}_{8}\times \mathrm{E}_{8}$ theory (\ref{HetroTicE8XE8 metric})}
    \label{Het E8XE8 tree diagram}
\end{figure}

[$\bullet$]  Dimensional reduction along the $\theta_{2}$ direction then give us the {heterotic ${\rm E}_{8}\times {\rm E}_{8}$} theory on $\mathbb{R}^{3,1}\times   \hat{\mathbf{S}}^{1}_{\theta_1}\times\mathcal{M}_{4}\times \tilde{\mathbf{S}}_{11}^{1}
$. The metric configuration of the same being
\begin{align}
ds^{2}_{\mathrm{Het~E}_8\times \mathrm{E}_8 }&=\tilde{g}_{s}^{-1}g_{s}^{-2}\mathrm{H}(y)\sqrt{{\rm F}_{3}(t)} \left(-dt^2+ \sum_{i=1}^3\left(dx^{i}\right)^2\right)+\tilde{g}_{s}\mathrm{H}(y)\sqrt{\mathrm{F}_{3}(t)~}(d\tilde{x}_{11})^2\nonumber\\
&+\tilde{g}_{s}^{-1}\mathrm{H}(y)\sqrt{{\rm F}_{3}(t)}\bigg[\frac{1}{\mathrm{H}^2(y)\mathrm{F}_{1}(t)~}(d\theta_{1})^2+\mathrm{H}^{2}(y)\mathrm{F}_{2}(t)~g_{mn}dy^m dy^n\bigg]\nonumber\\
&=a^2(t)\sqrt{\mathrm{F}_{1}(t)\mathrm{F}_{3}(t)}\left(-dt^2+ \sum_{i=1}^3\left(dx^{i}\right)^2\right)+\sqrt{\frac{{\rm F}_3}{{\rm F}_1}}\left((d\theta_1)^2+(d\tilde{x}_{11})^2\right)\nonumber\\
&+\mathrm{H}^{4}(y)\sqrt{\mathrm{F}_{1}(t)\mathrm{F}_{3}(t)}\mathrm{F}_{2}(t)~g_{mn}dy^{m}dy^{n}.\label{HetroTicE8XE8 metric}
    \end{align}
It has the NS-NS two-form fluxes $B_{\rm PQ}$. When $(\mathrm{P,Q})\in \mathbb{R}^{3,1}$ we obtain model-independent axions and $(\mathrm{P,Q})\in \mathcal{M}_4\times \left( \hat{\mathbf{S}}^1_{\theta_1}\times\widetilde{\mathbf{S}}^1_{11}\right)$ gives model-dependent axions in the heterotic $\mathrm{E}_8\times \mathrm{E}_8$ theory. The coupling of this theory is as follows 
\begin{align}
g_{\text{Het~E}_8\times \text{E}_8}&=\mathrm{H}^{2}(y)\mathrm{F}_{1}^{1/4}(t)\mathrm{F}_{3}^{3/4}(t).\label{CouplIngE8E8HeT}
\end{align}
 Now, demanding de Sitter isometry on the heterotic $\mathrm{E}_{8}\times \mathrm{E}_{8}$ theory gives the following form of the $\mathrm{IIA}$ coupling constant $g_{s}$  
\begin{align}
a^2(t){\sqrt{\mathrm{F}_{1}(t)\mathrm{F}_{3}(t)}}{}&\equiv g_{s}^{-2}\mathrm{H}^2(y)\sqrt{\mathrm{F}_{1}(t)\mathrm{F}_{3}(t)}=\frac{1}{\Lambda t^{2}}\label{dDSIsomHetE8E8thry}\\
\implies \frac{g_{s}}{\mathrm{H}(y)}\equiv\overline{g}_s=&\sqrt{\Lambda}|t|\left(\mathrm{F}_{1}(t)\mathrm{F}_{3}(t)\right)^{1/4}~~\text{with}~~-\infty <t<0.\label{gsinE8thEorY}
\end{align}
Further, the time independence of the four dimensional Newton's constant implying the time independence of the volume of the six dimensional internal space of $\mathrm{E}_{8}\times \mathrm{E}_{8}$ theory give rise to the following constraint between the three time dependent warp factors
    \begin{align}
\mathrm{F}_{1}(t)\mathrm{F}_{2}^{4}(t)\mathrm{F}_{3}^{3}(t)&=1.\label{ConstrF1F2F3}
    \end{align}

\subsubsection{Dynamical duality sequence:}
For the above described duality sequence (Fig. \ref{Het E8XE8 tree diagram}) to proceed dynamically we have to ensure that the time dependent warp factor $\mathrm{F}_{1}(t)$ goes to zero faster as compared to $\mathrm{F}_{3}(t)$. In order to quantify these let us consider 
\begin{align}
    \mathrm{F}_{1}(t)=(\sqrt{\Lambda}|t|)^{p},~~~\mathrm{F}_{3}(t)=(\sqrt{\Lambda}|t|)^{q}.\label{wArpFactorsF1AnDF3}
\end{align}
Substituting these in  (\ref{gsinE8thEorY}) we obtain 
\begin{align}
    \overline g_s&=(\sqrt{\Lambda}|t|)^{\frac{4+p+q}{4}}.\label{TypeIIAE8E8TheOrY}
\end{align}
Therefore, the warp factors \eqref{wArpFactorsF1AnDF3} take the following form
\begin{align}
    \mathrm{F}_{1}(t)\equiv \mathrm{F}_{1}\left( \overline g_s\right)&=\left( \overline g_s\right)^{\frac{4p}{4+p+q}},\quad\quad
        \mathrm{F}_{3}(t)\equiv \mathrm{F}_{3}\left( \overline g_s \right)=\left( \overline g_s\right)^{\frac{4q}{4+p+q}}.
\end{align}
Since, at the very first step of the duality sequence, the M-theory torus along the $3,11$ directions (\ref{MthThreewarpFac}) must shrink to zero size in order to reach the type IIB theory (\ref{DualIIBE8XE8}), we must impose the following conditions: 
\begin{align}
    0<\frac{4p}{4+p+q}<\frac{2}{3}~~,~~~0<\frac{4q}{4+p+q}<\frac{2}{3}.
\end{align}
Furthermore, to implement that $\mathrm{F}_1(t)$ decays faster than $\mathrm{F}_3(t)$ we  define 
\begin{align}
 \mathrm{F}_{1}(\overline{g}_s)=\left(\overline{g}_s\right)^{  \frac{4p}{4+p+q}}:=\left(\overline{g}_s\right)^{\frac{2\hat{\alpha}_o}{3}},\quad \mathrm{F}_3(\overline{g}_s)=(\overline{g}_s)^{\frac{4q}{4+p+q}}:=\left(\overline{g}_s\right)^{\frac{2\hat{\beta}_o}{3}},\label{pqinalphabeta}
\end{align}
with $1>\hat{\alpha}_{o}>\hat{\beta}_o>0$. Solving (\ref{pqinalphabeta}) for $p$ and $q$ we get 
\begin{align}
    p=\frac{4\hat{\alpha}_{o}}{6-\hat{\alpha}_o-\hat{\beta}_o},~~~~q=\frac{4\hat{\beta}_o}{6-\hat{\alpha}_o-\hat{\beta}_o}.
\end{align}
This implies the type IIA coupling constant (\ref{TypeIIAE8E8TheOrY}) becomes 
\begin{align}
    \overline g_s &=\left(\sqrt{\Lambda}|t|\right)^{\frac{6}{6-\hat{\alpha}_o-\hat{\beta}_o}}.\label{StrCCinE8E8}
\end{align}
Thus, the EFT criteria (\ref{EfTCriTeria2}) leads to the temporal domain
\begin{align}
\overline{g}_s<1\implies (\sqrt{\Lambda}|t|)^{\frac{6}{6-\hat{\alpha}_o-\hat{\beta}_o}}<1\implies -\frac{1}{\sqrt{\Lambda}}<t<0,\label{TempDomHetE8E8TheoRY}
\end{align}
within which the de Sitter GS state in heterotic $\mathrm{E}_8\times \mathrm{E}_8$ theory remains well defined. In fact this is identical to the domain obtained for the type IIB as well as the heterotic $\mathrm{SO}(32)$ theory \S\ref{sec:TCCconsTraints}.

Note that, even though the constraint (\ref{ConstrF1F2F3}) keeps the volume of the six dimensional internal space of the heterotic $\mathrm{E}_8\times \mathrm{E}_{8}$ theory (\ref{HetroTicE8XE8 metric}) time independent, in the limit $t\rightarrow 0$ both the circles denoted by coordinates $\theta_{1}$ and $\tilde{x}_{11}$ grow as 
\begin{align}
\left(\frac{\mathrm{F}_{3}}{\mathrm{F}_1}\right)^{\frac{1}{4}}\equiv (\overline{g}_s)^{\frac{\hat{\beta}_o-\hat{\alpha}_o}{6}}\rightarrow \infty,\quad(\because~ 1>\hat{\alpha}_o>\hat{\beta}_o>0)
\end{align}
 whereas the volume of the internal four manifold $\mathcal{M}_{4}$ shrinks as 
 \begin{align}
\left(\frac{\mathrm{F}_{1}}{\mathrm{F}_{3}}\right)^{\frac{1}{2}}\equiv (\overline{g}_s)^{\frac{\hat{\alpha}_o-\hat{\beta}_o}{3}}\rightarrow 0.   \label{VolM4E8E8ThrY}  
 \end{align}
To avoid such late time singularity of the internal manifolds we break the temporal domain (\ref{TempDomHetE8E8TheoRY}) into two parts and define: 
\begin{align}
&\quad\quad\mathrm{F}_{1}(t)=(\overline{g}_s)^{\frac{2\hat{\alpha}(t) }{3}},\quad\quad 
    \mathrm{F}_{3}(t)=(\overline{g}_s)^{\frac{2\hat{\beta}(t) }{3}}, \label{F1F3inHeteRE8E8}
    \\
&\text{where}\nonumber\\
   \hat{\alpha}(t)&=\begin{cases}   \hat{\alpha}_o,\quad-\frac{1}{\sqrt{\Lambda }}<t\leq -\epsilon\\
\hat{\gamma}_{o},\quad\quad-\epsilon< t<0
\end{cases},\quad\hat{\beta}(t)=\begin{cases}
\hat{\beta}_o,\quad\quad-\frac{1}{\sqrt{\Lambda }}<t\leq -\epsilon\\
\hat{\gamma}_{o},\quad\quad-\epsilon< t<0 
\end{cases},\label{hatAlphHatBetaHeteRE8E8}
\end{align}
where the IIA coupling $\overline{g}_s$ (\ref{StrCCinE8E8}) is of the form
\begin{align}
    \overline{g}_s&=(\sqrt{\Lambda}|t|)^{\frac{6}{6-\hat{\alpha}(t)-\hat{\beta}(t)}}.\label{TimeDepIIAcConstE8Th}
\end{align}
 That is, in the late time regime, $-\epsilon<t<0,$ once we identify $\mathrm{F}_1(t)$ with $\mathrm{F}_3(t)$, the metric (\ref{HetroTicE8XE8 metric}) takes the following structure 
\begin{align}
ds^2_{\mathrm{Het~E}_8\times \mathrm{E}_8 }&=   a^2(t)~\mathrm{F}_{1}(t)\left[-dt^2+ \sum_{i=1}^3\left(dx^{i}\right)^2\right]+(d\theta_{1})^2+(d\tilde{x}_{11})^2 \nonumber\\
&+\mathrm{H}^{4}(y)\mathrm{F}_{1}(t)\mathrm{F}_{2}(t)g_{mn}dy^{m}dy^{n},
\end{align}
and the constraint (\ref{ConstrF1F2F3}) becomes
 \begin{align}
     \mathrm{F}_{1}(t)\mathrm{F}_{2}(t)=1.
 \end{align}
Hence we obtain time independent internal space in the heterotic $\mathrm{E}_{8}\times \mathrm{E}_{8}$ theory having no late time singularity.    

\noindent
We determine the value of $\epsilon$ by letting the volume of the four manifold $\mathcal{M}_{4}$ (\ref{VolM4E8E8ThrY}) to attain size $\mathrm{V}_{1}>>\alpha^{\prime~2}$ at $|t|=\epsilon.$ If $\mathrm{V}_{o}(-1/\sqrt{\Lambda})>>\mathrm{V}_1$ is its stabilized volume at $|t|=1/\sqrt{\Lambda}$ then  
\begin{align}
\frac{\left({\overline{g}_s}\right)^{\frac{\hat{\alpha}_o-\hat{\beta}_o}{3}}\bigg|_{|t|=\epsilon}}{\left({\overline{g}_s}\right)^{\frac{\hat{\alpha}_o-\hat{\beta}_o}{3}}\bigg|_{|t|=1/\sqrt{\Lambda}}}
    &=\frac{\mathrm{V}_1}{\mathrm{V}_o(-1/\sqrt{\Lambda})}\nonumber\\
    \implies (\sqrt{\Lambda}\epsilon)^{\frac{2(\hat{\alpha}_o-\hat{\beta}_o)}{6-\hat{\alpha}_o-\hat{\beta}_o}}&=\frac{\mathrm{V}_1}{\mathrm{V}_{o}(-1/\sqrt{\Lambda})}\nonumber\\
    \implies \epsilon=\frac{1}{\sqrt{\Lambda}}&\left[\frac{\mathrm{V}_1}{\mathrm{V}_o(-1/\sqrt{\Lambda})} \right]^{\frac{6-\hat{\alpha}_o-\hat{\beta}_o}{2\hat{\alpha}_o-2\hat{\beta}_o}}.\label{EpSilonInStrFrE8E8}
\end{align}
\noindent
 Using the constraint (\ref{ConstrF1F2F3}), we find 
\begin{align}
    \mathrm{F}_{2}(t)=\left({\overline g_s}\right)^{-\frac{\hat{\alpha}(t)}{6}-\frac{\hat{\beta}(t)}{2}}=\left(\sqrt{\Lambda}|t|\right)^{-\left(\frac{\hat{\alpha}(t)+3\hat{\beta}(t)}{6-\hat{\alpha}(t)-\hat{\beta}(t)}\right)}. \label{ExpofF2inE8TheorY}
\end{align}
Now, the coupling (\ref{IIA2CouplinG}) of the type IIA configuration (\ref{TypeIIA2MetRic}) approaches the strong coupling limit as  
\begin{align}
\tilde{g}_{s}\propto \frac{1}{\sqrt{\mathrm{F}_{1}(t)}}=\left(\overline{g}_{s}\right)^{-\hat{\alpha}(t)/3}.
\end{align}
The above coupling becomes strong faster as compared to the rate at which the $\mathbf{S}^1_{\theta_2}$ circle shrinks to zero size, viz. $\sqrt{\mathrm{F}_{3}(t)}\equiv \left( {\overline{g}_s}\right)^{\hat{\beta}(t)/3},$ leading to an uplift to M-theory (\ref{Mth2ndE8E8dS}). A careful look into this metric structure of M-theory (\ref{Mth2ndE8E8dS}) tells us that in the late time all the internal directions are growing except $\mathbf{S}^1_{\theta_2}$ and $\mathcal{M}_{4}$. This is because the volume of the four manifold scales as 
\begin{align}
\sqrt{\left(\mathrm{F}_{1}^{1/3}\mathrm{F}_{2}\right)^4}=\left( \frac{g_s}{\mathrm{H}(y)}\right)^{\frac{\hat{\alpha}(t)}{9}-\hat{\beta}(t)}\rightarrow 0 ~~~(\mathrm{as~long~as}~\frac{\hat{\alpha}_o}{9}>  \hat{\beta}_o),
\end{align}
and the circle $\mathbf{S}^1_{\theta_2}$ scales as 
\begin{align}
\sqrt{\mathrm{F}_{1}^{1/3}\mathrm{F}_{3}}&=\left(\frac{g_s}{\mathrm{H}(y)}\right)^{\frac{\hat{\alpha}(t)+3\hat{\beta}(t)}{9}}\rightarrow 0. \label{HoravaWittenWAlls}
\end{align}
At this stage, since the duality sequence (Fig. \ref{Het E8XE8 tree diagram}) requires only the circle $\mathbf{S}^1_{\theta_2}$ to shrink, and not the four manifold, the two parameters $\hat{\alpha}_{o}$ and $\hat{\beta}_{o}$  must satisfy   
\begin{align}
1>\hat{\alpha}_{o}>\hat{\beta}_{o}>\frac{\hat{\alpha}_o}{9}>0.
\label{CondhAtAlpohatBetao}\end{align}
Therefore, if the above condition is satisfied by the parameters $\hat{\alpha}_o$ and $\hat{\beta}_o$, the duality sequence shown in Fig.~\ref{Het E8XE8 tree diagram} can be implemented consistently. Starting from the M-theory configuration (\ref{MthThreewarpFac}), the sequence evolves dynamically toward the heterotic $\mathrm{E}_8\times \mathrm{E}_8$ theory (\ref{HetroTicE8XE8 metric}). In the late-time limit of the flat-slicing coordinates, this leads to the emergence of a four-dimensional de Sitter Glauber--Sudarshan state in heterotic $\mathrm{E}_8\times \mathrm{E}_8$ theory.

Since the coupling \eqref{CouplIngE8E8HeT} in the heterotic $\mathrm{E}_8\times \mathrm{E}_8$ theory is not equal to $1$, the corresponding analysis in the Einstein frame will differ substantially from the string-frame analysis presented above. In the following subsection, we therefore analyse this scenario in detail.
\subsection{The string frame versus the Einstein frame}
\label{EinsVsStrE8E8Thry}
 In Einstein frame, the metric of heterotic $\mathrm{E}_8\times \mathrm{E}_8$ theory (\ref{HetroTicE8XE8 metric}) takes the following form 
\begin{align}
ds^2_{\mathrm{Het}~\mathrm{E_8\times E_8 ,E}}&= \frac{1}{\sqrt{g_{\mathrm{Het~E}_8\times \mathrm{E}_8} }}ds^2_{\mathrm{Het}~\mathrm{E}_8\times\mathrm{E}_8}=\frac{1}{\mathrm{H}(y) \mathrm{F}_{1}^{1/8}(t)\mathrm{F}_3^{3/8}(t) }ds^2_{\mathrm{Het}~\mathrm{E}_8\times\mathrm{E}_8}\nonumber\\
&=\frac{a^2(t)}{\mathrm{H}(y)} \mathrm{F}_1^{3/8}(t)\mathrm{F}_3^{1/8}(t)\left[-dt^2+\sum_{i=1}^3(dx^i)^2 \right]+\frac{1}{\mathrm{H}(y)}\mathrm{F}_1^{-5/8}(t)\mathrm{F}_3^{1/8}(t)\big[(d\tilde{x}_{11})^2\nonumber\\
&+(d\theta_1)^2\big]+\mathrm{H}^4(y)\mathrm{F}_1^{3/8}(t)\mathrm{F}_2(t)\mathrm{F}_3^{1/8}(t)g_{mn}dy^m dy^n ,\label{EinframeinE8Th}
\end{align} 
where we have substituted the coupling $g_{\mathrm{Het~E}_8\times \mathrm{E}_8}$  from Eq. (\ref{CouplIngE8E8HeT}).
Therefore, in this theory, we can have a four dimensional de Sitter isometry in flat slicing provided the following condition is satisfied 
\begin{align}
a^2(t)\mathrm{F}_1^{3/8}(t)\mathrm{F}_3^{1/8}(t)&\equiv g_{s}^{-2}\mathrm{H^2}(y)\mathrm{F}_1^{3/8}(t)\mathrm{F}_3^{1/8}(t)=\frac{1}{\Lambda t^2 }\nonumber\\
\implies \frac{g_s}{\mathrm{H}(y)}\equiv \overline{g}_s=&\sqrt{\Lambda }|t|( \mathrm{F}_1^3(t)\mathrm{F}_3(t))^{1/16}. \label{dSinHetE8tImesE8}
\end{align}
In the above expression, substituting $\mathrm{F}_1(t)$ and $\mathrm{F}_3(t)$ from Eq. (\ref{F1F3inHeteRE8E8}):
\begin{align}
\mathrm{F}_1(t)=(\overline{g}_s)^{\frac{2\hat{\alpha}(t)}3},\quad \text{and} \quad \mathrm{F}_3(t)=(\overline{g}_s)^{\frac {2\hat{\beta}(t)} 3},    \label{F1F3inEinStFraMeE8Th}
\end{align}
 we find 
\begin{align} \overline{g}_s=(\sqrt{\Lambda} |t|)^{\frac{24}{24-3\hat{\alpha}(t)-\hat{\beta}(t)}}.\label{ExpofgsinEinFrameE8Th}
\end{align}
Note however, in this case the form of $\hat{\alpha}(t)$ and $\hat{\beta}(t)$ will differ from those appearing in (\ref{hatAlphHatBetaHeteRE8E8}).

\noindent
Now, the time independence of the volume of the six dimensional internal space (\ref{EinframeinE8Th}) implies 
\begin{align}
\left( \mathrm{F}_1^{3/8}(t)\mathrm{F}_2(t)\mathrm{F}_3^{1/8}(t)\right)^2 (\mathrm{F}_1^{-5/8}(t) \mathrm{F}_3^{1/8}(t))=&1\nonumber\\
\implies  \mathrm{F}_1^{\frac{1}{16}}(t)\mathrm{F}_{2}(t)\mathrm{F}_3^{\frac{3}{16}}(t) =& 1\nonumber\\
\implies \mathrm{F}_{2}(t)=\left({ \overline{g}_s}\right)^{\frac{ -\hat{\alpha}(t) -3\hat{\beta}(t)}{24} }&. \label{ExpofF2(t)inEinstFrameE8Th}
\end{align}
Substituting (\ref{F1F3inEinStFraMeE8Th}), (\ref{ExpofgsinEinFrameE8Th}) and (\ref{ExpofF2(t)inEinstFrameE8Th}) in the metric (\ref{EinframeinE8Th}) we arrive at
\begin{align}
ds^2_{ \mathrm{Het}~\mathrm{E}_8\times \mathrm{E}_8,~{\rm E}}&=\frac{\mathrm{H}^{-1}(y)}{\Lambda t^2 } ds^2_{ \mathbb{R}^{3,1}}+ \mathrm{H}^3(y) \left(\Lambda t^2\right)^{ \frac{ 5 \hat{\alpha}(t)-\hat{\beta}(t)}{48- 6\hat{\alpha}(t)-2\hat{\beta}(t)}}ds^2_{\mathcal{M}_4 }+\mathrm{H}^{-1}(y)\left( \Lambda t^2 \right)^{\frac{\hat{\beta}(t)-5\hat{\alpha}(t)}{24- 3\hat{\alpha}(t)-\hat{\beta}(t)}}ds^2_{\mathbb{T}^2}.\label{EinsFrameMetDSisOE8Th} 
\end{align}

It shows that in the Einstein frame as $t\rightarrow 0^-$, $\hat{\beta}(t)$ has to be equal to $5 \hat{\alpha}(t)$ for there to be no late time singularities of the internal manifolds $\mathcal{M}_4$ and $\mathbb{T}^2=\hat{\mathbf{S}}^1_{\theta_1}\times\widetilde{\mathbf{S}}^1_{11}$. So, we define 
\begin{align}
\hat{\alpha}(t)&=\begin{cases}   \hat{\alpha}_o,\quad-\frac{1}{\sqrt{\Lambda }}<t\leq -\epsilon\\
\hat{\gamma}_{o},\quad\quad-\epsilon< t<0
\end{cases},\quad\hat{\beta}(t)=\begin{cases}
\hat{\beta}_o,\quad\quad-\frac{1}{\sqrt{\Lambda }}<t\leq -\epsilon\\
5\hat{\gamma}_{o},\quad\quad-\epsilon< t<0 
\end{cases}.\label{FuncAnsTzhatAlphahatBeta}
\end{align}
where $\hat{\alpha}_o$ and $\hat{\beta}_o$ should still satisfy the condition (\ref{CondhAtAlpohatBetao}). 
We can compactly express the type IIA coupling constant obtained in Eqs.  (\ref{TimeDepIIAcConstE8Th}) and (\ref{ExpofgsinEinFrameE8Th}) as
\begin{align}
\overline{g}_s&=( \sqrt{\Lambda} |t|)^{\frac{6v^2}{ 6v^2 -(2v-1)\hat{\alpha}(t)-\hat{\beta}(t) }},\label{IIAinString&EinsFrame}
\end{align}
and the warp factor $\mathrm{F}_2(t)$ obtained in Eqs. \eqref{ExpofF2inE8TheorY} and \eqref{ExpofF2(t)inEinstFrameE8Th} as
\begin{align}
\mathrm{F}_2(t)\equiv\mathrm{F}_2(\overline{g}_s)&=(\overline{g}_s)^{-\frac{(\hat{\alpha}(t)+3\hat{\beta}(t))}{4v^2}},\label{F2tHetE8thryinEitherFr}
\end{align}
where $v=1,2$ for the string frame, the Einstein frame respectively. Likewise, the parameters $\hat{\alpha}(t)$ and $\hat{\beta}(t)$ in Eqs. (\ref{hatAlphHatBetaHeteRE8E8}), (\ref{FuncAnsTzhatAlphahatBeta}) can be written for either frame as
\begin{align}
\hat{\alpha}(t)&=\begin{cases}   \hat{\alpha}_o,\quad-\frac{1}{\sqrt{\Lambda }}<t\leq -\epsilon\\
\hat{\gamma}_{o},\quad\quad-\epsilon< t<0
\end{cases},\quad\hat{\beta}(t)=\begin{cases}
\hat{\beta}_o,\quad\quad\quad\quad-\frac{1}{\sqrt{\Lambda }}<t\leq -\epsilon\\
(4v-3)\hat{\gamma}_{o},\quad\quad-\epsilon< t<0 
\end{cases}.\label{EitherfrFuncAnsTzhatAlphahatBeta}
\end{align}
Note that the time dependence of $\hat{\alpha}(t)$ and $\hat{\beta}(t)$ appearing in the expression for $\overline{g}_s$ in Eq.~(\ref{IIAinString&EinsFrame}) does not alter the temporal domain of validity of the de Sitter Glauber--Sudarshan state in heterotic $\mathrm{E}_8\times \mathrm{E}_8$ theory. Consequently, the allowed temporal interval remains identical to the TCC bound as obtained in Eq.~\eqref{TempDomHetE8E8TheoRY}.

Next, we investigate whether this de Sitter Glauber--Sudarshan state also  satisfies the other effective field theory constraint (\ref{EftCriTeriA1}). As discussed in \S\ref{sec:NECIIB}, this condition is equivalent to satisfying the null energy condition for a $(3+1)$-dimensional FLRW cosmology. We will follow the methodology developed therein.
\section{NEC in heterotic \texorpdfstring{$\mathrm{E}_{8}\times \mathrm{E}_{8}$}{E8XE8}  string theory }
\label{sec:NECE8XE8}
The metric coefficient of the 3+1 dimensional non-compact space for the heterotic $\mathrm{E}_8\times \mathrm{E}_8$ theory in string frame (\ref{HetroTicE8XE8 metric}) is
\begin{align} g_{\mu\nu}^{\mathrm{Het~E}_8\times \mathrm{E}_8}&=a^2(t)\sqrt{\mathrm{F}_1(t)\mathrm{F}_3(t)}\eta_{\mu\nu}
\end{align}
where $\eta_{\mu\nu}=\mathrm{diag}(-1,1,1,1).$ Following the procedure of  \S\ref{sec:NECIIB}, we consider
\begin{align}
\sqrt{\mathrm{F}_1(t)\mathrm{F}_3(t)} a^2(t)&= \left( \Lambda t^2\right)^n.
\end{align}
Substituting $\mathrm{F}_1(t)$ and $\mathrm{F}_3(t)$ from Eq. (\ref{pqinalphabeta}) we obtain 
\begin{align}   \left({\overline{g}_s}\right)^{\frac{\hat{\alpha}_o+\hat{\beta}_o}{3}-2}&=(\Lambda t^2)^n
    \implies \overline{g}_s= (\Lambda t^2)^{\frac{3n}{\hat{\alpha}_o+\hat{\beta}_o-6}}.\label{SframeinE8Th}
\end{align}
In Einstein frame, the metric coefficient of 3+1 dimensional non-compact space  (\ref{EinframeinE8Th}) takes the following form 
\begin{align}
g_{\mu\nu}^{\mathrm{Het}~\mathrm{E_8\times E_8 ,E}}
&=\frac{a^2(t)}{\mathrm{H}(y)} \mathrm{F}_1^{3/8}(t)\mathrm{F}_3^{1/8}(t)\eta_{\mu\nu}.
\end{align}
 Thus, the time dependent factors of this metric coefficient can be equated to 
\begin{align}
    \mathrm{F}_1^{3/8}(t)\mathrm{F}_3^{1/8}(t)a^2(t)&=(\Lambda t^2)^n\nonumber\\
    \implies \left(\frac{g_s}{\mathrm{H}(y)} \right)^{\frac{3\hat{\alpha}_o+\hat{\beta}_o}{12}-2 }&=(\Lambda t^2)^n\nonumber\\
    \implies \overline{g}_s =(\Lambda t^2&)^{\frac{12n }{3\hat{\alpha}_o+\hat{\beta}_o-24}},
\end{align}
where we have substituted $\mathrm{F}_1(t)$ and $\mathrm{F}_3(t)$ from Eq. (\ref{pqinalphabeta}). The above expression for $\overline{g}_s$ in the Einstein frame can be written together with its string-frame counterpart, Eq.~(\ref{SframeinE8Th}), in the compact form
\begin{align}
    \overline{g}_s&=(\Lambda t^2)^{\frac{n}{\frac{(5-v)\hat{\alpha}_o+(7-3v)\hat{\beta}_o}{12}-2}},
\end{align}
where $v=1,2$ for the string frame, the Einstein frame respectively. Computing its derivative with respect to $t$ gives 
\begin{align}
    \frac{\partial \overline{g}_s }{\partial t}&=\left(\frac{2n}{\frac{(5-v)\hat{\alpha}_o+(7-3v)\hat{\beta}_o}{12}-2}\right)\frac{\overline{g}_s}{t}\nonumber\\
    &\propto \overline{g}_s^{\left(\frac{2- \frac{(5-v)\hat{\alpha}_o+(7-3v)\hat{\beta}_o }{12}}{2n}+1\right)\geq 0}\nonumber\\
    \therefore~~~~ & \frac{2- \frac{(5-v)\hat{\alpha}_o+(7-3v)\hat{\beta}_o }{12}}{2n}+1\geq 0\implies \frac{1}{n}\geq -\frac{1}{1- \frac{(5-v)\hat{\alpha}_o+(7-3v)\hat{\beta}_o}{24} }.\label{EFTBound}
\end{align}
As explained in \S\ref{E8XE8ThrdistWarpf}, the range of $\hat{\alpha}_o$ and $\hat{\beta}_o$ is
$0<\frac{\hat{\alpha}_{o}}{9}<\hat{\beta}_o<\hat{\alpha}_{o}<1$.
For all admissible values of these parameters, taking the intersection of the allowed regions for $\frac{1}{n}$ implied by \eqref{EFTBound}, we obtain the null energy condition:
\begin{align}
    \frac{1}{n} \ge -1.
    \label{NECinHetE8ThrYTimeIndPar}
\end{align}

\vspace{5pt}
\noindent
If suppose we take into account the time dependence of the parameters $\hat{\alpha}(t)$ and $\hat{\beta}(t)$ in string frame \eqref{hatAlphHatBetaHeteRE8E8} then the string coupling in Eq.  (\ref{SframeinE8Th})   gets modified as  
\begin{align}
    \overline{g}_s&=({\Lambda}t^2)^{\frac{3n }{\hat{\alpha}(t) +\hat{\beta}(t)-6}}.\label{StringCouPStrFRaMEE8}
\end{align}
Taking its derivative with respect to $t$, we obtain an expression analogous to \eqref{tderofgsbar}, with the replacements $v=1$ and $\beta(t)\rightarrow \frac{\hat{\alpha}(t)+\hat{\beta}(t)}{3}$. Below we write down the result
\begin{align}
    \frac{ \partial \overline{g}_s}{\partial t}&=\overline{g}_s^{1+ {\mathrm{B}}(t)}- 6n \sqrt{ \Lambda} \overline{g}_s^{1+ \frac{1}{n}- \frac{\hat{\alpha}(t)+\hat{\beta}(t)}{6n}+{\mathrm{C}}(t)}\label{PartIalDerHetE8ThNeC}
\end{align}
where 

\begin{align}
{\mathrm{B}}(t)&=\frac{ \log\left[\left( \frac{ \dot{\hat{\alpha}}(t)+\dot{\hat{\beta}}(t) }{6-\hat{\alpha}(t)-\hat{\beta}(t)} \right) \log \overline{g}_s \right] }{\log\overline{g}_s},~~~\mathrm{C}(t)=-\frac{\log \left(6-\hat{\alpha}(t)-\hat{\beta}(t) \right) }{\log \overline{g}_s}.
\end{align}
Since $\overline{g}_s<<1$ and $\hat{\beta}(t)<\hat{\alpha}(t)<1$, the sign of $\mathrm{C}(t)$ is positive. As in the case of heterotic $\mathrm{SO}(32)$ theory, cf.~(\ref{PosI2ndTermNEC}), requiring the exponent of $\overline{g}_s$ in the second term of \eqref{PartIalDerHetE8ThNeC} to be non-negative yields
\begin{align}
    \frac{1}{n}\geq -\frac{1+C(t)}{1-\frac{\hat{\alpha}(t)+\hat{\beta}(t)}{6}}.
\end{align}
The null energy condition \eqref{NECforFLRWcos} is obtained by taking the intersection of the allowed ranges of $\frac{1}{n}$ corresponding to all permitted values of the parameters $\hat{\alpha}(t)$ and $\hat{\beta}(t)$:
\begin{align}
    \frac{1}{n}\geq -1.\label{NECinE8E8hEtTheory}
\end{align}
Further, for $\mathrm{B}(t)$ to make sense it is necessary that we have $(\dot{\hat{\alpha}}(t)+\dot{\hat{\beta}}(t))\leq 0$, i.e. 
\begin{align}
    \mathrm{B}(t)&=-\frac{ \log\left[ \frac{\vert \dot{\hat{\alpha}}(t)+\dot{\hat{\beta}}(t) \vert }{\left( 6-\hat{\alpha}(t)-\hat{\beta}(t)\right) }  \vert \log \overline{g}_s \vert \right] }{\vert\log\overline{g}_s \vert}.
\end{align}
As in the case of heterotic $\mathrm{SO}(32)$ theory \S\ref{sec:NEC}, the sign of $\mathrm{B}(t)$ would require us to know the precise value of $| \dot{\hat{\alpha}}(t)+\dot{\hat{\beta}}(t) |.$ However, if the functional forms of $\hat{\alpha}(t)$ and $\hat{\beta}(t)$ are such that the following condition holds
\begin{align}
    \frac{ |\dot{\hat{\alpha}}(t)+\dot{\hat{\beta}}(t) | }{(6 -\hat{\alpha}(t)-\hat{\beta}(t) )}<\frac{1}{|\log\overline{g}_s|}
\end{align}
then $\text{sign}~\mathrm{B}(t)>0$, which immediately implies the existence of well defined effective field theory description.

In addition to the NEC constraint that we looked at, one has to also respect the axion cosmology constraints for the heterotic $\mathrm{E}_8\times \mathrm{E}_8$ theory\cite{PeterSvrcek_2006,KeshavDasgupta_2008}. In fact the cosmological bound on the axion decay constant may modify the functional form of the time dependent parameters $\hat{\alpha}(t)$ and $\hat{\beta}(t)$ as obtained in (\ref{EitherfrFuncAnsTzhatAlphahatBeta}). We will elaborate on these aspects in the following section.  
\section{Axions in \texorpdfstring{$\mathrm{E}_{8}\times \mathrm{E}_{8}$}{E8XE8}  heterotic string theory}
\label{sec:AxionsinE8Thoery}

The axion is a pseudo-scalar field originally introduced by Roberto Peccei and Helen Quinn as a dynamical solution to the strong CP problem in quantum chromodynamics (QCD) \cite{PhysRevLett.38.1440}. 

The origin of this problem lies in the presence of a CP-violating term in the QCD Lagrangian of the form
\begin{align}
\mathcal{L}_{\overline{\theta}}&=\frac{\overline{\theta}}{8\pi^2}\int \mathbf{tr} \,(\mathrm{F}\wedge \mathrm{F})\nonumber\\
&=\frac{\overline{\theta}}{16\pi^2}\int d^4x\, \mathbf{tr}\,(\mathrm{F}_{\mu\nu}\widetilde{\rm F}^{\mu\nu})\nonumber\\
&=\frac{\overline{\theta}}{32\pi^2}\int d^4x\, \epsilon^{\mu\nu\rho\sigma}\,\mathbf{tr}(\mathrm{F}_{\mu\nu}\mathrm{F}_{\rho\sigma}) 
\end{align}
where $\mathbf{tr}$ is a trace in the three-dimensonal representation of $SU(3)$,  $\mathrm{F}=\frac{1}{2}\mathrm{F}_{\mu\nu}dx^\mu \wedge dx^\nu$ is the gauge two-form and $\widetilde{\rm F}_{\mu\nu}=\frac{1}{2}{\epsilon}_{\mu\nu \rho\sigma}\mathrm{F}^{\rho\sigma}$ is the dual field strength. Experimental bounds constrain the parameter $\overline{\theta}$ to be extremely small, $|\overline{\theta}| \lesssim 10^{-10}$. This poses a significant theoretical puzzle, as there is no \emph{a priori} reason for $\overline{\theta}$ to be so tiny. Explaining this extreme smallness of $\overline{\theta}$ is referred to as the strong CP problem \cite{TURNER199067,KIM19871,hook2023tasilecturesstrongcp}. 

The Peccei--Quinn mechanism resolves this issue by promoting $\overline{\theta}$ to a dynamical field. This is achieved by introducing a global $U(1)_{\mathrm{PQ}}$ symmetry, which is spontaneously broken at a scale $f_a$, giving rise to a pseudo-Nambu--Goldstone boson: the axion, $a$. The resulting low-energy effective Lagrangian contains a term
\begin{align}
\left(\overline{\theta} + \frac{a}{f_a}\right)\frac{1}{16\pi^2}\mathbf{tr}(\mathrm{F}^{\mu\nu}\tilde{\rm F}_{\mu\nu}),
\end{align}
where $f_a$ referred to as the axion decay constant, controls the strength of the axion coupling.  Non-perturbative QCD effects generate a potential for the axion whose minimum dynamically drives the effective parameter $\overline{\theta}_{\text{eff}} = \overline{\theta} + a/f_a$ to zero, thereby eliminating CP violation in the strong sector\cite{hook2023tasilecturesstrongcp}.

String theories compactified to four dimensions naturally contain axion-like fields already at the classical level\cite{PeterSvrcek_2006}. In heterotic $\mathrm{E}_8 \times \mathrm{E}_8$ string theory, such an axion arises from the Neveu--Schwarz two-form field $B_{\mathrm{A}\mathrm{B}}$. Upon compactification to four dimensions, the spacetime components of this two-form can be dualised to a pseudo-scalar field ${a}$, which plays the role of the axion-field: $d\mathbf{B}_2=\star_{4}da$. The effective four-dimensional interaction contains a term\cite{KeshavDasgupta_2008}
\begin{align}
\frac{\tilde{a}}{f_a}\frac{1}{16\pi^2}\mathrm{Tr}\left(\mathrm{F}_{\mathrm{A}\mathrm{B}}\mathrm{\tilde{F}}^{\mathrm{A}\mathrm{B}}\right)\quad\quad\text{where}~\widetilde{a}\propto a.
\end{align}
This reproduces precisely the structure required for the Peccei--Quinn mechanism. In this framework, the axion decay constant $f_a$ is not an arbitrary parameter but is determined by the underlying compactification data. In particular, it depends on the string scale, the Planck scale, and the time-dependent warp factors of the internal manifold\cite{10.1002/prop.70047,Archana_CC2026}. According to the astrophysical and cosmological bounds, $f_a$ should lie within the range \cite{PeterSvrcek_2006}:
\begin{align}
    10^9 \,\text{GeV} \lesssim f_a \lesssim 10^{12} \,\text{GeV}.\label{RanGefa(t)HasTosaT}
\end{align}
This section is devoted to analysing how such bounds on the axion decay constant $f_a$ further constrains the functional form of $\hat{\alpha}(t)$ and $\hat{\beta}(t)$ as obtained in Eq.  \eqref{EitherfrFuncAnsTzhatAlphahatBeta} \cite{10.1002/prop.70047,Archana_CC2026}.  

Recall, in string frame the metric of $\mathrm{E}_{8}\times \mathrm{E}_{8}$ heterotic string theory (\ref{HetroTicE8XE8 metric}) with de Sitter isometry is given as
\begin{align}    ds^2_{10}&=\frac{1}{\Lambda t^2} \eta_{\mu\nu}dx^\mu dx^\nu+\mathrm{H}^4(y)\mathrm{F}_{2}(t)\sqrt{\mathrm{F}_1(t)\mathrm{F}_3(t)}g_{mn}(y)dy^m dy^n+\sqrt{\frac{\mathrm{F}_3(t)}{\mathrm{F}_1(t)}}((d\theta_1)^2+(d\tilde{x}_{11})^2)\nonumber\\
&= {g}^{d \mathrm{S}}_{\mu\nu}dx^\mu dx^\nu+\mathrm{H}^4(y)\left(\Lambda t^2\right)^{\frac{\hat{\alpha}(t)-\hat{\beta}(t)}{12-2\hat{\alpha}(t)-2\hat{\beta}(t)}}g_{mn}(y)dy^m dy^n+
    \left(\Lambda t^2\right)^{\frac{\hat{\beta}(t)-\hat{\alpha}(t)}{6-\hat{\alpha}(t)-\hat{\beta}(t)}}((d\theta_1)^2+(d\tilde{x}_{11})^2),\label{SframeMetE8xE8ThrwithdS}
\end{align}
where to arrive at the second line we have substituted the expressions of $\mathrm{F}_1(t),~\mathrm{F}_3(t)$ and $\mathrm{F}_2(t)$ from (\ref{F1F3inHeteRE8E8}), and (\ref{ExpofF2inE8TheorY}), along with the value of $\overline{g}_{s}$ from (\ref{TimeDepIIAcConstE8Th}). Further we have called the 3+1 dimensional de Sitter metric coefficients in flat slicing to be 
\begin{align}
{g}^{d\mathrm{S}}_{\mu\nu}&=\frac{1}{\Lambda t^2}\eta_{\mu\nu},\quad\text{where}\quad \mu,\nu=0,1,2,3\quad\text{and}\quad \eta_{\mu\nu}=\text{diag}(-1,1,1,1).\label{DefOvErlInegMunU}
\end{align}
The coupling of this theory (\ref{CouplIngE8E8HeT}) can be written as 
\begin{align}
g_{\mathrm{het}}&=\mathrm{H}^2(y)\mathrm{F}_{1}^{{1}/{4}}(t)\mathrm{F}_{3}^{{3}/{4}}(t)=\mathrm{H}^2(y)\left( {\overline {g}_s}\right)^{\frac{\hat{\alpha}(t)+3\hat{\beta}(t)}{6}}=\mathrm{H}^2(y)\left( \Lambda t^2\right)^{\frac{\hat{\alpha}(t)+3\hat{\beta}(t)}{12-2\hat{\alpha}(t)-2\hat{\beta}(t)}}.\label{HetCoUStringFrame}
\end{align}  
\vspace{5pt}
\noindent
In string frame, the action of the heterotic $\mathrm{E}_8\times \mathrm{E}_8$ string theory is as follows 
\begin{align}
    S&=\frac{1}{2\kappa_{10}^2}\int d^{10}x \sqrt{-\mathrm{det}G}~e^{-2\phi}\left[\mathrm{R}+4 G^{\rm AB}\partial_{\rm A}\phi \partial_{\rm B}\phi-\frac{1}{2}\vert \mathbf{H}_{3}\vert^2-\frac{\alpha^\prime}{120}\mathrm{Tr}(\vert \mathbf{F}\vert^2) \right]
\end{align}
where $\mathrm{R}$ is the Ricci scalar, $G$ denotes the string frame metric, $\phi$ is the dilaton, the trace $\mathrm{Tr}$ is taken with respect to the adjoint representation of $\mathrm{E}_{8}\times \mathrm{E}_{8}$ theory and for any $p$ form $\mathrm{F}_p$, 
\begin{align}
|{\rm F}_p|^2=\frac{1}{p!}{\rm F}_{\rm A_1\cdots A_p}{\rm F}_{\rm B_1\cdots B_p}G^{\rm A_1B_1}\cdots G^{\rm A_pB_p}.    
\end{align}
 Taking into account the string frame metric (\ref{SframeMetE8xE8ThrwithdS}), the zero mode action of the graviton turns out to be 
\begin{align}
    S_{G}^{(0)}&=\frac{1}{2\kappa_{10}^2 }\int d^{10}x \sqrt{-\mathrm{det}G  }~e^{-2\phi} ~\mathrm{R}\nonumber\\
    &\rightarrow \frac{1}{2\kappa_{10}^2}\int d^4x d^6y\sqrt{(\mathrm{H}^4(y))^4}~ \sqrt{\mathrm{det}g_{\mathcal{M}_4}\mathrm{det}g_{\mathbb{T}^2}}\frac{1}{\mathrm{H}^4(y)\left( \Lambda t^2\right)^{\frac{\hat{\alpha}(t)+3\hat{\beta}(t)}{6-\hat{\alpha}(t)-\hat{\beta}(t)}}} \sqrt{-\mathrm{det}{g}^{d\mathrm{S}}}~{\mathrm{R}^{d\mathrm{ S}}}\nonumber\\
    &=\frac{1}{2\kappa_{10}^2}\int d^4x  \frac{1}{\left( \Lambda t^2\right)^{\frac{\hat{\alpha}(t)+3\hat{\beta}(t)}{6-\hat{\alpha}(t)-\hat{\beta}(t)}}}\sqrt{-\mathrm{det}{g}^{d\mathrm{S}}}~{\mathrm{R}^{d\mathrm{S}}}\int d^6y~ \mathrm{H}^4(y)\sqrt{\mathrm{det}g_{\mathcal{M}_4}\mathrm{det}g_{\mathbb{T}^2}}\nonumber\\
    &=\frac{\mathrm{M}_p^2}{2}\int d^4x  \frac{1}{\left( \Lambda t^2\right)^{\frac{\hat{\alpha}(t)+3\hat{\beta}(t)}{6-\hat{\alpha}(t)-\hat{\beta}(t)}}}\sqrt{-\mathrm{det}{g}^{d\mathrm{S}}}~{\mathrm{R}^{d\mathrm{S}}}
\end{align}
where in the second step we have used the coupling of heterotic theory in string frame as given in Eq. (\ref{HetCoUStringFrame}). $\mathrm{det}g_{\mathcal{M}_4}$ and $\mathrm{det}g_{\mathbb{T}^2}$ denote the determinant of the unwarped metric of $\mathcal{M}_4$ and $\mathbb{T}^2$. $\mathrm{det}g^{d\mathrm{S}}$ is determinant of the de Sitter metric \eqref{DefOvErlInegMunU}. ${\rm R}^{d{\rm S}}$ is the Ricci scalar computed with the de Sitter metric ${g}^{d{\rm S}}_{\mu\nu}$ (\ref{DefOvErlInegMunU}) and the Planck mass $\mathrm{M}_p$ is defined as 
\begin{align}
\mathrm{M}_{p}^2&=\frac{1}{\kappa_{10}^2}\int d^6y ~\mathrm{H}^4(y)\sqrt{\mathrm{det}g_{\mathcal{M}_4}\mathrm{det}g_{\mathbb{T}^2}}\nonumber\\
    &=4\pi \mathrm{M}_{s}^{8}\int d^6y ~\mathrm{H}^4(y)\sqrt{\mathrm{det}g_{\mathcal{M}_4}\mathrm{det}g_{\mathbb{T}^2}}.\label{Mp2ExpResSION}
\end{align}
Likewise, the action due to the zero mode of the NS-NS field turns out to be 
\begin{align}
    S_{\rm NS}^{(0)}&=   -\frac{1}{4\kappa_{10}^2}\int d^{10}x \sqrt{-\mathrm{det}G} ~e^{-2\phi}~ \vert \mathbf{H}_3\vert^2 
    \nonumber\\
    &\rightarrow-\frac{1}{4\kappa_{10}^2}\int d^4 x d^6y\sqrt{(\mathrm{H}^4(y))^4}~ \sqrt{\mathrm{det}g_{\mathcal{M}_4}\mathrm{det}g_{\mathbb{T}^2}}\frac{1}{\mathrm{H}^4(y)\left( \Lambda t^2\right)^{\frac{\hat{\alpha}(t)+3\hat{\beta}(t)}{6-\hat{\alpha}(t)-\hat{\beta}(t)}}}\sqrt{-\mathrm{det}{g}^{d\mathrm{S}}}~\vert{\mathbf{H}}_{3}\vert^2_{d{\rm S}}\nonumber\\
    &=-\frac{1}{4\kappa_{10}^2}\int d^4x \left( \Lambda t^2\right)^{\frac{\hat{\alpha}(t)+3\hat{\beta}(t)}{\hat{\alpha}(t)+\hat{\beta}(t)-6}}\sqrt{-\mathrm{det}{g}^{d\mathrm{S}}}~\vert{\mathbf{H}}_{3}\vert^2_{d{\rm S}}\int d^6y~{\rm H}^{4}(y)\sqrt{\mathrm{det}g_{\mathcal{M}_4}\mathrm{det}g_{\mathbb{T}^2}}\nonumber\\
    &=-\frac{\mathrm{M}_{p}^2}{4}\int d^4x ~\beta(t)~\sqrt{-\mathrm{det}{g}^{d\mathrm{S}}}~\vert{\mathbf{H}}_{3}\vert^2_{d{\rm S}}
\end{align}
where $\vert{\mathbf{H}}_{3}\vert^2_{d{\rm S}}$ is defined with respect to the metric ${g}^{d{\rm S}}_{\mu\nu}:$
\begin{align}
\vert{\mathbf{H}}_{3}\vert^2_{d{\rm S}}=\mathbf{H}_{\mu\nu\rho}\mathbf{H}_{\mu^\prime\nu^\prime\rho^\prime}{g}_{d{\rm S}}^{\mu\mu^\prime}{g}_{d{\rm S}}^{\nu\nu^\prime}{g}_{d{\rm S}}^{\rho\rho^\prime},
\end{align}
and $\beta(t)$ is defined as\footnote{Note that $\beta(t)$ appearing in this chapter should not be confused with the time dependent warp factor in heterotic $\mathrm{SO}(32)$ theory, which is also denoted as $\beta(t)$ in Chapter \ref{Chap:3}.}
\begin{align}
    \beta(t)&=\left(\Lambda t^2 \right)^{\frac{\hat{\alpha}(t)+3\hat{\beta}(t)}{\hat{\alpha}(t)+\hat{\beta}(t)-6}}.\label{AxiOnbetatStringFrame}
\end{align}
In Einstein frame, the ten-dimensional action of the heterotic string theory is given as 
\begin{align}
S_{\mathrm{E}}&=\frac{1}{2 \kappa_{10}^2} \int d^{10}x \sqrt{ -\mathrm{det}G_{\rm (E)}}\left[\mathrm{R}+ 4 G_{\rm (E)}^{\rm AB} \partial_{\rm A}\phi \partial_{\rm B}\phi -\frac{1}{2}e^{-\phi} \vert \mathbf{H}_3\vert^2- \frac{\alpha^\prime}{120}e^{-\phi /2}\mathrm{Tr}(\vert \mathbf{F}\vert^2) \right].
\end{align}
Here ${\rm A,B}$ are the 10-dimensional indices, $\mathrm{R}$ is the Ricci scalar, $G_{(\rm E)}$ denotes the Einstein frame metric and $\phi$ is the dilaton.  Now, we express the Einstein frame metric (\ref{EinsFrameMetDSisOE8Th}) as 
\begin{align}
ds^2_{ \mathrm{Het}~\mathrm{E}_8\times \mathrm{E}_8,~{\rm E}}&=\frac{1}{\mathrm{H}(y)\Lambda t^2 } \eta_{\mu\nu}dx^\mu dx^\nu+ \mathrm{H}^3(y) \left(\Lambda t^2\right)^{ \frac{ 5 \hat{\alpha}(t)-\hat{\beta}(t)}{48- 6\hat{\alpha}(t)-2\hat{\beta}(t)}}g_{mn}(y)dy^m dy^n\nonumber\\
&\quad\quad\quad+\mathrm{H}^{-1}(y)\left( \Lambda t^2 \right)^{\frac{\hat{\beta}(t)-5\hat{\alpha}(t)}{24- 3\hat{\alpha}(t)-\hat{\beta}(t)}}((d\theta_1)^2+(d\tilde{x}_{11})^2)\nonumber\\
&=\frac{1}{\mathrm{H}(y) } {g}^{d{\rm S}}_{\mu\nu}dx^\mu dx^\nu+ \mathrm{H}^3(y) \left(\Lambda t^2\right)^{ \frac{ 5 \hat{\alpha}(t)-\hat{\beta}(t)}{48- 6\hat{\alpha}(t)-2\hat{\beta}(t)}}g_{mn}(y)dy^m dy^n\nonumber\\
&\quad\quad\quad+\mathrm{H}^{-1}(y)\left( \Lambda t^2 \right)^{\frac{\hat{\beta}(t)-5\hat{\alpha}(t)}{24- 3\hat{\alpha}(t)-\hat{\beta}(t)}}((d\theta_1)^2+(d\tilde{x}_{11})^2), 
\end{align}
to write down the zero mode action due to the graviton in the following way 
\begin{align}
S_{G^\mathrm{E}}^{(0)}&=\frac{1}{2 \kappa_{10}^2} \int d^{10}x \sqrt{-\mathrm{det}G^{\rm E}} ~\mathrm{R}\nonumber\\
    &\rightarrow \frac{1}{2\kappa_{10}^2}\int d^4xd^6y \sqrt{ {(\mathrm{H}^{-1}(y) )^4} (\mathrm{H}^3(y))^4{(\mathrm{H}^{-1}(y))^2}}\sqrt{ \mathrm{det}g_{\mathcal{M}_4 } \mathrm{det}g_{\mathbb{T}^2}}~\mathrm{H}(y) \sqrt{-\mathrm{det}{g}^{d\mathrm{S}}}~{\mathrm{R}^{d\mathrm{S}}}\nonumber\\
    &=\frac{1}{2 \kappa_{10}^2}\int d^4x~\sqrt{-\mathrm{det}{g}^{d\mathrm{S}}}~{\mathrm{R}^{d\mathrm{S}}} \int d^6y~\mathrm{H}^4(y)\sqrt{ \mathrm{det}g_{\mathcal{M}_4 } \mathrm{det}g_{\mathbb{T}^2}}\nonumber\\
    &=\frac{ \mathrm{M}_p^2 }{2}\int d^4x~ \sqrt{-\mathrm{det}{g}^{d\mathrm{S}}}~{\mathrm{R}^{d\mathrm{S}}}
\end{align}
where $\mathrm{M}_p^2$ is as given in Eq. (\ref{Mp2ExpResSION}). In a similar way, the zero mode  action due to the NS-NS field gives 
\begin{align}
    S^{(0)}_{\mathrm{NS}, ~\mathrm{E}}&=-\frac{1}{4\kappa_{10}^2}\int d^{10}x \sqrt{ -\mathrm{det}G^{\rm E}} ~e^{-\phi} \vert \mathbf{H}_3\vert^2\nonumber\\
    &\rightarrow -\frac{1}{4 \kappa_{10}^2} \int d^4x d^6 y \sqrt{(\mathrm{H}^{-1}(y))^4(\mathrm{H}^3(y))^4(\mathrm{H}^{-1}(y))^2}\sqrt{ \mathrm{det} g_{\mathcal{M}_4 }\mathrm{det} g_{\mathbb{T}^2}} \nonumber\\
    &~~~~~~~~~~~~~~~~\times \frac{1}{ \mathrm{H}^2(y) (\Lambda t^2 )^{ \frac{2 \hat{\alpha}(t)+6\hat{\beta}(t)}{24-3 \hat{\alpha}(t)-\hat{\beta}(t) }} } \times (\mathrm{H}^{-1}(y))^{-3} \sqrt{-\mathrm{det}{g}^{d\mathrm{S}}}~\vert{\mathbf{H}}_{3}\vert^2_{d{\rm S}} \nonumber\\
    &= -\frac{1}{  4\kappa_{10}^2}\int d^4x ~(\Lambda t^2 )^{ \frac{2 \hat{\alpha}(t)+6\hat{\beta}(t)}{3 \hat{\alpha}(t)+\hat{\beta}(t)-24 }} \sqrt{-\mathrm{det}{g}^{d\mathrm{S}}}~\vert{\mathbf{H}}_{3}\vert^2_{d{\rm S}}\int d^6y~ \mathrm{H}^{4}(y)~ \sqrt{ \mathrm{det} g_{\mathcal{M}_4 }\mathrm{det} g_{\mathbb{T}^2}}\nonumber\\
    &=-\frac{ \mathrm{M}_p^2 }{4}\int d^4 x ~\beta(t) ~\sqrt{-\mathrm{det}{g}^{d\mathrm{S}}}~\vert{\mathbf{H}}_{3}\vert^2_{d{\rm S}} ,  
\end{align}
to arrive at the second step we have substituted the coupling of heterotic theory in Einstein frame:
\begin{align} g_{\mathrm{Het~E}_8\times \mathrm{E}_8,~\mathrm{E}}&=\mathrm{H}^2(y)\mathrm{F}_1^{1/4}(t)\mathrm{F}_3^{3/4}(t)=\mathrm{H}^2(y)\left( {\overline{g}_s }\right)^{\frac{\hat{\alpha}(t)+3\hat{\beta}(t)}{6}}=\mathrm{H}^2(y)\left(\Lambda t^2\right)^{\frac{ 2\hat{\alpha}(t)+6\hat{\beta}(t)}{24- 3\hat{\alpha}(t)-\hat{\beta}(t)}}.\label{HetCoUEinsFrameE8Th}
\end{align}
Thus $\beta(t)$ in Einstein frame is given as
\begin{align}
    \beta(t)=(\Lambda t^2)^{ \frac{2 \hat{\alpha}(t)+ 6\hat{\beta}(t) }{3 \hat{\alpha}(t)+\hat{\beta}(t)-24}}.\label{AxiOnbetatEinFrame}
\end{align}
The results of $\beta(t)$ in the string frame (\ref{AxiOnbetatStringFrame}) and the Einstein frame (\ref{AxiOnbetatEinFrame}) can be compactly  written as 
\begin{align}
    \beta(t)&=( \Lambda t^2 )^{ \frac{v \hat{\alpha}(t)+ 3 v \hat{\beta}(t) }{(2v-1)\hat{\alpha}(t)+\hat{\beta}(t) -6v^2 } }\label{ExPreSSionOfbeTA(t)}
\end{align}
where $v=1,2$ corresponds to the string frame and the Einstein frame respectively. We can rewrite the dimensionally reduced heterotic action due to the NS-NS field as
\begin{align}
    S_{4}&=-\frac{ \mathrm{M}_p^2}{4}\int \beta(t)  \mathbf{H}_3\wedge \star_{d{\rm S}}\mathbf{H}_3 + \int a \left[d\mathbf{H}_3 -\frac{\alpha^\prime}{4}\left(\mathrm{tr} \mathbf{R}\wedge \mathbf{R}- \frac{1}{30}\mathrm{Tr}\mathbf{F}\wedge \mathbf{F}\right) \right].\label{FourDimActIOnwLagMultiPlier}
\end{align}
Here $a$ is a Lagrange multiplier for the Bianchi identity which in fact is the axion field and $\star_{d\mathrm{S}}$ denotes the Hodge operation with respect to the de Sitter metric (\ref{DefOvErlInegMunU}). We first determine the equation of motion for $\mathbf{H}_3$ by varying the action (\ref{FourDimActIOnwLagMultiPlier}) with respect to $\mathbf{H}_3$
\begin{align}
    \delta S_{4}&=-\frac{{\rm M}_p^2}{4}\int \beta(t) 2~\delta\mathbf{H}_3\wedge\star_{d\mathrm{S}}\mathbf{H}_3+ \int a~d(\delta \mathbf{H}_3).
\end{align}
Integrating the second term by parts we get 
\begin{align}
    \int a~d(\delta \mathbf{H}_3)\equiv     \int a\wedge d(\delta \mathbf{H}_3) &=-\int da \wedge \delta \mathbf{H}_3,
\end{align}
upto boundary terms. Therefore,
\begin{align}
    \delta S_{4}&=-\frac{{\rm M}_p^2}{2}\int \beta(t) ~\delta\mathbf{H}_3\wedge\star_{d\mathrm{S}}\mathbf{H}_3-\int da \wedge \delta \mathbf{H}_3\nonumber\\
    &=\int \delta\mathbf{H}_3\wedge \left[-\frac{\mathrm{M}_p^2}{2}\beta(t) \star_{d{\rm S}}\mathbf{H}_3+da\right].
\end{align}
Upon setting the variation $\delta S_4$ to zero, for arbitrary $\delta\mathbf{H}_3$ we get
\begin{align}
    \star_{d{\rm S}}\mathbf{H}_3&=\left(\frac{ 2}{ {\rm M}_p^2\beta(t)}\right)da\implies     \mathbf{H}_3=\left(\frac{ 2}{ {\rm M}_p^2\beta(t)}\right)\star_{d{\rm S}}da.
\end{align}
This implies that in four dimensions the axion is Hodge dual of the anti-symmetric Kalb-Ramond field. Substituting it back into the action (\ref{FourDimActIOnwLagMultiPlier}) we get 
\begin{align}
    S_{4}&=-\frac{\mathrm{M}_p^2}{4}\int \beta(t) \left(\frac{ 2}{ {\rm M}_p^2\beta(t)}\right)^2 \left(\star_{d{\rm S}}da\wedge da\right) -\int da \wedge \left(\frac{ 2}{ {\rm M}_p^2\beta(t)}\right)\star_{d{\rm S}}da\nonumber\\
    &\quad\quad\quad -\frac{\alpha^\prime}{4}\int a\left(\mathrm{tr} \mathbf{R}\wedge \mathbf{R}- \frac{1}{30}\mathrm{Tr}\mathbf{F}\wedge \mathbf{F}\right)\nonumber\\
    &=-\frac{1}{\mathrm{M}_p^2}\int \frac{1}{\beta(t)}da\wedge \star_{d \mathrm{S}}da -\frac{\alpha^\prime}{4}\int a\left(\mathrm{tr} \mathbf{R}\wedge \mathbf{R}- \frac{1}{30}\mathrm{Tr}\mathbf{F}\wedge \mathbf{F}\right)\nonumber\\
    &=\frac{2}{\mathrm{M}_p^2}\int d^4 x \frac{1}{\beta(t)}\sqrt{-\mathrm{det}g^{d\mathrm{S}}} ~\left(-\frac{1}{2}g_{d{\rm S}}^{\mu\nu}\partial_{\mu}a\partial_{\nu}a \right)-\frac{\alpha^\prime}{4}\int a\left(\mathrm{tr} \mathbf{R}\wedge \mathbf{R}- \frac{1}{30}\mathrm{Tr}\mathbf{F}\wedge \mathbf{F}\right)
\end{align}
To obtain the canonical form of the action, we rescale $a$ as 
\begin{align}
    \tilde{a}&=\sqrt{ \frac{ 2}{ \mathrm{M}_p^2~ \beta(t)}}~a.
\end{align}
This implies 
\begin{align}
    S_{4}&=\int d^4 x \sqrt{-\mathrm{det}g^{d\mathrm{S}}} ~\left(-\frac{1}{2}g_{d{\rm S}}^{\mu\nu}\partial_{\mu}\widetilde{a}\partial_{\nu}\widetilde{a} \right)-\int \frac{\tilde{a} }{\sqrt{2/(\mathrm{M}_p^2 \beta(t) ) } }{ \frac{\alpha^\prime}{4}}\left(\mathrm{tr} \mathbf{R}\wedge \mathbf{R}-\frac{1}{30}\mathrm{Tr}\mathbf{F}\wedge \mathbf{F}\right)\nonumber\\
  &=\int d^4 x \sqrt{-\mathrm{det}g^{d\mathrm{S}}} ~\left(-\frac{1}{2}g_{d{\rm S}}^{\mu\nu}\partial_{\mu}\widetilde{a}\partial_{\nu}\widetilde{a} \right)-\int \frac{\tilde{a} }{f_{a} }\frac{1}{16\pi^2} \left(\mathrm{tr}\mathbf{R}\wedge \mathbf{R}-\frac{1}{30}\mathrm{Tr}\mathbf{F}\wedge \mathbf{F}\right)
\end{align}
where we have replaced $\alpha^\prime=\frac{1}{4\pi^2 \mathrm{M}_s^2}$, $\mathrm{M}_s$ denotes the string scale and $f_a$ referred to as the axion decay constant has the following form 
\begin{align}
    f_{a}(t)&=\sqrt{ \frac{ 2}{\beta(t)} }\frac{ \mathrm{M}_s^2 }{ \mathrm{M}_p }=\frac{1}{ {2\pi } } (\sqrt{\Lambda}|t|)^{\frac{ v\hat{\alpha}(t)+ 3 v \hat{\beta}(t)}{ 6v^2 -(2v-1)\hat{\alpha}(t)-\hat{\beta}(t)}} \left( \frac{2\pi \mathrm{M}_s^{-4}}{\int d^6y \,\mathrm{H}^4(y)\sqrt{ \mathrm{det}g_{\mathcal{M}_4}(y)\mathrm{det}g_{\mathbb{T}^2}(y)} }\right)^{1/2}.\label{AxioNicCoupConst}
\end{align}
To arrive at the second step, we have substituted $\mathrm{M}_p$ from \eqref{Mp2ExpResSION} and $\beta(t)$ from \eqref{ExPreSSionOfbeTA(t)}. As $t\rightarrow 0^-$ in the late time limit, $f_{a}(t)\rightarrow 0$. This is as opposed to the range \eqref{RanGefa(t)HasTosaT} that $f_{a}(t)$ has to satisfy. In order to set a lower bound of $f_{a}(t)$ we make $\hat{\gamma}_{o}$ in Eq. (\ref{EitherfrFuncAnsTzhatAlphahatBeta}) time dependent such that 
\begin{align}
    \left(\sqrt{\Lambda} |\epsilon|\right)^{\frac{v \hat{\alpha}(-\epsilon)+3 v\hat{\beta}(-\epsilon)}{6v^2 -(2v-1)\hat{\alpha}(-\epsilon)-\hat{\beta}(-\epsilon)}}&=\left(\sqrt{\Lambda}|t|\right)^{\frac{v \hat{\gamma}_o(t) +3v(4v-3)\hat{\gamma}_o(t)}{6v^2-(2v-1)\hat{\gamma}_o(t)-(4v-3)\hat{\gamma}_o(t)}},~(\mathrm{for}~-\epsilon<t<0 ).
\end{align}
Taking logarithm on both sides we get 
\begin{align}
c_{o}:=\frac{(v\hat\alpha_o + 3v\hat\beta_o)} {6v^2 - (2v-1)\hat\alpha_o -\hat\beta_o}  \big\vert {\rm log}(\sqrt{\Lambda} \vert\epsilon\vert)\big\vert &={\frac{ (v \hat{\gamma}_o(t) +3v(4v-3)\hat{\gamma}_o(t))}{6v^2-(2v-1)\hat{\gamma}_o(t)-(4v-3)\hat{\gamma}_o(t)}} \nonumber\\
&~~~\times \big\vert {\rm log}(\sqrt{\Lambda} \vert t\vert)\big\vert\nonumber\\
\implies c_{o}&=\frac{ (12 v^2 -8 v )\hat{\gamma}_o(t)}{6v^2 -(6v-4)\hat{\gamma}_{o}(t)}\big\vert {\rm log}(\sqrt{\Lambda} \vert t\vert)\big\vert \nonumber\\
\implies c_{o}&=\frac{ (3 v^2 - 2 v )\hat{\gamma}_o(t)}{3v^2 -(3v-2)\hat{\gamma}_{o}(t)}\big\vert {\rm log}({\Lambda}  t^2)\big\vert \nonumber\\
\implies\hat{\gamma}_o(t)&=\left( \frac{3v^2}{3v-2}\right)\left(\frac{c_o}{c_o+v\big\vert {\rm log}({\Lambda}  t^2)\big\vert} \right).\label{ExpRsco}
\end{align}
Incorporating the aforementioned modification we express (\ref{EitherfrFuncAnsTzhatAlphahatBeta}) as
\begin{align}
    \hat{\alpha}(t)&=\begin{cases}   \hat{\alpha}_o,~~~~-\frac{1}{\sqrt{\Lambda }}<t\leq -\epsilon\\
\hat{\gamma}_{o}(t),~~~~~-\epsilon< t<0
\end{cases},~~~~~\hat{\beta}(t)=\begin{cases}
\hat{\beta}_o,~~~~~~~~~~~~~~-\frac{1}{\sqrt{\Lambda }}<t\leq-\epsilon\\
(4v-3)\hat{\gamma}_{o}(t),~~~~~-\epsilon< t<0 
\end{cases}.\label{AxionicCCTimeDepFactors}
\end{align}
Thus, within the temporal domain $-\epsilon <t<0$, the axion decay constant \eqref{AxioNicCoupConst} becomes 
\begin{align}
    f_a(t)&=\frac{1}{{2\pi}}e^{-c_o}\left( \frac{2\pi \mathrm{M}_s^{-4} }{\int d^6y \mathrm{H}^4(y)\sqrt{ \mathrm{det}g_{\mathcal{M}_4}(y)\mathrm{det}g_{\mathbb{T}^2}(y)}} \right)^{1/2},\quad\text{where}\, -\epsilon<t<0\nonumber\\
    &=\frac{1}{2\pi}e^{-c_o}\left(\frac{2\pi \mathrm{M}_s^{-4}}{\mathrm{M}_p^2/(4\pi \mathrm{M}_s^8 )} \right)^{1/2}\nonumber\\
    &=\sqrt{2} e^{-c_o}\mathrm{M}_s^2\mathrm{M}_p^{-1}.
\end{align}
As a result, in the late time limit $f_a(t)$ no longer vanishes and can be made to lie well within the range \eqref{RanGefa(t)HasTosaT}. 

In this chapter, we have presented the construction of a transient de Sitter phase in heterotic $\mathrm{E}_8\times \mathrm{E}_8$ theory, providing a framework for studying axion cosmology with a time-dependent axion decay constant. Both the transient de Sitter background and the corresponding axion coupling are consistent with the relevant cosmological constraints. Nevertheless, a broader exploration of the cosmological implications of these models remains an interesting direction for future research.
In the following chapter, we summarize the constructions presented throughout this thesis and discuss several open problems and possible directions for future investigation.

\chapter{Conclusions and Outlook}
\label{sec:Summary&Con}

\section{Conclusions}

In this thesis, we have presented the realization of de Sitter space as an excited state in string theory. This excited state, referred to as the Glauber--Sudarshan (GS) state, is constructed over a time-independent supersymmetric Minkowski background. Consequently, such a construction circumvents the vacuum-based no-go theorems while simultaneously satisfying the null energy condition (NEC), the Trans-Planckian Censorship Conjecture (TCC) bound, and phenomenological constraints arising from axion cosmology.

In Chapter~\ref{Chap:Intro}, we first discussed the no-go theorems that obstruct de Sitter vacuum solutions in supergravity theories. We then outlined several proposals for evading these no-go conditions. Subsequently, we explained how realizing de Sitter space as a Glauber--Sudarshan (GS) state, rather than as a vacuum configuration, can circumvent the vacuum-based no-go theorems and provide a well-defined effective field theory description.
Following this, in \S\ref{dSspacePoincareCo}, we reviewed the basic properties of de Sitter space, describing it from the embedding-space perspective and parametrizing it using Poincar\'e (flat-slicing) coordinates. This led to the four-dimensional de Sitter metric presented in Eq.~(\ref{dSFlAtSlic}). In \S\ref{GS state}, we discussed the salient features of the GS state and highlighted the ways in which it differs from conventional coherent states.

In Chapter~\ref{Chapter:2}, we reviewed the realization of de Sitter space as a GS state in type IIB string theory following Refs.~\cite{Brahma:2020tak,brahma2021four,Bernardo:2021rul,Chakravarty_2024,Brahma:2022wdl}. In \S\ref{typeIIBdSisom}, we discussed the dynamical duality sequence shown in Fig.~\ref{Type IIB tree diagram}, which starts from an M-theory metric configuration and flows dynamically to a type IIB background exhibiting four-dimensional de Sitter isometry. In \S\ref{sec:PIinMth}, we reviewed the computation of the expectation value of the metric operator in M-theory using path-integral techniques. The analysis was carried out for scalar fields with the all-order derivative interaction term given in Eq.~(\ref{SinTeraCtiOn}). Employing functional integral methods, we evaluated the tree-level contribution together with the first-order quantum correction and expressed the resulting expectation value as the asymptotic series shown in Eq.~(\ref{ExpValMOp}). In \S\ref{BorelREsum}, we performed the Borel transform and subsequent Borel resummation of this asymptotic series, thereby obtaining the finite result quoted in Eq.~(\ref{BORELRSummed}). Finally, in \S\ref{sec:NECIIB}, we demonstrated that satisfying the EFT criterion in Eq.~(\ref{EftCriTeriA1}) is equivalent to satisfying the null energy condition (\ref{NECforFLRWcos}), for a four-dimensional FLRW cosmology.

Chapter~3 is devoted to the study of de Sitter space as a GS state in heterotic $\mathrm{SO}(32)$ string theory. In \S\ref{dSinHetSO(32)}, we presented a dynamical duality sequence (see Fig.~\ref{Het SO(32) tree diagram}) that leads to a heterotic $\mathrm{SO}(32)$ background exhibiting four-dimensional de Sitter isometry (\ref{dSIsometryHetSO}), in the late-time limit of flat-slicing coordinates. The resulting metric is given in Eq.~(\ref{HeTerotiCSO(32)MetrIc}). 

Unlike the de Sitter GS-state construction in type IIB theory, the heterotic $\mathrm{SO}(32)$ coupling is not fixed to unity \eqref{HeterCoupling}. Consequently, the Einstein-frame analysis yields results that differ significantly from those obtained in the string frame. These distinctions are discussed in detail in \S\ref{StrFraMEinstFramSO32}. \S\ref{sec:NEC} is devoted to the analysis of the null energy condition within the framework of heterotic $\mathrm{SO}(32)$ theory. In \S\ref{sec:TCCconsTraints}, we examined the temporal regime over which the effective field theory description remains valid, which is identical to the Trans-Planckian Censorship Conjecture (TCC) bound (\ref{TCCBdinSO(32)HeTThy}). By considering explicit functional forms \eqref{AnsatzofBetat1}, \eqref{2ndbetatAnsatZ} for the warp factor, we demonstrated how different ans$\ddot{\text {a}}$tze affect the duration of validity of the effective field theory. We found that the smoother functional form given in Eq.~(\ref{2ndbetatAnsatZ}) preserves the same domain of validity as that implied by the TCC bound.


In Chapter~\ref{Chap:4}, we presented the realization of the de Sitter GS state in heterotic $\mathrm{E}_8\times\mathrm{E}_8$ string theory. An important distinction of this construction is that the generic M-theory metric configuration, Eq.~(\ref{MthThreewarpFac}), contains three time-dependent warp factors, $\mathrm{F}_1(t)$, $\mathrm{F}_2(t)$, and $\mathrm{F}_3(t)$. This contrasts with the heterotic $\mathrm{SO}(32)$ and type IIB constructions, Eqs.~(\ref{MthHetSO32}) and (\ref{MthEorYmeTric}), respectively, which involve only two time-dependent warp factors, $\mathrm{F}_1(t)$ and $\mathrm{F}_2(t)$. In \S\ref{E8XE8ThrdistWarpf}, we presented the dynamical duality sequence shown in Fig.~\ref{Het E8XE8 tree diagram}, which leads to a heterotic $\mathrm{E}_8\times\mathrm{E}_8$ background exhibiting four-dimensional de Sitter isometry (\ref{dDSIsomHetE8E8thry}), with metric given in Eq.~(\ref{HetroTicE8XE8 metric}). \S\ref{EinsVsStrE8E8Thry} examined the differences between the Einstein-frame and string-frame descriptions. In \S\ref{sec:NECE8XE8}, we analyzed the null energy condition within the framework of heterotic $\mathrm{E}_8\times\mathrm{E}_8$ theory. Finally, in \S\ref{sec:AxionsinE8Thoery}, we investigated constraints arising from axion cosmology. We showed that the bound on the axion decay constant, $f_a(t)$, given in Eq.~(\ref{AxioNicCoupConst}), modifies the dominant scaling behavior of the time-dependent warp-factor exponents $\hat{\alpha}(t)$ and $\hat{\beta}(t)$ from the forms given in Eq.~(\ref{EitherfrFuncAnsTzhatAlphahatBeta}) to those in Eq.~(\ref{AxionicCCTimeDepFactors}).


A concise summary of the construction of de Sitter GS state in the above mentioned three different string theories has been presented in Table \ref{tab:dSsummary}.  

\begingroup
\footnotesize
\setlength{\tabcolsep}{1pt}
\renewcommand{\arraystretch}{1.1}
\begin{longtable}{|M{0.09\textwidth}|L{0.29\textwidth}|L{0.29\textwidth}|L{0.33\textwidth}|}

\hline
&
\parbox[t]{0.28\textwidth}{\centering
\textbf{Type IIB theory}
}
&
\parbox[t]{0.28\textwidth}{\centering
\textbf{Heterotic $\mathrm{SO}(32)$ theory}
}
&
\parbox[t]{0.28\textwidth}{\centering
\textbf{Heterotic $\mathrm{E}_8\times \mathrm{E}_8$ theory}
}
\\
\hline

\endfirsthead

\hline
&
\parbox[t]{0.28\textwidth}{\centering
\textbf{Type IIB theory}
}
&
\parbox[t]{0.28\textwidth}{\centering
\textbf{Heterotic $\mathrm{SO}(32)$ theory}
}
&
\parbox[t]{0.28\textwidth}{\centering
\textbf{Heterotic $\mathrm{E}_8\times \mathrm{E}_8$ theory}
}
\\
\hline

\endhead

\hline
\endfoot

\caption{Summary of the construction of four-dimensional de Sitter space as a
Glauber--Sudarshan state in type IIB, heterotic $\mathrm{SO}(32)$,
and heterotic $\mathrm{E}_8\times\mathrm{E}_8$ theories.
\label{tab:dSsummary}}
\\
\endlastfoot


\textbf{M-theory}
&
Starting point of the duality sequence in Fig.~\ref{Type IIB tree diagram}.
Compactified on: $
\mathbb{R}^{2,1}\times
\mathcal{M}_2
\times \mathcal{M}_4\times
\frac{\mathbb{T}^{2}_{3,11}}
{\mathcal{G}}$.

$\bullet$ Metric configuration \eqref{MthEorYmeTric}:

$
ds_{11}^{2}
=
g_s^{-\frac 8 3}
\big(
-dt^2+\sum_{i=1}^{2}(dx^i)^2
\big)
+g_s^{-\frac 2 3}\mathrm{H}^2(y)
\big[
\mathrm{F}_1(t)g_{\alpha\beta}dy^\alpha dy^\beta
+
\mathrm{F}_2(t)g_{mn}dy^m dy^n
\big]+g_s^{\frac 4 3}\delta_{ab}d\omega^a d\omega^b
$.

\vspace{0.2cm}
&
Starting point of the duality sequence in Fig.~\ref{Het SO(32) tree diagram}.
Compactified on: $
\mathbb{R}^{2,1}\times
\frac{\mathbb{T}^2_{4,5}}{\mathcal{I}_4\mathcal{I}_5}
\times \mathcal{M}_4\times
\frac{\mathbb{T}^{2}_{3,11}}
{\mathcal{I}_3 \mathcal{I}_{11} } $ (\ref{MThCompcTifedForSO(32)}).

$\bullet$ Metric configuration \eqref{MthHetSO32}:

$
ds_{11}^{2}
=
g_s^{-\frac 8 3}
\big(
-dt^2+\sum_{i=1}^{2}(dx^i)^2
\big)
+g_s^{-\frac 2 3}\mathrm{H}^2(y)
\big[
\mathrm{F}_1(t)\delta_{\alpha\beta}dy^\alpha dy^\beta
+
\mathrm{F}_2(t)g_{mn}dy^m dy^n
\big]+g_s^{\frac 4 3}\delta_{ab}d\omega^a d\omega^b
$.

\vspace{0.2cm}
&
Starting point of the duality sequence in Fig.~\ref{Het E8XE8 tree diagram}.
Compactified on: $
\mathbb{R}^{2,1}\times
\mathbf{S}^1_{\theta_1}\times \frac{\mathbf{S}^1_{\theta_2}}{\mathcal{I}_{\theta_2}}
\times \mathcal{M}_4\times
\frac{\mathbb{T}^{2}_{3,11}}
{\mathcal{G}} $ (\ref{CompInMTheoRY}).

$\bullet$ Metric configuration \eqref{MthThreewarpFac}:

$
ds^{2}_{11}=g_{s}^{-\frac 8 3} \big(-dt^2+ \sum_{i=1}^{2}\big(dx^{i}\big)^{2}\big)+g_{s}^{-\frac 2 3}\mathrm{H}^{2}(y)\big[\mathrm{F}_{1}(t)(d\theta_{1})^2+\mathrm{F}_{3}(t)(d\theta_{2})^2 +\mathrm{F}_{2}(t)~g_{mn}dy^m dy^n\big]+g_{s}^{\frac 4 3}\delta_{ab}d\omega^a d\omega^b
$.

\vspace{0.2cm}
\\
\hline


\textbf{Final stage of the duality sequence}
&
See Fig.~\ref{Type IIB tree diagram}. Type IIB
compactified on: $\underbrace{\mathbb{R}^{3,1}}_{d\mathrm{S}_4}\times \mathcal{M}_2\times \mathcal{M}_4,$ with 3+1 dimensional spacetime having de Sitter isometry in flat slicing (\ref{dSFlAtSlic}).

$\bullet$ Metric configuration (\ref{tyPeIIBMetrIc}):

$ds^2_{\mathrm{IIB}}=g_s^{-2}\big[-dt^2+\sum_{i=1}^3\big(dx^{i}\big)^2 \big] +\mathrm{H}^{2}(y)\big[\mathrm{F}_{1}(t)g_{\alpha\beta}dy^{\alpha}dy^{\beta}+\mathrm{F}_{2}(t)g_{mn}dy^{m}dy^{n}\big].$

$\bullet$ Type IIB coupling (\ref{tyPeIIbCoupLConstaNT}): $g_{b}=1.$

\vspace{0.2cm}
&
See Fig.~\ref{Het SO(32) tree diagram}. Heterotic $\mathrm{SO}(32)$
compactified on: $\underbrace{\mathbb{R}^{3,1}}_{d\mathrm{S}_4}\times \mathcal{M}_4\rtimes (\widetilde{\bf S}^1_{4}\times \widetilde{\bf S}^1_{5}).$

    $\bullet$ Metric in string frame (\ref{HeTerotiCSO(32)MetrIc}):

$ds^2_{\mathrm{Het~SO}(32)}=\mathrm{F}_{1}(t)g_s^{-2}\mathrm{H}^2(y)\big[-dt^2+\sum_{i=1}^3\big(dx^{i}\big)^2 \big] +\mathrm{H}^{4}(y)\mathrm{F}_{1}(t)\mathrm{F}_{2}(t)g_{mn}dy^{m}dy^{n}
+\delta_{\alpha\beta} \big(dy^{\alpha}+\mathcal{B}^{(1)\alpha}_{~~~m} dy^{m}\big)\big(dy^{\beta}+\mathcal{B}^{(1)\beta}_{~~~n} dy^{n}\big).$

$\bullet$ Coupling (\ref{HeterCoupling}): $g_{\rm Het}=\mathrm{H}^2(y)\mathrm{F}_1(t).$

$\bullet$ Metric in Einstein frame (\ref{EinsFrameSO32}):

$ds^2_{\mathrm{Het~SO(32),~E}}={\sqrt{\mathrm{F}_{1}(t)}}g_s^{-2}{\mathrm{H}(y)}\big[-dt^2+\sum_{i=1}^3\big(dx^{i}\big)^2 \big]+\mathrm{H}^{3}(y){\mathrm{F}^{\frac 1 2}_{1}(t)}\mathrm{F}_{2}(t)g_{mn}dy^{m}dy^{n}+{\mathrm{H}^{-1}(y){\mathrm{F}^{-\frac 1 2}_{1}(t)}}\delta_{\alpha\beta}\big(dy^{\alpha}+\mathcal{B}^{(1)\alpha}_{~~~m} dy^{m}\big)\big(dy^{\beta}+\mathcal{B}^{(1)\beta}_{~~~n} dy^{n}\big)$.

\vspace{0.2cm}
&
See Fig.~\ref{Het E8XE8 tree diagram}. Heterotic $\mathrm{E}_8\times \mathrm{E}_8$
compactified on: $\underbrace{\mathbb{R}^{3,1}}_{d\mathrm{S}_4}\times \mathcal{M}_4\times \hat{\bf S}^1_{\theta_1}\times \tilde{\bf S}^1_{11}.$

    $\bullet$ Metric in string frame (\ref{HetroTicE8XE8 metric}):

$ds^2_{\mathrm{Het~E}_8\times \mathrm{E}_8}=g_s^{-2}\mathrm{H}^2(y)\sqrt{\mathrm{F}_{1}(t)\mathrm{F}_{3}(t)}\big[-dt^2+ \sum_{i=1}^3\big(dx^{i}\big)^2\big]+\sqrt{\frac{{\rm F}_3(t)}{{\rm F}_1(t)}}\big((d\theta_1)^2+(d\tilde{x}_{11})^2\big)+\mathrm{H}^{4}(y){\mathrm{F}^{\frac 1 2}_{1}(t)\mathrm{F}^{\frac 1 2}_{3}(t)}\mathrm{F}_{2}(t)~g_{mn}dy^{m}dy^{n}.$

$\bullet$ Coupling (\ref{CouplIngE8E8HeT}): $g_{\text{Het~E}_8\times \text{E}_8}=\mathrm{H}^{2}(y)\mathrm{F}_{1}^{1/4}(t)\mathrm{F}_{3}^{3/4}(t).$

$\bullet$ Metric in Einstein frame (\ref{EinframeinE8Th}):

$ds^2_{\mathrm{Het~E_8\times E_8,~E}}=g_s^{-2}{\mathrm{H}(y)} \mathrm{F}_1^{3/8}(t)\mathrm{F}_3^{1/8}(t)\big[-dt^2+\sum_{i=1}^3(dx^i)^2 \big]+\frac{1}{\mathrm{H}(y)}\mathrm{F}_1^{-\frac 58}(t)\mathrm{F}_3^{\frac 18}(t)\big[(d\tilde{x}_{11})^2+(d\theta_1)^2\big]+\mathrm{H}^4(y)\mathrm{F}_1^{\frac 38}(t)\mathrm{F}_2(t)\mathrm{F}_3^{\frac 18}(t)g_{mn}dy^m dy^n$.

\vspace{0.2cm}
\\
\hline


\textbf{Type IIA coupling $(g_s)$}
&
The coupling \eqref{IIAcouPlingdeSitterIIB} of dimensionally reduced type IIA theory (second stage of the duality sequence in Fig.~\ref{Type IIB tree diagram}):
$\overline{g}_s\equiv\frac{g_{s}}{\mathrm{H}(y)}=|t|\sqrt{ \Lambda }$.
&
The coupling of dimensionally reduced type IIA theory (second stage of the duality sequence in Fig.~\ref{Het SO(32) tree diagram}):

    $\bullet$ In string frame (\ref{TYpe IIA HetSO(32)}): $\overline{g}_s\equiv\frac{g_{s}}{\mathrm{H}(y)}=|t|\sqrt{ \Lambda \mathrm{F}_1(t)}$.
    
    $\bullet$ In Einstein frame (\ref{CCinEinsFrameHetSO32thr}): $\overline{g}_s=|t|\sqrt{\Lambda \sqrt{\mathrm{F}_1(t)}}.$

&
The coupling of dimensionally reduced type IIA theory (second stage of the duality sequence in Fig.~\ref{Het E8XE8 tree diagram}):

    $\bullet$ In string frame (\ref{gsinE8thEorY}): $\overline{g}_s=|t|\sqrt{\Lambda}\big(\mathrm{F}_{1}(t)\mathrm{F}_{3}(t)\big)^{1/4}$.
    
    $\bullet$ In Einstein frame (\ref{dSinHetE8tImesE8}): $\overline{g}_s=|t|\sqrt{\Lambda}(\mathrm{F}_1^3(t)\mathrm{F}_3(t))^{1/16}.$

\\
\hline


\textbf{Time-dep warp factors}
&
Dominant scalings of the two time-dependent warp factors  are \eqref{ScalingOfFi's}, \eqref{UbdofdomFactor}:

$ \mathrm{F}_1(t)\equiv \mathrm{F}_1(\overline{g}_s)=(\overline{g}_s)^{\beta_o},\, \beta_o<\frac{2}{3}$

$ \mathrm{F}_2(t)\equiv \mathrm{F}_2(\overline{g}_s)=(\overline{g}_s)^{\alpha_o},\, \alpha_o<\frac{2}{3}.$

In fact $\alpha_o=\beta_o=0$ \eqref{TheValofalphAoBetAo}, for the type IIB configuration to be the final stage of the duality sequence.
\vspace{5pt}
&
Dominant scalings of the two time-dependent warp factors are \eqref{F1(t)withbeta(t)}, \eqref{F2tinEinstFrSO(32)}:

    $\mathrm{F}_1(t)\equiv \mathrm{F}_1(\overline{g}_s)=(\overline{g}_s)^{\beta(t)},$ where 
    $\beta(t)=\begin{cases}
       {\beta_o},\quad -{ \frac{1}{ \sqrt{\Lambda}}} < t \leq -\epsilon \\
 0,\quad\quad -\epsilon < t < 0
    \end{cases}$
    
   $\mathrm{F}_2(t)\equiv \mathrm{F}_2(\overline{g}_s)=(\overline{g}_s)^{\alpha(t)},$ where $\alpha(t)=-\frac{\beta(t)}{v^{2}}.$ $v=1,2$ in string, Einstein frame respectively.

Parameter $\beta_o$ satisfies $0<\beta_o<\frac{2}{3}$ (\ref{beta0Range}) and $\epsilon$ is as given in Eq.~(\ref{ValueofEpsilonHetSO}).
\vspace{5pt}
&
Dominant scalings of the three time-dependent warp factors   are  \eqref{F1F3inHeteRE8E8}, \eqref{F2tHetE8thryinEitherFr}:

     $\mathrm{F}_1(t)=(\overline{g}_s)^{\frac{2\hat{\alpha}(t)}{3}},\quad \mathrm{F}_3(t)=(\overline{g}_s)^{\frac{2\hat{\beta}(t)}{3}},$
     
     $ \mathrm{F}_2(t)=(\overline{g}_s)^{-\frac{(\hat{\alpha}(t)+3\hat{\beta}(t))}{4v^2}}$,

    where \eqref{EitherfrFuncAnsTzhatAlphahatBeta}:

    $\hat{\alpha}(t)=\begin{cases}   \hat{\alpha}_o,\quad-\frac{1}{\sqrt{\Lambda }}<t\leq -\epsilon\\
\hat{\gamma}_{o},\quad-\epsilon< t<0
\end{cases}$ 

$\hat{\beta}(t)=\begin{cases}
\hat{\beta}_o,\quad\quad\quad-\frac{1}{\sqrt{\Lambda }}<t\leq -\epsilon\\
(4v-3)\hat{\gamma}_{o},\quad-\epsilon< t<0
\end{cases},$

with $1>\hat{\alpha}_{o}>\hat{\beta}_{o}>\frac{\hat{\alpha}_o}{9}>0$ \eqref{CondhAtAlpohatBetao}, and $\epsilon$ is as given in Eq.~(\ref{EpSilonInStrFrE8E8}).
\vspace{5pt}
\\
\hline


\textbf{TCC bound}
&
Arising due to the EFT criteria \eqref{TCCinTypeIIBStrINgTh}: $\overline{g}_s<1.$

$\overline{g}_s=\sqrt{\Lambda}|t|<1~\implies -\frac{1}{\sqrt{\Lambda}}<t<0$.
&
Arising due to the EFT criteria \eqref{TCCBdinSO(32)HeTThy}: $\overline{g}_s<1.$

$\overline{g}_s=(\sqrt{\Lambda}|t|)^{\frac{2v}{2v-\beta(t)}}<1~\implies -\frac{1}{\sqrt{\Lambda}}<t<0$.
&
Arising due to the EFT criteria \eqref{TempDomHetE8E8TheoRY}: $\overline{g}_s<1$.

$\overline{g}_s=( \sqrt{\Lambda} |t|)^{\frac{6v^2}{ 6v^2 -(2v-1)\hat{\alpha}(t)-\hat{\beta}(t) }}<1~\implies -\frac{1}{\sqrt{\Lambda}}<t<0$.
\\
\hline


\textbf{NEC}
&
Arising due to the EFT criteria \eqref{NECinTypeIIBThry}: $
\frac{\partial\overline{g}_s}{\partial t}=\overline{g}_s^{q},$ where $q\geq 0$.

With $\overline{g}_s=(\sqrt{\Lambda}|t|)^{-n}$ (\ref{ValUeofgsForNECinTyPeIIB}), we have

$\frac{\partial \overline{g}_s}{\partial t}\propto \overline{g}_s^{(1+1/n)}\,\implies \frac{1}{n}\geq -1.$
&
Arising due to the EFT criteria \eqref{NECinHetSO32Thry}: $
\frac{\partial\overline{g}_s}{\partial t}=\overline{g}_s^{q},$ where $q\geq 0$.

With $\overline{g}_s=(\sqrt{\Lambda}|t|)^{\frac{2nv}{\beta(t)-2v}}$ (\ref{gsbarToObtainNEC}), we have

$\frac{\partial \overline{g}_s}{\partial t}=\overline{g}_s^{1+\mathrm{B}(t)}-2 nv \sqrt{\Lambda}\overline{g}_s^{1+\frac{1}{n}-\frac{\beta(t)}{2nv}+\mathrm{C}(t)} \eqref{tderofgsbar},$

where $\mathrm{B}(t)=\frac{\log\big[\big(\frac{\dot{\beta}(t) }{2v-\beta(t)}\big) \log \overline{g}_s\big]}{\log \overline{g}_s},\, C(t)=-\frac{\log(2v-\beta(t))}{\log\overline{g}_s}.$

The second term gives the NEC \eqref{NuEnConDiSO32TimDep}: $\frac{1}{n}\geq -1$.
&
Arising due to the EFT criteria \eqref{EFTBound}: $
\frac{\partial\overline{g}_s}{\partial t}=\overline{g}_s^{q},$ where $q\geq 0$.

With $\overline{g}_s=(\sqrt{\Lambda}|t|)^{\frac{6n}{\hat{\alpha}(t)+\hat{\beta}(t)-6}}$ (\ref{StringCouPStrFRaMEE8}):

$\frac{ \partial \overline{g}_s}{\partial t}=\overline{g}_s^{1+ {\mathrm{B}}(t)}- 6n \sqrt{ \Lambda} \overline{g}_s^{1+ \frac{1}{n}- \frac{\hat{\alpha}(t)+\hat{\beta}(t)}{6n}+{\mathrm{C}}(t)}\eqref{PartIalDerHetE8ThNeC},$

where $\mathrm{B}(t)=\frac{ \log\big[\big( \frac{ \dot{\hat{\alpha}}(t)+\dot{\hat{\beta}}(t) }{6-\hat{\alpha}(t)-\hat{\beta}(t)} \big) \log \overline{g}_s \big] }{\log\overline{g}_s},\, C(t)=\frac{\log \big(6-\hat{\alpha}(t)-\hat{\beta}(t) \big) }{|\log \overline{g}_s|}.$

The second term gives the NEC (\ref{NECinE8E8hEtTheory}): $\frac{1}{n}\geq -1$.
\\
\hline




\end{longtable}
\endgroup

\section{Outlook}
\begin{itemize}
\item Our analysis in \S\ref{BorelREsum} allowed us to write a non-perturbative definition of the cosmological constant as in Eq. \eqref{NpeRDefCosMCons}. A detailed proof for the positivity of this cosmlogical constant has been given in \cite{Brahma:2022wdl}. However, showing that it is a small quantity is a very interesting open problem to investigate in future.
\item Recent DESI results \cite{w4c6-1r5j} have provided evidence for dynamical dark energy. In light of these developments, it would be interesting to investigate how a time-varying dark energy may emerge within the framework of the GS-state construction presented in this thesis. A first step towards this direction was taken in Ref.~\cite{Dasgupta:2025ypg}, where the possibility of temporally varying dark energy was incorporated into the analysis.
\item The isometry group of the underlying time-independent supersymmetric Minkowski background is the super-Poincar\'e group acting along the external spacetime directions. It would be interesting to investigate the isometry group associated with the de Sitter excited state constructed over this supersymmetric Minkowski background. Since our state breaks supersymmetry spontaneously and is transient in nature its isometry group  may not be the exact $SO(1,4)$ group of a de Sitter space.
    \item In \cite{Brahma:2024zuk, Dasgupta:2024aif}, attempts have been made to relate the Glauber--Sudarshan state to the wavefunctional of the Universe satisfying the Wheeler--DeWitt (WDW) equation. This turned out to be an {\it emergent} equation, although an explicit solution to this equation was not obtained therein. It would therefore be interesting to explore how the technique used in solving the WDW equation in \cite{chakraborty2023holographyinformationsitterspace},
that uses information theoretic approaches to de Sitter space, may be applied to our case for solving the corresponding {\it emergent} WDW equation.

\end{itemize}

\appendix
\renewcommand{\chaptermark}[1]{%
  \markboth{APPENDIX \thechapter.\ #1}{}}

\chapter{Duality rules}
\label{DualitYRulES}
\textbf{T-duality:}
Let us consider a closed string theory where a string is moving in 9-dimensional Minkowski space-time ($\mathbb{R}^{8,1}$) times a circle ($S^1$) of radius $R$. There is a $\mathbb{Z}_{2}$  $\mathrm{T}-$duality transformation that inverts the radius of the circle $R\rightarrow\tilde{R}=\frac{\alpha^\prime}{R}$. This mapping of $R\rightarrow \tilde{R}$ leaves the mass formula of the string invariant provided the Kaluza-Klein excitation number is exchanged with the string winding number. 
The above transformation is usually referred to as Buscher's transformation. There is a generalization of it which transforms solution of type $\mathrm{IIA}$ string on a background with one isometry to a type $\mathrm{IIB}$ string solution on a background with one isometry. For the Neveu-Schwarz/Neveu-Schwarz sector this transformation is essentially that of Buscher's but it interchanges the R-R fields of type $\mathrm{IIA}$ with those of the R-R fields of type $\mathrm{IIB}$ string theory. 

We denote the isometry direction to be $y$, the remaining directions as $x^{\mu}$ and all the directions as $x^{\alpha}=(x^{\mu},y)$. All fields have to be independent of $y$ coordinate. Below we list how the fields of type $\mathrm{IIA}$ superstring theory get mapped to that of the fields of type $\mathrm{IIB}$ superstring theory under a $\mathrm{T}-$duality performed along $y$ direction  \cite{bergshoeff1995duality}:
    \begin{align}
 \tilde{{g}}_{\mu \nu}&=\hat{g}_{\mu \nu}-\left(\hat{g}_{y\mu}\hat{g}_{y\nu}-{B}^{}_{y\mu}{B}^{}_{y\nu}\right)/\hat{g}_{yy}\\
\tilde{g}_{y\mu}&={B}^{}_{y\mu}/\hat{g}_{yy}\\
\tilde{g}_{yy}&=1/\hat{g}_{yy}\\
{D}_{y\mu \nu \rho}&=\frac{3}{8}\bigg[{C}_{\mu \nu \rho}-{A}_{[\mu}{B}^{}_{\nu \rho ]}+\hat{g}_{y[\mu}{B}^{}_{\nu \rho]}{A}_{y}/\hat{g}_{yy}-\frac{3}{2}\hat{g}_{y[\mu}{C}_{\nu \rho]y}/\hat{g}_{yy}\bigg] \\
{\mathcal{B}}^{(2)}_{\mu \nu}&=\frac{3}{2}{C}_{\mu \nu y}-2 {A}_{[\mu}{B}^{}_{\nu]y}+2\hat{g}_{y[\mu}{B}^{}_{\nu]y}{A}_{y}/\hat{g}_{yy}\\
{\mathcal{B}}^{(2)}_{y\mu}&=-{A}_{\mu}+{A}_{y}\hat{g}_{y\mu}/\hat{g}_{yy}\\
{\mathcal{B}}^{(1)}_{\mu \nu}&={B}^{}_{\mu \nu}+2 \hat{g}_{y[\mu}{B}^{}_{\nu]y}/\hat{g}_{yy}\\
{\mathcal{B}}^{(1)}_{y\mu}&=\hat{g}_{y\mu}/\hat{g}_{yy}\\
{l}&={A}_{y}\\
{\varphi}&={\phi}-\frac{1}{2}\log (\hat{g}_{yy})
    \end{align}
Likewise, a  $\mathrm{T}-$duality performed along $y$ direction maps the fields of type $\mathrm{IIB}$ superstring theory onto the fields of type $\mathrm{IIA}$ superstring theory in the following way: 
    \begin{align}
        \hat{{g}}_{\mu \nu}&=\tilde{g}_{\mu \nu}-\left(\tilde{g}_{y\mu}\tilde{g}_{y\nu}-{\mathcal{B}}^{(1)}_{y\mu}{\mathcal{B}}^{(1)}_{y\nu}\right)/\tilde{g}_{yy}\\
    \hat{g}_{y\mu}&={\mathcal{B}}^{(1)}_{y\mu}/\tilde{g}_{yy}\\
\hat{g}_{yy}&=1/\tilde{g}_{yy}\\
{C}_{y\mu \nu}&=\frac{2}{3}\left[{\mathcal{B}}^{(2)}_{\mu\nu}+2 {\mathcal{B}}^{(2)}_{y[\mu}\tilde{g}_{\nu]y}/\tilde{g}_{yy}\right]\\
{C}_{\mu \nu \rho}&=\frac{8}{3}{D}_{y\mu \nu \rho}+\epsilon^{ij}{\mathcal{B}}^{(i)}_{y[\mu}{\mathcal{B}}^{(j)}_{\nu \rho]}+\epsilon^{ij}{\mathcal{B}}^{(i)}_{y[\mu}{\mathcal{B}}^{(j)}_{|y|\nu}\tilde{g}_{\rho]y}/\tilde{g}_{yy}\\
{B}^{}_{\mu \nu}&={\mathcal{B}}^{(1)}_{\mu \nu}+2 {\mathcal{B}}^{(1)}_{y[\mu}\tilde{g}_{\nu]y}/\tilde{g}_{yy}\\
{B}^{}_{y\mu}&=\tilde{g}_{y\mu}/\tilde{g}_{yy}\\
{A}_{\mu}&=-{\mathcal{B}}^{(2)}_{y\mu}+{l}{\mathcal{B}}^{(1)}_{y\mu}\\
{A}_{y}&={l}\\
{\phi}&={\varphi}-\frac{1}{2}\log (\tilde{g}_{yy})
    \end{align} 
In the above rules, fields of type $\mathrm{IIA}$ and $\mathrm{IIB}$ theory are denoted as
\begin{align}
    \mathrm{NS-NS~sector ~of ~type~ IIA :}~&(\hat{g}_{\alpha\beta},~B^{}_{\alpha\beta},~\phi)\nonumber\\
    \mathrm{R-R ~sector~ of ~type~ IIA :}~&(A_{\alpha},~C_{\alpha\beta\gamma})\nonumber\\
        \mathrm{NS-NS~sector ~of ~type~ IIB :}~&(\tilde{g}_{\alpha\beta},~\mathcal{B}^{(1)}_{\alpha\beta},~\varphi)\nonumber\\
    \mathrm{R-R ~sector~ of ~type~ IIB :}~&(l,~\mathcal{B}^{(2)}_{\alpha\beta},~D_{\alpha\beta\gamma\delta}).\nonumber
\end{align}

\noindent
    \textbf{S duality:} S-duality is a transformation that relates a strongly coupled theory to a weakly coupled one. Unlike T-duality, it is a non-perturbative symmetry. For example, the type $\mathrm{I}$ and $SO(32)$ superstring theories are $\mathrm{S}$ dual to each other as they are equivalent when the following conditions hold
    \begin{align}\Phi_{\mathrm{Het}}=-&\Phi_{\mathrm{I}},~~~g_{\alpha\beta}^{\mathrm{Het}}=e^{-\Phi_{\mathrm{I}}}g^{\mathrm{I}}_{\alpha\beta},~B_{\alpha\beta}^{\mathrm{Het}}=C_{\alpha\beta}^{\mathrm{I}},~~A_{\alpha}^{\mathrm{Het}}=A_{\alpha}^{\mathrm{I}}.
    \end{align}
It implies that the coupling constant of type $\mathrm{I}$ superstring theory is the reciprocal of the coupling constant of heterotic $SO(32)$ theory $g^{\mathrm{I}}=1/g^{\mathrm{Het}},$ in other words when type $\mathrm{I}$ theory is strongly coupled, the $\mathrm{S}$ dual $SO(32)$ heterotic theory is weakly coupled and vice versa and the Weyl rescaled metric of type $\mathrm{I}$ theory is the metric of the dual heterotic $SO(32)$ theory 
\begin{align}
ds^{2}_{\text{Het}}&=e^{-\phi_{\mathrm{I}}}ds^{2}_{\text{I}}=\frac{1}{g^{\mathrm{I}}}ds^2_{\mathrm{I}}.
    \end{align}
  \noindent 
  \textbf{Dimensional reduction:} Low-energy limit of the type $\mathrm{IIA}$ superstring theory is the type $\mathrm{IIA}$ supergravity theory and it can be obtained from $\mathcal{N}=1,~d=11$ supergravity theory by dimensional reduction.  Bosonic fields of the 11 dimensional theory consist of the metric $\mathbf{g}_{\mathrm{MN}}$ and a three-form potential $\mathbf{C}_{\mathrm{MNP}}$. The coordinates are denoted as $x^{\mathrm{M}}=(x^\alpha, z),~\alpha=0,1,...9.$ In order to dimensionally reduce the 11th direction (viz. the $z$ coordinate) one has to compactify that direction on a circle of radius $R_{11}$ and assume all fields to be independent of $z$ and consequently the fields of ten-dimensional type $\mathrm{IIA}$ supergravity can be expressed in terms of the eleven-dimensional ones as
    \begin{align}
    \hat{g}_{\alpha\beta}&=(\mathbf{g}_{zz})^{1/2}~\left(\mathbf{g}_{\alpha\beta}-\mathbf{g}_{\alpha z } \mathbf{g}_{\beta z }/\mathbf{g}_{zz}\right),\\
{A}_{\alpha}&=\mathbf{g}_{\alpha z}/\mathbf{g}_{zz}\\
    {\phi}&=\frac{3}{4}\log \left(\mathbf{g}_{zz}\right)\label{DilatonIIA}\\
{C}_{\alpha\beta\gamma}&=\mathbf{C}_{\alpha\beta\gamma}\\
{B}^{}_{\alpha\beta}&=\frac{3}{2}\mathbf{C}_{\alpha\beta z}.
    \end{align}
That is, dimensional reduction of the metric gives rise to a ten -dimensional metric $\hat{g}_{\alpha \beta}$, a vector field $A_{\alpha}$, and a scalar field (dilaton) $\phi$ while the dimensional reduction of the three-form potential gives rise to a ten-dimensional three form $C_{\alpha\beta\gamma}$, and a two-from $B_{\alpha\beta}$. From (\ref{DilatonIIA}) it turns out that 
\begin{align}
R_{11}=\sqrt{\mathbf{g}_{zz}}l_{p}=e^{2\phi/3}l_{p}=l_{s}g_{s},\label{R_{11}}    
\end{align}
 where $g_{s}=e^{\phi}$ is the type $\mathrm{IIA}$ coupling constant. Thus the perturbative regime ($g_{s}\rightarrow 0$) of type $\mathrm{IIA}$ theory corresponds to the $R_{11}\rightarrow 0$ limit. 

\noindent
        \textbf{Dimensional Uplift:} The strong coupling limit of type $\mathrm{IIA}$ theory, that is the limit $g_{s}\rightarrow \infty$, leads to decompactification of the 11th circle of radius $R_{11}$ (\ref{R_{11}}). Hence in this limit the ten dimensional type $\mathrm{IIA}$ supergravity theory gets uplifted to the eleven dimensional supergravity theory. The bosonic fields of the eleven dimensional theory can be expressed in terms of the ten dimensional ones as 
    \begin{align}
\mathbf{g}_{\alpha\beta}&= e^{-\frac{2}{3}{\phi}} \hat{g}_{\alpha\beta} + e^{\frac{4}{3}{\phi}}{A}_{\alpha}{A}_{\beta}\\
\mathbf{g}_{\alpha z}&=e^{\frac{4}{3}{\phi}}{A}_{\alpha}\\
\mathbf{g}_{zz}&=e^{\frac{4}{3}{\phi}}\\
\mathbf{C}_{\alpha\beta\gamma}&=C_{\alpha\beta\gamma}\\
\mathbf{C}_{\alpha\beta z}&=\frac{2}{3}{B}_{\alpha\beta}.
\end{align}

\chapter{Functional integral approach for \texorpdfstring{$\varphi^4$}{phi4}  theory}
\label{FIntApproachPhi4}
\vskip.1in
In this appendix, we elaborate the two point correlator of $\varphi^{4}$ theory in 3+1 dimensional quantum field theory for a Klein Gordon field using the method of functional integral. We follow closely the method given in Chap. 9 of Peskin \& Schroeder \cite{Peskin:1995ev}:
\begin{align}
    \langle \varphi(x_1)\varphi(x_2)\rangle &= \frac{\int [\mathcal{D}\varphi(x)]e^{i\mathbf{S}_{\mathrm{tot}}}\varphi(x_1)\varphi(x_{2})}{\int [\mathcal{D}\varphi(x)]e^{i\mathbf{S}_{\mathrm{tot}}}}\label{2ptCorrelator}
\end{align}
where 
\begin{align}
\mathbf{S}_{\mathrm{tot}}&=\mathbf{S}_{0}+\mathbf{S}_{\mathrm{int}}=\int d^{4}x \left(\frac{1}{2}\partial_{\mu}\varphi(x) \partial^{\mu} \varphi (x)-\frac{1}{2}m^2\varphi^2(x)\right)+\frac{\lambda}{4!} \int d^{4}x \varphi^{4}(x).\label{TotalAction}
\end{align}
The first step is to map the space-time into a square lattice having discrete points and thus the field $\varphi(x)$ defined on a continuum of points has to be mapped to $\varphi(x_{i})$ defined at the points of the aforementioned lattice. Consequently, the integral $\mathcal{D}\varphi$ over the space of all field configurations can be replaced in terms of a finite but large number of integrals over the lattice points 
\begin{align}
    \int [\mathcal{D}\varphi(x)]\rightarrow \prod_{i}\int d\varphi(x_{i})\label{Dvarphi}
\end{align}
The discrete Fourier transform of the field $\varphi(x_i)$ is given by
\begin{align}
    \varphi(x_{i})&=\frac{1}{\mathrm{V}}\sum_{n} e^{-ik_{n}\cdot x_i}\widetilde{\varphi}(k_{n}),
\end{align}
where the four dimensional space-time volume $\mathrm{V}=L^{4}$. The periodic boundary condition $\varphi(x_{i})=\varphi(x_{i}+L)$ imposed on the fields implies $k_{n}^{\mu}=2\pi n^{\mu}/L$, $n^\mu$ are the integers.  Since we are considering a real Klein Gordon field 
\begin{align}
    \varphi^{\star}(x_{i})&=\frac{1}{\mathrm{V}}\sum_{n}e^{ik_{n}\cdot x_{i}}\widetilde{\varphi}^{\star}(k_{n})=\varphi(x_{i}),
\end{align}
the Fourier coefficients have to obey the constraint $\widetilde{\varphi}^{\star}(k_{n})=\widetilde{\varphi}(-k_{n})$. Therefore, we can consider $\mathbf{Re}\widetilde{\varphi}(k_n)$ and $\mathbf{Im}\widetilde{\varphi}(k_{n})$ with $k_{n}^{0}>0$ as independent variables. In terms of these fourier coefficients we can rewrite the integral in (\ref{Dvarphi}) as 
\begin{align}
    \int[\mathcal{D}\varphi(x)]\rightarrow  \prod_{k^{0}
_{n}>0}\int d~ \mathbf{Re}\widetilde{\varphi}(k_{n}) d ~\mathbf{Im}\widetilde{\varphi}(k_{n})
\end{align}
Likewise, the free part of the action (\ref{TotalAction}) can be expressed as
\begin{align}
  \mathbf{S}_{0}&=  \int d^{4}x \frac{1}{2}\left(\partial_{\mu}\varphi(x)\partial^{\mu}\varphi(x)-m^2 \varphi^2(x)\right)\nonumber\\
  &= -\frac{1}{\mathrm{V}^2}\sum_{m,n} \int d^{4}x \frac{1}{2}~e^{-i(k_m+k_n)\cdot x} (k_{m\mu}k_{n}^{\mu} +m^2)\widetilde{\varphi}(k_{m})\widetilde{\varphi}(k_{n})\nonumber\\
    &=-\frac{1}{\mathrm{V}}\sum_{n}\frac{1}{2}(m^2-k_{n}^2) |\widetilde{\varphi}(k_{n})|^2=-\frac{1}{\mathrm{V}}\sum_{k_{n}^{0}>0}(m^2-k_{n}^{2})|\widetilde{\varphi}(k_{n})|^2,
\end{align}
and the interaction part can be expressed as 
\begin{align}
    \mathbf{S}_{\mathrm{int}}=\frac{\lambda}{4!}\int d^{4}x \varphi^{4}(x)&=\frac{\lambda}{4!}\frac{1}{\mathrm{V}^4}\sum_{m,n,p,q}\int d^4 x e^{-i(k_{m}+k_{n}+k_{p}+k_{q})\cdot x} \widetilde{\varphi}(k_{m})\widetilde{\varphi}(k_{n})\widetilde{\varphi}(k_{p})\widetilde{\varphi}(k_{q})\nonumber\\
    &=\frac{\lambda}{4!}\frac{1}{\mathrm{V}^3}\sum_{m,n,p}\widetilde{\varphi}(k_{m})\widetilde{\varphi}(k_{n})\widetilde{\varphi}(k_{p})\widetilde{\varphi}^{\star}(k_{m}+k_{n}+k_{p}).
\end{align}
With these ingredients we now analyse the numerator of (\ref{2ptCorrelator}) 
\begin{align}
&\mathrm{Num}[\langle \varphi(x_1)\varphi(x_2)\rangle]=\int [\mathcal{D}\varphi(x)] e^{i\mathbf{S}_{\mathrm{tot}}}\varphi(x_1)\varphi(x_2)\nonumber\\
&=\left(\prod_{k_{n}^{0}>0}\int d~\mathbf{Re}\widetilde{\varphi}(k_{n}) d~\mathbf{Im}\widetilde{\varphi}(k_{n})  \right) \left(\frac{1}{\mathrm{V}^2}\sum_{q,r}e^{-i(k_{q}\cdot x_1+k_{r}\cdot x_2)}\widetilde{\varphi}(k_{q})\widetilde{\varphi}(k_{r}) \right)\nonumber\\
&\times \exp \bigg[-\frac{i}{\mathrm{V}}\sum_{k_{n}^0>0}(m^2-k_{n}^2) |\widetilde{\varphi}(k_n)|^2+\frac{\lambda}{4!}\frac{i}{\mathrm{V}^3}\sum_{m,n,p}\widetilde{\varphi}(k_{m})\widetilde{\varphi}(k_{n})\widetilde{\varphi}(k_{p})\widetilde{\varphi}^{\star}(k_{m}+k_{n}+k_{p}) \bigg]\nonumber\\
    &=\left(\prod_{k_{n}^{0}>0}\int d~\mathbf{Re}\widetilde{\varphi}(k_{n}) d~\mathbf{Im}\widetilde{\varphi}(k_{n})  \right)\exp \bigg[-\frac{i}{\mathrm{V}}\sum_{k_{n}^0>0}(m^2-k_{n}^2) \left[(\mathbf{Re}\widetilde{\varphi}(k_{n}))^2+(\mathbf{Im}\widetilde{\varphi}(k_{n}))^2\right]\bigg]\nonumber\\
    &~~~~~\times \left(\frac{1}{\mathrm{V}^2}\sum_{q,r}e^{-i(k_{q}\cdot x_1+k_{r}\cdot x_2)} (\mathbf{Re}\widetilde{\varphi}(k_q)+i\mathbf{Im}\widetilde{\varphi}(k_q))(\mathbf{Re}\widetilde{\varphi}(k_r)+i\mathbf{Im}\widetilde{\varphi}(k_r))\right)\nonumber\\
    &~~~~~\times \bigg(1+\frac{\lambda}{4!}\frac{i}{\mathrm{V}^3}\sum_{m,n,p}\left(\mathbf{Re}\widetilde{\varphi}(k_{m})+i\mathbf{Im}\widetilde{\varphi}(k_{m})\right)\left(\mathbf{Re}\widetilde{\varphi}(k_{n})+i\mathbf{Im}\widetilde{\varphi}(k_{n})\right)\nonumber\\
&~~~~~\left(\mathbf{Re}\widetilde{\varphi}(k_{p})+i\mathbf{Im}\widetilde{\varphi}(k_{p})\right)\left(\mathbf{Re}\widetilde{\varphi}(k_{m}+k_{n}+k_{p})-i\mathbf{Im}\widetilde{\varphi}(k_{m}+k_{n}+k_{p})\right)+\mathcal{O}(\lambda^2)\bigg)
\end{align}
The tree level contribution to the above integral is given by 
\begin{align}
    &\mathrm{Num}[\langle \varphi(x_1)\varphi(x_{2})\rangle]\bigg|_{\mathcal{O}(\lambda^0)}\nonumber\\
    &=\left(\prod_{k_{n}^{0}>0}\int d~\mathbf{Re}\widetilde{\varphi}(k_{n}) d~\mathbf{Im}\widetilde{\varphi}(k_{n})  \exp \bigg[-\frac{i}{\mathrm{V}}(m^2-k_{n}^2) \left[(\mathbf{Re}\widetilde{\varphi}(k_{n}))^2+(\mathbf{Im}\widetilde{\varphi}(k_{n}))^2\right]\bigg]\right)\nonumber\\
    &~~~~~\times \left(\frac{1}{\mathrm{V}^2}\sum_{q,r}e^{-i(k_{q}\cdot x_1+k_{r}\cdot x_2)} (\mathbf{Re}\widetilde{\varphi}(k_q)+i\mathbf{Im}\widetilde{\varphi}(k_q))(\mathbf{Re}\widetilde{\varphi}(k_r)+i\mathbf{Im}\widetilde{\varphi}(k_r))\right)
\end{align}
From the structure of the integrand it is obvious that for the values of $k_{q}\neq k_{r}$ the integral evaluates to zero. The result may become non vanishing for $k_{q}=\pm k_{r}$. Below we analyse these two cases separately.
\vskip.2in
\noindent{\textbf{Case 1:} {$\boldsymbol{k_{q} =+ k_{r}}$}
\begin{align}
    &\mathrm{Num}[\langle \varphi(x_1)\varphi(x_{2})\rangle]\bigg|^{\mathrm{Case ~1}}_{\mathcal{O}(\lambda^0)}\nonumber\\
    &=\left(\prod_{k_{n}^{0}>0}\int d~\mathbf{Re}\widetilde{\varphi}(k_{n}) d~\mathbf{Im}\widetilde{\varphi}(k_{n})  \exp \bigg[-\frac{i}{\mathrm{V}}(m^2-k_{n}^2) \left[(\mathbf{Re}\widetilde{\varphi}(k_{n}))^2+(\mathbf{Im}\widetilde{\varphi}(k_{n}))^2\right]\bigg]\right)\nonumber\\
    &~~\times \left(\frac{1}{\mathrm{V}^2}\sum_{q}e^{-ik_{q}\cdot (x_1+x_2)} ((\mathbf{Re}\widetilde{\varphi}(k_q))^2-(\mathbf{Im}\widetilde{\varphi}(k_q))^2+2i\mathbf{Re}\widetilde{\varphi}(k_q)\mathbf{Im}\widetilde{\varphi}(k_q))\right)
\end{align}
The third term $\mathbf{Re}\widetilde{\varphi}(k_q)\mathbf{Im}\widetilde{\varphi}(k_{q})$ makes the integrand odd and hence doesn't contribute to the integral. The remaining two terms viz. $(\mathbf{Re}\widetilde{\varphi}(k_{q}))^2$ and $(\mathbf{Im}\widetilde{\varphi}(k_{q}))^2$ contribute equally to the integral and hence the net result vanishes. 
\vskip.2in
\noindent{\textbf{Case 2:} {$\boldsymbol{k_{q} =- k_{r}}$}
\begin{align}
    &\mathrm{Num}[\langle \varphi(x_1)\varphi(x_{2})\rangle]\bigg|^{\mathrm{Case ~2}}_{\mathcal{O}(\lambda^0)}\nonumber\\
    &=\left(\prod_{k_{n}^{0}>0}\int d~\mathbf{Re}\widetilde{\varphi}(k_{n}) d~\mathbf{Im}\widetilde{\varphi}(k_{n})  \exp \bigg[-\frac{i}{\mathrm{V}}(m^2-k_{n}^2) \left[(\mathbf{Re}\widetilde{\varphi}(k_{n}))^2+(\mathbf{Im}\widetilde{\varphi}(k_{n}))^2\right]\bigg]\right)\nonumber\\
    &~~~~~~~~~~~~\times \left(\frac{1}{\mathrm{V}^2}\sum_{q}e^{-ik_{q}\cdot (x_1-x_2)} ((\mathbf{Re}\widetilde{\varphi}(k_q))^2+(\mathbf{Im}\widetilde{\varphi}(k_q))^2)\right)\nonumber\\
    &=\frac{1}{\mathrm{V}^2}\sum_{q}e^{-ik_{q}\cdot (x_1-x_2)}\left(\prod_{k_{n}^0>0} \frac{-i\pi \mathrm{V}}{m^2-k_{n}^2}\right) \frac{-i \mathrm{V}}{m^2-k_{q}^2}
\end{align}
The factor similar to that in the parentheses of last line also appears from the denominator of the path integral (\ref{2ptCorrelator}) and hence gets canceled out once we take into account the denominator as well. Now, if we take the continuum limit (of the remaining terms), $\mathrm{V}\rightarrow \infty$ we get the tree level contribution to be
\begin{align}
    \mathrm{Num}[\langle \varphi(x_1)\varphi(x_2)\rangle]\bigg|_{\mathcal{O}(\lambda^0)}&=\left(\prod_{k_{n}^0>0} \frac{-i\pi \mathrm{V}}{m^2-k_{n}^2}\right) \int \frac{d^4 k}{(2\pi)^4}\frac{i e^{-ik\cdot(x_1-x_2)}}{k^2-m^2}. 
\end{align}
which is nothing but the Feynman propagator upto an overall factor. In a similar vein one can analyse the quantum corrections. We follow this method extensively in Section \ref{sec:PIinMth} to evaluate one point function with respect to an excited state.

\chapter{Volume of the internal manifold}
\label{sec:AppendixD}
Consider the six-dimensional internal space metric (\ref{HeTerotiCSO(32)MetrIc}) written in Kaluza--Klein (KK) form
\begin{align}
   ds^2_{6}&= \delta_{\alpha\beta}\left(dy^{\alpha}+\mathcal{B}^{(1)\alpha}_{~~~m} dy^{m}\right)\left(dy^{\beta}+\mathcal{B}^{(1)\beta}_{~~~n} dy^{n}\right)+\mathrm{H}^{4}(y)\mathrm{F}_{1}(t)\mathrm{F}_{2}(t)g_{mn}dy^{m}dy^{n}
\end{align}
where $\alpha,\beta=4,5$ and $m,n=6,\ldots,9$.
The vectors $\mathcal{B}^{(1)\alpha}{}_m$ are the KK gauge fields describing the twisting of
the two-dimensional fiber ($\tilde{\mathbf{S}}^1_{4}\times \tilde{\mathbf{S}}^1_5$) over the four-dimensional base $\mathcal{M}_4$. Let us define
$$
A(y,t) \equiv \mathrm{H}^4(y)\,\mathrm{F}_1(t)\, \mathrm{F}_2(t),~~~\mathrm{where}~y\equiv y^m\in \mathcal{M}_4.
$$
The six dimensional metric can be written in block form as
$$
g_{MN}
=
\begin{pmatrix}
\delta_{\alpha \beta} &~~ \mathcal{B}^{(1)\gamma}{}_n \delta_{\alpha \gamma}\\
\mathcal{B}^{(1)~\gamma}_{~~m} \delta_{\gamma \beta} & ~~A g_{mn} + \mathcal{B}^{(1)}_{~~m\gamma }\delta^{\gamma \tilde{\gamma} } \mathcal{B}^{(1)}_{~~\tilde{\gamma} n} 
\end{pmatrix},~~\mathrm{where}~M,N=4,5,6,\cdots ,9.
$$
Using the standard block determinant identity
\begin{align}
\det
\begin{pmatrix}
\mathrm{P} & \mathrm{Q} \\
\mathrm{Q}^{\mathrm{T}} &~~ \mathrm{S}
\end{pmatrix}= 
\det(\mathrm{P} )\,\det( \mathrm{S} - \mathrm{Q}^{\mathrm{T}} \mathrm{P}^{-1} \mathrm{Q}),
\end{align}
for invertible $\mathrm{P}$ and identifying 
\begin{align}
&\mathrm{P}_{2\times 2}=\delta_{\alpha \beta},~~\mathrm{Q}_{2\times 4}=\mathcal{B}^{(1)}_{~~\alpha n},~~\mathrm{S}_{4\times 4}=A g_{mn} + \mathcal{B}^{(1)}_{~~m\gamma }\delta^{\gamma \tilde{\gamma} } \mathcal{B}^{(1)}_{~~\tilde{\gamma} n}
\end{align}
we find 
\begin{align}
    \mathrm{det}(g_{MN})&=\mathrm{det}(\delta_{\alpha \beta }) \mathrm{det}(A g_{mn} + \mathcal{B}^{(1)}_{~~m\gamma }\delta^{\gamma \tilde{\gamma} } \mathcal{B}^{(1)}_{~~\tilde{\gamma} n}-\mathcal{B}^{(1)}_{~~m\alpha }\delta^{\alpha {\beta} } \mathcal{B}^{(1)}_{~~{\beta} n} )\nonumber\\
    &=\mathrm{det}(\delta_{\alpha \beta } ) \mathrm{det}(A g_{mn})=A^4 \mathrm{det}(g_{mn}).
\end{align}
It shows that the contribution due to the KK vector cancels out. Therefore, the internal six-dimensional volume is
$$
\mathrm{Vol}_6(t)
= \int d^6y\,\sqrt{g_6(y,t)}
= \int d^6y\;\,\big(\mathrm{H}^4(y) \mathrm{F}_1(t)\mathrm{F}_2(t)\big)^2\,\sqrt{g_4(y)}.
$$
\subsubsection*{Differential-form interpretation}
With the introduction of the shifted one-forms
$$
\eta^\alpha = dy^\alpha + \mathcal{B}^{(1)\alpha}_{~~~m} dy^m,
$$
the metric takes the form
$$
ds_6^2 =\delta_{\alpha\beta}\,\eta^\alpha \eta^\beta+ A(y,t) g_{mn}(y) dy^m dy^n 
$$
The six-dimensional volume form is proportional to 
\begin{align}
    \mathrm{Vol}_6 &\sim  \eta^4 \wedge \eta^5 \wedge dy^6\wedge dy^7 \wedge dy^8 \wedge dy^9 \nonumber\\
    &= (dy^4 + \mathcal{B}^{(1)\alpha}_{~~~m} dy^m )\wedge (dy^5 + \mathcal{B}^{(1)5}_{~~~n} dy^n)\wedge dy^6\wedge dy^7 \wedge dy^8\wedge dy^9\nonumber\\
    &=dy^4 \wedge dy^5 \wedge dy^6\wedge dy^7 \wedge dy^8 \wedge dy^9,
\end{align}
since the wedge product is completely anti-symmetric, the terms involving the KK vectors do not contribute to the volume form.

\subsubsection*{Geometric interpretation}

The KK vectors $B^\alpha{}_m$ act as connection one-forms on the fiber bundle.
They describe how the internal two-torus is twisted over the four-dimensional
base manifold. A connection affects parallel transport and curvature, but not
local volumes. Consequently, the internal volume depends only on the base metric
$A g_{mn}$ and the fiber metric $\delta_{\alpha\beta}$, and not on the KK
connection.

\bibliographystyle{JHEP}
\bibliography{references.bib}

\end{document}